\documentclass[12pt]{article}

\usepackage[utf8]{inputenc}
\usepackage{textcomp}
\usepackage{amsmath}
\usepackage[shortlabels]{enumitem}
\usepackage{chetnew}
\usepackage{tikz}
\usepackage{amssymb,amsfonts,mathrsfs,dsfont,yfonts,bbold}
\usetikzlibrary{calc}
\usepackage{xcolor}
\usepackage{braket}
\usepackage[normalem]{ulem}
\usepackage{import}
\usepackage{booktabs}
\usepackage{varioref}
\usepackage{bm}

\usepackage{tikz-cd}
\usepgflibrary{patterns} 
\usepgflibrary[patterns] 
\usetikzlibrary{patterns} 
\usetikzlibrary[patterns] 
\usetikzlibrary{decorations.pathmorphing}

\usepackage{tabularx}

\newcommand{\xmathpalette}[2]{\mathchoice
{#1\displaystyle\textfont{#2}}%
{#1\textstyle\textfont{#2}}%
{#1\scriptstyle\scriptfont{#2}}%
{#1\scriptscriptstyle\scriptscriptfont{#2}}%
}

\makeatletter
\newcommand{\twistedproduct}{%
\mathbin{\xmathpalette\twisted@product\relax}%
}

\newcommand{\twisted@product}[3]{%
\vbox{%
\ialign{\hfil##\hfil\cr
\scalebox{0.7}{$\m@th#1\sim$}\cr
\noalign{\nointerlineskip\kern-0.3\fontdimen5 #2 2}
$\m@th#1\otimes$\cr
}%
}%
}
\makeatother

\usepackage{amsthm}

\newcommand{\DD}{\mathbb{D}}

\newcommand{\be}{\begin{equation}}
\newcommand{\ee}{\end{equation}}
\newcommand{\Hom}{{\rm Hom}}
\newcommand{\bea}{\begin{eqnarray}}
\newcommand{\eea}{\end{eqnarray}}

\newcommand{\DZ}{\mathbb{Z}}
\newcommand{\CL}{\mathcal{L}}
\newcommand{\CA}{\mathcal{A}}

\newcommand{\CD}{\mathcal{D}}

\newcommand{\CB}{\mathcal{B}}
\newcommand{\CC}{\mathcal{C}}
\newcommand{\CZ}{\mathcal{Z}}

\newcommand{\CT}{\mathcal{T}}
\newcommand{\CI}{\mathcal{I}}

\newcommand{\CM}{\mathcal{M}}

\newcommand{\zz}{\mathbb{Z}}

\newcommand{\nn}{\nonumber}

\newcommand{\trl}{1}

\newcommand{\msf}{\mathsf}
\newcommand{\mc}{\mathcal}
\newcommand{\Vect}{\msf{Vec}}
\newcommand{\TVect}{\msf{2Vec}}
\newcommand{\Mod}{\msf{Mod}}
\newcommand{\Rep}{\msf{Rep}}
\newcommand{\TC}{\mc D(\mathbb Z_2)}
\newcommand{\QDDih}{\mc D(\mathbb D_6)}
\newcommand{\TQDDih}{\mc D^{\alpha_p}(\mathbb D_6)}
\newcommand{\rU}{{\rm U}}

\newcommand{\q}{\quad}
\newcommand{\act}{\triangleright}
\newcommand{\cat}{\triangleleft}
\newcommand{\DAS}{\mc A_S}

\usepackage{amsmath}	
\title{Gauging Non-Invertible Symmetries in \\[5mm] (2+1)d Topological Orders}
\author{Mahesh~K.~N.~Balasubramanian,$^{\between}$ Matthew~Buican,$^{\between}$ Clement~Delcamp,$^{\boxtimes}$ and~Rajath~Radhakrishnan$^{\theta}$}

\date{July 2025}
\affiliation{$^{\between}$\smallskip CTP and Department of Physics and Astronomy\\
Queen Mary University of London, London E1 4NS, UK\\[1mm]$^{\boxtimes}$ Laboratoire Alexander Grothendieck, Institut des Hautes \'Etudes Scientifiques \& CNRS, Bures-sur-Yvette, France \\[1mm] $^{\theta}$International Centre for Theoretical Physics, Strada Costiera 11, Trieste 34151, Italy}

\abstract{We present practical and formal methods for gauging non-invertible symmetries in (2+1)d topological quantum field theories. Along the way, we generalize various aspects of invertible 0-form gauging, including symmetry fractionalization, discrete torsion, and the fixed point theorem for symmetry action on lines. Our approach involves two complementary strands: the fusion of topological interfaces and Morita theory of fusion 2-categories. We use these methods to derive constraints on gaugeable symmetries and their duals while unifying the prescription for gauging non-invertible 0-form and 1-form symmetries and various higher structures. With a view toward recent advances in creating non-Abelian topological orders from Abelian ones, we give a simple recipe for non-invertible 0-form gauging that takes large classes of the latter to the former. We also describe conditions under which iterated gauging of invertible 0-form symmetries is equivalent to a single-step gauging of a non-invertible symmetry. We conclude with a set of concrete examples illustrating these various phenomena involving gauging symmetries of the infrared limit of the toric code.\\
}

\begin{document}
\setcounter{tocdepth}{2}
\maketitle
\toc

\section{Introduction}

\noindent
Quantum field theories (QFTs) in (2+1)d are both constrained and enriched by a wild zoo of $0$-form symmetries, $1$-form symmetries, and higher structures that combine them. Gauging these symmetries---whenever the relevant anomalies vanish---gives a natural way to (partially) explore the space of (2+1)d QFTs and exhibit new phenomena. When the symmetries in question are invertible, this gauging procedure is well-understood: loosely speaking, we couple ($p$+1)-form gauge fields to the QFT we are studying and perform an appropriate sum over these fields. Via 
Poincar\'e duality, we can equivalently understand gauging of invertible $p$-form symmetries in terms of insertions of networks of (2$-p$)-dimensional defects. 

The gauging procedure is somewhat murkier when the symmetries are non-invertible (although see \cite{Carqueville:2018sld,cui2019generalized,Mulevicius:2020tgg,Mulevicius:2022gce,Delcamp:2023kew,Hsin:2024aqb,Wen:2025thg,Hsin:2025ria,Inamura:2025cum} for recent progress). An exception to this statement is the well-understood gauging of 1-form symmetries\footnote{Note that throughout this paper, the terms ``1-form symmetries'' and ``0-form symmetries'' without further modifiers can refer either to invertible symmetries or to non-invertible symmetries.} via networks of lines (see App.~\ref{sec:Morita1} for a review). In this case, we gauge a 1-form symmetry by first choosing a ``condensable'' algebra with an underlying object, $A_L$, built from a sum of the simple bosonic lines that generate the 1-form symmetry (here the ``$L$'' subscript in $A_L$ reminds us that the underlying object is built from lines). Gauging amounts to constructing an appropriate network or mesh of $A_L$ in spacetime that implements a sum over 1-form symmetry lines.\footnote{In App.~\ref{sec:Morita1}, we also review the well-understood related procedure for gauging 0-form symmetries in (1+1)d. See \cite{Seifnashri:2025fgd} for recent progress understanding such (continuum) gauging in the non-invertible case from the dual perspective of gauge fields.}

In the present paper we are mostly interested in better understanding the gauging of non-invertible symmetries in non-spin (2+1)d topological QFTs (TQFTs).\footnote{Hereafter, when we say TQFT, the reader should understand a non-spin TQFT (i.e., a TQFT that does not depend on a spin structure).} Gauging 1-form symmetries of TQFTs associated with condensable algebras of lines produces ``simpler'' theories, in the sense that it generates TQFTs with less topological entanglement entropy (or, more precisely, with smaller total quantum dimension \cite{Kim:2023ydi}). In particular, this procedure can produce Abelian TQFTs from non-Abelian ones. Importantly, the resulting theories possess dual 0-form symmetry, which can be gauged to recover the initial theory. This procedure produces more ``complicated'' theories from simpler ones (including cases where it produces non-Abelian topological orders from Abelian ones).

This latter possibility is now of practical interest and has received considerable attention. Indeed, several protocols using adaptive constant depth quantum circuits have recently been proposed to prepare so-called string-net models, giving rise to non-Abelian anyons, built from sequences of Abelian group extensions  (e.g., see \cite{Verresen:2021wdv,PhysRevX.14.021040,PRXQuantum.4.020339,Ren:2024ayb}). Such theories include quantum double models built from solvable groups. More generally, consider a quantum double model for which a $\Rep(G)$ 1-form subsymmetry can be gauged so as to produce an Abelian topological order. In this case, the resulting theory always possesses an invertible dual 0-form (possibly non-Abelian) $G$ symmetry that can be gauged so as to recover the initial model. A prototypical example of this phenomenon is the gauging of $\Rep(\mathbb{Z}_2)$ Wilson lines in the quantum double model, $\mc D(\mathbb D_6)$, based on the dihedral group of order six to produce $\CD(\mathbb Z_3)$ (here we have $\mathbb D_6\cong S_3$, where $S_3$ is the group of permutations on three letters). The dual 0-form symmetry is the charge conjugation symmetry of the Abelian $\mathbb Z_3$ theory, and it can be gauged to produce the non-Abelian $\CD(\mathbb D_6)$.

One of the main motivations of this paper is to better understand the above discussion when the 1-form symmetry in question no longer corresponds to a fusion subcategory but instead corresponds to a more general condensable algebra, $A_L$. In this case, we show that it is still possible to recover the initial model by gauging a symmetry related to an ``algebra of surfaces'', $\CA_S$, describing a non-invertible 0-form symmetry (see Sec.~\ref{OPE} and App.~\ref{DetSimple}). As in the case of 1-form symmetry gauging associated with an algebra $A_L$, we can gauge the non-invertible 0-form symmetry related to $\CA_S$ by inserting a network or mesh of $\CA_S$ in spacetime. In practice, explicitly constructing such a network can be complicated, and so we provide practical tools in this paper that allow us to understand the corresponding gauged topological phase. As a particularly simple example (among many others in Sec.~\ref{ToricCode}), we show how to gauge a non-invertible 0-form symmetry of $\TC$ to produce $\CD(\mathbb D_6)$.

Although it might naively seem that 0-form and 1-form symmetries are on different footing---indeed, while the former are associated with codimension-one defects, the latter are associated with codimension-two defects---they are both encoded in the surfaces that make up $\CA_S$. The heuristic reason for this fact is that the surfaces in $\CA_S$ are condensation surfaces. In particular, we generate such surfaces by ``higher-gauging'' a 1-form symmetry on a co-dimension one surface in the (2+1)d bulk \cite{Roumpedakis_2023} (i.e., we build the surface as a mesh of lines for some algebra, $A_L$, that is not necessarily condensable in the full (2+1)d bulk).

As a pedestrian example to illustrate this point, consider the case of $\TC$ on the three-torus $\mathbb T^3$. This theory has two non-trivial bosons, $e$ and $m$, and a non-trivial fermion, $f$, all of order two under fusion and all with non-trivial mutual braiding. We can then consider gauging the symmetry corresponding to the Algebra object
\begin{equation}
\CA_S=S_{1 \oplus e}\, ,
\end{equation}
where $S_{1 \oplus e}$ is the surface obtained from higher gauging $A_L=1 \oplus e$ (i.e., the algebra corresponding to the $\mathbb Z_2$ 1-form symmetry generated by $e$). To sum over a network of $\CA_S$ on $\mathbb T^3$, we can insert $S_{1 \oplus e}$ on each of the three 2-cycles $\Sigma_i\in H_2(\mathbb T^3,\mathbb Z)$. Following the above discussion, we can define $S_{1 \oplus e}$ wrapping $\Sigma$ as \cite{Roumpedakis_2023}
\begin{equation}
S_{1 \oplus e}(\Sigma)={1\over\sqrt{H_1(\Sigma,\mathbb{Z}_2)}}\sum_{\gamma\in H_1(\Sigma,\mathbb{Z}_2)}e(\gamma)\, ,
\end{equation}
where we understand $e(\gamma)$ as the $e$ line wrapping the corresponding 1-cycle $\gamma\in H_1(\Sigma,\mathbb Z_2)$. Then, up to an overall normalization, gauging $\CA_S$ amounts to inserting a copy of $S_{1 \oplus e}$ along each of the three $\Sigma_i$ (with one-cycles $\gamma_i$ and $\gamma_j$ for $\Sigma_k$ and $i,j,k$ distinct). It is easy to check we obtain
\begin{equation}\label{gaugeSeT3}
S_{1 \oplus e}(\Sigma_1)S_{1 \oplus e}(\Sigma_2)S_{1 \oplus e}(\Sigma_3)=\prod_{i=1}^3({\rm id} + e(\gamma_i))\, .
\end{equation}
Therefore, we see from this procedure that gauging $\CA_S$ is the same as $1 \oplus e$ anyon condensation on the entire $\mathbb T^3$ spacetime (see also the discussion in \cite{Kong:2024ykr}). 

Gaining a better understanding of more general symmetries and the corresponding gauging procedure requires elucidating the underlying category theory. From this point of view, $\CA_S$ is a separable algebra in the 2-category $\Mod(\mc B)$, where $\mc B$ is the modular tensor category (MTC) that encodes the anyonic excitations of the TQFT we are studying. The objects of $\Mod(\mc B)$ are the condensation surfaces we have discussed above.\footnote{More generally, a (2+1)d TQFT can contain surfaces that are not condensation surfaces (and therefore not in $\Mod(\mc B)$). These are elements of $\msf{2Vec}_G^{\omega}$ for some $\omega\in H^4(G,\rU(1))$. Unlike the surfaces in $\Mod(\mc B)$, these surfaces do not admit topological boundary conditions and can, for example, arise when we couple TQFTs to non-topological QFTs as in \cite{Balasubramanian:2024nei}. Note that in contrast to such surfaces, condensation surfaces are porous---they correspond to a mesh of lines---and hence act trivially on local operators.} The underlying object of the separable algebra, $\CA_S$, is typically a decomposable condensation surface in $\Mod(\mc B)$. By abuse of notation, we often refer to both this object and the corresponding algebra as $\CA_S$. However, the reader should be aware that the algebra comes with a fusion 1-category structure and that this notation can potentially be ambiguous (see Sec.~\ref{sec:Morita2} for more details; in ambiguous contexts, we use more precise notation).

In fact, defining this additional fusion 1-category structure for $\CA_S$ leads to notions of discrete torsion and symmetry fractionalization (see Sec.~\ref{sec:simple lines after gauging}) that are crucial for defining the gauging procedure and generalize the corresponding notions in the invertible case \cite{Barkeshli:2014cna}. We illustrate these concepts in several examples in Sec.~\ref{ToricCode} and also use them to study various constraints on the gaugeability of symmetries in Sec.~\ref{GenConstraints}.

More abstractly, given the fact that algebras of surfaces or, equivalently, separable algebras in $\Mod(\mc B)$ encode 0-form and 1-form symmetries in such closely related ways, it is reasonable to ask how to disentangle 0-form and 1-form symmetries from various higher structures in this construction. Building on work in \cite{Huston:2022utd,Buican:2023bzl}, we derive several general results in this direction in Sec.~\ref{OPE} and Sec.~\ref{GenConstraints}. Another related issue we confront in this context is how to generalize the duality between invertible 0-form and 1-form symmetries to the most general classes of (non-invertible) symmetries in (2+1)d.

One of our main tools for understanding this question is via fusion of interfaces of the form $\CI\otimes\CI^{\dagger}$ and $\CI^{\dagger}\otimes\CI$ (see Sec.~\ref{OPEdetails}), where $\CI$ is a general topological interface between two TQFTs and $\CI^{\dagger}$ is its orientation reversal. Indeed, we argue that such fusion processes can be used to recover the algebras of surfaces corresponding to the dual symmetries we gauge to go between $\Mod(\CB_1)$ and $\Mod(\CB_2)$. In particular
\begin{equation}
\CA_S \simeq \CI^{\dagger}\otimes\CI\ , \q \widehat\CA_S \simeq \CI\otimes\CI^{\dagger}\, .
\end{equation}
From the associativity of the fusion of surfaces we then obtain the relation
\begin{equation}\label{AAhRel}
\CA_S\otimes\CI \simeq \CI\otimes\widehat\CA_S\, ,
\end{equation}
which gives rise to various constraints we elaborate further upon in Sec.~\ref{OPE} and Sec.~\ref{GenConstraints}.

In Sec.~\ref{OPE}, we also develop the dictionary between the fusion 1-categories underlying the dual $\CA_S$ and $\widehat\CA_S$ algebras of surfaces and the theory on the $\CI$ interface, $\mc C_{\CI}$. Developing this dictionary motivates us in turn to reveal a non-invertible generalization of the fixed point theorem for invertible symmetry action on MTCs \cite{Barkeshli:2014cna}. In other words, we generalize the equality between the rank of the twisted sector of a given invertible surface (the number of symmetry defects in a given $g\in G$ sector in the language of \cite{Barkeshli:2014cna}) and the number of fixed points of the surface's action on anyons.\footnote{Our method of proof is quite different from that in \cite{Barkeshli:2014cna}, which involves a $G$-crossed generalization of modularity. It would be interesting to understand a generalized version of $G$-crossed modularity that captures non-invertible symmetries for a variety of reasons (e.g., better understanding anomalies in the case of non-invertible symmetries).}

In order to apply these results to concrete theories and also to better understand constraints arising from the above discussion, we review Morita theory for 2-categories in Sec.~\ref{sec:Morita2}. Using this formalism, we give a simple prescription for understanding how to gauge generalized symmetries in (2+1)d topological orders. We then use these results to find further constraints on gaugeable symmetries and to understand cases in which non-invertible 0-form symmetry gauging can be broken up into invertible steps (again with an eye toward constructions similar to those in \cite{Verresen:2021wdv,PhysRevX.14.021040,PRXQuantum.4.020339,Ren:2024ayb}). With these results in hand, we then apply our formalism to a series of examples involving gauging of various symmetries in $\TC$ as a means of illustration and clarification.

The broad plan of the paper is as follows. In the next section, we review the gauging of invertible 0-form symmetries of (2+1)d TQFTs. We highlight relevant aspects of the formalism detailed in \cite{Barkeshli:2014cna} that we then generalize to the non-invertible case in Sec.~\ref{OPE}. This latter section forms the heart of the paper: we describe non-invertible symmetries and their gauging via the fusion of interfaces in Sec.~\ref{OPEdetails} and via Morita theory in Sec.~\ref{sec:Morita2}. Along the way we prove various other results in this section. We provide a roadmap for these results in the introduction to Sec.~\ref{OPE}. In Sec.~\ref{GenConstraints}, we apply the results of Sec.~\ref{OPE} to derive constraints on gaugeable symmetries. Then, in Sec.~\ref{sec:oneshotNshot} we tackle the question of when non-invertible gauging can be broken up into steps involving invertible symmetry gauging. Interestingly, we will see that invertible gauging is not closed under composition. In Sec.~\ref{ToricCode} we apply the results of the previous sections to various concrete examples involving gauging symmetries of $\TC$. We conclude with a discussion of some open problems.

We also collect various results in our Appendices. In particular, in App.~\ref{sec:Morita1}, we review 0-form symmetry gauging in (1+1)d and the closely related (higher) gauging of 1-form symmetry in (2+1)d. Then, in App.~\ref{DetSimple}, we explain how to derive the full set of surfaces that appear in the fusion of interfaces for the special case of Abelian theories. We use this discussion as an opportunity to also show that, by gauging non-invertible symmetries, we can obtain any topological order given by a non-Abelian discrete gauge theory from an Abelian one. Finally, in App.~\ref{FPabelian}, we give a more direct proof of the (generalized) fixed point theorem for symmetry action in the context of Abelian TQFTs (the general case is proven in Sec.~\ref{sec:number of twisted sectors A_S and hat A_S}). This proof gives all quantities in terms of simple group theoretical data.

\newsec{Invertible 0-form gauging}\label{InvertRev}
In this section, we briefly review the data determining the action of an invertible 0-form symmetry, $G$, on an MTC,  $\mc B$, corresponding to a (2+1)d TQFT, $\CT$, and we summarize the process of gauging $G$.\footnote{We refer the reader to \cite{Barkeshli:2014cna} for a detailed exposition.} Along the way, we highlight various aspects of this discussion that we will generalize in Sec.~\ref{OPE} when considering non-invertible symmetries.

Let us begin by considering the topological surfaces $S_g$, with $g\in G$, that implement the 0-form symmetry. Their action is given by the following data:
\begin{itemize}
\item Permutations, $\rho_g$, capturing the action of $S_g$ on the simple lines.
\item Matrices, $U_g(a,b,c)$, describing the action of $S_g$ on the $V_{ab}^c$ fusion spaces associated with simple lines $a,b,c \in \text{Irr}(\mc B)$.
\item Phases, $\eta_a(g_1,g_2)$, encoding the fractionalization of the symmetry action on lines $a$ in $\mc B$.
\item The Postnikov class, $\beta \in H^3_{\rho}(G,A_{\rm ab})$, where $A_{\rm ab}$ is the invertible 1-form symmetry associated with $\mc B$.\footnote{In many cases (e.g., the Fibonacci theory), $A_{\rm ab}$ is trivial and therefore the Postnikov class is also, by construction, trivial. More generally, when $A_{\rm ab}$ is non-trivial, the Postnikov class may or may not be trivial.} If $\beta$ is non-trivial, then the 0-form symmetry $G$ forms a (non-split) 2-group with $A_{\rm ab}$.
\item The anomaly $\varpi \in H^4(G,\rU(1))$ of the 0-form symmetry. 
\end{itemize}
When the $S_g$ are condensation surfaces, they admit non-trivial boundary conditions. We can think of these boundary conditions as non-genuine / twisted lines bounding $S_g$ and consider the category
\be
\mc B_G:=\boxplus_{g \in G} \mc B_g\, ,
\ee
of all twisted lines (here $\mc B_1\simeq\mc B$ corresponds to genuine lines, but we can also think of them as bounding the trivial surface). The number of simple lines in $\mc B_g$, $|\mc B_g|$, is the number of simple boundary conditions of the surface $S_g$. Consistency conditions relate the $S_g$ action on $\mc B$ to the non-genuine lines in $\mc B_{g\ne1}$. For example, from the fixed-point theorem for group action on MTCs, $|\mc B_g|$ is given by the number of fixed points of the $G$-action on the genuine lines \cite{Barkeshli:2014cna}.\footnote{We will generalize this result to non-invertible surfaces in Sec.~\ref{sec:number of twisted sectors A_S and hat A_S}.} Moreover, the action of $S_g$ on the genuine lines can be extended to $\mc B_{G}$.

Upon gauging the invertible symmetry $G$, the surfaces implementing the symmetry become trivial. Therefore, the non-genuine lines in $\mc B_G$ become genuine lines in the TQFT, $\CT_G$, obtained from gauging $G$. These lines must have a fusion 1-category structure. This fact requires $\mc B_G$ itself to form a consistent fusion 1-category. For this constraint to be obeyed, both the Postnikov class $\beta$ and anomaly $\varpi$ must be trivial \cite{Barkeshli:2014cna,Cui_2016}. Indeed, if either $\beta$ or $\varpi$ is non-trivial, then $G$ cannot be consistently gauged on its own. When $\beta$ is trivial, there is a freedom in choosing the fusion rules for the twisted sector lines. Indeed, consider the fusion of two twisted sector lines $a_{g_1}$ and $b_{g_2}$:
\be\label{fusionFree}
a_{g_1} \otimes b_{g_2} \cong  c_{g_1g_2}\, .
\ee
We have the freedom to modify \eqref{fusionFree} by an invertible line operator, $\gamma(g_1,g_2)$. Moreover, when $\varpi$ is trivial, the associator of the fusion 1-category $\mc B_G$ can be modified by a 3-cocycle, $\alpha \in H^3(G,\rU(1))$. This modification can be understood as the freedom to stack a $G$-SPT on the TQFT. Note that this stacking only changes the associator of the twisted sector lines, leaving the MTC, $\mc B$, of genuine lines unchanged. In Sec.~\ref{OPE}, we will discuss a generalization of symmetry fractionalization and SPT stacking to non-invertible symmetries. 

\begin{figure}[h!]

\tikzset{every picture/.style={line width=0.75pt}} 

\begin{tikzpicture}[x=0.75pt,y=0.75pt,yscale=-0.94,xscale=0.94]

\draw  [color={rgb, 255:red, 245; green, 166; blue, 35 }  ,draw opacity=1 ][fill={rgb, 255:red, 255; green, 224; blue, 187 }  ,fill opacity=1 ] (93.55,173) -- (93.57,74.86) -- (184.58,30.06) -- (184.57,128.19) -- cycle ;
\draw [color={rgb, 255:red, 139; green, 6; blue, 24 }  ,draw opacity=1 ]   (139.07,101.53) -- (241.26,101.4) ;
\draw [color={rgb, 255:red, 139; green, 6; blue, 24 }  ,draw opacity=1 ] [dash pattern={on 4.5pt off 4.5pt}]  (98.71,102.26) -- (139.07,101.53) ;
\draw [color={rgb, 255:red, 139; green, 6; blue, 24 }  ,draw opacity=1 ]   (30.86,101.79) -- (94.42,102.26) ;
\draw  [color={rgb, 255:red, 139; green, 6; blue, 24 }  ,draw opacity=1 ][fill={rgb, 255:red, 139; green, 6; blue, 24 }  ,fill opacity=1 ] (136.23,101.53) .. controls (136.23,100.74) and (136.87,100.11) .. (137.65,100.11) .. controls (138.43,100.11) and (139.07,100.74) .. (139.07,101.53) .. controls (139.07,102.31) and (138.43,102.95) .. (137.65,102.95) .. controls (136.87,102.95) and (136.23,102.31) .. (136.23,101.53) -- cycle ;
\draw  [color={rgb, 255:red, 245; green, 166; blue, 35 }  ,draw opacity=1 ][fill={rgb, 255:red, 255; green, 224; blue, 187 }  ,fill opacity=0 ][dash pattern={on 4.5pt off 4.5pt}] (497.55,169) -- (497.57,70.86) -- (588.58,26.06) -- (588.57,124.19) -- cycle ;
\draw [color={rgb, 255:red, 139; green, 6; blue, 24 }  ,draw opacity=1 ]   (543.07,97.53) -- (645.26,97.4) ;
\draw [color={rgb, 255:red, 139; green, 6; blue, 24 }  ,draw opacity=1 ]   (434.86,97.79) -- (540.23,97.53) ;
\draw  [color={rgb, 255:red, 139; green, 6; blue, 24 }  ,draw opacity=1 ][fill={rgb, 255:red, 139; green, 6; blue, 24 }  ,fill opacity=1 ] (540.23,97.53) .. controls (540.23,96.74) and (540.87,96.11) .. (541.65,96.11) .. controls (542.43,96.11) and (543.07,96.74) .. (543.07,97.53) .. controls (543.07,98.31) and (542.43,98.95) .. (541.65,98.95) .. controls (540.87,98.95) and (540.23,98.31) .. (540.23,97.53) -- cycle ;

\draw (162.86,39.93) node [anchor=north west][inner sep=0.75pt]    {$S_g$};
\draw (242.99,92.28) node [anchor=north west][inner sep=0.75pt]    {$a$};
\draw (-11.01,91.42) node [anchor=north west][inner sep=0.75pt]    {$S_g(a)$};
\draw (565.86,37.93) node [anchor=north west][inner sep=0.75pt]    {$S_g$};
\draw (646.99,88.28) node [anchor=north west][inner sep=0.75pt]    {$a$};
\draw (392.01,87.42) node [anchor=north west][inner sep=0.75pt]    {$S_g(a)$};
\draw (290,88.4) node [anchor=north west][inner sep=0.75pt]    {$\xrightarrow{{\rm Gauging}\ S_g}$};

\end{tikzpicture}
\caption{Gauging $S_g$ identifies the lines $a$ and $\rho_g(a)$.}
\label{fig:gauging Sg gives point operators}
\end{figure}
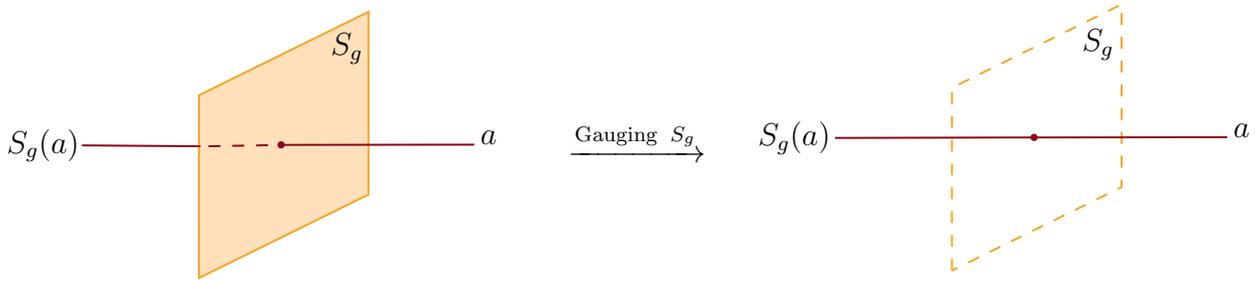

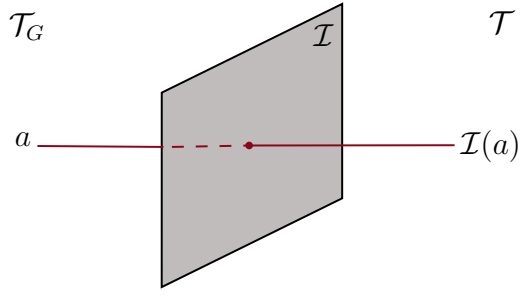
\begin{figure}
    \centering

\tikzset{every picture/.style={line width=0.75pt}} 

\begin{tikzpicture}[x=0.75pt,y=0.75pt,yscale=-1,xscale=1]

\draw  [color={rgb, 255:red, 0; green, 0; blue, 0 }  ,draw opacity=1 ][fill={rgb, 255:red, 192; green, 188; blue, 188 }  ,fill opacity=1 ] (270.55,176) -- (270.57,77.86) -- (361.58,33.06) -- (361.57,131.19) -- cycle ;
\draw [color={rgb, 255:red, 139; green, 6; blue, 24 }  ,draw opacity=1 ]   (316.07,104.53) -- (418.26,104.4) ;
\draw [color={rgb, 255:red, 139; green, 6; blue, 24 }  ,draw opacity=1 ] [dash pattern={on 4.5pt off 4.5pt}]  (275.71,105.26) -- (316.07,104.53) ;
\draw [color={rgb, 255:red, 139; green, 6; blue, 24 }  ,draw opacity=1 ]   (207.86,104.79) -- (271.42,105.26) ;
\draw  [color={rgb, 255:red, 139; green, 6; blue, 24 }  ,draw opacity=1 ][fill={rgb, 255:red, 139; green, 6; blue, 24 }  ,fill opacity=1 ] (313.23,104.53) .. controls (313.23,103.74) and (313.87,103.11) .. (314.65,103.11) .. controls (315.43,103.11) and (316.07,103.74) .. (316.07,104.53) .. controls (316.07,105.31) and (315.43,105.95) .. (314.65,105.95) .. controls (313.87,105.95) and (313.23,105.31) .. (313.23,104.53) -- cycle ;

\draw (345.86,41.93) node [anchor=north west][inner sep=0.75pt]    {$\CI$};
\draw (419.99,95.28) node [anchor=north west][inner sep=0.75pt]    {$\CI( a)$};
\draw (195.01,96.42) node [anchor=north west][inner sep=0.75pt]    {$a$};
\draw (192,36.4) node [anchor=north west][inner sep=0.75pt]    {$\CT_{G}$};
\draw (434,35.4) node [anchor=north west][inner sep=0.75pt]    {$\CT$};

\end{tikzpicture}
    \caption{Action of $\CI$ on line operators.}
    \label{fig:action of I}
\end{figure}

Supposing the Postnikov class, $\beta$, and the anomaly, $\varpi$, vanish, then $\mc B_G$ is a consistent fusion 1-category with a $G$-action.\footnote{$\mc B_G$ does not have  a braiding in the usual sense, because it contains non-genuine lines. However, it does have a $G$-crossed braiding that captures how general non-genuine lines braid with each other by passing through the topological surface they are attached to.} We can then consistently gauge $G$ to produce a new TQFT, $\CT_{G}$. The lines of $\CT_{G}$ are orbits of the genuine and non-genuine line in $\mc B_G$ under the $G$-action. To understand this statement, consider the action of $S_g$ on a line $a$ in  $\mc B$. The lines $a$ and $\rho_g(a) \equiv S_g(a)$ admit a junction on the surface $S_g$. Gauging $S_g$ trivializes it. Therefore, in the gauged theory the line $a$ and $\rho_g(a)$ must be identified (see Fig.~\ref{fig:gauging Sg gives point operators}). 
This discussion shows that the simple lines related by the action of the surfaces being gauged are identified by gauging the surfaces.  Consider the junction of two $a$ lines on the surface $S := \boxplus_{g \in G}S_g$. Let $V_{S(a)}^a$ be the vector space of local operators at this junction. Note that $\dim V_{S(a)}^a>0$, because $S$ contains the trivial surface. Upon gauging the symmetry $G$, the surface $S$ is trivialized. In fact, $\dim  V_{S(a)}^a =|K_a|>0$, where $K_a \leq G$ is the subgroup of surfaces that act trivially on $a$. Note that the point junction at the intersection of $S$ with $a$ is a topological operator, because it can be moved along the line by moving the surface in the ambient spacetime. Upon gauging $G$, we can think of the line $a$ as a quantum mechanical system with symmetry $K_a$. Therefore, in the gauged theory, $a$ is a non-simple line operator, and it can be decomposed into simple lines $a_\pi$ as follows 
\be
a \xrightarrow[]{\text{Gauging } G} \bigoplus_{\pi \in \text{Irr}(\Rep(K_a))} a_{\pi}\, .
\ee
In particular, $S$ acts trivially on the trivial line $1$. Therefore, we have
\be
1 \xrightarrow[]{\text{Gauging } G} \bigoplus_{\pi \in \text{Irr}(\Rep(G))} 1_{\pi}\, .
\ee
This logic shows that the TQFT, $\CT_G$, contains a subset of lines that form the category $\Rep(G)$ under fusion. This category is the dual 1-form symmetry whose gauging takes us back to the TQFT, $\CT$. The corresponding condensable algebra is the algebra $\mathbb C^G$ equipped with a $G$-action so that, as an object in $\Rep(G)$, it decomposes as $\bigoplus_{a \in \text{Irr}(\Rep(G))} d_a \cdot a$ (see App.~\ref{sec:Morita1}). Gauging $\Rep(G)$ in half-space gives an interface, $\CI$, between $\CT_{G}$ and $\CT$. Mathematically, $\CI$ can be understood as a monoidal functor that maps the category of line operators in $\CT_G$ to $\CB_G$ (see Fig.~\ref{fig:action of I}). 
Note that, upon gauging the 1-form symmetry $\Rep(G)$, the lines in $\Rep(G)$ become trivial in $\CT$. More precisely, we have $\CI(a) \cong d_a \cdot 1$, where $a\in \Rep(G)$.

We close this section by reviewing that under gauging a 1-form symmetry with algebra object, $A_L \cong \bigoplus_{a \in \text{Irr}(\mc B)} n_a\cdot a$, in a TQFT,  the dual 0-form symmetry is invertible if and only if 
\be\label{invCond2}
\CI(A_L) \cong d_{A_L} \cdot 1\, .
\ee
Here the interface $\CI$ is thought as a map that implements the 1-form symmetry gauging, $1$ is the identity line of the gauged theory, and $d_{A_L}$ is the quantum dimension of $A_L$ as an object. Note that, since quantum dimensions are positive in a unitary TQFT, the above equation holds if and only if
\be\label{aCondComp}
\CI(a) \cong d_a\cdot1 \, , \q \forall \; a\in A_L \q \text{and} \q  d_a\in \mathbb{N}\, .
\ee
This equation is equivalent to requiring that $n_a=d_a$\footnote{This statement follows from the fact that $A_L$ can end on the domain wall separating the gauged and ungauged phases, and $n_a$ gives the multiplicity of the corresponding junction of $a\in A_L$ with the trivial line. Note that, since $n_a\le d_a$ by conservation of quantum dimension, the case of invertible 0-form symmetry corresponds to a ``maximal'' algebra object. Clearly, the non-invertible case is the generic scenario. \label{extremalA}} and is tantamount to saying that all $a\in A_L$ are trivialized under gauging.

Suppose instead that $\CI(a) \cong  n_a\cdot1 \oplus x$ for some non-trivial line, $x$, in the gauged theory. Then, under the action of the dual symmetry, $1 \to x$. That is, $1$ and $x$ must be in the same orbit under the action of the dual symmetry. But an invertible symmetry cannot map $1$ to a non-trivial line. Therefore, the dual symmetry must be non-invertible. This argument shows that the condition \eqref{invCond2} is necessary for the dual symmetry to be invertible.

Now, suppose that \eqref{invCond2} holds. Then, by \eqref{aCondComp}, we have
\be
\CI(a) \otimes \CI(b) \cong d_a d_b\cdot 1\, . 
\ee
Since the condensation map preserves fusion rules, we have 
\be
\CI(a \otimes b) \cong \sum_c N_{ab}^c \, \CI(c) \cong  d_a d_b\cdot 1\, . 
\ee
Moreover, since quantum dimensions are positive real numbers in unitary TQFTs, for the above equality to hold, we must have 
\be
\CI(c) \cong d_c \cdot1 \, , \q   \forall \; N_{ab}^c \neq 0\, .
\ee
Therefore, we find that if $a$ and $b$ are condensed, then all lines in $a \otimes b$ are also condensed. In other words, the simple lines in $A_L$ are closed under fusion. This fact implies that the simple lines in $A_L$ form a fusion subcategory isomorphic to $\Rep(G)$ for some group $G$. As a result, we see that the dual of \eqref{invCond2} is gauging the invertible 0-form symmetry $G$. 

\newsec{Generalized symmetry gauging}\label{OPE}

In the previous section, we reviewed the special case of invertible 0-form symmetry gauging. We argued that the dual 1-form symmetry gauging involves an algebra of lines, $A_L$, that satisfies
\be\label{condenseComp2}
\CI(A_L) \cong n_{A_L}\cdot 1 \, .
\ee
Indeed, we learned that \eqref{condenseComp2} is a necessary and sufficient condition for the dual 0-form symmetry to be invertible. Equivalently, using the perspective of footnote \ref{extremalA}, such an $A_L \cong \sum_{a \in \text{Irr}(\mc B_1)} n_a \cdot a$ is maximal in the sense that each coefficient satisfies $n_a=d_a$ for all $a\in A_L$. It is then natural to consider the more general situation in which 1-form symmetry gauging is associated with an algebra of lines satisfying
\be\label{genCondLO}
\CI(A_L) \cong n_{A_L,1}\cdot 1 \oplus n_{A_L,x}\cdot x \oplus \cdots\, , \q x\ne1\, .
\ee
By our above discussion, the presence of  a non-trivial line operator, $x$, implies that the dual symmetry must be non-invertible.

In fact, the dual symmetry is a non-invertible 0-form symmetry in the following sense. Let us consider the interface $\CI:\mc B_1\to\mc B_2$ constructed from performing the 1-form symmetry gauging corresponding to summing over $A_L$ in half the spacetime. Then, the orientation reversed interface $\CI^{\dagger}:\mc B_2\to\mc B_1$ implements the dual gauging. Note that for any non-trivial simple line $a$ in $\mc B_2$, we have $\CI^{\dagger}(a) \cong b$ with $\dim \Hom_{\mc B_1}(b,1)=0$. Indeed, this statement follows from the fact that 1-form symmetry gauging maps the trivial line to the trivial line. Physically, this discussion shows that the dual symmetry we gauge does not involve a 1-form symmetry and is therefore a ``pure'' 0-form symmetry.

More generally, we will consider gauging an abstract symmetry in a TQFT associated with a collection of surfaces that requires a choice of fusion 1-category satisfying compatibility conditions. We will refer to this symmetry structure as an ``algebra of surfaces''. Given an underlying MTC, $\CB$, an algebra of surfaces mathematically amounts to a separable algebra in the fusion 2-category $\Mod(\CB)$ (see Sec.~\ref{sec:Morita2}). We will denote such an algebra of surfaces as $\CA_S$, or variations thereof, (the subcript is to remind us that the underlying object is built from surfaces instead of lines). As in the case of algebras in 1-categories, we denote the algebra of surfaces and its underlying collection of surfaces, i.e. the underlying object in $\Mod(\mc B)$, in the same way. Although this conflation is ambiguous in general---since the same underlying object can be equipped with multiple algebraic structures via distinct choices of fusion 1-categories (see Sec.~\ref{sec:simple lines after gauging})---our meaning should be clear from the context.

In principle, gauging an abstract symmetry is performed by summing over a spacetime network built from $\CA_S$  in such a way that the tensor product of the underlying fusion 1-category is implemented at every interface involving three surfaces, and that its associator encodes the junction of four such interfaces.
In the introduction, we commented that 1-form symmetry gauging can also be rephrased in terms of surfaces. Therefore, it should come as no surprise that the most general symmetries corresponding to algebras of surfaces are combinations of (non-)invertible 0-form and 1-form symmetries organized into  higher structures. In this sense we can think of summing over $\CA_S$ as gauging generalized symmetries in (2+1)d. To distinguish the more general case from the 0-form gauging described previously, we will, as in the discussion above, sometimes refer to the latter as gauging a ``pure'' 0-form symmetry (although we will drop the ``pure'' modifier when no confusion arises). 
\begin{table}
\centering
\small
\begin{tabularx}{\textwidth}{|X|X|X|}
\hline
Concept & Invertible 0-form symmetry & Generalized symmetry
\\ \hline
Underlying surface& $\boxplus_{g\in G} S_g$ & $\boxplus_{i} S_i$
\\ \hline
Fusion category of twisted sector lines  & $\CB_G = \boxplus_{g\in G } \CB_g$ & $\DAS$
\\ \hline
Number of non-genuine lines & Fixed points of $S_g$ & Fixed points of $S_i$ 
\\ \hline
Symmetry fractionalization & $\gamma \in H^2_{\rho}(G,A_{\rm ab})$  & Choice of fusion rules for $\DAS$ 
\\ \hline
Discrete torsion & $\alpha \in H^3(G,\rU(1))$ &  Choice of $F$-symbols for $\DAS$ 
\\ \hline
Anomaly free symmetry & Both $\beta$ and $\varpi$ vanish & $\exists$ a braided tensor functor, \newline $\CB\to\CZ(\DAS)$
\\\hline
\end{tabularx}
\caption{Some properties of non-invertible symmetries we discuss in this section, building on corresponding properties of invertible 0-form symmetries discussed in Sec. \ref{InvertRev}. As we will describe in more detail below (see also the discussion in the introduction), summing over an algebra of surfaces, $\CA_S$, can also be used to gauge more general (non-)invertible symmetries (0-form and 1-form) and various higher structures. In Sec.~\ref{sec:Morita2} we define the algebra of surfaces as a separable algebra in a 2-category and formalize the gauging of the corresponding symmetry via Morita theory.}
\label{GeneralizationTable}
\end{table}
Therefore, we are motivated to study various properties of non-invertible surfaces and their gauging. In particular, we will generalize several aspects of invertible 0-form symmetry gauging to more general symmetries.

The plan of the rest of this section is as follows. In Sec.~\ref{OPEdetails}, we explain how to relate the $\CA_S$ algebras of surfaces (and the dual $\widehat\CA_S$ algebras) to fusion of gapped interfaces between topological phases. We also explain how the lines on $\CI$ are related to the twisted sector lines of the algebras of surfaces. In Sec.~\ref{sec:number of twisted sectors A_S and hat A_S} we further constrain the twisted sector-lines of dual algebras of surfaces via a generalization of the invertible fixed point theorem mentioned in Sec.~\ref{InvertRev}. Next, in Sec.~\ref{sec:simple lines after gauging} we describe generalized notions of symmetry fractionalization and discrete torsion. Then, in Sec.~\ref{Int1form} we discuss the special case of 1-form symmetry gauging and dual 0-form symmetry gauging and explain distinct notions of 1-form symmetry 't Hooft anomalies. In Sec.~\ref{IntExt}, we describe how non-trivial braiding of the 1-form symmetry we gauge affects the dual surface algebra. Finally, in Sec.~\ref{sec:Morita2} we introduce additional formalism for non-invertible gauging that will be useful when we analyze particular examples later in the paper. Table \ref{GeneralizationTable} summarizes a few of the main results of this section.

\subsection{Fusion of interfaces, algebras of surfaces, and (non-)invertible gauging}\label{OPEdetails}
\begin{figure}[h!]
\centering

\tikzset{every picture/.style={line width=0.75pt}} 

\begin{tikzpicture}[x=0.75pt,y=0.75pt,yscale=-1,xscale=1]

\draw  [color={rgb, 255:red, 0; green, 0; blue, 0 }  ,draw opacity=1 ][fill={rgb, 255:red, 192; green, 188; blue, 188 }  ,fill opacity=1 ] (123.47,169) -- (123.49,91.37) -- (209.99,41.79) -- (209.98,119.43) -- cycle ;
\draw [color={rgb, 255:red, 139; green, 6; blue, 24 }  ,draw opacity=1 ]   (63,68) -- (63,166) ;
\draw    (329,110) -- (362,110) ;
\draw [shift={(364,110)}, rotate = 180] [color={rgb, 255:red, 0; green, 0; blue, 0 }  ][line width=0.75]    (10.93,-3.29) .. controls (6.95,-1.4) and (3.31,-0.3) .. (0,0) .. controls (3.31,0.3) and (6.95,1.4) .. (10.93,3.29)   ;
\draw  [color={rgb, 255:red, 0; green, 0; blue, 0 }  ,draw opacity=1 ][fill={rgb, 255:red, 192; green, 188; blue, 188 }  ,fill opacity=1 ] (489.47,177) -- (489.49,99.37) -- (575.99,49.79) -- (575.98,127.43) -- cycle ;
\draw [color={rgb, 255:red, 0; green, 0; blue, 0 }  ,draw opacity=1 ]   (535,74) -- (536,151) ;

\draw (52.67,40.45) node [anchor=north west][inner sep=0.75pt]    {$\CB_{1}$};
\draw (267.65,37.31) node [anchor=north west][inner sep=0.75pt]    {$\CB_{2}$};
\draw (195.6,47.45) node [anchor=north west][inner sep=0.75pt]    {$\CI$};
\draw (418.67,48.45) node [anchor=north west][inner sep=0.75pt]    {$\CB_{1}$};
\draw (633.65,45.31) node [anchor=north west][inner sep=0.75pt]    {$\CB_{2}$};
\draw (561.6,55.45) node [anchor=north west][inner sep=0.75pt]    {$\CI$};
\draw (536.6,94.45) node [anchor=north west][inner sep=0.75pt]    {$\CI_{L}( x)$};
\draw (67,97.4) node [anchor=north west][inner sep=0.75pt]    {$x$};

\end{tikzpicture}
\caption{Fusion of a line $x$ in $\CB_1$ from the left on the interface produces a line operator, $\CI_L(x)$, on it. One can similarly consider the action of a line in $\CB_2$ from the right.}
\label{fig:enter-label}
\end{figure}
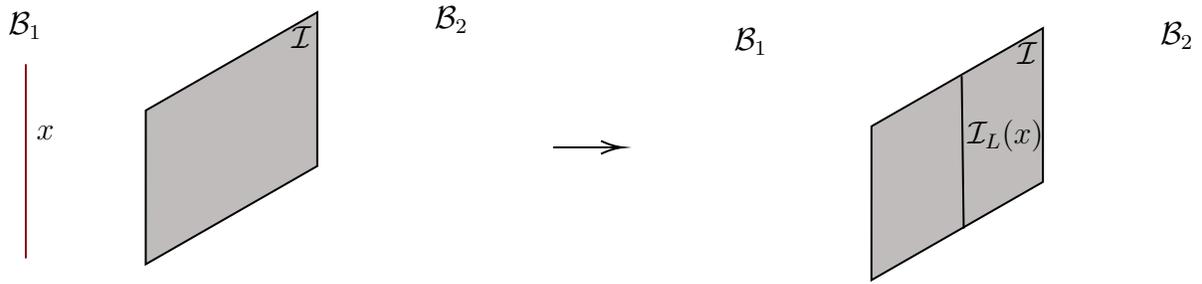

To better understand non-invertible symmetry and its gauging, let us study some general properties of topological interfaces between TQFTs. To that end,
consider two (2+1)d TQFTs with MTCs $\mc B_1$ and $\mc B_2$ that admit a topological interface $\CI$ between them. There is a manifest action of lines in $\CB_1$ and $\CB_2$ on this interface arising via parallel fusion (from either side) with lines on the interface (see Fig.~\ref{fig:enter-label}).

More precisely, let $\CC_{\CI}$ be the category of lines on the interface. Then, for $x \in \CC_{\CI}$, there is an action of $\CB_1$ and $\CB_2$ on $\CC_{\CI}$ by fusion on the interface from the left and right, respectively. Of course, the fusion of lines does not change the interface itself, but it does generally change the lines on the interface.\footnote{The situation in Fig.~\ref{fig:enter-label} can be thought of as fusion with the trivial line in $\CC_{\CI}$.}

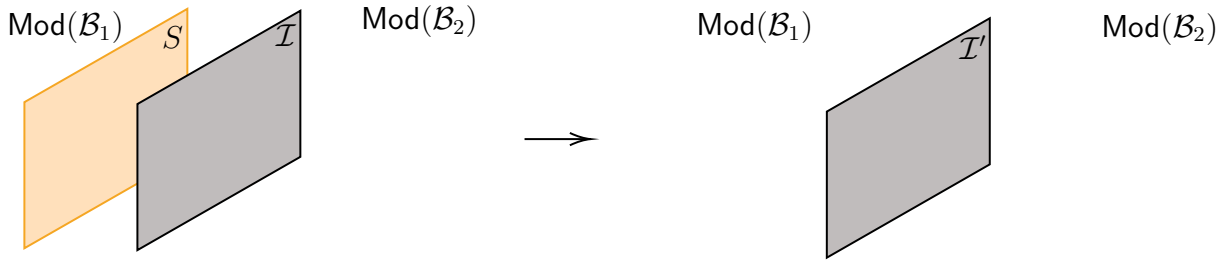
\begin{figure}[h!]

\tikzset{every picture/.style={line width=0.75pt}} 

\begin{tikzpicture}[x=0.75pt,y=0.75pt,yscale=-0.95,xscale=0.95]

\draw  [color={rgb, 255:red, 245; green, 166; blue, 35 }  ,draw opacity=1 ][fill={rgb, 255:red, 255; green, 224; blue, 187 }  ,fill opacity=1 ] (63.47,168) -- (63.49,90.37) -- (149.99,40.79) -- (149.98,118.43) -- cycle ;
\draw  [color={rgb, 255:red, 0; green, 0; blue, 0 }  ,draw opacity=1 ][fill={rgb, 255:red, 192; green, 188; blue, 188 }  ,fill opacity=1 ] (123.47,169) -- (123.49,91.37) -- (209.99,41.79) -- (209.98,119.43) -- cycle ;
\draw    (329,110) -- (362,110) ;
\draw [shift={(364,110)}, rotate = 180] [color={rgb, 255:red, 0; green, 0; blue, 0 }  ][line width=0.75]    (10.93,-3.29) .. controls (6.95,-1.4) and (3.31,-0.3) .. (0,0) .. controls (3.31,0.3) and (6.95,1.4) .. (10.93,3.29)   ;
\draw  [color={rgb, 255:red, 0; green, 0; blue, 0 }  ,draw opacity=1 ][fill={rgb, 255:red, 192; green, 188; blue, 188 }  ,fill opacity=1 ] (489.47,173) -- (489.49,95.37) -- (575.99,45.79) -- (575.98,123.43) -- cycle ;
\textbf{}
\draw (52.67,40.45) node [anchor=north west][inner sep=0.75pt]    {$\Mod(\CB_{1})$};
\draw (240.65,37.31) node [anchor=north west][inner sep=0.75pt]    {$\Mod(\CB_{2})$};
\draw (195.6,47.45) node [anchor=north west][inner sep=0.75pt]    {$\CI$};
\draw (418.67,40.45) node [anchor=north west][inner sep=0.75pt]    {$\Mod(\CB_{1})$};
\draw (633.65,41.31) node [anchor=north west][inner sep=0.75pt]    {$\Mod(\CB_{2})$};
\draw (558.6,53.45) node [anchor=north west][inner sep=0.75pt]    {$\CI'$};
\draw (135,48.4) node [anchor=north west][inner sep=0.75pt]    {$S$};
\end{tikzpicture}
\caption{Fusion of a surface, $S\in \Mod(\CB_1)$, with the interface $\CI$ from the left produces a new interface, $\CI'$.}
\label{fig:surface-interface fusion}
\end{figure}

Now, let us also consider the surfaces in the two TQFTs. These operators are contained in the fusion 2-categories $\Mod(\CB_1)$ and $\Mod(\CB_2)$ (see Sec.~\ref{sec:Morita2} for a more detailed mathematical description of these 2-categories). Now, the surfaces in either TQFT can fuse with the interface $\CI$ to produce new interfaces (see Fig.~\ref{fig:surface-interface fusion}). 
Let $\mathfrak{I}$ be the set of all interfaces between $\Mod(\CB_{1})$ and $\Mod(\CB_{2})$. As shown in \cite[Theorem 5.3.4]{Decoppet:2021skp}, any interface $\CI \in \mathfrak{I}$ can be obtained by gauging an algebra $\CA_S$ in $\Mod(\CB_{1})$ in order to obtain $\Mod(\CB_{2})$. Physically, this gauging corresponds to obtaining $\Mod(\CB_{2})$ from inserting a network of $\CA_S$ surfaces in $\Mod(\CB_{1})$. Upon half-gauging, we get an interface, $\CI$, between $\Mod(\CB_{1})$ and $\Mod(\CB_{2})$. 

\begin{figure}[h!]
\centering

\tikzset{every picture/.style={line width=0.75pt}} 

\begin{tikzpicture}[x=0.75pt,y=0.75pt,yscale=-1,xscale=1]

\draw  [color={rgb, 255:red, 245; green, 166; blue, 35 }  ,draw opacity=1 ][fill={rgb, 255:red, 255; green, 224; blue, 187 }  ,fill opacity=1 ] (240.5,179.95) -- (238.69,91.25) -- (331.73,66.02) -- (333.54,154.72) -- cycle ;
\draw  [color={rgb, 255:red, 0; green, 0; blue, 0 }  ,draw opacity=1 ][fill={rgb, 255:red, 192; green, 188; blue, 188 }  ,fill opacity=0.75 ] (286.45,184) -- (286.47,95.68) -- (375.75,39.31) -- (375.74,127.62) -- cycle ;
\draw    (331.73,66.02) -- (333.54,154.73) ;

\draw (192.04,38.97) node [anchor=north west][inner sep=0.75pt]    {$\Mod(\CB_{1})$};
\draw (436.63,42.22) node [anchor=north west][inner sep=0.75pt]    {$\Mod(\CB_{2})$};
\draw (358.09,53.75) node [anchor=north west][inner sep=0.75pt]    {$\CI$};
\draw (240.99,96.15) node [anchor=north west][inner sep=0.75pt]    {$\CA_S$};
\draw (290.89,95.05) node [anchor=north west][inner sep=0.75pt]    {$\CI$};

\end{tikzpicture}
\caption{Unlike the case of fusion of general surfaces in $\Mod(\CB_1)$ with $\CI$, the fusion of $\CA_S$ with the interface $\CI$ leaves it unchanged.}
\label{fig:fusion of A with I}
\end{figure}
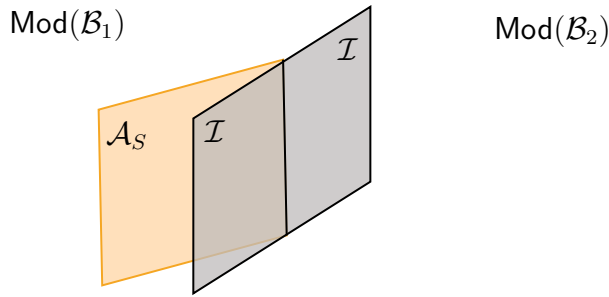

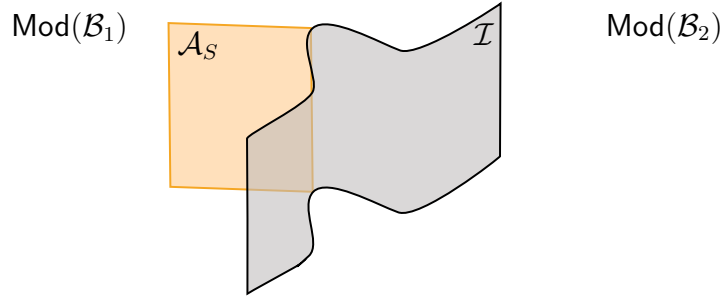
\begin{figure}[h!]
\centering
\tikzset{every picture/.style={line width=0.75pt}} 

\begin{tikzpicture}[x=0.75pt,y=0.75pt,yscale=-1,xscale=1]

\draw  [color={rgb, 255:red, 245; green, 166; blue, 35 }  ,draw opacity=1 ][fill={rgb, 255:red, 255; green, 224; blue, 187 }  ,fill opacity=1 ] (241.53,63.47) -- (313.32,65.98) -- (314.24,148.64) -- (242.45,146.13) -- cycle ;
\draw  [fill={rgb, 255:red, 192; green, 188; blue, 188 }  ,fill opacity=0.58 ] (281.32,200.08) .. controls (281.35,200.3) and (281.34,124.25) .. (281.12,121.88) .. controls (280.89,119.5) and (311.32,104.98) .. (314.32,97.98) .. controls (317.32,90.98) and (306.32,72.98) .. (314.32,65.98) .. controls (322.32,58.98) and (348.13,73.53) .. (358.22,77.25) .. controls (368.32,80.98) and (409,53.86) .. (409,53.57) .. controls (409,53.29) and (408.71,130.14) .. (408.71,130.71) .. controls (408.71,131.29) and (368.32,162.98) .. (357.32,158.98) .. controls (346.32,154.98) and (321.79,141.25) .. (314.55,148.61) .. controls (307.32,155.98) and (319.06,174.67) .. (312.32,180.98) .. controls (305.57,187.29) and (307.27,185.42) .. (307.57,185.35) .. controls (307.86,185.29) and (281.29,199.86) .. (281.32,200.08) -- cycle ;

\draw (160.67,56.45) node [anchor=north west][inner sep=0.75pt]    {$\Mod(\CB_{1})$};
\draw (460.65,56.31) node [anchor=north west][inner sep=0.75pt]    {$\Mod(\CB_{2})$};
\draw (394.17,62.04) node [anchor=north west][inner sep=0.75pt]    {$\CI$};
\draw (244.53,66.87) node [anchor=north west][inner sep=0.75pt]    {$\CA_S$};

\end{tikzpicture}
\caption{Pinching $\CI$ to the left shows that there is a non-trivial interface between $\CA_S$ and $\CI^{\dagger} \otimes \CI$. }
\label{fig:pinching I}
\end{figure}

Since $\Mod(\CB_{2})$ is obtained from a network of $\CA_S$ surfaces ending on $\CI$,  we can consider the configuration in Fig.~\ref{fig:fusion of A with I} (recall that, as described in the introduction to Sec.~\ref{OPE}, we are freely conflating the algebra of surfaces, $\CA_S$, with the underlying object). Pinching the interface $\CI$ to the left, we find that there is a 1-dimensional interface between $\CA_S$ and $\CI^{\dagger} \otimes \CI$ (see Fig.~\ref{fig:pinching I}).\footnote{This maneuver is closely related to the ``pinching trick'' described in the context of entanglement bootstrap partons in \cite{Buican:2025zpm}.} In fact, 
\be
\label{eq:A_S from I}
\CA_S \simeq \CI^{\dagger}\otimes \CI\, . 
\ee
Mathematically, this derivation can be motivated via an internal hom construction for module 2-categories \cite[Proposition 4.1.1]{Decoppet:2021skp}. This argument shows that, given the interface $\CI$, we can recover the surface $\CA_S$ that can be gauged in half-space to create $\CI$ using \eqref{eq:A_S from I}. In particular, the action of $\CA_S$ on the lines in $\Mod(\CB_1)$ can be determined from the action of $\CI^{\dagger} \otimes \CI$ on the lines in $\CB_1$. 

\begin{figure}[h!]
\centering

\tikzset{every picture/.style={line width=0.75pt}} 

\begin{tikzpicture}[x=0.75pt,y=0.75pt,yscale=-0.95,xscale=0.95]

\draw  [color={rgb, 255:red, 245; green, 166; blue, 35 }  ,draw opacity=1 ][fill={rgb, 255:red, 255; green, 224; blue, 187 }  ,fill opacity=1 ] (71.55,173) -- (71.57,74.86) -- (162.58,30.06) -- (162.57,128.19) -- cycle ;
\draw  [color={rgb, 255:red, 139; green, 6; blue, 24 }  ,draw opacity=1 ][fill={rgb, 255:red, 139; green, 6; blue, 24 }  ,fill opacity=1 ] (114.23,101.53) .. controls (114.23,100.74) and (114.87,100.11) .. (115.65,100.11) .. controls (116.43,100.11) and (117.07,100.74) .. (117.07,101.53) .. controls (117.07,102.31) and (116.43,102.95) .. (115.65,102.95) .. controls (114.87,102.95) and (114.23,102.31) .. (114.23,101.53) -- cycle ;
\draw  [color={rgb, 255:red, 0; green, 0; blue, 0 }  ,draw opacity=1 ][fill={rgb, 255:red, 192; green, 188; blue, 188 }  ,fill opacity=1 ] (360.55,169) -- (360.57,70.86) -- (451.58,26.06) -- (451.57,124.19) -- cycle ;
\draw  [color={rgb, 255:red, 0; green, 0; blue, 0 }  ,draw opacity=1 ][fill={rgb, 255:red, 192; green, 188; blue, 188 }  ,fill opacity=1 ] (488.55,170) -- (488.57,71.86) -- (579.58,27.06) -- (579.57,125.19) -- cycle ;
\draw  [color={rgb, 255:red, 139; green, 6; blue, 24 }  ,draw opacity=1 ][fill={rgb, 255:red, 139; green, 6; blue, 24 }  ,fill opacity=1 ] (404.65,98.95) .. controls (404.65,98.16) and (405.28,97.53) .. (406.07,97.53) .. controls (406.85,97.53) and (407.48,98.16) .. (407.48,98.95) .. controls (407.48,99.74) and (406.85,100.37) .. (406.07,100.37) .. controls (405.28,100.37) and (404.65,99.74) .. (404.65,98.95) -- cycle ;
\draw  [color={rgb, 255:red, 139; green, 6; blue, 24 }  ,draw opacity=1 ][fill={rgb, 255:red, 139; green, 6; blue, 24 }  ,fill opacity=1 ] (534.07,98.53) .. controls (534.07,97.74) and (534.7,97.11) .. (535.48,97.11) .. controls (536.26,97.11) and (536.9,97.74) .. (536.9,98.53) .. controls (536.9,99.31) and (536.26,99.95) .. (535.48,99.95) .. controls (534.7,99.95) and (534.07,99.31) .. (534.07,98.53) -- cycle ;
\draw [color={rgb, 255:red, 139; green, 6; blue, 24 }  ,draw opacity=1 ]   (406.07,99.37) -- (534.07,98.53) ;

\draw (136.86,38.93) node [anchor=north west][inner sep=0.75pt]    {$\CA_{S}{}$};
\draw (277,93.4) node [anchor=north west][inner sep=0.75pt]    {$=$};
\draw (366.57,70.26) node [anchor=north west][inner sep=0.75pt]    {$\CI$};
\draw (494.57,71.26) node [anchor=north west][inner sep=0.75pt]    {$\CI^{\dagger}$};
\draw (7,37.4) node [anchor=north west][inner sep=0.75pt]    {Mod($\CB_{1})$};
\draw (192,36.4) node [anchor=north west][inner sep=0.75pt]    {$\Mod(\CB_{1})$};
\draw (303,39.4) node [anchor=north west][inner sep=0.75pt]    {$\Mod( \CB_{1})$};
\draw (589.58,38.46) node [anchor=north west][inner sep=0.75pt]    {$\Mod(\CB_{1})$};
\draw (455,37.4) node [anchor=north west][inner sep=0.75pt]    {$\Mod(\CB_{2})$};
\draw (459,101.4) node [anchor=north west][inner sep=0.75pt]    {$x $};
\draw (117.65,106.35) node [anchor=north west][inner sep=0.75pt]    {$O_{x }$};
\end{tikzpicture}
\caption{Local operators on $\CA_{S}$ correspond to lines that can end on both $\CI$ and $\CI^{\dagger}$.}
\label{fig: local operators on A_S}
\end{figure}
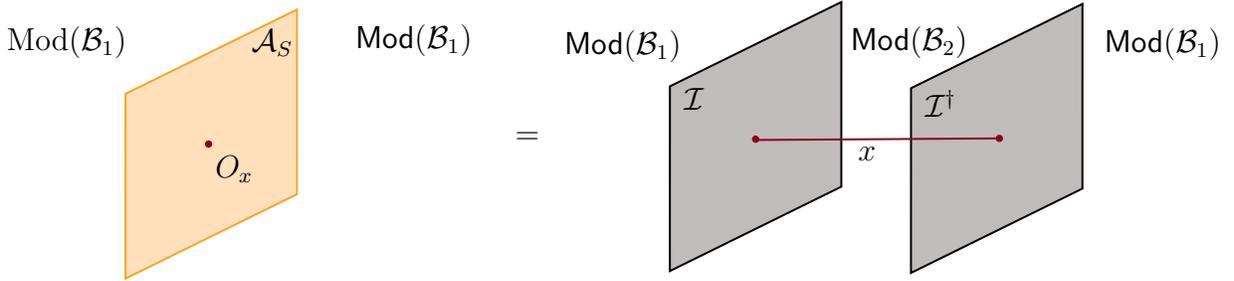

Note that $\CA_S$ typically does not correspond to an indecomposable surface. In general $\CA_S$ will decompose into a number of indecomposable surfaces determined by the number of local operators it hosts. Local operators on the surface $\CA_S$ can be obtained from a line operator, $x \in \CB_2$, that ends on both $\CI$ and $\CI^{\dagger}$ as in Fig.~\ref{fig: local operators on A_S}. In this case, $x$ decomposes into as many local operators as the dimension of the vector space of local operators on $\CA_S$.

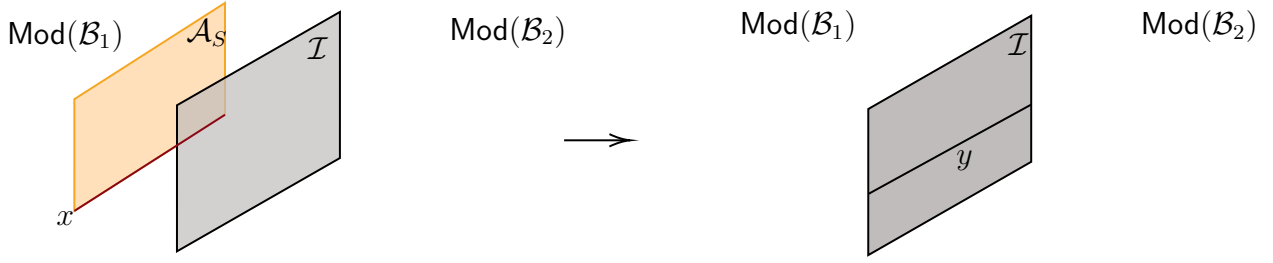
\begin{figure}[h!]
\centering

\tikzset{every picture/.style={line width=0.75pt}} 

\begin{tikzpicture}[x=0.75pt,y=0.75pt,yscale=-0.95,xscale=0.95]

\draw  [color={rgb, 255:red, 245; green, 166; blue, 35 }  ,draw opacity=1 ][fill={rgb, 255:red, 255; green, 224; blue, 187 }  ,fill opacity=1 ] (69,147.54) -- (69.01,88.12) -- (148.99,36.99) -- (148.98,96.41) -- cycle ;
\draw  [color={rgb, 255:red, 0; green, 0; blue, 0 }  ,draw opacity=1 ][fill={rgb, 255:red, 192; green, 188; blue, 188 }  ,fill opacity=0.7 ] (123.47,169) -- (123.49,91.37) -- (209.99,41.79) -- (209.98,119.43) -- cycle ;
\draw    (329,110) -- (362,110) ;
\draw [shift={(364,110)}, rotate = 180] [color={rgb, 255:red, 0; green, 0; blue, 0 }  ][line width=0.75]    (10.93,-3.29) .. controls (6.95,-1.4) and (3.31,-0.3) .. (0,0) .. controls (3.31,0.3) and (6.95,1.4) .. (10.93,3.29)   ;
\draw  [color={rgb, 255:red, 0; green, 0; blue, 0 }  ,draw opacity=1 ][fill={rgb, 255:red, 192; green, 188; blue, 188 }  ,fill opacity=1 ] (490.47,171) -- (490.49,93.37) -- (576.99,43.79) -- (576.98,121.43) -- cycle ;
\draw [color={rgb, 255:red, 139; green, 6; blue, 24 }  ,draw opacity=1 ]   (148.98,96.41) -- (69,147.54) ;
\draw    (490.73,138.4) -- (577,91) ;

\draw (31.67,44.45) node [anchor=north west][inner sep=0.75pt]    {$\Mod(\CB_{1})$};
\draw (266.65,42.31) node [anchor=north west][inner sep=0.75pt]    {$\Mod(\CB_{2})$};
\draw (191.6,53.45) node [anchor=north west][inner sep=0.75pt]    {$\CI$};
\draw (420.67,39.45) node [anchor=north west][inner sep=0.75pt]    {$\Mod(\CB_{1})$};
\draw (633.65,39.31) node [anchor=north west][inner sep=0.75pt]    {$\Mod(\CB_{2})$};
\draw (563.6,51.45) node [anchor=north west][inner sep=0.75pt]    {$\CI$};
\draw (127,44.4) node [anchor=north west][inner sep=0.75pt]    {$\CA_S$};
\draw (535.73,113.8) node [anchor=north west][inner sep=0.75pt]    {$y$};
\draw (58,146.4) node [anchor=north west][inner sep=0.75pt]    {$x$};

\end{tikzpicture}
\caption{Fusing a twisted sector line operator $x$ of $\DAS$, on $\CI$ produces a line on it.} 
\label{fig:fusing twisted x on I}
\end{figure}

\begin{figure}[h!]
    \centering

\tikzset{every picture/.style={line width=0.75pt}} 

\begin{tikzpicture}[x=0.75pt,y=0.75pt,yscale=-1,xscale=1]

\draw    (328,114) -- (359,114) ;
\draw [shift={(361,114)}, rotate = 180] [color={rgb, 255:red, 0; green, 0; blue, 0 }  ][line width=0.75]    (10.93,-3.29) .. controls (6.95,-1.4) and (3.31,-0.3) .. (0,0) .. controls (3.31,0.3) and (6.95,1.4) .. (10.93,3.29)   ;
\draw  [color={rgb, 255:red, 245; green, 166; blue, 35 }  ,draw opacity=1 ][fill={rgb, 255:red, 255; green, 224; blue, 187 }  ,fill opacity=1 ] (437.5,176.95) -- (435.69,88.25) -- (528.73,63.02) -- (530.54,151.72) -- cycle ;
\draw  [color={rgb, 255:red, 0; green, 0; blue, 0 }  ,draw opacity=1 ][fill={rgb, 255:red, 192; green, 188; blue, 188 }  ,fill opacity=0.75 ] (483.45,181) -- (483.47,92.68) -- (572.75,36.31) -- (572.74,124.62) -- cycle ;
\draw    (528.73,63.02) -- (530.54,151.73) ;
\draw [color={rgb, 255:red, 139; green, 6; blue, 24 }  ,draw opacity=1 ]   (435.69,88.25) -- (437.5,176.95) ;
\draw  [fill={rgb, 255:red, 192; green, 188; blue, 188 }  ,fill opacity=0.58 ] (108.32,183) .. controls (108.35,183.22) and (108.34,107.17) .. (108.12,104.8) .. controls (107.89,102.43) and (134,92) .. (137,83) .. controls (140,74) and (133.32,55.9) .. (141.32,48.9) .. controls (149.32,41.9) and (175.13,56.46) .. (185.22,60.18) .. controls (195.32,63.9) and (236,36.78) .. (236,36.5) .. controls (236,36.21) and (235.71,113.07) .. (235.71,113.64) .. controls (235.71,114.21) and (195.32,145.9) .. (184.32,141.9) .. controls (173.32,137.9) and (148.79,124.17) .. (141.55,131.54) .. controls (134.32,138.9) and (146.06,157.59) .. (139.32,163.9) .. controls (132.57,170.21) and (134.27,168.34) .. (134.57,168.27) .. controls (134.86,168.21) and (108.29,182.78) .. (108.32,183) -- cycle ;
\draw [color={rgb, 255:red, 74; green, 144; blue, 226 }  ,draw opacity=1 ]   (141.32,48.9) -- (142.38,130.51) ;

\draw (391.04,37.97) node [anchor=north west][inner sep=0.75pt]    {$\Mod(\CB_{1})$};
\draw (557.09,50.75) node [anchor=north west][inner sep=0.75pt]    {$\CI$};
\draw (441.69,87.65) node [anchor=north west][inner sep=0.75pt]    {$\CA_S$};
\draw (426,175.4) node [anchor=north west][inner sep=0.75pt]    {$x$};
\draw (624.65,38.31) node [anchor=north west][inner sep=0.75pt]    {$\Mod(\CB_{2})$};
\draw (60.67,34.38) node [anchor=north west][inner sep=0.75pt]    {$\Mod(\CB_{1})$};
\draw (268.65,38.24) node [anchor=north west][inner sep=0.75pt]    {$\Mod(\CB_{2})$};
\draw (224.17,44.96) node [anchor=north west][inner sep=0.75pt]    {$\CI$};
\draw (125,105.4) node [anchor=north west][inner sep=0.75pt]    {$y$};

\end{tikzpicture}
\caption{Pinching the interface and fusing in the presence of a line operator, $y\in \CC_{\CI}$, produces a twisted sector line operator $x$ bounding a surface in $\mc A_S$.}
\label{fig:twisted sector lines from the interface}
\end{figure}
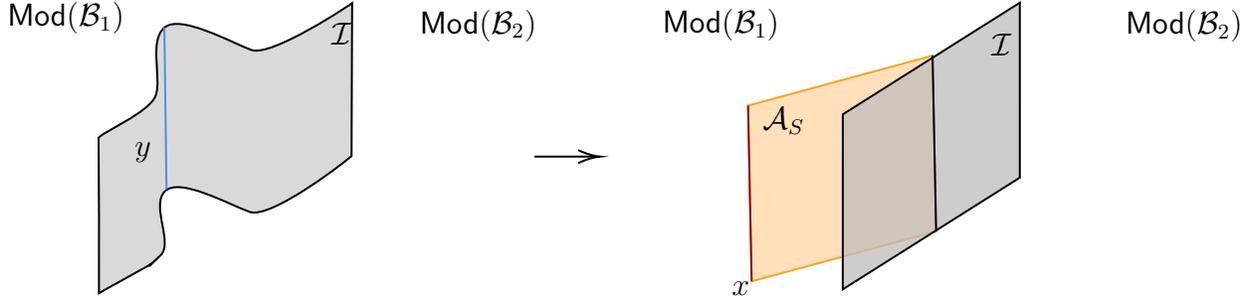

As commented above, a choice of an algebra of surfaces, $\mc A_S$, requires a choice of fusion 1-category. Physically, this is the fusion 1-category formed by the lines bounding the surfaces $\mc A_S$ decomposes into. This fusion 1-category is closely related to the category of lines on $\CI$, $\CC_{\CI}$. Since $\CA_S$ acts trivially on $\CI$, fusing some line $x$ bounding a surface in $\CA_S$ with $\CI$ produces a line on the interface $\CI$ as in Fig.~\ref{fig:fusing twisted x on I}. Conversely, given $y\in \CC_{\CI}$. Then pinching the interface to the left as before produces $\CA_S$ along with a twisted sector line operator $x$, as in Fig.~\ref{fig:twisted sector lines from the interface}.

We can repeat the above argument by starting with $\Mod(\CB_2)$ and constructing $\Mod(\CB_1)$ by inserting a network built from the surface $\widehat \CA_S$ in $\Mod(\CB_2)$. This surface is then determined by the fusion of interfaces
\be
\label{eq:hat A_S from I}
\widehat \CA_S \simeq \CI \otimes \CI^{\dagger}\, .
\ee
In analogy with our previous discussion, we can fuse a twisted sector line of $\widehat \CA_S$ with the interface to produce a line on the interface. Moreover, pinching the interface to the right, produces twisted sector lines of $\widehat\CA_S$. Therefore, we find that the fusion 1-categories of twisted sector lines of $\DAS$ and $\widehat{\mc A}_S$, as well as the fusion 1-category $\mc C_\mc I$, 
are related to each other under fusion from the left and right of the interface $\CI$. Moreover, using associativity of the fusion of interfaces, we have that 
\be
\CA_S \otimes \CI \simeq \CI \otimes \widehat \CA_S\, .
\ee
Therefore, the twisted sector lines of $\CA_S$ and $\widehat \CA_S$ are mapped to each other by the interface $\CI$. In Sec.~\ref{sec:number of twisted sectors A_S and hat A_S}, we will show that both surfaces $\CA_S$ and $\widehat\CA_S$ admit the same number of twisted sector lines bounding the corresponding surfaces.

\begin{figure}[h!]
\centering

\tikzset{every picture/.style={line width=0.75pt}} 

\begin{tikzpicture}[x=0.75pt,y=0.75pt,yscale=-0.95,xscale=0.95]

\draw  [color={rgb, 255:red, 0; green, 0; blue, 0 }  ,draw opacity=1 ][fill={rgb, 255:red, 192; green, 188; blue, 188 }  ,fill opacity=1 ] (137.55,177) -- (137.57,78.86) -- (228.58,34.06) -- (228.57,132.19) -- cycle ;
\draw [color={rgb, 255:red, 139; green, 6; blue, 24 }  ,draw opacity=1 ]   (97,62) -- (97,158) ;
\draw  [color={rgb, 255:red, 0; green, 0; blue, 0 }  ,draw opacity=1 ][fill={rgb, 255:red, 192; green, 188; blue, 188 }  ,fill opacity=1 ] (437.55,176) -- (437.57,77.86) -- (528.58,33.06) -- (528.57,131.19) -- cycle ;
\draw  [color={rgb, 255:red, 245; green, 166; blue, 35 }  ,draw opacity=1 ][fill={rgb, 255:red, 255; green, 224; blue, 187 }  ,fill opacity=0.93 ] (486.55,152.33) -- (486.57,54.5) -- (577.58,53.17) -- (577.56,151) -- cycle ;
\draw [color={rgb, 255:red, 139; green, 6; blue, 24 }  ,draw opacity=1 ]   (577.58,53.17) -- (577.56,151) ;

\draw (144.57,79.26) node [anchor=north west][inner sep=0.75pt]    {$\CI$};
\draw (338,96.4) node [anchor=north west][inner sep=0.75pt]    {$=$};
\draw (447.57,80.26) node [anchor=north west][inner sep=0.75pt]    {$\CI$};
\draw (92,158.4) node [anchor=north west][inner sep=0.75pt]    {$a$};
\draw (573,151.4) node [anchor=north west][inner sep=0.75pt]    {$x$};
\draw (554.57,55.26) node [anchor=north west][inner sep=0.75pt]    {$\widehat \CA_S$};
\draw (17,40.4) node [anchor=north west][inner sep=0.75pt]    {$\Mod( \CB_{1})$};
\draw (253,40.4) node [anchor=north west][inner sep=0.75pt]    {$\Mod( \CB_{2})$};
\draw (386,38.4) node [anchor=north west][inner sep=0.75pt]    {$\Mod( \CB_{1})$};
\draw (591,37.4) node [anchor=north west][inner sep=0.75pt]    {$\Mod( \CB_{2})$};

\end{tikzpicture}
\caption{Passing a line $a\in \CB_1$ through the interface $\CI$ produces a line $x$ bounding a surface in $\widehat\CA_S$. If $x$ is supported on the trivial surface in $\widehat\CA_S$, then $x$ is a genuine line in $\CB_2$.}
\label{fig:action of I on lines}
\end{figure}
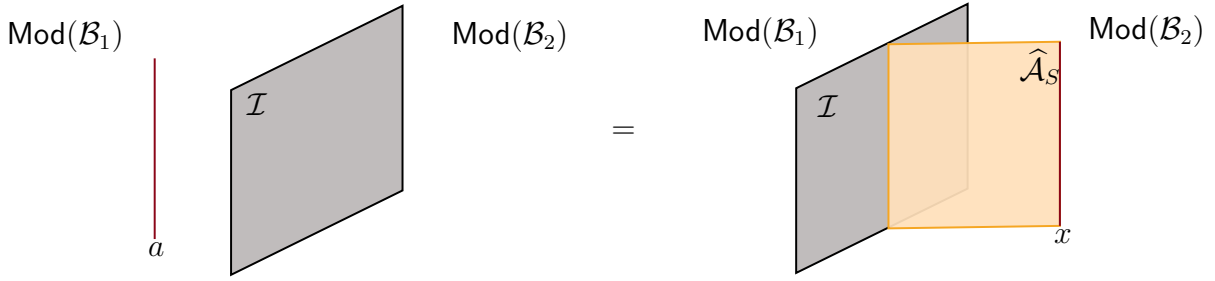

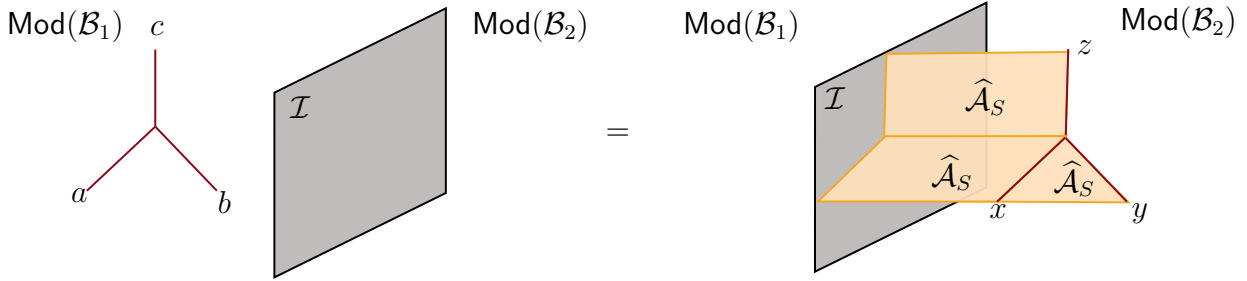
\begin{figure}[h!]
\centering

\tikzset{every picture/.style={line width=0.75pt}} 

\begin{tikzpicture}[x=0.75pt,y=0.75pt,yscale=-0.95,xscale=0.95]

\draw  [color={rgb, 255:red, 245; green, 166; blue, 35 }  ,draw opacity=1 ][fill={rgb, 255:red, 255; green, 224; blue, 187 }  ,fill opacity=0.93 ] (600,136.18) -- (503.73,135.62) -- (469.6,100.72) -- (565.87,101.28) -- cycle ;
\draw  [color={rgb, 255:red, 0; green, 0; blue, 0 }  ,draw opacity=1 ][fill={rgb, 255:red, 192; green, 188; blue, 188 }  ,fill opacity=1 ] (146.55,176) -- (146.57,77.86) -- (237.58,33.06) -- (237.57,131.19) -- cycle ;
\draw [color={rgb, 255:red, 139; green, 6; blue, 24 }  ,draw opacity=1 ]   (83.32,55) -- (83.32,95.91) ;
\draw  [color={rgb, 255:red, 0; green, 0; blue, 0 }  ,draw opacity=1 ][fill={rgb, 255:red, 192; green, 188; blue, 188 }  ,fill opacity=1 ] (433.55,173) -- (433.57,74.86) -- (524.58,30.06) -- (524.57,128.19) -- cycle ;
\draw  [color={rgb, 255:red, 245; green, 166; blue, 35 }  ,draw opacity=1 ][fill={rgb, 255:red, 255; green, 224; blue, 187 }  ,fill opacity=0.93 ] (470.19,101.98) -- (471.48,57.25) -- (567.98,56.25) -- (566.68,100.99) -- cycle ;
\draw [color={rgb, 255:red, 139; green, 6; blue, 24 }  ,draw opacity=1 ]   (83.32,95.91) -- (47,130) ;
\draw [color={rgb, 255:red, 139; green, 6; blue, 24 }  ,draw opacity=1 ]   (83.32,95.91) -- (116,130) ;
\draw [color={rgb, 255:red, 139; green, 6; blue, 24 }  ,draw opacity=1 ]   (566.46,101.73) -- (599.15,135.82) ;
\draw [color={rgb, 255:red, 139; green, 6; blue, 24 }  ,draw opacity=1 ]   (568,54.64) -- (566.68,100.98) ;
\draw  [color={rgb, 255:red, 245; green, 166; blue, 35 }  ,draw opacity=1 ][fill={rgb, 255:red, 255; green, 224; blue, 187 }  ,fill opacity=0.93 ] (471.01,101) -- (566.69,101) -- (530.82,135.7) -- (435.14,135.7) -- cycle ;
\draw [color={rgb, 255:red, 139; green, 6; blue, 24 }  ,draw opacity=1 ]   (566.46,101.73) -- (530.15,135.82) ;

\draw (153.57,78.26) node [anchor=north west][inner sep=0.75pt]    {$\CI$};
\draw (321,95.4) node [anchor=north west][inner sep=0.75pt]    {$=$};
\draw (437.57,73.26) node [anchor=north west][inner sep=0.75pt]    {$\CI$};
\draw (37,127.4) node [anchor=north west][inner sep=0.75pt]    {$a$};
\draw (115,128.4) node [anchor=north west][inner sep=0.75pt]    {$b$};
\draw (79,38.4) node [anchor=north west][inner sep=0.75pt]    {$c$};
\draw (524.57,135.59) node [anchor=north west][inner sep=0.75pt]    {$x$};
\draw (600.15,135.22) node [anchor=north west][inner sep=0.75pt]    {$y$};
\draw (571,48.4) node [anchor=north west][inner sep=0.75pt]    {$z$};
\draw (493.57,109.26) node [anchor=north west][inner sep=0.75pt]    {$\widehat\CA_S$};
\draw (511.57,69.26) node [anchor=north west][inner sep=0.75pt]    {$\widehat\CA_S$};
\draw (558.57,112.4) node [anchor=north west][inner sep=0.75pt]    {$\widehat\CA_S$};
\draw (3,32.4) node [anchor=north west][inner sep=0.75pt]    {$\Mod(\CB_{1})$};
\draw (250,32.4) node [anchor=north west][inner sep=0.75pt]    {$\Mod(\CB_{2})$};
\draw (362,32.4) node [anchor=north west][inner sep=0.75pt]    {$\Mod(\CB_{1})$};
\draw (594,30.4) node [anchor=north west][inner sep=0.75pt]    {$\Mod(\CB_{2})$};
\end{tikzpicture}
\caption{Passing a point operator at the junction of three lines through the interface $\CI$ produces a point operator at the junction of three generically non-genuine lines bounding surfaces in $\widehat\CA_S$.  We have suppressed the label of the point operator at the point junction for clarity. The line at the junction of the three $\widehat\CA_S$ surfaces is defined by the choice of fusion 1-category underlying defining the algebra $\widehat\CA_S$.}
\label{fig:action of I on fusion spaces}
\end{figure}

In general, passing genuine lines in $\mc B_1$ through the interface $\CI$ may produce non-genuine lines in $\mc B_2$ (see Fig.~\ref{fig:action of I on lines}). In particular, we get lines bounding surfaces in $\widehat\CA_S$, because the twisted sector lines of $\widehat \CA_S$ are precisely those that become genuine lines in $\CB_1$ upon gauging $\CA_S$. Moreover, passing a fusion vertex consisting of three lines in $\CB_2$ through $\CI$  produces twisted sector lines attached to a trivalent junction of the $\widehat\CA_S$ surface as in Fig.~\ref{fig:action of I on fusion spaces}.

It will be useful for us to keep track of how the simple genuine lines in $\CB_1$ map to those in $\CB_2$ through the interface $\CI$. To that end, let $W$ be the matrix whose rows are labeled by the simple genuine lines in $\CB_2$ and whose columns are labeled by simple genuine lines in $\CB_1$ (e.g., See \cite{Lan:2014uaa}). The matrix elements of $W$ are defined by
\be\label{Wdefn}
W_{ij}:= N_{\CI(a_j)}^{x_i}\, ,
\ee
where $N_{\CI(a_i)}^{x_j}$ is the multiplicity of the simple line $x_j \in \CB_2$ in $\CI(a_i)$, and $a_i \in \CB_1$.
Note that some columns of this matrix can be zero. This phenomenon arises when a simple line in $\CB_1$ passes through the interface $\CI$ to become a non-genuine line operator. The actions of $\CA_S$ and $\widehat\CA_S$ on simple lines in $\CB_1$ and $\CB_2$ are given by $W^{\dagger}W$ and $WW^{\dagger}$ respectively.

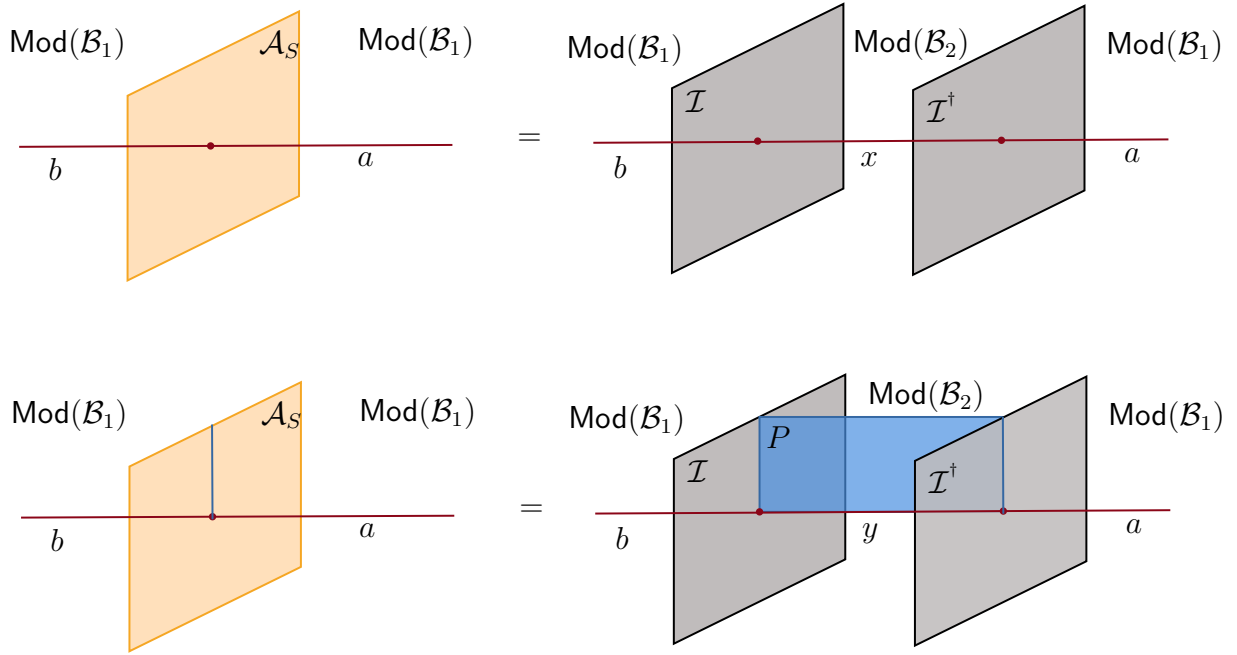
\begin{figure}[h!]
\centering

\tikzset{every picture/.style={line width=0.75pt}} 

\begin{tikzpicture}[x=0.75pt,y=0.75pt,yscale=-0.95,xscale=0.95]

\draw  [color={rgb, 255:red, 0; green, 0; blue, 0 }  ,draw opacity=1 ][fill={rgb, 255:red, 192; green, 188; blue, 188 }  ,fill opacity=1 ] (361.55,366) -- (361.57,267.86) -- (452.58,223.06) -- (452.57,321.19) -- cycle ;
\draw  [color={rgb, 255:red, 53; green, 104; blue, 163 }  ,draw opacity=1 ][fill={rgb, 255:red, 74; green, 144; blue, 226 }  ,fill opacity=0.7 ] (407.04,245.53) -- (536.23,245.53) -- (536.26,295.95) -- (407.07,295.95) -- cycle ;
\draw  [color={rgb, 255:red, 245; green, 166; blue, 35 }  ,draw opacity=1 ][fill={rgb, 255:red, 255; green, 224; blue, 187 }  ,fill opacity=1 ] (71.55,173) -- (71.57,74.86) -- (162.58,30.06) -- (162.57,128.19) -- cycle ;
\draw  [color={rgb, 255:red, 139; green, 6; blue, 24 }  ,draw opacity=1 ][fill={rgb, 255:red, 139; green, 6; blue, 24 }  ,fill opacity=1 ] (114.23,101.53) .. controls (114.23,100.74) and (114.87,100.11) .. (115.65,100.11) .. controls (116.43,100.11) and (117.07,100.74) .. (117.07,101.53) .. controls (117.07,102.31) and (116.43,102.95) .. (115.65,102.95) .. controls (114.87,102.95) and (114.23,102.31) .. (114.23,101.53) -- cycle ;
\draw  [color={rgb, 255:red, 0; green, 0; blue, 0 }  ,draw opacity=1 ][fill={rgb, 255:red, 192; green, 188; blue, 188 }  ,fill opacity=1 ] (360.55,169) -- (360.57,70.86) -- (451.58,26.06) -- (451.57,124.19) -- cycle ;
\draw  [color={rgb, 255:red, 0; green, 0; blue, 0 }  ,draw opacity=1 ][fill={rgb, 255:red, 192; green, 188; blue, 188 }  ,fill opacity=1 ] (488.55,170) -- (488.57,71.86) -- (579.58,27.06) -- (579.57,125.19) -- cycle ;
\draw  [color={rgb, 255:red, 139; green, 6; blue, 24 }  ,draw opacity=1 ][fill={rgb, 255:red, 139; green, 6; blue, 24 }  ,fill opacity=1 ] (404.65,98.95) .. controls (404.65,98.16) and (405.28,97.53) .. (406.07,97.53) .. controls (406.85,97.53) and (407.48,98.16) .. (407.48,98.95) .. controls (407.48,99.74) and (406.85,100.37) .. (406.07,100.37) .. controls (405.28,100.37) and (404.65,99.74) .. (404.65,98.95) -- cycle ;
\draw  [color={rgb, 255:red, 139; green, 6; blue, 24 }  ,draw opacity=1 ][fill={rgb, 255:red, 139; green, 6; blue, 24 }  ,fill opacity=1 ] (534.07,98.53) .. controls (534.07,97.74) and (534.7,97.11) .. (535.48,97.11) .. controls (536.26,97.11) and (536.9,97.74) .. (536.9,98.53) .. controls (536.9,99.31) and (536.26,99.95) .. (535.48,99.95) .. controls (534.7,99.95) and (534.07,99.31) .. (534.07,98.53) -- cycle ;
\draw [color={rgb, 255:red, 139; green, 6; blue, 24 }  ,draw opacity=1 ]   (406.07,99.37) -- (534.07,98.53) ;
\draw [color={rgb, 255:red, 139; green, 6; blue, 24 }  ,draw opacity=1 ]   (117.07,101.53) -- (244,101) ;
\draw [color={rgb, 255:red, 139; green, 6; blue, 24 }  ,draw opacity=1 ]   (14,102) -- (114.23,101.53) ;
\draw [color={rgb, 255:red, 139; green, 6; blue, 24 }  ,draw opacity=1 ]   (318.83,99.85) -- (406.07,99.37) ;
\draw [color={rgb, 255:red, 139; green, 6; blue, 24 }  ,draw opacity=1 ]   (536.9,98.53) -- (624.13,98.06) ;
\draw  [color={rgb, 255:red, 245; green, 166; blue, 35 }  ,draw opacity=1 ][fill={rgb, 255:red, 255; green, 224; blue, 187 }  ,fill opacity=1 ] (72.55,370) -- (72.57,271.86) -- (163.58,227.06) -- (163.57,325.19) -- cycle ;
\draw  [color={rgb, 255:red, 139; green, 6; blue, 24 }  ,draw opacity=1 ][fill={rgb, 255:red, 139; green, 6; blue, 24 }  ,fill opacity=1 ] (115.23,298.53) .. controls (115.23,297.74) and (115.87,297.11) .. (116.65,297.11) .. controls (117.43,297.11) and (118.07,297.74) .. (118.07,298.53) .. controls (118.07,299.31) and (117.43,299.95) .. (116.65,299.95) .. controls (115.87,299.95) and (115.23,299.31) .. (115.23,298.53) -- cycle ;
\draw  [color={rgb, 255:red, 0; green, 0; blue, 0 }  ,draw opacity=1 ][fill={rgb, 255:red, 192; green, 188; blue, 188 }  ,fill opacity=0.86 ] (489.55,367) -- (489.57,268.86) -- (580.58,224.06) -- (580.57,322.19) -- cycle ;
\draw  [color={rgb, 255:red, 139; green, 6; blue, 24 }  ,draw opacity=1 ][fill={rgb, 255:red, 139; green, 6; blue, 24 }  ,fill opacity=1 ] (405.65,295.95) .. controls (405.65,295.16) and (406.28,294.53) .. (407.07,294.53) .. controls (407.85,294.53) and (408.48,295.16) .. (408.48,295.95) .. controls (408.48,296.74) and (407.85,297.37) .. (407.07,297.37) .. controls (406.28,297.37) and (405.65,296.74) .. (405.65,295.95) -- cycle ;
\draw  [color={rgb, 255:red, 139; green, 6; blue, 24 }  ,draw opacity=1 ][fill={rgb, 255:red, 139; green, 6; blue, 24 }  ,fill opacity=1 ] (535.07,295.53) .. controls (535.07,294.74) and (535.7,294.11) .. (536.48,294.11) .. controls (537.26,294.11) and (537.9,294.74) .. (537.9,295.53) .. controls (537.9,296.31) and (537.26,296.95) .. (536.48,296.95) .. controls (535.7,296.95) and (535.07,296.31) .. (535.07,295.53) -- cycle ;
\draw [color={rgb, 255:red, 139; green, 6; blue, 24 }  ,draw opacity=1 ]   (407.07,296.37) -- (535.07,295.53) ;
\draw [color={rgb, 255:red, 139; green, 6; blue, 24 }  ,draw opacity=1 ]   (118.07,298.53) -- (245,298) ;
\draw [color={rgb, 255:red, 139; green, 6; blue, 24 }  ,draw opacity=1 ]   (15,299) -- (115.23,298.53) ;
\draw [color={rgb, 255:red, 139; green, 6; blue, 24 }  ,draw opacity=1 ]   (319.83,296.85) -- (407.07,296.37) ;
\draw [color={rgb, 255:red, 139; green, 6; blue, 24 }  ,draw opacity=1 ]   (537.9,295.53) -- (625.13,295.06) ;
\draw [color={rgb, 255:red, 53; green, 104; blue, 163 }  ,draw opacity=1 ]   (536.23,245.53) -- (536.48,296.95) ;
\draw [color={rgb, 255:red, 53; green, 104; blue, 163 }  ,draw opacity=1 ]   (116.4,249.53) -- (116.65,299.95) ;

\draw (139.86,38.93) node [anchor=north west][inner sep=0.75pt]    {$\CA_{S}{}$};
\draw (277,93.4) node [anchor=north west][inner sep=0.75pt]    {$=$};
\draw (366.57,70.26) node [anchor=north west][inner sep=0.75pt]    {$\CI$};
\draw (494.57,71.26) node [anchor=north west][inner sep=0.75pt]    {$\CI^{^{\dagger }}$};
\draw (7,37.4) node [anchor=north west][inner sep=0.75pt]    {$\Mod( \mc B_1)$};
\draw (192,36.4) node [anchor=north west][inner sep=0.75pt]    {$\Mod( \mc B_1)$};
\draw (303,40.4) node [anchor=north west][inner sep=0.75pt]    {$\Mod( \mc B_1)$};
\draw (589.58,38.46) node [anchor=north west][inner sep=0.75pt]    {$\Mod( \mc B_1)$};
\draw (455,37.4) node [anchor=north west][inner sep=0.75pt]    {$\Mod( \mc B_2)$};
\draw (192,103.4) node [anchor=north west][inner sep=0.75pt]    {$a$};
\draw (28,106.4) node [anchor=north west][inner sep=0.75pt]    {$b$};
\draw (599,101.4) node [anchor=north west][inner sep=0.75pt]    {$a$};
\draw (328,105.4) node [anchor=north west][inner sep=0.75pt]    {$b$};
\draw (459,103.4) node [anchor=north west][inner sep=0.75pt]    {$x$};
\draw (140.86,235.93) node [anchor=north west][inner sep=0.75pt]    {$\CA_{S}{}$};
\draw (278,290.4) node [anchor=north west][inner sep=0.75pt]    {$=$};
\draw (367.57,267.26) node [anchor=north west][inner sep=0.75pt]    {$\CI$};
\draw (495.57,268.26) node [anchor=north west][inner sep=0.75pt]    {$\CI^{^{\dagger }}$};
\draw (8,234.4) node [anchor=north west][inner sep=0.75pt]    {$\Mod( \mc B_1)$};
\draw (193,233.4) node [anchor=north west][inner sep=0.75pt]    {$\Mod( \mc B_1)$};
\draw (304,237.4) node [anchor=north west][inner sep=0.75pt]    {$\Mod( \mc B_1)$};
\draw (590.58,235.46) node [anchor=north west][inner sep=0.75pt]    {$\Mod( \mc B_1)$};
\draw (463,224.4) node [anchor=north west][inner sep=0.75pt]    {$\Mod( \mc B_2)$};
\draw (193,300.4) node [anchor=north west][inner sep=0.75pt]    {$a$};
\draw (29,303.4) node [anchor=north west][inner sep=0.75pt]    {$b$};
\draw (600,298.4) node [anchor=north west][inner sep=0.75pt]    {$a$};
\draw (329,302.4) node [anchor=north west][inner sep=0.75pt]    {$b$};
\draw (460,300.4) node [anchor=north west][inner sep=0.75pt]    {$y$};
\draw (409.04,248.93) node [anchor=north west][inner sep=0.75pt]    {$P$};

\end{tikzpicture}
\caption{The top figure shows that two lines $a,b\in\CB_1$ can form a junction on the surface $\CA_S$ if and only if there exists some intermediate line $x \in \CB_2$ such that $a$ and $b$ are connected through the interfaces $\CI$ and $\CI^{\dagger}$. The bottom figure shows that, to find point junctions of two genuine lines on the surface $\CA_S$, the intermediate line must also be a genuine line operator.}
\label{fig:A_S action from I}
\end{figure}

Note that $W$ only captures the action of the interface on genuine lines to given genuine lines. However, this data is enough to capture the action of $\CA_S$ and $\widehat\CA_S$ on genuine lines: the intermediate non-genuine lines do not contribute to this action. Indeed, the action of $\CA_S$ on genuine lines is given by the possible junctions of two simple lines on the interface $\CI$. On the other hand, intermediate non-genuine lines give trivalent junctions of simple lines at the interface (see Fig.~\ref{fig:A_S action from I}). We will see in many examples that, given an interface $\CI$, computing $W^{\dagger}W$ and $WW^{\dagger}$ is often sufficient to identify the surfaces $\CA_S$ and $\widehat\CA_S$, respectively.

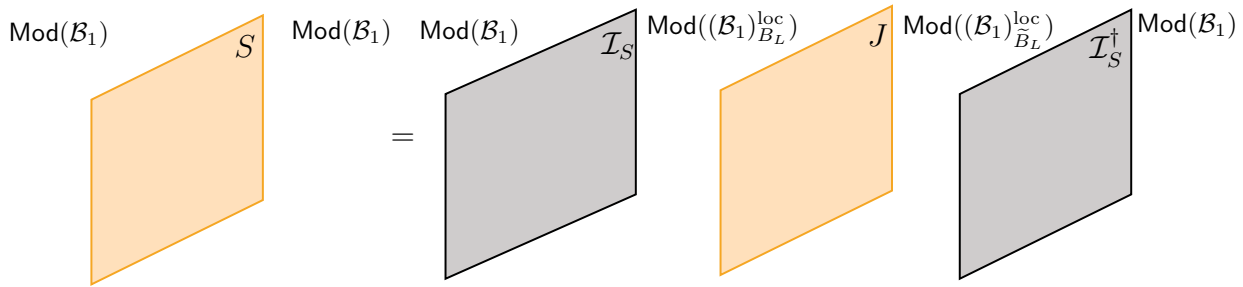
\begin{figure}[h!]
\centering

\tikzset{every picture/.style={line width=0.75pt}} 

\begin{tikzpicture}[x=0.75pt,y=0.75pt,yscale=-0.95,xscale=0.95]

\draw  [color={rgb, 255:red, 0; green, 0; blue, 0 }  ,draw opacity=1 ][fill={rgb, 255:red, 192; green, 188; blue, 188 }  ,fill opacity=0.75 ] (265.55,173) -- (265.57,74.86) -- (366.58,30.06) -- (366.57,128.19) -- cycle ;
\draw  [color={rgb, 255:red, 245; green, 166; blue, 35 }  ,draw opacity=1 ][fill={rgb, 255:red, 255; green, 224; blue, 187 }  ,fill opacity=1 ] (77.55,176) -- (77.57,77.86) -- (168.58,33.06) -- (168.57,131.19) -- cycle ;
\draw  [color={rgb, 255:red, 245; green, 166; blue, 35 }  ,draw opacity=1 ][fill={rgb, 255:red, 255; green, 224; blue, 187 }  ,fill opacity=1 ] (411.55,171) -- (411.57,72.86) -- (502.58,28.06) -- (502.57,126.19) -- cycle ;
\draw  [color={rgb, 255:red, 0; green, 0; blue, 0 }  ,draw opacity=1 ][fill={rgb, 255:red, 192; green, 188; blue, 188 }  ,fill opacity=0.75 ] (538.55,173) -- (538.57,74.86) -- (629.58,30.06) -- (629.57,128.19) -- cycle ;

\draw (151.86,41.93) node [anchor=north west][inner sep=0.75pt]    {$S$};
\draw (32,34.4) node [anchor=north west][inner sep=0.75pt]    {\footnotesize  $\Mod(\CB_1)$};
\draw (182,33.4) node [anchor=north west][inner sep=0.75pt]    {\footnotesize    $\Mod(\CB_1)$};
\draw (235,94.4) node [anchor=north west][inner sep=0.75pt]    {$=$};
\draw (250,33.4) node [anchor=north west][inner sep=0.75pt]    {\footnotesize  $\Mod(\CB_1)$};
\draw (631,29.9) node [anchor=north west][inner sep=0.75pt]    {\footnotesize   $\Mod(\CB_1)$};
\draw (347,40.4) node [anchor=north west][inner sep=0.75pt]    {$\CI_S$};
\draw (606,38.4) node [anchor=north west][inner sep=0.75pt]    {$\CI_S^{\dagger }$};
\draw (370.58,29.46) node [anchor=north west][inner sep=0.75pt]    {\footnotesize   $\Mod((\CB_1)_{B_L}^{\rm loc})$};
\draw (506.58,29.46) node [anchor=north west][inner sep=0.75pt]    {\footnotesize $\Mod((\CB_1)_{\widetilde B_L}^{\rm loc})$};
\draw (488,36.4) node [anchor=north west][inner sep=0.75pt]    {$J$};

\end{tikzpicture}
\caption{When we are interested in characterizing the action of a non-invertible condensation surface, $S$, on a genuine line, $a\in\CB_1$, it is often useful to expand the surface into a ``sandwich'' \cite{davydov2013structure,Huston:2022utd} as in the above figure and send $a$ through it (by the logic of Fig.~\ref{fig:A_S action from I}, we can ignore intermediate non-genuine lines). This sandwich involves an interface, $\CI_S$, that gauges a 1-form symmetry corresponding to a condensable algebra $B_L\in\CB_1$, a surface $J$ in $\Mod((\CB_1)_{\widetilde B_L}^{\rm loc})$, with invertible action on genuine lines, and a surface $\CI_S^{\dagger}$ that gauges a 0-form symmetry dual to the 1-form symmetry corresponding to $\widetilde B_L\in\CB_2$.}
\label{fig:sandwich}
\end{figure}

More generally, we would like to classify the types of indecomposable surfaces, $S_i$, that can appear in $\CA_S$ and $\widehat\CA_S$.\footnote{We will have more to say on this point in Sec.~\ref{GenConstraints} after introducing the formalism of Sec.~\ref{sec:Morita2}. See also App.~\ref{DetSimple} for a complete solution to this problem in the case of Abelian TQFTs.} We have partly explained one aspect of this problem around \eqref{genCondLO}: in general, we can have both invertible and non-invertible $S_i$ appearing (see also the examples in Sec.~\ref{ToricCode}). We can further refine the classification of surfaces appearing in $\CA_S$ and $\widehat\CA_S$ by characterizing their individual actions on genuine lines (i.e., their individual contributions to $W^{\dagger}W$ and $WW^{\dagger}$). For example, in the case of non-invertible surfaces, we will see in Sec.~\ref{GenConstraints} that it is often useful to follow \cite{davydov2013structure,Huston:2022utd} and decompose such surfaces into interfaces implementing condensation and dual 0-form gauging sandwiching a surface with invertible action on genuine lines as in Fig.~\ref{fig:sandwich}. 

More broadly, the symmetries associated with the $S_i$ are of two types:
\begin{enumerate}
\item {\it Faithful symmetries} corresponding to $S_i\in\Mod(\CB)$ that act non-trivially on genuine lines or their fusion spaces.
\item {\it Unfaithful symmetries} corresponding to $S_i\in\Mod(\CB)$ that leave all the genuine lines and their fusion spaces invariant. As an object in $\Mod(\CB)$, such an $S_i$ is a copy of the trivial surface: $S_i\simeq S_1$. If we consider extending the action of $S_i$ to twisted sector lines and their corresponding surfaces, we have a further refinement: the surface may or may not remain unfaithful. Examples of the former case include the surfaces implementing an Abelian 0-form symmetry, $G$, acting on an Abelian TQFT obtained from gauging the $\Rep(G)$ 1-form symmetry of another Abelian TQFT (e.g., see Sec.~\ref{TCDZ4}). Examples of the latter case include the case of a surface obtained by gauging a 1-form symmetry corresponding to a Lagrangian algebra in $\CZ(\msf{Fib})$ (e.g., see Sec.~\ref{ZFib}).
\end{enumerate}
We will see various examples involving each type of symmetry and mixtures thereof in Sec.~\ref{ToricCode}.

\subsec{Fixed point theorem and a consistency condition on dual symmetries}
\label{sec:number of twisted sectors A_S and hat A_S}
In the previous section, we showed that $\CI$, the interface between $\Mod(\CB_1)$ and $\Mod(\CB_2)$, relates the fusion 1-categories of lines bounding surfaces in $\DAS$ and ${\widehat\CA_S}$, respectively. In this section, we derive an important consistency condition on these fusion 1-categories: they possess the same number of simple objects.

To understand the above statements, we proceed in two steps:
\begin{enumerate}
\item We first show that the actions of the surfaces $\CA_S$ and $\widehat\CA_S$ on genuine lines have the same number of fixed points.
\item We then show that the number of fixed points of the action of a surface on genuine lines is equal to the number of simple boundary conditions of the surface.  
\end{enumerate}
This result is a generalization of the fixed point theorem for invertible 0-form symmetries proved in \cite{Barkeshli:2014cna}.\footnote{In \cite{Barkeshli:2014cna}, this property was derived for invertible surfaces implementing a $G$ symmetry by using $G$-crossed modularity.} To that end, let us first show that the action of the surfaces $\CA_S$ and $\widehat\CA_S$ have the same number of fixed points. Let $S$ be a surface whose action on lines is given by 
\be
    S(a) \cong \bigoplus_{b\in \text{Irr}(\CB)} N_{S(a)}^{b} \cdot b\, ,
\ee
where $N_{S(a)}^b$ are non-negative integers, and $a,b$ are genuine lines. The number of fixed points of the action of $S$ is 
\be
    X_S:= \sum_{a\in \text{Irr}(\CB)} N_{S(a)}^a\, .
\ee
The relevant surfaces in our setting are
\be
\CA_S \simeq \CI^{\dagger} \otimes \CI \, , \q \widehat\CA_S \simeq \CI \otimes \CI^{\dagger}\, .  
\ee
The action of these surfaces on the genuine lines is given by the $W^{\dagger}W$ and $WW^{\dagger}$ matrices introduced around \eqref{Wdefn}. Clearly, 
\be
\label{eq:equal fixed points of A_S and hat A_S}
X_{\CA_S}= \sum_{i} (W^{\dagger}W)_{ii}= \sum_{i,j} W_{ji}W_{ji}= \sum_{j} (WW^{\dagger})_{jj}=X_{\widehat\CA_S}\, .
\ee

Now, we will show that, for a general surface in a (2+1)d TQFT, the number of fixed points of its action on genuine lines is equal to the number of simple boundary conditions admitted by the surface. For the special case of Abelian TQFTs, an alternate explicit argument is given in App.~\ref{FPabelian} in terms of group theoretical data. For a general surface $S$, consider the category of twisted sector lines of $S$. We will show that the rank of this category, defined as its number of simple objects, is equal to $X_S$.

\begin{figure}[h!]
\centering

\tikzset{every picture/.style={line width=0.75pt}} 

\begin{tikzpicture}[x=0.75pt,y=0.75pt,yscale=-1,xscale=1]

\draw  [color={rgb, 255:red, 245; green, 166; blue, 35 }  ,draw opacity=1 ][fill={rgb, 255:red, 255; green, 224; blue, 187 }  ,fill opacity=1 ] (128.55,176) -- (128.57,77.86) -- (219.58,33.06) -- (219.57,131.19) -- cycle ;
\draw [color={rgb, 255:red, 139; green, 6; blue, 24 }  ,draw opacity=1 ]   (174.07,104.53) -- (276.26,104.4) ;
\draw [color={rgb, 255:red, 139; green, 6; blue, 24 }  ,draw opacity=1 ] [dash pattern={on 4.5pt off 4.5pt}]  (133.71,105.26) -- (174.07,104.53) ;
\draw [color={rgb, 255:red, 139; green, 6; blue, 24 }  ,draw opacity=1 ]   (65.86,104.79) -- (129.42,105.26) ;
\draw  [color={rgb, 255:red, 139; green, 6; blue, 24 }  ,draw opacity=1 ][fill={rgb, 255:red, 139; green, 6; blue, 24 }  ,fill opacity=1 ] (171.23,104.53) .. controls (171.23,103.74) and (171.87,103.11) .. (172.65,103.11) .. controls (173.43,103.11) and (174.07,103.74) .. (174.07,104.53) .. controls (174.07,105.31) and (173.43,105.95) .. (172.65,105.95) .. controls (171.87,105.95) and (171.23,105.31) .. (171.23,104.53) -- cycle ;
\draw  [color={rgb, 255:red, 0; green, 0; blue, 0 }  ,draw opacity=1 ][fill={rgb, 255:red, 192; green, 188; blue, 188 }  ,fill opacity=1 ] (520.54,174) -- (520.55,75.86) -- (611.57,31.06) -- (611.56,129.19) -- cycle ;
\draw [color={rgb, 255:red, 139; green, 6; blue, 24 }  ,draw opacity=1 ] [dash pattern={on 4.5pt off 4.5pt}]  (525.7,103.26) -- (566.06,102.53) ;
\draw [color={rgb, 255:red, 139; green, 6; blue, 24 }  ,draw opacity=1 ]   (457.85,102.79) -- (521.41,103.26) ;
\draw  [color={rgb, 255:red, 139; green, 6; blue, 24 }  ,draw opacity=1 ][fill={rgb, 255:red, 139; green, 6; blue, 24 }  ,fill opacity=1 ] (563.22,102.53) .. controls (563.22,101.74) and (563.86,101.11) .. (564.64,101.11) .. controls (565.42,101.11) and (566.06,101.74) .. (566.06,102.53) .. controls (566.06,103.31) and (565.42,103.95) .. (564.64,103.95) .. controls (563.86,103.95) and (563.22,103.31) .. (563.22,102.53) -- cycle ;
\draw    (338,103) -- (373,103) ;
\draw [shift={(375,103)}, rotate = 180] [color={rgb, 255:red, 0; green, 0; blue, 0 }  ][line width=0.75]    (10.93,-3.29) .. controls (6.95,-1.4) and (3.31,-0.3) .. (0,0) .. controls (3.31,0.3) and (6.95,1.4) .. (10.93,3.29)   ;
\draw  [dash pattern={on 4.5pt off 4.5pt}]  (409.54,174) -- (520.54,174) ;
\draw  [dash pattern={on 4.5pt off 4.5pt}]  (409.55,75.86) -- (520.55,75.86) ;
\draw  [dash pattern={on 4.5pt off 4.5pt}]  (500.57,31.06) -- (611.57,31.06) ;
\draw  [dash pattern={on 4.5pt off 4.5pt}]  (500.56,129.19) -- (611.56,129.19) ;

\draw (205.86,39.93) node [anchor=north west][inner sep=0.75pt]    {$S$};
\draw (277.99,95.28) node [anchor=north west][inner sep=0.75pt]    {$a$};
\draw (53.01,94.42) node [anchor=north west][inner sep=0.75pt]    {$a$};
\draw (590.85,42.93) node [anchor=north west][inner sep=0.75pt]    {$\CB_{S}$};
\draw (417,92.42) node [anchor=north west][inner sep=0.75pt]    {$a \boxtimes \overline a$};
\draw (63,42.4) node [anchor=north west][inner sep=0.75pt]    {$\CB$};
\draw (268,41.4) node [anchor=north west][inner sep=0.75pt]    {$\CB$};
\draw (417,51.4) node [anchor=north west][inner sep=0.75pt]    {$\CB\boxtimes \overline{\CB}$};

\end{tikzpicture}
\caption{Folding the fixed point action of $S$ on $a$.}
\label{fig:folding fixed point}
\end{figure}
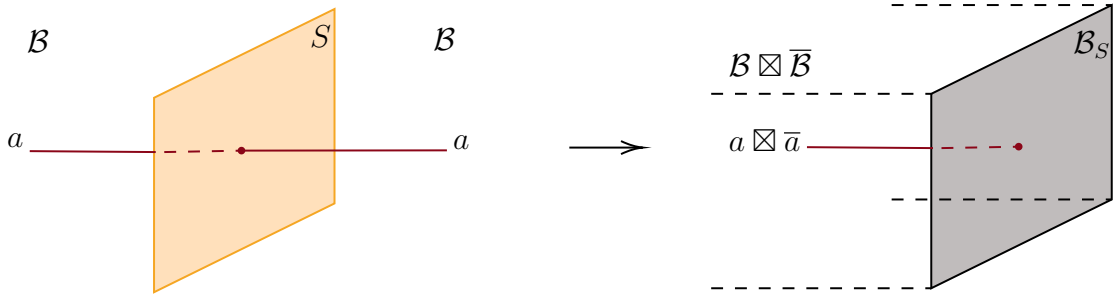

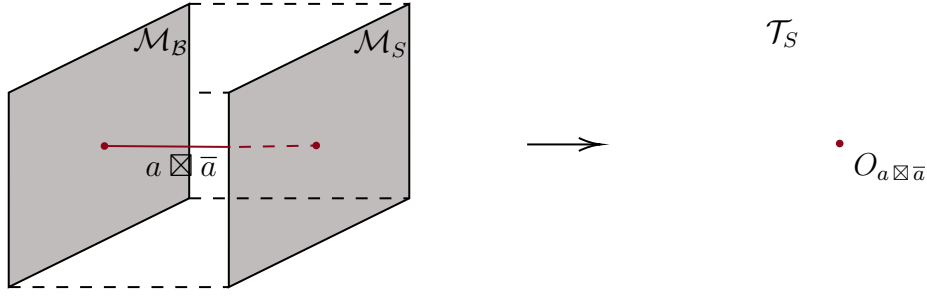
\begin{figure}[h!]
\centering

\tikzset{every picture/.style={line width=0.75pt}} 

\begin{tikzpicture}[x=0.75pt,y=0.75pt,yscale=-1,xscale=1]

\draw    (338,103) -- (373,103) ;
\draw [shift={(375,103)}, rotate = 180] [color={rgb, 255:red, 0; green, 0; blue, 0 }  ][line width=0.75]    (10.93,-3.29) .. controls (6.95,-1.4) and (3.31,-0.3) .. (0,0) .. controls (3.31,0.3) and (6.95,1.4) .. (10.93,3.29)   ;
\draw  [color={rgb, 255:red, 0; green, 0; blue, 0 }  ,draw opacity=1 ][fill={rgb, 255:red, 192; green, 188; blue, 188 }  ,fill opacity=1 ] (187.54,175) -- (187.55,76.86) -- (278.57,32.06) -- (278.56,130.19) -- cycle ;
\draw [color={rgb, 255:red, 139; green, 6; blue, 24 }  ,draw opacity=1 ] [dash pattern={on 4.5pt off 4.5pt}]  (192.7,104.26) -- (233.06,103.53) ;
\draw  [color={rgb, 255:red, 139; green, 6; blue, 24 }  ,draw opacity=1 ][fill={rgb, 255:red, 139; green, 6; blue, 24 }  ,fill opacity=1 ] (230.22,103.53) .. controls (230.22,102.74) and (230.86,102.11) .. (231.64,102.11) .. controls (232.42,102.11) and (233.06,102.74) .. (233.06,103.53) .. controls (233.06,104.31) and (232.42,104.95) .. (231.64,104.95) .. controls (230.86,104.95) and (230.22,104.31) .. (230.22,103.53) -- cycle ;
\draw  [dash pattern={on 4.5pt off 4.5pt}]  (76.54,175) -- (187.54,175) ;
\draw  [dash pattern={on 4.5pt off 4.5pt}]  (76.55,76.86) -- (187.55,76.86) ;
\draw  [dash pattern={on 4.5pt off 4.5pt}]  (167.57,32.06) -- (278.57,32.06) ;
\draw  [dash pattern={on 4.5pt off 4.5pt}]  (167.56,130.19) -- (278.56,130.19) ;
\draw  [color={rgb, 255:red, 0; green, 0; blue, 0 }  ,draw opacity=1 ][fill={rgb, 255:red, 192; green, 188; blue, 188 }  ,fill opacity=1 ] (76.54,175) -- (76.55,76.86) -- (167.57,32.06) -- (167.56,130.19) -- cycle ;
\draw [color={rgb, 255:red, 139; green, 6; blue, 24 }  ,draw opacity=1 ]   (124.85,103.79) -- (188.41,104.26) ;
\draw  [color={rgb, 255:red, 139; green, 6; blue, 24 }  ,draw opacity=1 ][fill={rgb, 255:red, 139; green, 6; blue, 24 }  ,fill opacity=1 ] (123.43,103.79) .. controls (123.43,103.01) and (124.07,102.37) .. (124.85,102.37) .. controls (125.63,102.37) and (126.26,103.01) .. (126.26,103.79) .. controls (126.26,104.58) and (125.63,105.21) .. (124.85,105.21) .. controls (124.07,105.21) and (123.43,104.58) .. (123.43,103.79) -- cycle ;
\draw  [color={rgb, 255:red, 139; green, 6; blue, 24 }  ,draw opacity=1 ][fill={rgb, 255:red, 139; green, 6; blue, 24 }  ,fill opacity=1 ] (494.22,102.53) .. controls (494.22,101.74) and (494.86,101.11) .. (495.64,101.11) .. controls (496.42,101.11) and (497.06,101.74) .. (497.06,102.53) .. controls (497.06,103.31) and (496.42,103.95) .. (495.64,103.95) .. controls (494.86,103.95) and (494.22,103.31) .. (494.22,102.53) -- cycle ;

\draw (456.85,38.93) node [anchor=north west][inner sep=0.75pt]    {$\CT_{S}$};
\draw (248.85,43.93) node [anchor=north west][inner sep=0.75pt]    {$\mc M_{S}$};
\draw (144,106.42) node [anchor=north west][inner sep=0.75pt]    {$a \boxtimes \overline a$};
\draw (137.85,42.93) node [anchor=north west][inner sep=0.75pt]    {$\mc M_{\CB}$};
\draw (501.06,105.93) node [anchor=north west][inner sep=0.75pt]    {$O_{a \, \boxtimes \, \overline a}$};

\end{tikzpicture}
\caption{If $a$ is a fixed point of $S$, then the line $a \boxtimes \overline a$ can end on both $\mc M_{\CB}$ and $\mc M_{S}$.}
\label{fig:a ending on two boundaries}
\end{figure}

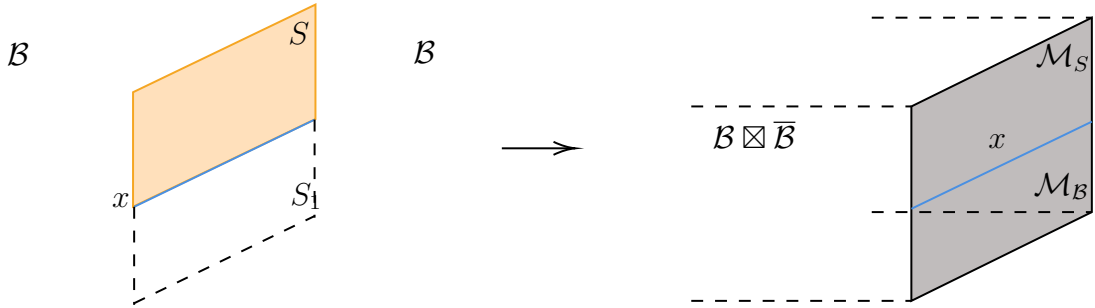
\begin{figure}[h!]
\centering

\tikzset{every picture/.style={line width=0.75pt}} 

\begin{tikzpicture}[x=0.75pt,y=0.75pt,yscale=-1,xscale=1]

\draw  [color={rgb, 255:red, 245; green, 166; blue, 35 }  ,draw opacity=1 ][fill={rgb, 255:red, 255; green, 224; blue, 187 }  ,fill opacity=1 ] (115,132.49) -- (115.01,74.69) -- (207.01,30.44) -- (207,88.24) -- cycle ;
\draw  [color={rgb, 255:red, 0; green, 0; blue, 0 }  ,draw opacity=1 ][fill={rgb, 255:red, 192; green, 188; blue, 188 }  ,fill opacity=1 ] (507.54,180) -- (507.55,81.86) -- (598.57,37.06) -- (598.56,135.19) -- cycle ;
\draw    (301,103) -- (336,103) ;
\draw [shift={(338,103)}, rotate = 180] [color={rgb, 255:red, 0; green, 0; blue, 0 }  ][line width=0.75]    (10.93,-3.29) .. controls (6.95,-1.4) and (3.31,-0.3) .. (0,0) .. controls (3.31,0.3) and (6.95,1.4) .. (10.93,3.29)   ;
\draw  [dash pattern={on 4.5pt off 4.5pt}]  (396.54,180) -- (507.54,180) ;
\draw  [dash pattern={on 4.5pt off 4.5pt}]  (396.55,81.86) -- (507.55,81.86) ;
\draw  [dash pattern={on 4.5pt off 4.5pt}]  (487.57,37.06) -- (598.57,37.06) ;
\draw  [dash pattern={on 4.5pt off 4.5pt}]  (487.56,135.19) -- (598.56,135.19) ;
\draw [color={rgb, 255:red, 74; green, 144; blue, 226 }  ,draw opacity=1 ]   (115.57,132.53) -- (206.57,88.53) ;
\draw  [dash pattern={on 4.5pt off 4.5pt}]  (115.57,132.53) -- (115.55,182) ;
\draw  [dash pattern={on 4.5pt off 4.5pt}]  (206.57,88.53) -- (206.55,138) ;
\draw  [dash pattern={on 4.5pt off 4.5pt}]  (115.57,181.19) -- (206.57,137.19) ;
\draw [color={rgb, 255:red, 74; green, 144; blue, 226 }  ,draw opacity=1 ]   (507.56,133.53) -- (598.56,89.53) ;

\draw (191.86,37.93) node [anchor=north west][inner sep=0.75pt]    {$S$};
\draw (568.85,48.93) node [anchor=north west][inner sep=0.75pt]    {$\mc M_{S}$};
\draw (50,48.4) node [anchor=north west][inner sep=0.75pt]    {$\CB$};
\draw (255,47.4) node [anchor=north west][inner sep=0.75pt]    {$\CB$};
\draw (406,87.4) node [anchor=north west][inner sep=0.75pt]    {$\CB\boxtimes \overline{\CB}$};
\draw (192.86,118.93) node [anchor=north west][inner sep=0.75pt]    {$S_1$};
\draw (103,124.4) node [anchor=north west][inner sep=0.75pt]    {$x$};
\draw (567.85,113.93) node [anchor=north west][inner sep=0.75pt]    {$\mc M_{\CB}$};
\draw (545,94.4) node [anchor=north west][inner sep=0.75pt]    {$x$};

\end{tikzpicture}
\caption{On folding, a boundary condition of the surface $S$ becomes an interface between the gapped boundaries $\mc M_{S}$ and $\mc M_{\CB}$.}
\label{fig:folding twisted sector}
\end{figure}

\begin{figure}[h!]
\centering
\tikzset{every picture/.style={line width=0.75pt}} 

\begin{tikzpicture}[x=0.75pt,y=0.75pt,yscale=-1,xscale=1]

\draw  [fill={rgb, 255:red, 74; green, 74; blue, 74 }  ,fill opacity=0.15 ] (254.4,80.83) -- (253.65,191.73) -- (169.81,143.32) -- (170.56,32.42) -- cycle ;
\draw  [fill={rgb, 255:red, 74; green, 74; blue, 74 }  ,fill opacity=0.15 ] (254.4,80.83) -- (254.21,190.56) -- (167.94,271.36) -- (168.13,161.63) -- cycle ;
\draw [color={rgb, 255:red, 74; green, 144; blue, 226 }  ,draw opacity=1 ][line width=1.5]    (254.3,192.22) -- (254.47,80.89) ;
\draw    (317,158) -- (397,158) ;
\draw [shift={(399,158)}, rotate = 180] [color={rgb, 255:red, 0; green, 0; blue, 0 }  ][line width=0.75]    (10.93,-3.29) .. controls (6.95,-1.4) and (3.31,-0.3) .. (0,0) .. controls (3.31,0.3) and (6.95,1.4) .. (10.93,3.29)   ;
\draw [color={rgb, 255:red, 74; green, 144; blue, 226 }  ,draw opacity=1 ][line width=1.5]    (525.98,209.86) -- (526,92) ;
\draw    (526,102) -- (538,96) ;
\draw    (526,148) -- (538,142) ;
\draw    (527,158) -- (539,152) ;
\draw    (526,168) -- (538,162) ;
\draw    (525.99,177.93) -- (537.99,171.93) ;
\draw    (526,189) -- (538,183) ;
\draw    (526,199) -- (538,193) ;
\draw    (525.98,207.86) -- (537.98,201.86) ;
\draw    (526,119) -- (538,113) ;
\draw    (526,138) -- (538,132) ;
\draw    (526,128) -- (538,122) ;
\draw    (526,111) -- (538,105) ;

\draw (174,161.4) node [anchor=north west][inner sep=0.75pt]    {$\mc M_{S}$};
\draw (172,48.4) node [anchor=north west][inner sep=0.75pt]    {$\mc M_{\CB}$};
\draw (260,127.4) node [anchor=north west][inner sep=0.75pt]    {$x$};
\draw (459.41,94.01) node [anchor=north west][inner sep=0.75pt]  [font=\small]  {$\CT_{S}$};
\draw (500,139.4) node [anchor=north west][inner sep=0.75pt]    {$m_{x}$};
\draw (92,64.4) node [anchor=north west][inner sep=0.75pt]    {$\CB\boxtimes \overline{\CB}$};

\end{tikzpicture}
\caption{The interface $x$ between the gapped boundaries $\mc M_{S}$ and $\mc M_{\CB}$ gives a 1d gapped boundary of the (1+1)d TQFT $\CT_{S}$ upon folding.}
\label{fig:1d gapped boundary from interface}
\end{figure}
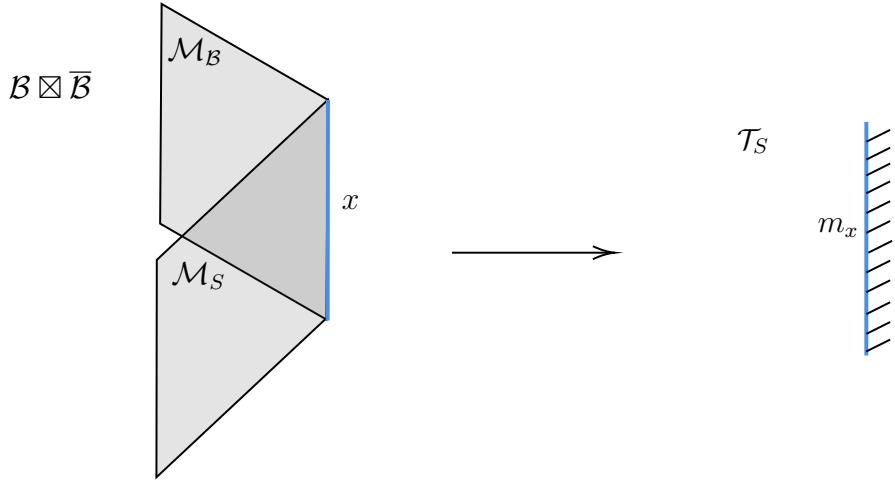

To that end, consider a genuine line $a$ for which the action of $S$ has a fixed point, $S(a) \ni a$.
In the case of an invertible surface, a fixed point means that $S(a)\cong a$, but, more generally, $S(a)$ can also contain other simple genuine lines (or even higher multiplicity of $a$). Now, consider a configuration in which the line remains unchanged as it passes through $S$. Folding this configuration, we get a gapped boundary, $\mc M_{S}$, on which $a \boxtimes \overline a \in \CB \boxtimes \overline \CB$ ends (see Fig.~\ref{fig:folding fixed point}). 
The dimension of the vector space associated with the junction between $a \boxtimes \overline a$ and the boundary is precisely $N_{S(a)}^a$. Note that the trivial surface yields the canonical boundary $\mc M_{\CB}$, of the folded theory. Clearly, all lines of the form $a \boxtimes \overline a$ can end on this gapped boundary, and $N_{S_1(a)}^a=1$ for all $a\in \CB$, where $S_1$ is the trivial surface. Next, consider the configuration in Fig.~\ref{fig:a ending on two boundaries}.
Upon interval compactification, we get a (1+1)d TQFT, $\CT_{S}$. Every line $a \boxtimes \overline a$ with $N_{S(a)}^a\neq 0$ becomes a local operator of $\CT_S$.\footnote{Conversely, all local operators can be extended to lines satisfying this condition upon ``expanding'' the 2d theory into a 3d TQFT.} Therefore, the dimension of the vector space of local operators in $\CT_{S}$ is precisely $X_S$. Now, consider a twisted sector line bounding a surface in $\mc A_S$. In the folded theory, $x$ becomes an interface between the two gapped boundaries $\mc M_{\CB}$ and $\mc M_{S}$ as in Fig.~\ref{fig:folding twisted sector}. 
We can fold along the interface $x$ and compactify to get a (1+1)d TQFT with each line $x\in \mc A_S$ specifying a simple gapped boundary condition of $\CT_{S}$ (see Fig.~\ref{fig:1d gapped boundary from interface}). Therefore, the number of simple gapped boundaries of $\CT_{S}$ is the rank of the category of lines bounding $S$. 
Since in a (1+1)d TQFT the dimension of the vector space of local operators is equal to the number of simple boundary conditions \cite{Moore:2006dw,Huang:2021zvu}, we obtain that this rank equals $X_S$.
In other words, we have arrived at the following result:

\bigskip
\noindent
{\bf Fixed point theorem:} {\it The number of twisted sector lines of a condensation surface in a $(2+1)d$ TQFT is equal to the number of fixed points of its action on simple lines.}

\bigskip
\noindent
Applying this theorem to the algebra of surfaces $\CA_S$ and $\widehat\CA_S$ and using \eqref{eq:equal fixed points of A_S and hat A_S}, we find that the fusion 1-categories of lines bounding surfaces in $\mc A_S$ and $\widehat{\mc A}_S$, respectively, have the same rank.
Therefore, as promised, we have the following corollary:

\bigskip
\noindent
{\bf Constraint on boundary conditions for dual symmetries:} {\it The number of twisted sector lines associated with dual surface algebras $\CA_S$ and $\widehat\CA_S$ coincide.}

\bigskip
\noindent
In Sec.~\ref{sec:Morita2}, we will see how this important consistency condition that dual symmetries must satisfy readily follows from the Morita theory of fusion 2-category.

\subsection{Gauging: simple lines, generalized symmetry fractionalization, and generalized discrete torsion}\label{sec:simple lines after gauging}
In Sec.~\ref{OPEdetails}, we learned that the interface, $\CI$, between the fusion 2-categories $\Mod(\CB_1)$ and $\Mod(\CB_2)$ can be created by starting from $\Mod(\CB_1)$ and summing over an algebra of surfaces, $\CA_S$, in half-space; or, alternatively, by starting with $\Mod(\CB_2)$ and summing over a dual algebra of surfaces, $\widehat\CA_S$, in half-space. For invertible symmetries, the simple lines of the gauged theory can be systematically computed using the action of the invertible surface on the lines in the ungauged theory. In this section, we will describe a generalization for gauging non-invertible symmetries.

Upon gauging $\CA_S$, the lines bounding the surfaces in $\DAS$, both genuine and non-genuine, become genuine lines in the gauged theory, $\Mod(\CB_2)$. Therefore, as in the case of gauging invertible 0-form symmetry, they must form a consistent fusion 1-category, which we also denote by $\mc A_S$. Recall that, by definition, specifying such a fusion 1-category requires specifying a monoidal structure as well as a solution to the corresponding pentagon axioms (see App.~\ref{sec:Morita1}). In other words, we have to make the following two choices in order to gauge $\CA_S$:
\begin{itemize}
\item \textbf{Generalized symmetry fractionalization:} A choice of fusion ring for the lines bounding surfaces in $\DAS$.\footnote{If rank finiteness holds for general (unitary) fusion 1-categories, then there are a finite number of choices of generalized symmetry fractionalization. For example, at rank 2, there are two choices: the $\mathbb{Z}_2$ fusion ring or the Fibonacci fusion ring \cite{ostrik2002fusion}. See also the interesting recent discussions on fractionalization for generalized symmetries in \cite{Hsin:2024aqb}. \label{rank2ring}} 
\item \textbf{Generalized discrete torsion:} A solution to the pentagon axiom associated with the choice of fusion ring. By Ocneanu rigidity, there are finitely-many such choices \cite[Section 9.1]{etingof2016tensor}. This freedom generalizes the notion of stacking with a $G$-SPT in the case of invertible 0-form symmetry.\footnote{For example, at rank 2 we can list all the choices of generalized discrete torsion given the discussion in footnote \ref{rank2ring}. In the case of the $\mathbb{Z}_2$ fusion ring, the two choices are the category with trivial and non-trivial associator. In the case of Fibonacci symmetry fractionalization, there is a unique generalized discrete torsion \cite{ostrik2002fusion}. We will come across all three of these possibilities in our examples below. \label{rank2}}\footnote{The authors in \cite{Hsin:2024aqb,Hsin:2025ria} define symmetry fractionalization and anomalies of coset non-invertible symmetries. It will be interesting to understand the relation to generalized symmetry fractionalization and generalized discrete torsion introduced in this paper.}
\end{itemize}
Naturally, symmetry fractionalization and generalized discrete torsion cannot be chosen arbitrarily. Indeed, the fusion 1-category associated with $\DAS$ must be Morita equivalent to the fusion 1-category, $\CC_{\CI}$, of lines on the interface. Upon folding, we find that the Drinfeld center of $\CC_{\CI}$ is (see also the mathematical discussion in Sec.~\ref{sec:Morita2})
\be
\CZ(\CC_{\CI})\simeq \mc Z(\mc A_S) \simeq  \CB_1 \boxtimes \overline \CB_2\, .
\ee
Therefore, the fusion 1-category underlying $\mc A_S$ must be chosen such that there is a braided tensor functor from $\CB_1$ to $\CZ(\DAS)$. If this condition is satisfied, then we can consistently gauge the symmetry related to $\CA_S$ and obtain $\Mod(\CB_2)$ \cite{DECOPPET2023108967,Decoppet:2023uoy}. In the examples that we consider in this paper, the different choices of braided tensor functors correspond to different interfaces between the same two theories, $\Mod(\CB_1)$ and $\Mod(\CB_2)$. 

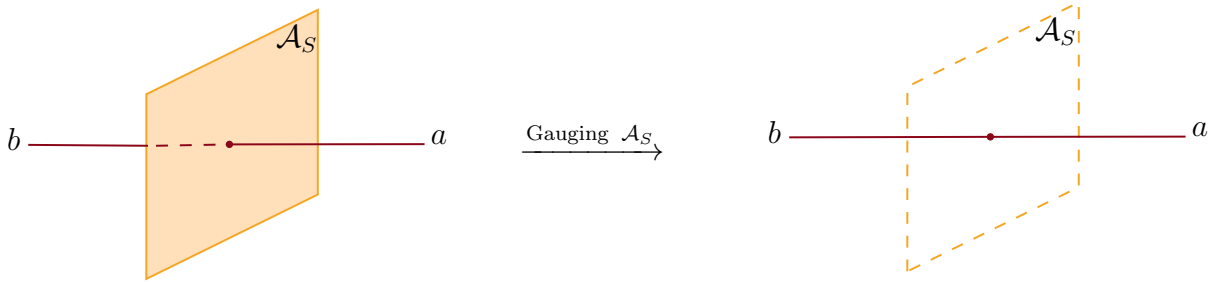
\begin{figure}[h!]

\tikzset{every picture/.style={line width=0.75pt}} 

\begin{tikzpicture}[x=0.75pt,y=0.75pt,yscale=-0.95,xscale=0.95]

\draw  [color={rgb, 255:red, 245; green, 166; blue, 35 }  ,draw opacity=1 ][fill={rgb, 255:red, 255; green, 224; blue, 187 }  ,fill opacity=1 ] (93.55,173) -- (93.57,74.86) -- (184.58,30.06) -- (184.57,128.19) -- cycle ;
\draw [color={rgb, 255:red, 139; green, 6; blue, 24 }  ,draw opacity=1 ]   (139.07,101.53) -- (241.26,101.4) ;
\draw [color={rgb, 255:red, 139; green, 6; blue, 24 }  ,draw opacity=1 ] [dash pattern={on 4.5pt off 4.5pt}]  (98.71,102.26) -- (139.07,101.53) ;
\draw [color={rgb, 255:red, 139; green, 6; blue, 24 }  ,draw opacity=1 ]   (30.86,101.79) -- (94.42,102.26) ;
\draw  [color={rgb, 255:red, 139; green, 6; blue, 24 }  ,draw opacity=1 ][fill={rgb, 255:red, 139; green, 6; blue, 24 }  ,fill opacity=1 ] (136.23,101.53) .. controls (136.23,100.74) and (136.87,100.11) .. (137.65,100.11) .. controls (138.43,100.11) and (139.07,100.74) .. (139.07,101.53) .. controls (139.07,102.31) and (138.43,102.95) .. (137.65,102.95) .. controls (136.87,102.95) and (136.23,102.31) .. (136.23,101.53) -- cycle ;
\draw  [color={rgb, 255:red, 245; green, 166; blue, 35 }  ,draw opacity=1 ][fill={rgb, 255:red, 255; green, 224; blue, 187 }  ,fill opacity=0 ][dash pattern={on 4.5pt off 4.5pt}] (497.55,169) -- (497.57,70.86) -- (588.58,26.06) -- (588.57,124.19) -- cycle ;
\draw [color={rgb, 255:red, 139; green, 6; blue, 24 }  ,draw opacity=1 ]   (543.07,97.53) -- (645.26,97.4) ;
\draw [color={rgb, 255:red, 139; green, 6; blue, 24 }  ,draw opacity=1 ]   (434.86,97.79) -- (540.23,97.53) ;
\draw  [color={rgb, 255:red, 139; green, 6; blue, 24 }  ,draw opacity=1 ][fill={rgb, 255:red, 139; green, 6; blue, 24 }  ,fill opacity=1 ] (540.23,97.53) .. controls (540.23,96.74) and (540.87,96.11) .. (541.65,96.11) .. controls (542.43,96.11) and (543.07,96.74) .. (543.07,97.53) .. controls (543.07,98.31) and (542.43,98.95) .. (541.65,98.95) .. controls (540.87,98.95) and (540.23,98.31) .. (540.23,97.53) -- cycle ;

\draw (160.86,36.93) node [anchor=north west][inner sep=0.75pt]    {$\CA_S$};
\draw (242.99,92.28) node [anchor=north west][inner sep=0.75pt]    {$a$};
\draw (18.01,91.42) node [anchor=north west][inner sep=0.75pt]    {$b$};
\draw (563.86,32.93) node [anchor=north west][inner sep=0.75pt]    {$\CA_S$};
\draw (646.99,88.28) node [anchor=north west][inner sep=0.75pt]    {$a$};
\draw (422.01,87.42) node [anchor=north west][inner sep=0.75pt]    {$b$};
\draw (290,88.4) node [anchor=north west][inner sep=0.75pt]    {$\xrightarrow{{\rm Gauging}\ \CA_S}$};

\end{tikzpicture}
\caption{Gauging surfaces introduces new junctions between simple lines.}
\label{fig:gauging S gives point operators}
\end{figure}

Upon gauging $\CA_S$, all lines bounding surfaces in $\DAS$ become genuine lines in $\CB_2$. The spectrum of lines get put into orbits under the action of $\CA_S$. Moreover, lines that are simple in $\mc B_1$ can become non-simple lines in $\CB_2$. To understand this point, consider the action of $\CA_S$ on the simple lines in $\Mod(\CB_1)$. If a simple line operator, $b$, is contained in the action of $\CA_S$ on $a$, then $a$ and $b$ can form a junction on $\CA_S$. Gauging $\CA_S$ involves trivializing this surface. Therefore, after gauging, $a$ and $b$ are not independent lines but instead have a point operator connecting them (see Fig \ref{fig:gauging S gives point operators}) and so
\be
\label{eq:inverse Frobenius reciprocity relation}
\Hom_{\CB_2}(b,a) \cong \Hom_{\CB_1}(b,\CA_S(a))\, .
\ee
As a special case, suppose $a \in \CB_1$ is a fixed point of the $\CA_S$ action. Then, two $a$ lines can form a junction on the surface $\CA_S$. Suppose $N_{\CA_S(a)}^a$ is the dimension of the vector space of local operators at the junction of $\CA_S$ with the two $a$ lines. After gauging $\CA_S$, the line $a$ hosts this vector space of local operators on it. Indeed, using \eqref{eq:inverse Frobenius reciprocity relation} with $a=b$, we get
\be
\Hom_{\CB_2}(a,a) \cong \Hom_{\CB_1}(a,\CA_S(a))\, .
\ee
Therefore, $a$ splits into a direct sum of simple lines in $\CB_2$. The topological spin of the resulting lines is given by $\theta_a$. This construction can be extended to the action of $\CA_S$ on the twisted sector lines of $\CA_S$. We will apply the discussion above to various explicit examples and determine the simple lines in the gauged theory. 

\subsection{Topological interfaces from 1-form symmetry gauging}\label{Int1form}
So far, we have mostly studied properties of general topological interfaces between two TQFTs. In this section, we will restrict ourselves to interfaces created by gauging 1-form symmetry in half the spacetime. A 1-form symmetry can be gauged if it is non-anomalous. However, as in the case of non-invertible 0-form symmetries in (1+1)d \cite{Choi:2023xjw}, there are two notions of anomalies for non-invertible 1-form symmetries in (2+1)d (as in \cite{Choi:2023xjw}, these notions coincide in the invertible case).

To understand the distinction, let us begin by considering a 1-form symmetry of a QFT described by a braided fusion 1-category, $\CB$, that is preserved under an RG flow. If the existence of such a $\CB$ forbids a trivially gapped phase, then we can say that $\CB$ is anomalous. This statement holds when the lines in $\CB$ braid non-trivially with each other. On the other hand, if all lines in $\CB$ braid trivially with each other, then $\CB$ is compatible with a trivially gapped phase. Therefore, it is non-anomalous. In this case, the anomaly is completely captured by the modular data of $\CB$.\footnote{Different consistent braidings on a fixed fusion category is a classification of possible anomalies of the 1-form symmetry \cite{Putrov:2025xmw}.} 

In this paper, we are interested in a more general notion of an anomaly. We will say that the symmetry associated with a line operator, $x \in \CB$, is anomalous if it cannot be gauged. A necessary and sufficient condition for it to be possible to gauge $x$ is that it admits the structure of a connected commutative separable algebra (see the discussion in App.~\ref{sec:Morita1}). Let $A_L$ be such an algebra. We say that $A_L$ is ``gaugeable'' or ``condensable''.

Note that gaugeability of $A_L$ is not in general captured by the $S$-matrix element $\msf S_{A_LA_L}$.\footnote{Here $\msf S_{A_LA_L}$ is sufficient to capture the full modular data associated with $A_L$, because $A_L$ is built from simple bosons.} Indeed, gauging $A_L$ involves decorating the dual of the triangulation of the spacetime 3-manifold with the corresponding object line and implementing its multiplication $\mu$ at every trivalent junction. For this gauging to be consistent, we only need braidings of $A_L$ lines with each other in the $\mu$ fusion channel to be trivial. This triviality can hold even if the S-matrix element $\msf S_{A_LA_L}$, which captures the braiding across all fusion channels of $A_L$ with $A_L$, is non-trivial. Indeed, suppose $\mu_{ab}^c$, where $a,b,c\in A_L$, is the multiplication of the algebra $A_L$. Then commutativity of the algebra is the constraint \cite{Cong:2017ffh}
\be
\label{eq:commutativity condition}
\mu_{ba}^c R_{ab}^c= \mu_{ab}^c\, .
\ee
A multiplication, $\mu$, which satisfies this constraint can exist even if $\msf S_{A_LA_L}\neq \frac{1}{\mc D} d_{A_L}^2$ where $\mc D:=\sqrt{\sum_{a\in \text{Irr}(\CB)} d_a^2}$ (we will give one such example in Sec.~\ref{ZFib}). This analysis shows that, in general,  $A_L$ can be gauged even if $A_L$ braids non-trivially with itself. In analogy with the 0-form case in (1+1)d \cite{Choi:2023xjw}, for invertible 1-form symmetry the two notions of anomaly discussed above coincide.\footnote{Indeed, taking equation \eqref{eq:commutativity condition} and multiplying by the same relation with $a\leftrightarrow b$, we see that the braiding satisfies $M_{ab}=R_{a\bar b}^{a+b}R_{\bar ba}^{a+b}=1$ and so $\msf S_{A_LA_L}=\frac{1}{\CD} d_{A_L}^2$.}

Let us now suppose we have a gaugeable 1-form symmetry corresponding to a commutative algebra $A_L$ in $\CB_1$. Suppose $\CB_2$ is obtained from gauging this 1-form symmetry in half the spacetime. In this case, the interface $\CI$ implements the gauging or condensation map from $\Mod(\CB_1)$ to $\Mod(\CB_2)$. It is natural to wonder if $\CA_S$ and $\widehat\CA_S$ have any special properties in this somewhat simpler scenario.

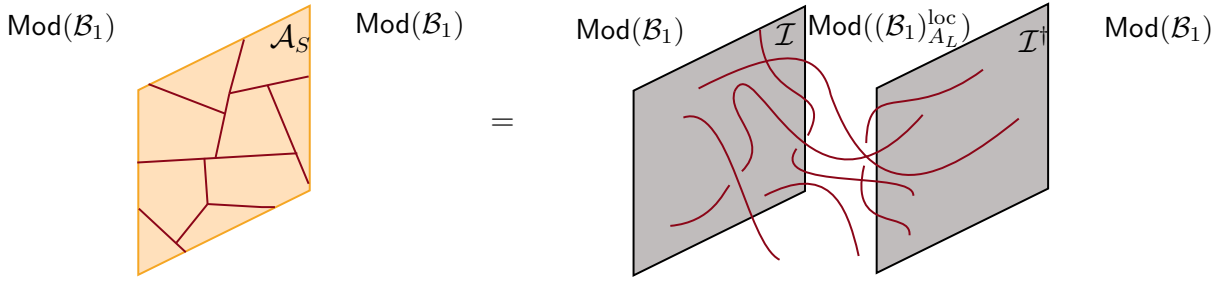
\begin{figure}
\centering

\tikzset{every picture/.style={line width=0.75pt}} 

\begin{tikzpicture}[x=0.75pt,y=0.75pt,yscale=-0.95,xscale=0.95]

\draw  [color={rgb, 255:red, 245; green, 166; blue, 35 }  ,draw opacity=1 ][fill={rgb, 255:red, 255; green, 224; blue, 187 }  ,fill opacity=1 ] (78.55,178) -- (78.57,79.86) -- (169.58,35.06) -- (169.57,133.19) -- cycle ;
\draw  [color={rgb, 255:red, 0; green, 0; blue, 0 }  ,draw opacity=1 ][fill={rgb, 255:red, 192; green, 188; blue, 188 }  ,fill opacity=1 ] (341.48,178.31) -- (341.5,80.17) -- (432.52,35.36) -- (432.5,133.5) -- cycle ;
\draw  [color={rgb, 255:red, 0; green, 0; blue, 0 }  ,draw opacity=1 ][fill={rgb, 255:red, 192; green, 188; blue, 188 }  ,fill opacity=1 ] (470.55,177) -- (470.57,78.86) -- (561.58,34.06) -- (561.57,132.19) -- cycle ;
\draw [color={rgb, 255:red, 139; green, 6; blue, 24 }  ,draw opacity=1 ]   (367.95,94.44) .. controls (392.55,86.05) and (401.12,160.92) .. (419,170) ;
\draw [color={rgb, 255:red, 139; green, 6; blue, 24 }  ,draw opacity=1 ]   (361,152) .. controls (383,150) and (389.01,134.57) .. (392.41,130.37) ;
\draw [color={rgb, 255:red, 139; green, 6; blue, 24 }  ,draw opacity=1 ]   (399,123) .. controls (415,107) and (390.72,93.74) .. (395.18,80.46) ;
\draw [color={rgb, 255:red, 139; green, 6; blue, 24 }  ,draw opacity=1 ]   (395.18,80.46) .. controls (406.42,56.01) and (435,162) .. (495,93) ;
\draw [color={rgb, 255:red, 139; green, 6; blue, 24 }  ,draw opacity=1 ]   (428,111) .. controls (412.09,133.99) and (491.01,122.03) .. (490,136) ;
\draw [color={rgb, 255:red, 139; green, 6; blue, 24 }  ,draw opacity=1 ]   (411,136) .. controls (444,119) and (454.39,137.96) .. (461,168) ;
\draw [color={rgb, 255:red, 139; green, 6; blue, 24 }  ,draw opacity=1 ]   (408.62,47.83) .. controls (410.18,89.75) and (448.65,76.05) .. (434,104) ;
\draw [color={rgb, 255:red, 139; green, 6; blue, 24 }  ,draw opacity=1 ]   (465,108) .. controls (466,72) and (487,99) .. (527,69) ;
\draw [color={rgb, 255:red, 139; green, 6; blue, 24 }  ,draw opacity=1 ]   (463.91,120.01) .. controls (455.08,157.04) and (489.01,143.03) .. (488,157) ;
\draw [color={rgb, 255:red, 139; green, 6; blue, 24 }  ,draw opacity=1 ]   (134.66,53) -- (126.93,81.59) ;
\draw [color={rgb, 255:red, 139; green, 6; blue, 24 }  ,draw opacity=1 ]   (169,72.6) -- (126.93,81.59) ;
\draw [color={rgb, 255:red, 139; green, 6; blue, 24 }  ,draw opacity=1 ]   (126.93,81.59) -- (119.64,115.48) ;
\draw [color={rgb, 255:red, 139; green, 6; blue, 24 }  ,draw opacity=1 ]   (84.01,76.69) -- (124.5,92.31) ;
\draw [color={rgb, 255:red, 139; green, 6; blue, 24 }  ,draw opacity=1 ]   (147.11,77.1) -- (168.98,129.63) ;
\draw [color={rgb, 255:red, 139; green, 6; blue, 24 }  ,draw opacity=1 ]   (115.34,140.4) -- (144.1,142.03) -- (151.1,142.03) ;
\draw [color={rgb, 255:red, 139; green, 6; blue, 24 }  ,draw opacity=1 ]   (78,118.34) -- (162.99,113.44) ;
\draw [color={rgb, 255:red, 139; green, 6; blue, 24 }  ,draw opacity=1 ]   (113.63,115.89) -- (115.34,140.4) ;
\draw [color={rgb, 255:red, 139; green, 6; blue, 24 }  ,draw opacity=1 ]   (115.34,140.4) -- (98.75,161) ;
\draw [color={rgb, 255:red, 139; green, 6; blue, 24 }  ,draw opacity=1 ]   (78.86,142.85) -- (103.75,166) ;
\draw [color={rgb, 255:red, 139; green, 6; blue, 24 }  ,draw opacity=1 ]   (446.18,82.46) .. controls (469,145) and (504,129) .. (546,95) ;
\draw [color={rgb, 255:red, 139; green, 6; blue, 24 }  ,draw opacity=1 ]   (376,79) .. controls (409,62) and (439.57,52.42) .. (446.18,82.46) ;

\draw (146.86,43.93) node [anchor=north west][inner sep=0.75pt]    {$\CA_{S}{}$};
\draw (264,93.4) node [anchor=north west][inner sep=0.75pt]    {$=$};
\draw (415.5,42.57) node [anchor=north west][inner sep=0.75pt]    {$\CI$};
\draw (544.57,44.26) node [anchor=north west][inner sep=0.75pt]    {$\CI^{\dagger}$};
\draw (7,37.4) node [anchor=north west][inner sep=0.75pt]    {\small $\Mod(\CB_{1})$};
\draw (192,36.4) node [anchor=north west][inner sep=0.75pt]    {\small $\Mod(\CB_{1})$};
\draw (309,39.4) node [anchor=north west][inner sep=0.75pt]    {\small $\Mod(\CB_{1})$};
\draw (589.58,38.46) node [anchor=north west][inner sep=0.75pt]    {\small $\Mod(\CB_{1})$};
\draw (432,35.4) node [anchor=north west][inner sep=0.75pt]    {\small $\Mod((\CB_{1})^{\rm loc}_{A_L})$};

\end{tikzpicture}
\caption{When $\CB_2 \simeq (\CB_1)_{A_L}^{\rm loc}$ is obtained from gauging a line $A_L\in\CB_1$, then shrinking a slab of $\CB_2$ theory within $\CB_1$ gives a surface created from a network of $A_L$ lines. This construction shows that $\CA_S \simeq S_{A_L}$ is the condensation surface created by higher-gauging $A_L$. In fact, we can think of this figure as a special example of Fig.~\ref{fig:sandwich} where $J$ is absent (or, if one prefers, completely trivial), and $B_L \cong \widetilde B_L \cong A_L$.}
\label{fig:condensation surface A_S from slab theory}
\end{figure}

To answer this question, consider a slab of $\Mod(\CB_2)$ inside $\Mod(\CB_1)$. In other words, $\Mod(\CB_2)$ is separated from $\Mod(\CB_1)$ by the interfaces $\CI$ and $\CI^{\dagger}$. Effectively, we have performed the 1-form symmetry gauging within the slab, and $\CB_2$ is created from a network of $A_L$ lines in $\CB_1$. Upon shrinking the slab, we get a surface in $\Mod(\CB_1)$ that is created from a network of $A_L$ lines (see Fig.~\ref{fig:condensation surface A_S from slab theory}).
Therefore, we have (see also the discussion in \cite{Buican:2023bzl}) that $\mc A_S \simeq \mc I^\dagger \otimes \mc I \simeq S_{A_L}$
is the condensation surface obtained from higher-gauging $A_L$. This logic supplements our discussion in the introduction and shows that gauging the 1-form symmetry corresponding to the algebra $A_L$ can be equivalently understood as the gauging a 0-form symmetry obtained by endowing the indecomposable surface $S_{A_L}$ with an algebra structure $\mc A_S$. This algebra structure is specified by choosing the underlying fusion 1-category of twisted sector lines to be the category $(\mc B_1)_{A_L}$ of modules over $A_L$ in $\CB_1$ (see App.~\ref{sec:Morita1}), where the fusion structure descends from the commutativity of $A_L$. Moreover, as is well known, lines on $\CI$ are also labelled by elements of this fusion 1-category. In other words, as fusion 1-categories, we have the following equivalences $\DAS\simeq \CC_{\CI}\simeq (\CB_1)_{A_L}$. 

A few comments are in order:
\begin{itemize}
\item As a surface, $\CA_S$ is indecomposable. In Sec.~\ref{sec:Morita2} and Sec. \ref{GenConstraints}, we will argue this statement holds if and only if we are gauging a 1-form symmetry.
\item The generalized symmetry fractionalization and generalized discrete torsion associated with $\CA_S$, whose underlying surface is $S_{A_L}$, is fixed by $A_L$ (this statement is natural, because fixing $A_L$ fixes the 1-form symmetry we gauge). Essentially, this statement holds because of the identification $\DAS\simeq\CC_{\CI}\simeq(\CB_1)_{A_L}$ (or, more loosely, because $\CI$ is fixed once we choose an $A_L\in\CB_1$ to sum over).\footnote{A more elaborate argument is as follows. Consider a fusion 1-category structure  for $\DAS$. In particular, specify all trivalent junctions of twisted sector lines with three $\CA_S$ surfaces attached. Now, for each such junction, consider a configuration where these three surfaces also end on the interface $\CI$. Pushing the trivalent junction of twisted sector lines onto the interface gives us a trivalent junction of lines on the interface. We know that $\DAS\simeq \CC_{\CI}\simeq (\CB_1)_{A_L}$. Moreover, the fusion of lines on the interface is fixed by the tensor product of $A_L$-modules which is, in turn, fixed by $A_L$. Therefore, in this case $\DAS$ is completely fixed by $A_L$.} This fact is important for the general consistency of our construction. 

\item By considering Fig.~\ref{fig:condensation surface A_S from slab theory}, it is clear that the action of $\CA_S$ on the identity line is given by $(\mc I \otimes \mc I^\dagger)(1)$, which is isomorphic to $A_L$. In other words
\begin{equation}\label{SlabOut}
\CA_S (1) \equiv S_{A_L}(1) \cong A_L\, .
\end{equation}
In particular, the output does not contain any twisted sector lines.
\end{itemize}
From general considerations, we know there is a dual symmetry corresponding to $\widehat\CA_S$ in $\Mod(\CB_2)$ that can be gauged to obtain $\Mod(\CB_1)$. As described in the introduction to Sec.~\ref{OPE}, given that this symmetry is dual to a 1-form symmetry, it is a (non-)invertible 0-form symmetry. The surface algebra, $\widehat\CA_S$ satisfies $\widehat \CA_S \cong  \CI \otimes \CI^{\dagger}$ and its action on the lines of $\Mod(\CB_2)$ is given by the matrix $WW^{\dagger}$ described below \eqref{Wdefn}. Moreover, since the interface maps the non-genuine lines bounding $\CA_S$ to those of $\widehat\CA_S$, we can determine the full set of boundary conditions of $\widehat \CA_S$ from the category $\DAS$.

We saw above that $\CA_S$ has special properties in the case at hand (e.g., it is an indecomposable surface). Does $\widehat\CA_S$ have any corresponding special properties? Interpreting results of \cite{Huston:2022utd} in our language, an $\widehat\CA_S$ corresponding to a (non-)invertible 0-form symmetry must have the trivial surface as an indecomposable summand.\footnote{In the case of an invertible $G$ 0-form symmetry (the dual of ${\Rep}(G)$ 1-form symmetry), this statement holds because there is a trivial surface corresponding to the trivial element $1\in G$.} In other words, $S_1\in\widehat\CA_S$. In Sec.~\ref{GenConstraints}, we will find conditions under which $S_1\in\widehat\CA_S$ is sufficient to conclude that the symmetry associated with $\widehat\CA_S$ is a 0-form symmetry.

\subsec{Algebras of surfaces from gauging 1-form symmetry with non-trivial braiding}\label{IntExt}
As discussed in the previous section, in order to gauge a 1-form symmetry it must be associated with a commutative algebra, $A_L$. In particular, this requirement does not imply that $A_L$ has trivial braiding (i.e., we can have $\msf S_{A_L,A_L}\ne\CD^{-1}d_{A_L}^2$). 

In this section, we will explore the consequences on the dual 0-form symmetry of gauging a 1-form symmetry with non-trivial braiding. More precisely, we consider the setup of Fig.~\ref{fig:condensation surface A_S from slab theory} with $A_L$ braiding non-trivially with itself. We begin by noting that, to preserve the braiding of lines across the interface $\CI$, $\CI(A_L)$ must contain non-trivial lines. In Sec.~\ref{InvertRev}, we showed that if $\CI(A_L)$ contains non-trivial lines, then $\widehat\CA_S$ must contain simple non-invertible surfaces. To better understand the corresponding physics, let us distinguish between the following two cases:\footnote{An example of the first type arises in the gauging of the $\mathbb{Z}_2$ symmetry generated by $\psi \boxtimes \overline \psi \in \msf{Ising} \boxtimes \overline{\msf{Ising}}$ to get $\TC$ or the gauging of the $\mathbb{Z}_2$ symmetry generated by $e\in\TC$ to produce the trivial TQFT.  An example of the second type arises from summing over the algebra object $A_L=1\boxtimes \overline 1 \oplus \tau\boxtimes\overline \tau \in \msf{Fib} \boxtimes \overline{\msf{Fib}}$ to get the trivial TQFT (we will describe this example in detail in Sec.~\ref{ZFib}).}
\begin{enumerate}
\item All lines in $\CI(A_L)$ are genuine.
\item There is a non-genuine line in $\CI(A_L)$. 
\end{enumerate}
Let us first establish that case 2 occurs if $A_L$ has non-trivial braiding. To that end, consider gauging the 1-form symmetry associated with $A_L\in\CB_1$ to get a new MTC, $\CB_2$. Let us study the braiding of two lines $a,b\in A_L$ 
\be
\msf S_{ab}=\frac{1}{\mc D} \sum_c N_{a\bar b}^c \frac{\theta_c}{\theta_a \theta_b} d_c= \frac{1}{\mc D} \sum_{c} N_{ab}^c \theta_c d_c\, .
\ee
Suppose $a$ and $b$ braid trivially so that 
\be\label{abBraid}
\msf S_{ab}= \frac{d_ad_b}{\mc D}\, .
\ee
In a unitary MTC, the quantum dimensions are all positive. Therefore, for \eqref{abBraid} to hold, we must have $\theta_c=1$ for all $N_{ab}^c\neq 0$. Conversely, if $a,b\in A_L$ braid non-trivially with each other, $a\otimes b$ must contain non-bosons.

With this groundwork, it is easy to show that if $A_L$ contains non-trivially braiding bosons, its image under condensation includes twisted sector lines in $\CB_2$. Indeed, we can arrive at this conclusion by first noting that under condensation
\be
\Hom_{\CB_2}(a,b)\cong \Hom_{\CB_1}(a,A_L\otimes b)\, .
\ee
Now, suppose $b_1,b_2\in A_L$ braid non-trivially. Then, we have a non-boson, say $c \in b_1 \otimes b_2$, and
\be
\Hom_{\CB_2}(c,b_2)\cong \Hom_{\CB_1}(c,A_L\otimes b_2)\, .
\ee
The product $A_L \otimes b_2$ contains the fusion $b_1 \otimes b_2$ which in turn contains the line $c$. Therefore, $\Hom_{\CB_2}(c,b_2)$ is non-trivial. This logic shows that there is a genuine or twisted sector line $x\in\CB_2$ such that 
\be
\CI(b_2) \cong 1 \oplus x  \oplus \cdots \q \text{and} \q \CI(c) \cong x \oplus \cdots\, . 
\ee
As a result, the simple line $x \in \CB_2$ lifts to a direct sum of lines of the form $c \oplus b_2 \oplus \cdots\in\CB_1$. If $x$ is a genuine line in $\CB_2$, it must have a well-defined spin given by the spin of the simple lines $b_2$ and $c$. Mathematically, this fact follows from the result that a simple $A_L$-module, $m$, as an object in $A_L$ is local if and only if $\theta_m = \xi$, where $\xi$ is some root of unity \cite[Corollary 3.18]{Frohlich:2003hm} (see also \cite{Kong:2013aya}). However, $b_2$ is a boson while $c$ is non-boson. This discussion shows that $x$ must be a twisted sector line operator. Therefore, we have
\be
\CI(b_2) \cong 1  \oplus x \oplus \cdots\, .
\ee
To summarize, we see that if $A_L$ contains bosons that braid non-trivially, their images under 1-form symmetry gauging contain twisted sector lines.

What can we say about the dual 0-form symmetry surface algebra, $\widehat\CA_S$? From the above discussion, we know it should act on the identity as
\be
\widehat\CA_S (1) \cong x \oplus \cdots\, .
\ee
We therefore see that $\widehat\CA_S$ maps the trivial line to a twisted sector line operator. As a result, $\widehat\CA_S$ cannot be made up purely of operators, $S_{B_L}$, built as in Fig.~\ref{fig:condensation surface A_S from slab theory}, from gauging a 1-form symmetry in a slab and compressing it (see the discussion around \eqref{SlabOut}).

\subsubsection{Example: $\msf{Fib} \boxtimes \overline{\msf{Fib}}$}\label{ZFib}
In this section, we work out an explicit example of 1-form symmetry gauging of the type discussed in Sec.~\ref{IntExt}. Along the way, we illustrate the distinction we described in Sec.~\ref{Int1form} between gaugeability of a 1-form symmetry and non-trivial braiding.

To that end, consider $(G_2)_1$ Chern-Simons theory and its associated Fibonacci (Fib) MTC. The simple lines $1, \tau$ have topological spins $1$ and $e^{\frac{4  \pi i}{5}}$ respectively. Now consider the TQFT with lines forming the category $\CZ(\msf{Fib})\cong \msf{Fib}\boxtimes \overline{\msf{Fib}}$
\be
1\boxtimes\overline1\, , \q 1\boxtimes\overline\tau\, , \q \tau\boxtimes\overline 1\, , \q \tau\boxtimes\overline\tau\, .
\ee
Consider gauging the 1-form symmetry corresponding to the following commutative algebra
\be
A_L=1\boxtimes\overline1 \oplus \tau\boxtimes\overline\tau\, .
\ee
Note that for this algebra object the commutativity condition \eqref{eq:commutativity condition} reduces to
\be
\mu_{\tau\boxtimes\overline\tau,\tau\boxtimes\overline\tau}^{c} R_{\tau\boxtimes\overline\tau,\tau\boxtimes\overline\tau}^{c} =\mu_{\tau\boxtimes\overline\tau,\tau\boxtimes\overline\tau}^c~ \implies R_{\tau\boxtimes\overline\tau,\tau\boxtimes\overline\tau}^c=1\, , \q   \forall 
 \; c\in A_L\, .
\ee
From the definition of the Deligne product, $\msf{Fib} \boxtimes \overline{\msf{Fib}}$, we find that the condition $R_{\tau\boxtimes\overline\tau,\tau\boxtimes\overline\tau}^c=|R_{\tau,\tau}^c|^2=1$ is indeed satisfied for $c\in A_L$. Therefore, $A_L$ is commutative for any choice of multiplication. However, 
\begin{equation}
\msf S_{A_L,A_L} = \msf S_{1\boxtimes\overline1,1\boxtimes\overline1} +  \msf S_{1\boxtimes\overline1,\tau\boxtimes\overline\tau} + \msf S_{\tau\boxtimes\overline\tau,1\boxtimes\overline1} + \msf S_{\tau\boxtimes\overline\tau,\tau\boxtimes\overline\tau}= \frac{1}{\mc D} (1+2d_{\tau}^2 \oplus 1) \neq \frac{1}{\mc D} \left(1+d_{\tau}^2\right)^2\, .
\end{equation}
This is an example of a 1-form symmetry that can be gauged, even though its S-matrix is non-trivial. Note that the commutativity of the algebra only requires $R_{\tau\boxtimes\overline\tau,\tau\boxtimes\overline\tau}^{1\boxtimes\overline1}=R_{\tau\boxtimes\overline\tau,\tau\boxtimes\overline\tau}^{\tau\boxtimes\overline\tau}=1$, but the $R$-matrix elements $R_{\tau\boxtimes\overline\tau,\tau\boxtimes\overline\tau}^{1\boxtimes\overline\tau}$ and $R_{\tau\boxtimes\overline\tau,\tau\boxtimes\overline\tau}^{\tau\boxtimes\overline1}$ that also contribute to the S-matrix element $\msf S_{\tau\boxtimes\overline\tau,\tau\boxtimes\overline\tau}$, are non-trivial. Indeed, the algebra only captures the $1\boxtimes\overline1,\tau\boxtimes\overline\tau\in\tau\boxtimes\overline\tau \otimes \tau\boxtimes\overline\tau$ fusion channels, while the S-matrix also receives contribution from the $1\boxtimes\overline\tau$ and $\tau\boxtimes\overline1$ fusion channels. This fact is the underlying reason why the algebra $A_L$ is commutative (and hence the corresponding symmetry is gaugeable) even though $A_L$ braids non-trivially with itself.

Upon gauging $A_L$, we get an interface, $\CI$, between $\msf{Fib} \boxtimes \overline{\msf{Fib}}$ and the trivial TQFT. The trivial line $1\boxtimes\overline 1 \in \msf{Fib} \boxtimes \overline{\msf{Fib}}$ becomes the trivial line $\trl$ in the trivial TQFT. On the other hand, $\tau \boxtimes\overline\tau$ becomes a line of the form $\trl \oplus \rho$. Note that $\rho$ must be non-trivial for the gauging to preserve quantum dimension. However, since $\trl$ is the only genuine line in the trivial TQFT, $\rho$ must be a non-genuine line operator. Since, $1\boxtimes\overline\tau$ and $\tau\boxtimes\overline1$ are related to each other by fusion with $\tau\boxtimes\overline\tau$, they are identified under condensation. Moreover, since there are no fixed points under the fusion of $A$ with $1\boxtimes\overline\tau$ and $\tau\boxtimes\overline1$, they do not split into multiple simple lines. Let us assume $\CI(1\boxtimes\overline\tau) \cong \CI(\tau\boxtimes\overline1) \cong x$ for some line $x$ in the trivial TQFT. In fact, for consistency with fusion rules, we must have $x=\rho$. Therefore, the action of the interface $\CI$ is given by 
\be
\CI(1\boxtimes \overline 1) \cong \trl\, , \q
\CI(1\boxtimes \overline \tau) \cong  \CI(\tau \boxtimes \overline 1) \cong \rho\, , \q 
\CI(\tau \boxtimes \overline \tau) \cong  \trl \oplus \rho\, .
\ee
The dual algebra of surfaces is 
\be
\widehat\CA_S \simeq \CI \otimes \CI^{\dagger} \simeq S_1\boxplus S_1\, .
\ee
In particular, the corresponding symmetry acts unfaithfully on the trivial TQFT with associated MTC $\CB_2\simeq\Vect$ and hence, as described in Sec.~\ref{OPEdetails}, $\CA_S$ consists of multiple copies of $S_1$. Without loss of generality, we can think of the second copy of $S_1$ as an ``extrinsic'' symmetry in the sense that its bounding line is $\rho$ as opposed to $\trl$ (here $\rho\otimes\rho \cong \trl \oplus \rho$, and our generalized discrete torsion is fixed by the uniqueness theorem described in footnote \ref{rank2}).\footnote{We will encounter a closely related example in Sec.~\ref{TCDZ4} that has a different generalized symmetry fractionalization and discrete torsion corresponding to $\mathbb{Z}_2$ fusion rules with and without non-trivial associator. Note that, from the perspective of microscopic lattice realizations, the question of which surfaces are extrinsic is not well defined, because different microscopic realizations may lead to different answers (e.g., see the discussion of Ex.~A in Sec.~V of \cite{Huston:2022utd}). Here we define intrinsic and extrinsic with respect to the IR TQFT.\label{MicroExt}} Moreover, when we extend the action of the second copy of $S_1$ to twisted sectors, the surface becomes non-invertible because $S_1 (\trl) \cong \trl \oplus \rho$. Therefore, the action of $\widehat\CA_S$ on the trivial line is given by 
\be
\widehat\CA_S(\trl) \cong 2\cdot \trl \oplus \rho\, ,
\ee
and $\widehat\CA_S$ is also an extrinsic surface.

\subsection{2-category symmetry, gauging, and Morita duality\label{sec:Morita2}}
We now formalize aspects of our previous discussion. In particular, we give 2-categorical definitions and background for the gauging process we described earlier. This formalism also allows us to explain some basic constraints on algebra objects corresponding to gaugeable 2-categorical symmetries, thereby setting up our explanation of further constraints in Sec.~\ref{GenConstraints}. We will also use the formalism developed in this section to analyze explicit examples in Sec.~\ref{ToricCode}.

Let us begin by reviewing the symmetries of (2+1)d TQFTs from the 2-categorical perspective. To that end, consider the Hamiltonian realisation of a (2+1)d state-sum TQFT with input data a spherical fusion 1-category, $\mc C$. One can think of the bulk lines of $\mc Z(\mc C)$ as generating an emergent, (non-)invertible, 't Hooft anomalous 1-form symmetry that is spontaneously broken in the ground state subspace. In particular, the number of degenerate ground states on the torus agrees with the number of simple objects in $\mc Z(\mc C)$ \cite{Levin:2004mi,Kirillov:2011mk}. As is well-known, and as we shall comment on further below, the mixed anomaly of the 1-form $\mc Z(\mc C)$-symmetry reflects the impossibility of simultaneously gauging all the anyonic excitations in the (2+1)d bulk (recall that Lagrangian algebras correspond to the ``maximal'' 1-form symmetry we can gauge in the bulk). More generally, one can consider a 1-form symmetry whose topological lines are encoded in any MTC, $\mc B$.

By analogy with the (1+1)d scenario reviewed in App.~\ref{sec:Morita1}, we would like to encode the 1-form symmetry associated with an MTC, $\mc B$, and more general associated symmetries in a fusion 2-category. Formally, this fusion 2-category is obtained via the so-called ``Karoubi completion'' of the delooping of $\mc B$, resulting in the 2-category, $\Mod(\mc B)$, of $\mc B$-module 1-categories, $\mc B$-module functors, and $\mc B$-natural transformations \cite{Douglas:2018qfz,Gaiotto:2019xmp}.  Physically, this construction amounts to our discussion in previous sections: we go beyond 1-form symmetry and add condensation defects that are obtained from gauging the 1-form $\mc B$-symmetry along co-dimension one submanifolds of spacetime \cite{Roumpedakis:2022aik,Lin:2022xod,Choi:2022zal,Delcamp:2023kew,Choi:2024rjm,Bartsch:2022mpm,Bartsch:2022ytj,Bhardwaj:2022yxj}. More precisely, a condensation defect associated with a simple object, $\mc M\in\Mod(\mc B)$, is constructed by inserting a network of lines labeled by an algebra $A_L\in\mc B$ along a co-dimension one submanifold such that $\mc B_{A_L} \simeq \mc M$.

As in the case of gauging symmetries in (1+1)d theories, it is also possible to construct condensation defects in (2+1)d theories directly in terms of the $\mc B$-module 1-category $\mc M$ \cite{Delcamp:2021szr,Delcamp:2023kew,Vancraeynest-DeCuiper:2025wkh}. The fusion 2-structure of $\Mod(\mc B)$ is then provided by the relative Deligne tensor product, $\boxtimes_{\mc B}$, of $\mc B$-module 1-categories over the braided fusion 1-category, $\mc B$, in such a way that the monoidal unit is given by $\mc B$ itself \cite{etingof2009fusioncategorieshomotopytheory,GREENOUGH20101818,D_coppet_2023}. One recovers the braided fusion 1-category encoding the topological lines of the 1-form symmetry as the endocategory of the monoidal unit. 

For concreteness, suppose $\mc B$ is given by the Drinfeld center of the fusion 1-category, $\Vect_H$, of $H$-graded vector spaces, where $H$ is a finite Abelian group. As a braided fusion 1-category, we have $\mc D(H) := \mc Z(\Vect_H) \simeq \Vect_{H \oplus H^\vee}^{R}$, where $H^\vee := \Hom(H,\rU(1))$ is the Pontryagin dual of $H$ and the braiding, $R$, is given by 
\begin{equation}
R_{\mathbb C_{(h_1 , \chi_1)}, \mathbb C_{(h_2 ,\chi_2)}} := \big( \chi_2(h_1) \otimes {\rm id} \big) \circ \text{swap} \, ,
\end{equation}
for every $h_1,h_2 \in H$ and $\chi_1,\chi_2 \in H^\vee$.
Let $\mc M(K,\phi)$ and $\mc M(J,\varphi)$ be two simple objects in $\Mod(\mc Z(\Vect_H))$. Then, we have the following equivalence of finite semisimple $\mc Z(\Vect_H)$-module 1-categories:
\begin{equation}
\mc M(K,\phi) \boxtimes_{\mc Z(\Vect_H)} \mc M(J,\varphi) \simeq \Mod_{\mc Z(\Vect_H)}(\mathbb C[K]^\phi \otimes \mathbb C[J]^\varphi) \, ,
\end{equation}
where, as described in App.~\ref{sec:Morita1}, $\mathbb C[K]^{\phi}$ and $C[J]^\varphi$ are twisted group algebras. The algebra structure on the tensor product, $\mathbb C[K]^\phi \otimes \mathbb C[J]^\varphi$, is provided by
\begin{align}
\nn
\mathbb C[K]^\phi \otimes \mathbb C[J]^\varphi \otimes \mathbb C[K]^\phi \otimes \mathbb C[J]^\varphi 
&\xrightarrow{{\rm id}_{\mathbb C[K]^\phi} \otimes R_{\mathbb C[K]^\phi, \mathbb C[J]^\varphi} \otimes {\rm id}_{\mathbb C[J]^\varphi}} 
\mathbb C[K]^\phi \otimes \mathbb C[K]^\phi \otimes \mathbb C[J]^\varphi \otimes \mathbb C[J]^\varphi 
\\[-.5em]
&\xrightarrow{\mu_{\mathbb C[K]^\phi} \otimes \mu_{\mathbb C[J]^\varphi}}
\mathbb C[K]^\phi \otimes \mathbb C[J]^\varphi \, .
\end{align}
More concretely, consider the 2-cochain, $\psi \in C^2(K \oplus J, \rU(1))$, defined by
\begin{equation}
\psi \big((k_1,j_1),(k_2,j_2) \big) := \phi(k_1,k_2) \, R(j_1,k_2) \, \varphi(j_1,j_2) \, ,
\end{equation}
for every $k_1,k_2 \in K$ and $j_1,j_2 \in J$. Via the homomorphism $K \oplus J \to H \oplus H^\vee, (k,j) \to kj$, the twisted group algebra, $\mathbb C[K \oplus J]^\psi$, defines an algebra in $\mc Z(\Vect_H)$ so that
\begin{equation}
\mc M(K,\phi) \boxtimes_{\mc Z(\Vect_H)} \mc M(J,\varphi) \simeq \Mod_{\mc Z(\Vect_H)}(\mathbb C[K \oplus J]^\psi) \, ,
\end{equation}
which one can then decompose into indecomposable $\mc Z(\Vect_H)$-module 1-categories by applying the results of \cite{etingof2009fusioncategorieshomotopytheory}. 

Analogously to the case of (1+1)d, (right) indecomposable module 2-categories over $\Mod(\mc B)$ encode the various ways of gauging the $\Mod(\mc B)$-symmetry. Succinctly, a right module 2-category over $\Mod(\mc B)$ consists of a finite semisimple 2-category, $\mc M$, equipped with a module structure, $(\cat,\alpha^{\cat},\pi^{\cat})$, consisting of a binary action 2-functor, $\cat : \mc M \otimes \Mod(\mc B) \to \mc M$, and an adjoint natural 2-equivalence, $\alpha^{\cat} : (- \cat -) \cat - \xrightarrow{\sim} - \cat (- \otimes -)$, satisfying  a ``pentagon axiom'' up to an invertible modification, $\pi^{\cat}$, which is itself required to satisfy an ``associahedron axiom''.

One can construct such a right module 2-category as the 2-category of left modules in $\Mod(\mc B)$ over a separable algebra in $\Mod(\mc B)$, where one defines a separable algebra in a fusion 2-category in the sense of \cite{DECOPPET2023108967}. In previous sections, we have referred to this separable algebra as ``an algebra of surfaces'' to emphasize that it is distinct from an algebra of objects (lines) in $\CB$. Succinctly, a separable algebra, $\CA_S$, in a fusion 2-category consists of an object together with a unit 1-morphism, $\eta_{\mc A_S} : 1 \to \mc A_S$, and a multiplication 1-morphism, $\mu_{\mc A_S}: \mc A_S \otimes \mc A_S \to \mc A_S$, satisfying coherence relations up to 2-isomorphisms that must themselves satisfy coherence axioms. Separability is the statement that the multiplication morphism, $\mu_{\mc A_S}$, admits a (right) adjoint, $\mu_{\mc A_S}^*$, as a morphism of $(\mc A_S,\mc A_S)$-bimodules as well as a 2-morphism of the form $1_{\mc A_S} \Rightarrow \mu_{\mc A_S} \circ \mu_{\mc A_S^*}$, which is a 2-morphism of $(\mc A_S,\mc A_S)$-bimodules playing the role of a section for the counit associated with $\mu_{\mc A_S}^*$.

Any fusion 1-category, $\mc A_S$, equipped with a braided tensor functor, $\lambda: \mc B \to \mc Z(\mc A_S)$, defines such a separable algebra\footnote{One can think of this statement as a categorification of the notion that an algebra with a module structure over a commutative algebra, $B$, is specified by an algebra, $A$, and a homomorphism $B \to Z(A)$.} in such a way that the left $\mc B$-module structure on $\mc A_S$ is provided by the composition $\msf{Forg} \circ \lambda: \mc B \to \mc Z(\mc A_S) \to \mc A_S$ of tensor functors, where $\msf{Forg}:\mc Z(\mc A_S) \to \mc A_S$ is the forgetful functor \cite{Brochier_Jordan_Snyder_2021,decoppet2021finite,DECOPPET2023108967}.\footnote{Note that for convenience, we abuse notation and allow $\CA_S$ to denote both the separable algebra and the associated fusion 1-category in the sense we are describing.} Notice that since $\mc B$ is an MTC, the braided tensor functor $\mc B \to \mc Z(\mc A_S)$ is fully faithful.
Left $\mc A_S$-modules in $\Mod(\mc B)$ are found to be finite semisimple left $\mc A_S$-module 1-categories, and so $\Mod(\mc A_S)$ has the structure of a right finite semisimple module 2-category over $\Mod(\mc B)$.

Generally speaking, given a theory with a fusion 2-category symmetry, one expects the symmetry of the theory resulting from performing the gauging operation associated with a given indecomposable module 2-category to be encoded in the Morita dual of the fusion 2-category with respect to the module 2-category. This notion has already been demonstrated for large classes of examples in \cite{Bartsch:2022mpm,Bartsch:2022ytj,Bhardwaj:2022yxj,Delcamp:2023kew,Bhardwaj:2024xcx,Vancraeynest-DeCuiper:2025wkh,Eck:2025ldx,Inamura:2025cum}. Therefore, given a theory with symmetry $\Mod(\mc B)$, gauging $\mc A_S$ results in a theory whose symmetry is encoded in the Morita dual of $\Mod(\mc B)$ with respect to $\Mod(\mc A_S)$, which is given by \cite{Decoppet:2023uoy}
\begin{equation}
\label{eq:Morita_ModB}
\Mod(\mc B)^*_{\Mod(\mc A_S)} \simeq \Mod(\overline{\mc Z^{\mc B}(\mc A_S)}) \, .
\end{equation}
Here $\mc Z^{\mc B}(\mc A_S)$ is the centralizer of the image of $\mc B$ in $\CZ(\mc A_S)$ under $\lambda$. 

As discussed in the previous subsections, this gauging operation can be performed physically by inserting a network of topological surfaces labeled by the condensation defects appearing in the object $\mc A_S \in \Mod(\mc B)$ in such a way that the tensor product of the fusion 1-category $\mc A_S$ is implemented at every interface involving three surfaces, and that the associator of $\mc A_S$ encodes the junction of four such interfaces. As a special case of \eqref{eq:Morita_ModB}, for any fusion 1-category, $\mc A_S$, one has
\begin{equation}
\label{eq:trivialGauging}
\TVect^*_{\Mod(\mc A_S)} \simeq \Mod(\overline{\mc Z(\mc A_S)}) \, ,
\end{equation}
where we have defined $\TVect := \Mod(\Vect)$. This formula encodes the notion that one can obtain the Hamiltonian realization of a (2+1)d TQFT with anyonic excitations in $\mc Z(\mc A_S)$ by performing an appropriate twisted gauging of a 0-form symmetry acting trivially on a theory with no anyonic excitations (see also the discussion in \cite{Kawagoe:2024tgv}).

Let us now consider the following situation: let $A_L\in\mc B$ be a Lagrangian algebra. By definition, $\mc B_{A_L}$ has both the structure of a fusion 1-category and that  of a left module 1-category over $\mc B$. Moreover, we have $\mc B_{A_L}^\text{loc} \simeq \Vect$ and $\mc Z(\mc B_{A_L}) \simeq \mc B$. Therefore, there is a braided tensor functor, $\lambda : \mc B \to \mc Z(\mc B_{A_L})$, such that $\msf{Forg} \circ \lambda$ recovers the left $\mc B$-module structure of $\mc B_{A_L}$, thereby endowing $\mc B_{A_L}$ with the structure of a separable algebra in $\mc B$. It follows from $\mc Z^{\mc B}(\mc B_{A_L}) \simeq \Vect$ that $\Mod(\mc B)^*_{\Mod(\mc B_{A_L})} \simeq \TVect$. Physically, this is the statement that condensing the Lagrangian algebra, $A_L$, amounts to gauging the condensation defect associated with $\mc B_{A_L}$ in $\Mod(\CZ(\CB))$ (which we first discussed around \eqref{gaugeSeT3} in the special case of $\TC$).

More generally, for any condensable algebra $A_L$ in $\mc B$, the fusion 1-category $\mc B_{A_L}$ is equipped with a braided tensor functor, $\mc B \to \mc Z(\mc B_{A_L})$, so that $\mc B_{A_L}$ has the structure of a separable algebra $\CA_S$ in $\Mod(\mc B)$ \cite{DECOPPET2023108967}.  Using physical reasoning, we argued that we could associate this algebra with the indecomposable surface, $S_{A_L}\in\Mod(\mc B)$, in the discussion around Fig.~\ref{fig:condensation surface A_S from slab theory} (and also in \cite{Buican:2023bzl}) and that gauging the corresponding symmetry amounts to gauging the 1-form symmetry corresponding to $A_L$. Most generally, for any separable algebra, $\mc A_S$, as constructed above, it must contain, as  a (decomposable) condensation defect, a topological surface obtained from the higher gauging of a condensable algebra in $\mc B$. Indeed, consider the smallest fusion subcategory in $\mc A_S$ containing the image of $\mc B$ under $\msf{Forg} \circ \lambda$. By construction, it has the structure of a $\mc B$-module 1-category and it appears in the decomposition of $\mc A_S$ as a $\mc B$-module 1-category. It then follows from Lemma 3.5. of \cite{DMNO09} that this fusion subcategory is monoidally equivalent to the category of modules in $\mc B$ over a condensable algebra.

Specializing to $\Mod(\mc Z(\Vect_H)) := \Mod(\Vect_{H \oplus H^\vee}^R)$, it follows from comments above that choosing the Lagrangian algebra $\mathbb C[H]$ in $\mc Z(\Vect_H)$ (physically, this corresponds to an algebra built from the magnetic fluxes) gives rise to a 1-category, $\mc Z(\Vect_H)_{\mathbb C[H]} \simeq \Vect_{H^\vee}$, endowed with the structure of an algebra in $\Mod(\mc Z(\Vect_H))$. In particular, $\Vect_{H^\vee} \simeq \mc M(H,1)$ is a simple object in $\Mod(\mc Z(\Vect_H))$. It encodes the condensation defect obtained by gauging all the magnetic fluxes along a co-dimension one surface. We could instead repeat the same procedure with the Lagrangian algebra $\mathbb C^H \cong \mathbb C[H^\vee]$ in $\mc Z(\Vect_H)$. Then we obtain a fusion 1-category, $\mc Z(\Vect_H)_{\mathbb C[H^\vee]} \simeq \Vect_{H}$, endowed with the structure of an algebra in $\Mod(\mc Z(\Vect_H))$. The simple object $\Vect_{H} \simeq \mc M(H^\vee,1)\in\Mod(\mc Z(\Vect_H))$ encodes the condensation defect obtained by gauging all the electric excitations along a co-dimension one surface.

As another example, consider the simple object $\Vect \simeq \mc M(H \oplus H^\vee,1)\in\Mod(\mc Z(\Vect_H))$, which physically encodes the condensation defect obtained by gauging all the anyonic excitations in $\mc Z(\Vect_H)$ along a co-dimension one surface. Can we endow the object $\Vect$ with the structure of an algebra in $\Mod(\mc Z(\Vect_H))$? No, because $\mc Z(\Vect_H)$ possesses a non-trivial braiding, and one cannot construct a braided tensor functor $\mc Z(\Vect_H) \to \mc Z(\Vect)$. This statement corresponds to the fact that it is not possible to simultaneously gauge all the electric and magnetic excitations in the $\mc Z(\Vect_H)$ bulk due to the mixed 't Hooft anomaly of the 1-form symmetry.\footnote{In general, the 't Hooft anomaly of a 1-form symmetry $H$, with $H$ an Abelian group, is captured by a cohomology class in $H^4(K(H,2),\rU(1))$ of the Eilenberg--Maclane space $K(H,2)$ of $H$, which is known to be isomorphic to the group of quadratic forms on $H$. Furthermore, $H^4(K(H,2),\rU(1))$ is also isomorphic to the cohomology group $H^3_\text{ab}(H,\rU(1))$ of Abelian 3-cocycles, consisting of group 3-cocycles and functions $R : H \otimes H \to \rU(1)$ satisfying ``hexagon equations'' involving the 3-cocycle. Such an Abelian 3-cocycle can be used to specify a braided monoidal structure on $\Vect_H$.} More generally, consider a simple object of the form $\mc Z(\Vect_H)_{A_L}$ in $\Mod(\mc Z(\Vect_H))$ such that the indecomposable algebra $A_L$ is not commutative. Then, $A_L$ cannot be endowed with the structure of a condensable algebra in $\mc Z(\Vect_H)$. As a result, it is not possible to endow $\mc Z(\Vect_H)_{A_L}$ with the structure of an algebra in $\Mod(\mc Z(\Vect_H))$. Indeed, we established above that any separable algebra must include, as a condensation defect, a topological surface resulting from the higher gauging of a condensable algebra. 

Motivated by the above observations, let us now explain, using the language of this section, one class of gaugings that produce a ``larger'' symmetry structure. Let $\mc B$ be an MTC, and let $A_L\in\mc B$ be a condensable algebra.\footnote{This class of gaugings was also discussed in Sec.~\ref{Int1form} from a slightly different perspective.} By construction, we have $\mc Z(\mc B_{A_L}) \simeq \mc B \boxtimes \overline{\mc B_{A_L}^\text{loc}}$ so that we have a braided tensor functor $\mc B_{A_L}^\text{loc} \to \overline{\mc Z(\mc B_{A_L})} \simeq \mc Z(\mc B_{A_L}^\text{op})$, thereby endowing the fusion 1-category $\mc B_{A_L}^\text{op}$ with the structure of a separable algebra $\widehat\CA_S$ in $\Mod(\mc B_{A_L}^\text{loc})$. Applying the formula given in \eqref{eq:Morita_ModB}, one finds
\begin{equation}
{\Mod(\mc B_{A_L}^\text{loc})}^*_{\Mod(\widehat\CA_S)} \simeq \Mod(\mc B) \, . 
\end{equation}
In this case, it turns out that $\widehat\CA_S$ is not an indecomposable module 1-category over $\mc B_{A_L}^\text{loc}$ (we expect this statement to hold from the discussion at the end of Sec.~\ref{Int1form}, and we will give a more rigorous argument in Sec.~\ref{GenConstraints}). Physically, one interprets this computation as the statement that gauging the symmetries corresponding to the algebra of surfaces $\widehat\CA_S$ in $\Mod(\mc B_{A_L}^\text{loc})$ is the dual of condensing the algebra $A_L$ in $\mc B$ and that $\widehat\CA_S$ contains a sum over at least two indecomposable surfaces. As discussed in the introduction to Sec.~\ref{OPE}, we think of the symmetry corresponding to $\CB_{A_L}^{\rm op}$ as a (non-)invertible 0-form symmetry.

In Sec.~\ref{GenConstraints}, we will further characterize such symmetries and also discuss more general ones. Moreover, we will use aspects of the formalism we have outlined above to derive constraints on the gaugeability of symmetries corresponding to certain topological surfaces. In the language of this section, we ask when such surfaces can be given the structure of an algebra in the corresponding fusion 2-category. We also partially classify the corresponding symmetries. In Sec.~\ref{ToricCode} , we provide numerous explicit examples of the phenomena discussed here and in the other parts of Sec.~\ref{OPE} for the case of $\TC$.

\newsec{Classifying and constraining gaugeable symmetries}\label{GenConstraints}

In this section, we discuss and collect several constraints on the gaugeability of non-invertible symmetries of TQFTs through the prism of the corresponding algebras of surfaces. However, before getting to these constraints, we would like to more precisely explain what we mean by gauging a (non-)invertible 0-form symmetry as opposed to a more general symmetry.

\begin{figure}[h!]
\centering

\tikzset{every picture/.style={line width=0.75pt}} 

\begin{tikzpicture}[x=0.75pt,y=0.75pt,yscale=-0.95,xscale=0.95]

\draw  [color={rgb, 255:red, 0; green, 0; blue, 0 }  ,draw opacity=1 ][fill={rgb, 255:red, 192; green, 188; blue, 188 }  ,fill opacity=0.75 ] (265.55,173) -- (265.57,74.86) -- (366.58,30.06) -- (366.57,128.19) -- cycle ;
\draw  [color={rgb, 255:red, 245; green, 166; blue, 35 }  ,draw opacity=1 ][fill={rgb, 255:red, 255; green, 224; blue, 187 }  ,fill opacity=1 ] (77.55,176) -- (77.57,77.86) -- (168.58,33.06) -- (168.57,131.19) -- cycle ;
\draw  [color={rgb, 255:red, 245; green, 166; blue, 35 }  ,draw opacity=1 ][fill={rgb, 255:red, 255; green, 224; blue, 187 }  ,fill opacity=1 ] (411.55,171) -- (411.57,72.86) -- (502.58,28.06) -- (502.57,126.19) -- cycle ;
\draw  [color={rgb, 255:red, 0; green, 0; blue, 0 }  ,draw opacity=1 ][fill={rgb, 255:red, 192; green, 188; blue, 188 }  ,fill opacity=0.75 ] (538.55,173) -- (538.57,74.86) -- (629.58,30.06) -- (629.57,128.19) -- cycle ;

\draw (151.86,41.93) node [anchor=north west][inner sep=0.75pt]    {$\CI$};
\draw (32,34.4) node [anchor=north west][inner sep=0.75pt]    {\footnotesize$\Mod(\CB_1)$};
\draw (182,33.4) node [anchor=north west][inner sep=0.75pt]    {\footnotesize$\Mod(\CB_2)$};
\draw (232,94.4) node [anchor=north west][inner sep=0.75pt]    {$=$};
\draw (504.86,36.93) node [anchor=north west][inner sep=0.75pt]    {$$};
\draw (250,33.4) node [anchor=north west][inner sep=0.75pt]    {\footnotesize $\Mod(\CB_1)$};
\draw (631,29.9) node [anchor=north west][inner sep=0.75pt]    {\footnotesize$\Mod(\CB_2)$};
\draw (339,40.4) node [anchor=north west][inner sep=0.75pt]    {$\widetilde\CI_{A_L}$};
\draw (604,41.4) node [anchor=north west][inner sep=0.75pt]    {$\widetilde\CI^{\dagger }_{B_L}$};
\draw (370.58,29.46) node [anchor=north west][inner sep=0.75pt]    {\footnotesize $\Mod((\CB_1)_{A_L}^{\rm loc})$};
\draw (505.58,29.46) node [anchor=north west][inner sep=0.75pt]    {\footnotesize $\Mod((\CB_2)_{B_L}^{\rm loc})$};
\draw (488,36.4) node [anchor=north west][inner sep=0.75pt]    {$J$};

\end{tikzpicture}
\caption{We can decompose a gapped domain wall, $\CI$, between $\CB_1$ and $\CB_2$ in terms of surfaces $\widetilde\CI_{A_L}$ and $\widetilde\CI_{B_L}^{\dagger}$ that implement 1-form symmetry gauging corresponding to condensable algebras $A_L$ in $\CB_1$ and $B_L$ in $\CB_2$, respectively. These surfaces sandwich a third surface, $J$, with an invertible action on genuine lines.}
\label{fig:Gensandwich}
\end{figure}
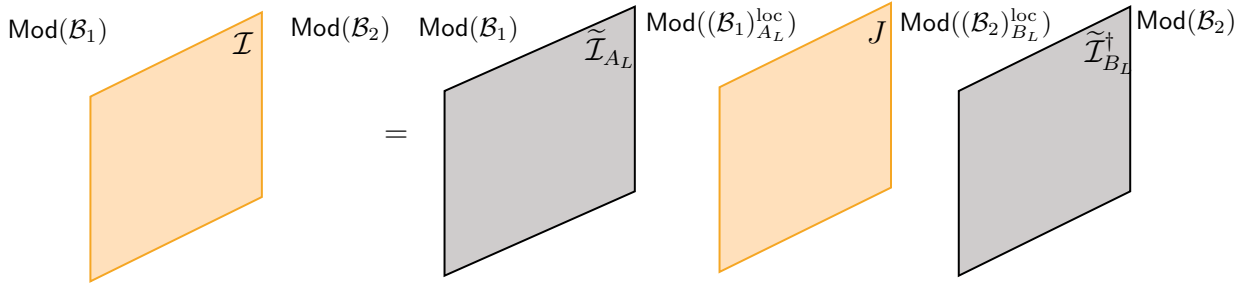
To that end, it is useful to follow \cite{davydov2013structure,Huston:2022utd} and decompose a gapped interface, $\CI$, between TQFTs associated with MTCs $\CB_1$ and $\CB_2$ into a ``sandwich'' as in Fig.~\ref{fig:Gensandwich}. There, $\widetilde\CI_{A_L}$ is an interface implementing 1-form symmetry gauging corresponding to a condensable algebra $A_L$ in $\CB_1$, $J$ is a surface with an invertible action on genuine lines, while $\widetilde\CI^{\dagger}_{B_L}$ implements 0-form symmetry gauging corresponding to the dual of 1-form symmetry gauging by the condensable algebra $B_L$ in $\CB_2$.

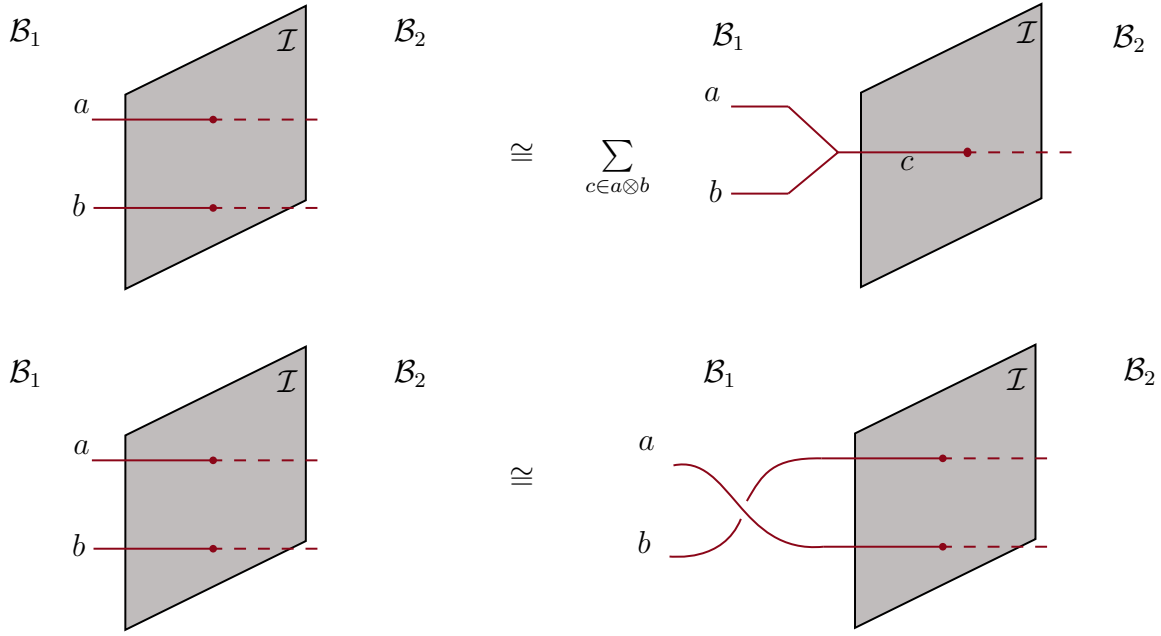
\begin{figure}[h!]
\centering

\tikzset{every picture/.style={line width=0.75pt}} 

\begin{tikzpicture}[x=0.75pt,y=0.75pt,yscale=-1,xscale=1]

\draw  [color={rgb, 255:red, 0; green, 0; blue, 0 }  ,draw opacity=1 ][fill={rgb, 255:red, 192; green, 188; blue, 188 }  ,fill opacity=1 ] (485.55,172) -- (485.57,73.86) -- (576.58,29.06) -- (576.57,127.19) -- cycle ;
\draw  [color={rgb, 255:red, 0; green, 0; blue, 0 }  ,draw opacity=1 ][fill={rgb, 255:red, 192; green, 188; blue, 188 }  ,fill opacity=1 ] (114.55,173) -- (114.57,74.86) -- (205.58,30.06) -- (205.57,128.19) -- cycle ;
\draw [color={rgb, 255:red, 139; green, 6; blue, 24 }  ,draw opacity=1 ]   (97.68,87.45) -- (157.42,87.45) ;
\draw [color={rgb, 255:red, 139; green, 6; blue, 24 }  ,draw opacity=1 ]   (98.5,132.05) -- (157.42,132.05) ;
\draw [color={rgb, 255:red, 139; green, 6; blue, 24 }  ,draw opacity=1 ]   (420.07,80.86) -- (448.9,80.86) ;
\draw [color={rgb, 255:red, 139; green, 6; blue, 24 }  ,draw opacity=1 ]   (420.07,124.82) -- (448.9,124.82) ;
\draw  [color={rgb, 255:red, 139; green, 6; blue, 24 }  ,draw opacity=1 ][fill={rgb, 255:red, 139; green, 6; blue, 24 }  ,fill opacity=1 ] (537.73,103.99) .. controls (537.73,102.94) and (538.42,102.09) .. (539.27,102.09) .. controls (540.12,102.09) and (540.8,102.94) .. (540.8,103.99) .. controls (540.8,105.05) and (540.12,105.9) .. (539.27,105.9) .. controls (538.42,105.9) and (537.73,105.05) .. (537.73,103.99) -- cycle ;
\draw [color={rgb, 255:red, 139; green, 6; blue, 24 }  ,draw opacity=1 ]   (448.9,80.86) -- (474,103.99) ;
\draw [color={rgb, 255:red, 139; green, 6; blue, 24 }  ,draw opacity=1 ]   (448.9,124.82) -- (474,103.99) ;
\draw [color={rgb, 255:red, 139; green, 6; blue, 24 }  ,draw opacity=1 ]   (473,103.99) -- (540.8,103.99) ;
\draw [color={rgb, 255:red, 139; green, 6; blue, 24 }  ,draw opacity=1 ]   (391,262) .. controls (420,256) and (424,307) .. (466.5,303.05) ;
\draw [color={rgb, 255:red, 139; green, 6; blue, 24 }  ,draw opacity=1 ]   (428,280) .. controls (436,266) and (441,258) .. (465.68,258.45) ;
\draw [color={rgb, 255:red, 139; green, 6; blue, 24 }  ,draw opacity=1 ]   (389,308) .. controls (391,308) and (416,311) .. (425,289) ;
\draw [color={rgb, 255:red, 139; green, 6; blue, 24 }  ,draw opacity=1 ] [dash pattern={on 4.5pt off 4.5pt}]  (157.42,87.45) -- (190,87.45) -- (217.15,87.45) ;
\draw [color={rgb, 255:red, 139; green, 6; blue, 24 }  ,draw opacity=1 ] [dash pattern={on 4.5pt off 4.5pt}]  (157.42,132.05) -- (190,132.05) -- (217.15,132.05) ;
\draw [color={rgb, 255:red, 139; green, 6; blue, 24 }  ,draw opacity=1 ] [dash pattern={on 4.5pt off 4.5pt}]  (537.73,103.99) -- (570.32,103.99) -- (597.47,103.99) ;
\draw  [color={rgb, 255:red, 139; green, 6; blue, 24 }  ,draw opacity=1 ][fill={rgb, 255:red, 139; green, 6; blue, 24 }  ,fill opacity=1 ] (157.42,87.45) .. controls (157.42,86.66) and (158.05,86.03) .. (158.83,86.03) .. controls (159.61,86.03) and (160.25,86.66) .. (160.25,87.45) .. controls (160.25,88.23) and (159.61,88.87) .. (158.83,88.87) .. controls (158.05,88.87) and (157.42,88.23) .. (157.42,87.45) -- cycle ;
\draw  [color={rgb, 255:red, 139; green, 6; blue, 24 }  ,draw opacity=1 ][fill={rgb, 255:red, 139; green, 6; blue, 24 }  ,fill opacity=1 ] (157.42,132.05) .. controls (157.42,131.27) and (158.05,130.63) .. (158.83,130.63) .. controls (159.61,130.63) and (160.25,131.27) .. (160.25,132.05) .. controls (160.25,132.84) and (159.61,133.48) .. (158.83,133.48) .. controls (158.05,133.48) and (157.42,132.84) .. (157.42,132.05) -- cycle ;
\draw  [color={rgb, 255:red, 0; green, 0; blue, 0 }  ,draw opacity=1 ][fill={rgb, 255:red, 192; green, 188; blue, 188 }  ,fill opacity=1 ] (114.55,345) -- (114.57,246.86) -- (205.58,202.06) -- (205.57,300.19) -- cycle ;
\draw [color={rgb, 255:red, 139; green, 6; blue, 24 }  ,draw opacity=1 ]   (97.68,259.45) -- (157.42,259.45) ;
\draw [color={rgb, 255:red, 139; green, 6; blue, 24 }  ,draw opacity=1 ]   (98.5,304.05) -- (157.42,304.05) ;
\draw [color={rgb, 255:red, 139; green, 6; blue, 24 }  ,draw opacity=1 ] [dash pattern={on 4.5pt off 4.5pt}]  (157.42,259.45) -- (190,259.45) -- (217.15,259.45) ;
\draw [color={rgb, 255:red, 139; green, 6; blue, 24 }  ,draw opacity=1 ] [dash pattern={on 4.5pt off 4.5pt}]  (157.42,304.05) -- (190,304.05) -- (217.15,304.05) ;
\draw  [color={rgb, 255:red, 139; green, 6; blue, 24 }  ,draw opacity=1 ][fill={rgb, 255:red, 139; green, 6; blue, 24 }  ,fill opacity=1 ] (157.42,259.45) .. controls (157.42,258.66) and (158.05,258.03) .. (158.83,258.03) .. controls (159.61,258.03) and (160.25,258.66) .. (160.25,259.45) .. controls (160.25,260.23) and (159.61,260.87) .. (158.83,260.87) .. controls (158.05,260.87) and (157.42,260.23) .. (157.42,259.45) -- cycle ;
\draw  [color={rgb, 255:red, 139; green, 6; blue, 24 }  ,draw opacity=1 ][fill={rgb, 255:red, 139; green, 6; blue, 24 }  ,fill opacity=1 ] (157.42,304.05) .. controls (157.42,303.27) and (158.05,302.63) .. (158.83,302.63) .. controls (159.61,302.63) and (160.25,303.27) .. (160.25,304.05) .. controls (160.25,304.84) and (159.61,305.48) .. (158.83,305.48) .. controls (158.05,305.48) and (157.42,304.84) .. (157.42,304.05) -- cycle ;
\draw  [color={rgb, 255:red, 0; green, 0; blue, 0 }  ,draw opacity=1 ][fill={rgb, 255:red, 192; green, 188; blue, 188 }  ,fill opacity=1 ] (482.55,344) -- (482.57,245.86) -- (573.58,201.06) -- (573.57,299.19) -- cycle ;
\draw [color={rgb, 255:red, 139; green, 6; blue, 24 }  ,draw opacity=1 ]   (465.68,258.45) -- (525.42,258.45) ;
\draw [color={rgb, 255:red, 139; green, 6; blue, 24 }  ,draw opacity=1 ]   (466.5,303.05) -- (525.42,303.05) ;
\draw [color={rgb, 255:red, 139; green, 6; blue, 24 }  ,draw opacity=1 ] [dash pattern={on 4.5pt off 4.5pt}]  (525.42,258.45) -- (558,258.45) -- (585.15,258.45) ;
\draw [color={rgb, 255:red, 139; green, 6; blue, 24 }  ,draw opacity=1 ] [dash pattern={on 4.5pt off 4.5pt}]  (525.42,303.05) -- (558,303.05) -- (585.15,303.05) ;
\draw  [color={rgb, 255:red, 139; green, 6; blue, 24 }  ,draw opacity=1 ][fill={rgb, 255:red, 139; green, 6; blue, 24 }  ,fill opacity=1 ] (525.42,258.45) .. controls (525.42,257.66) and (526.05,257.03) .. (526.83,257.03) .. controls (527.61,257.03) and (528.25,257.66) .. (528.25,258.45) .. controls (528.25,259.23) and (527.61,259.87) .. (526.83,259.87) .. controls (526.05,259.87) and (525.42,259.23) .. (525.42,258.45) -- cycle ;
\draw  [color={rgb, 255:red, 139; green, 6; blue, 24 }  ,draw opacity=1 ][fill={rgb, 255:red, 139; green, 6; blue, 24 }  ,fill opacity=1 ] (525.42,303.05) .. controls (525.42,302.27) and (526.05,301.63) .. (526.83,301.63) .. controls (527.61,301.63) and (528.25,302.27) .. (528.25,303.05) .. controls (528.25,303.84) and (527.61,304.48) .. (526.83,304.48) .. controls (526.05,304.48) and (525.42,303.84) .. (525.42,303.05) -- cycle ;

\draw (409,37.4) node [anchor=north west][inner sep=0.75pt]    {$\CB_{1}$};
\draw (611,38.4) node [anchor=north west][inner sep=0.75pt]    {$\CB_{2}$};
\draw (86.9,75.91) node [anchor=north west][inner sep=0.75pt]    {$a$};
\draw (86.29,123.36) node [anchor=north west][inner sep=0.75pt]    {$b$};
\draw (503.81,104.7) node [anchor=north west][inner sep=0.75pt]    {$c$};
\draw (306.84,96.4) node [anchor=north west][inner sep=0.75pt]    {$\cong ~~~ \sum\limits_{c\in a \otimes b} $};
\draw (54.44,35.4) node [anchor=north west][inner sep=0.75pt]    {$\CB_{1}$};
\draw (248.44,35.4) node [anchor=north west][inner sep=0.75pt]    {$\CB_{2}$};
\draw (563,36.4) node [anchor=north west][inner sep=0.75pt]    {$\CI$};
\draw (190,40.4) node [anchor=north west][inner sep=0.75pt]    {$\CI$};
\draw (404.9,69.91) node [anchor=north west][inner sep=0.75pt]    {$a$};
\draw (407.29,116.36) node [anchor=north west][inner sep=0.75pt]    {$b$};
\draw (306.84,261.4) node [anchor=north west][inner sep=0.75pt]    {$\cong $};
\draw (86.9,247.91) node [anchor=north west][inner sep=0.75pt]    {$a$};
\draw (86.29,295.36) node [anchor=north west][inner sep=0.75pt]    {$b$};
\draw (54.44,207.4) node [anchor=north west][inner sep=0.75pt]    {$\CB_{1}$};
\draw (248.44,207.4) node [anchor=north west][inner sep=0.75pt]    {$\CB_{2}$};
\draw (190,212.4) node [anchor=north west][inner sep=0.75pt]    {$\CI$};
\draw (371.9,245.91) node [anchor=north west][inner sep=0.75pt]    {$a$};
\draw (371.29,293.36) node [anchor=north west][inner sep=0.75pt]    {$b$};
\draw (404.44,207.4) node [anchor=north west][inner sep=0.75pt]    {$\CB_{1}$};
\draw (616.44,206.4) node [anchor=north west][inner sep=0.75pt]    {$\CB_{2}$};
\draw (558,211.4) node [anchor=north west][inner sep=0.75pt]    {$\CI$};

\end{tikzpicture}
\caption{Line operators which admit a junction with the trivial line on the interface $\CI$ form a condensable algebra that is gauged in $\CB_2$.}
\label{fig:commutative algebra from I}
\end{figure}

When we say that $\CI^{\dagger}$ implements 0-form gauging (or, equivalently, that $\CI$ implements 1-form gauging), we have in mind a situation in which $A_L\not\cong1$ and $B_L\cong1$. Indeed, for any $a\in A_L$, we see that
\begin{equation}\label{MapLRa}
    A_L\ni a\xrightarrow{\widetilde\CI_{A_L}} n_a\cdot1 \oplus x_a \xrightarrow{J} n_a\cdot1 \oplus y_a\xrightarrow{\widetilde\CI_{B_L}^{\dagger}}n\cdot B_L \oplus Y_a\, ,
\end{equation}
where $n_a,n>0$, and $\dim\Hom_{\CB_2}(Y_a,1)=0$ where $x_a$, $y_a$, and $Y_a$ are some (possibly non-genuine) lines in $\Mod((\CB_1)_{A_L}^{\rm loc})$, $\Mod((\CB_2)_{B_L}^{\rm loc})$, and $\Mod(\CB_2)$ respectively. In particular, the 1-form symmetry corresponding to $A_L$ is gauged in $\CB_2$, because $\dim \Hom_{\CB_2}(\CI(a),1)=n>0$. Conversely, any $a\not\in A_L$ is not gauged.\footnote{Here we use the same logic in Fig.~\ref{fig:A_S action from I} to argue that any intermediate twisted sector states do not affect our conclusions, because, at best, they lead to trivalent junctions on $\CI$ involving $a$, $1$, and a non-trivial $x\in\CC_{\CI}$.} Therefore, $A_L$ corresponds to the 1-form symmetry that is gauged in going from $\CB_1$ to $\CB_2$ and, by a similar argument, $B_L$ corresponds to the 1-form symmetry that is gauged in going from $\CB_2$ to $\CB_1$. Note also from the general arguments illustrated in Fig.~\ref{fig:commutative algebra from I} that operators forming a junction on $\CI$ with the trivial line form a condensable algebra that is gauged when crossing $\CI$. Therefore, without loss of generality, in order for $\CI^{\dagger}$ to implement 0-form gauging (and $\CI$ to implement 1-form gauging) we take $B_L \cong 1$ and $A_L\ncong 1$.

We would now like to understand when such a symmetry is gaugeable. A necessary condition for any symmetry to be gaugeable is that we can associate a corresponding algebra of surfaces with it. Therefore, recalling the discussion in Sec.~\ref{sec:Morita2}, we arrive at the following constraint:

\bigskip
\noindent
{\bf Constraint 4.1:} {\it Any gaugeable symmetry must have a corresponding algebra of surfaces that satisfies $\CA_S\ni S_{A_L}$, where $S_{A_L}$ is the surface obtained by higher gauging the condensable algebra $A_L$ (i.e., $A_L$ is associated with a 1-form symmetry that can be gauged in the full spacetime).}

\bigskip
\noindent
Combining this constraint with our discussion above, we arrive at the following characterization of gaugeable (non-)invertible 0-form symmetries:

\bigskip
\noindent
{\it A gaugeable 0-form symmetry has a corresponding algebra of surfaces $\widehat\CA_S$ in $\Mod(\CB_2)$ whose dual algebra, $\CA_S$ in $\Mod(\CB_1)$, corresponds to an indecomposable surface, $S_{A_L}$, built from higher gauging the condensable algebra $A_L$ in $\CB_1$.}

\bigskip
\noindent
Equivalently, such an $\CA_S$, whose underlying object is an indecomposable surface, corresponds to a gaugeable 1-form symmetry. Indeed, the fact that $A_L$ is condensable means it is  gaugeable in all of spacetime (i.e., it lacks a 1-form symmetry 't Hooft anomaly), and our discussion is consistent.  We have therefore arrived at a generalized notion of the duality between 0-form and 1-form symmetries.

More generally, we have $A_L, B_L \ncong 1$, and we are gauging a higher structure that combines (non-)invertible 0-form and 1-form symmetry. In what follows, we wish to constrain both this more general case as well as the case of 0-form gauging. In particular, we will study the following questions:
\begin{enumerate}
\item What additional constraints are there on the surfaces corresponding to a gaugeable 0-form symmetry?
\item Can gaugeable symmetries have corresponding algebras of surfaces satisfying $\widehat\CA_S\ni S_{\widetilde A_L}$ with $\widetilde A_L$ a non-commutative 1-form symmetry algebra (i.e., a 1-form symmetry that cannot be gauged in all of spacetime)?
\item Can gaugeable symmetries have corresponding algebras of surfaces satisfying $\widehat\CA_S\ni S_A, S_B$, where $A$ and $B$ are condensable but have non-trivial mutual or individual braiding?
\end{enumerate}
Although we will not answer all these questions in full generality, we shed some light on them in the remainder of this section.

In order to make this section slightly more self contained, let us recall some of the more precise notions of gaugeability discussed in Sec.~\ref{InvertRev}  and Sec.~\ref{OPE}. In the case of an invertible 0-form symmetry, the surfaces in $\widehat\CA_S$ implement the symmetry transformations of the corresponding group, and the
algebra of surfaces can be written as
\begin{equation}
\widehat\CA_S \simeq \boxplus_{g\in G}S_g\, ,
\end{equation}
where the relation (and all similar relations below) should be understood as holding at the level of the underlying object. As a fusion 1-category, $\widehat\CA_S$ is a $G$-crossed extension of $\CB_2$, and we say that it is gaugeable if
\begin{equation}\label{Zcond}
\CZ(\widehat\CA_S)\simeq\CB_2\boxtimes\overline\CB_1\, .
\end{equation}
Gaugeability in this case is not the same thing as vanishing 't Hooft anomaly. Indeed, as described in Sec.~\ref{InvertRev}, the $G$ symmetry should have vanishing 't Hooft anomaly $\varpi$ in $H^4(G,\rU(1))$ and vanishing Postnikov class in $H^3_{\rho}(G,A)$. Otherwise $G$ is part of a non-split 2-group, and we must gauge the full 2-group.  

Let us now consider more general algebras of surfaces for non-invertible symmetries. In this case, the fusion 1-category $\widehat\CA_S$ is no longer necessarily a $G$-crossed category. Following the discussion in Sec.~\ref{OPE}, we require the analog of \eqref{Zcond} to hold in order for the corresponding symmetry to be gaugeable. More specifically, we require 
\begin{equation}\label{ZcondGen}
    \CZ(\widehat\CA_S)\simeq\CB_2\boxtimes\overline\CB_1\, , \q \CZ(\CA_S)\simeq\CB_1\boxtimes\overline\CB_2\, .
\end{equation}
We are now ready to tackle the first question in our above list. In particular, consider an Abelian TQFT, and suppose that we can gauge a symmetry corresponding to
\begin{equation}\label{Apnoncomm}
\widehat\CA_S \simeq S_1\boxplus S_{A'_L}\, ,
\end{equation}
where $A'_L$ is an algebra that is not necessarily condensable. We claim \eqref{Apnoncomm} describes a 0-form symmetry. To understand this statement, we note that for genuine lines \cite{Buican:2023bzl}
\begin{equation}
S_{A'_L}(1) \cong \bigoplus_x n_{S_{A'_L}(1)}^{x}\cdot x\, , \q n_{S_{A'_L}(1)}^{x}\in\left\{0,1\right\}\, ,
\end{equation}
and so
\begin{equation}
    \label{ASoutcome}
    \widehat\CA_S(1) \cong \sum_{x} n_{\widehat\CA_S(1)}^{x}\cdot x\, , \q n_{\widehat\CA_S(1)}^{x}=
    \begin{cases} 
        2\, , &\text{if} \; x=1\\
        0 \; \text{or} \; 1\, , &\text{if} \; x\ne1\, . 
    \end{cases}
\end{equation}

\begin{figure}[h]
    \centering

\tikzset{every picture/.style={line width=0.75pt}} 

\begin{tikzpicture}[x=0.75pt,y=0.75pt,yscale=-0.95,xscale=0.95]

\draw  [color={rgb, 255:red, 0; green, 0; blue, 0 }  ,draw opacity=1 ][fill={rgb, 255:red, 192; green, 188; blue, 188 }  ,fill opacity=0.75 ] (42.55,206) -- (42.57,107.86) -- (133.58,63.06) -- (133.57,161.19) -- cycle ;
\draw  [color={rgb, 255:red, 245; green, 166; blue, 35 }  ,draw opacity=1 ][fill={rgb, 255:red, 255; green, 224; blue, 187 }  ,fill opacity=0.87 ] (109.55,206) -- (109.57,107.86) -- (200.58,63.06) -- (200.57,161.19) -- cycle ;
\draw  [color={rgb, 255:red, 0; green, 0; blue, 0 }  ,draw opacity=1 ][fill={rgb, 255:red, 192; green, 188; blue, 188 }  ,fill opacity=0.75 ] (178.55,207) -- (178.57,108.86) -- (269.58,64.06) -- (269.57,162.19) -- cycle ;
\draw  [color={rgb, 255:red, 0; green, 0; blue, 0 }  ,draw opacity=1 ][fill={rgb, 255:red, 192; green, 188; blue, 188 }  ,fill opacity=0.75 ] (378.55,206) -- (378.57,107.86) -- (469.58,63.06) -- (469.57,161.19) -- cycle ;
\draw  [color={rgb, 255:red, 245; green, 166; blue, 35 }  ,draw opacity=1 ][fill={rgb, 255:red, 255; green, 224; blue, 187 }  ,fill opacity=0.87 ] (445.55,206) -- (445.57,107.86) -- (536.58,63.06) -- (536.57,161.19) -- cycle ;
\draw  [color={rgb, 255:red, 0; green, 0; blue, 0 }  ,draw opacity=1 ][fill={rgb, 255:red, 192; green, 188; blue, 188 }  ,fill opacity=0.75 ] (514.55,207) -- (514.57,108.86) -- (605.58,64.06) -- (605.57,162.19) -- cycle ;
\draw [color={rgb, 255:red, 139; green, 6; blue, 24 }  ,draw opacity=1 ]   (12.92,135.32) -- (43,135) ;
\draw [color={rgb, 255:red, 139; green, 6; blue, 24 }  ,draw opacity=1 ]   (224.07,135.53) -- (378,136) ;
\draw [color={rgb, 255:red, 139; green, 6; blue, 24 }  ,draw opacity=1 ]   (560.07,135.53) -- (647,136) ;
\draw  [color={rgb, 255:red, 139; green, 6; blue, 24 }  ,draw opacity=1 ][fill={rgb, 255:red, 139; green, 6; blue, 24 }  ,fill opacity=1 ] (94.23,134.53) .. controls (94.23,133.74) and (94.87,133.11) .. (95.65,133.11) .. controls (96.43,133.11) and (97.07,133.74) .. (97.07,134.53) .. controls (97.07,135.31) and (96.43,135.95) .. (95.65,135.95) .. controls (94.87,135.95) and (94.23,135.31) .. (94.23,134.53) -- cycle ;
\draw [color={rgb, 255:red, 139; green, 6; blue, 24 }  ,draw opacity=1 ] [dash pattern={on 4.5pt off 4.5pt}]  (43,135) -- (95.65,134.53) ;
\draw [color={rgb, 255:red, 139; green, 6; blue, 24 }  ,draw opacity=1 ] [dash pattern={on 4.5pt off 4.5pt}]  (378,136) -- (422.65,135.95) ;
\draw  [color={rgb, 255:red, 139; green, 6; blue, 24 }  ,draw opacity=1 ][fill={rgb, 255:red, 139; green, 6; blue, 24 }  ,fill opacity=1 ] (152.23,135.53) .. controls (152.23,134.74) and (152.87,134.11) .. (153.65,134.11) .. controls (154.43,134.11) and (155.07,134.74) .. (155.07,135.53) .. controls (155.07,136.31) and (154.43,136.95) .. (153.65,136.95) .. controls (152.87,136.95) and (152.23,136.31) .. (152.23,135.53) -- cycle ;
\draw  [color={rgb, 255:red, 139; green, 6; blue, 24 }  ,draw opacity=1 ][fill={rgb, 255:red, 139; green, 6; blue, 24 }  ,fill opacity=1 ] (224.07,135.53) .. controls (224.07,134.74) and (224.7,134.11) .. (225.48,134.11) .. controls (226.26,134.11) and (226.9,134.74) .. (226.9,135.53) .. controls (226.9,136.31) and (226.26,136.95) .. (225.48,136.95) .. controls (224.7,136.95) and (224.07,136.31) .. (224.07,135.53) -- cycle ;
\draw  [color={rgb, 255:red, 139; green, 6; blue, 24 }  ,draw opacity=1 ][fill={rgb, 255:red, 139; green, 6; blue, 24 }  ,fill opacity=1 ] (421.23,135.53) .. controls (421.23,134.74) and (421.87,134.11) .. (422.65,134.11) .. controls (423.43,134.11) and (424.07,134.74) .. (424.07,135.53) .. controls (424.07,136.31) and (423.43,136.95) .. (422.65,136.95) .. controls (421.87,136.95) and (421.23,136.31) .. (421.23,135.53) -- cycle ;
\draw  [color={rgb, 255:red, 139; green, 6; blue, 24 }  ,draw opacity=1 ][fill={rgb, 255:red, 139; green, 6; blue, 24 }  ,fill opacity=1 ] (489.65,135.11) .. controls (489.65,134.32) and (490.28,133.68) .. (491.07,133.68) .. controls (491.85,133.68) and (492.48,134.32) .. (492.48,135.11) .. controls (492.48,135.89) and (491.85,136.53) .. (491.07,136.53) .. controls (490.28,136.53) and (489.65,135.89) .. (489.65,135.11) -- cycle ;
\draw  [color={rgb, 255:red, 139; green, 6; blue, 24 }  ,draw opacity=1 ][fill={rgb, 255:red, 139; green, 6; blue, 24 }  ,fill opacity=1 ] (557.23,135.53) .. controls (557.23,134.74) and (557.87,134.11) .. (558.65,134.11) .. controls (559.43,134.11) and (560.07,134.74) .. (560.07,135.53) .. controls (560.07,136.31) and (559.43,136.95) .. (558.65,136.95) .. controls (557.87,136.95) and (557.23,136.31) .. (557.23,135.53) -- cycle ;
\draw [color={rgb, 255:red, 139; green, 6; blue, 24 }  ,draw opacity=1 ]   (95.65,134.53) -- (109.8,134.6) ;
\draw [color={rgb, 255:red, 139; green, 6; blue, 24 }  ,draw opacity=1 ] [dash pattern={on 4.5pt off 4.5pt}]  (109.8,134.6) -- (153.65,135.53) ;
\draw [color={rgb, 255:red, 139; green, 6; blue, 24 }  ,draw opacity=1 ]   (155.07,135.53) -- (178.2,135.8) ;
\draw [color={rgb, 255:red, 139; green, 6; blue, 24 }  ,draw opacity=1 ] [dash pattern={on 4.5pt off 4.5pt}]  (178.2,135.8) -- (224.07,135.53) ;
\draw [color={rgb, 255:red, 139; green, 6; blue, 24 }  ,draw opacity=1 ]   (492.48,135.11) -- (515,135.4) ;
\draw [color={rgb, 255:red, 139; green, 6; blue, 24 }  ,draw opacity=1 ]   (424.07,135.53) -- (445,135.8) ;
\draw [color={rgb, 255:red, 139; green, 6; blue, 24 }  ,draw opacity=1 ] [dash pattern={on 4.5pt off 4.5pt}]  (446.42,135.58) -- (491.07,135.53) ;
\draw [color={rgb, 255:red, 139; green, 6; blue, 24 }  ,draw opacity=1 ] [dash pattern={on 4.5pt off 4.5pt}]  (515.42,135.58) -- (558.34,135.53) -- (560.07,135.53) ;

\draw (15,64.4) node [anchor=north west][inner sep=0.75pt]    {$\CB_2$};
\draw (323,66.4) node [anchor=north west][inner sep=0.75pt]    {$\CB_1$};
\draw (107,74.4) node [anchor=north west][inner sep=0.75pt]    {$\CI_{B_L}$};
\draw (241,75.4) node [anchor=north west][inner sep=0.75pt]    {$\CI^{\dagger }_{A_L}$};
\draw (182,74.4) node [anchor=north west][inner sep=0.75pt]    {$J^\dagger$};
\draw (518.86,71.93) node [anchor=north west][inner sep=0.75pt]    {$$};
\draw (441,74.4) node [anchor=north west][inner sep=0.75pt]    {$\CI_{A_L}$};
\draw (577,75.4) node [anchor=north west][inner sep=0.75pt]    {$\CI^{\dagger }_{B_L}$};
\draw (518,74.4) node [anchor=north west][inner sep=0.75pt]    {$J$};
\draw (637,63.4) node [anchor=north west][inner sep=0.75pt]    {$\CB_2$};
\draw (18,142.4) node [anchor=north west][inner sep=0.75pt]    {$1$};

\end{tikzpicture}
    \caption{We resolve the $\CI^{\dagger}$ and $\CI$ surfaces appearing in $\widehat\CA_S \simeq \CI^{\dagger}\otimes\CI$ in terms of invertible surfaces sandwiched between surfaces implementing 0-form and 1-form symmetry gauging.}
    \label{idMap}
\end{figure}
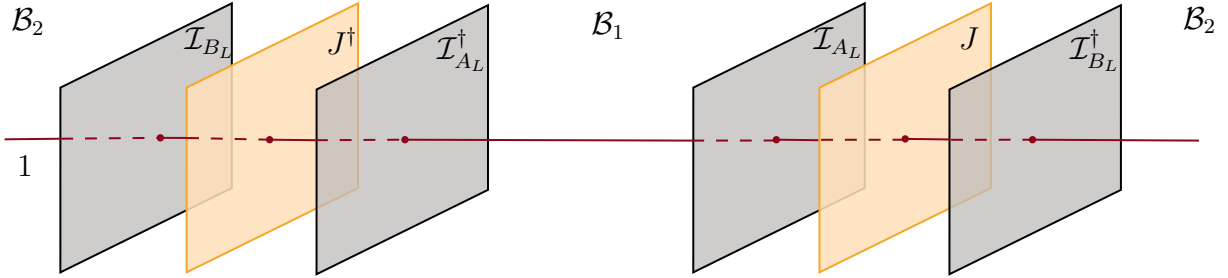

Now, let us consider the configuration shown in Fig.~\ref{idMap} coming from composing the surface in Fig.~\ref{fig:Gensandwich} and its orientation reversal (i.e., resolving $\widehat\CA_S \simeq \CI^{\dagger}\otimes\CI$ into two sandwiches). We are interested in counting junctions on $\widehat\CA_S$ of the trivial line with all genuine lines and comparing with \eqref{ASoutcome}. Mapping from left to right in Fig.~\ref{idMap}, we have
\begin{equation}\label{MapLR}
    \CB_2\ni 1
    \xrightarrow{\CI_{B_L}} 1
    \xrightarrow{J^{\dagger}} 1 \oplus x_1
    \xrightarrow{\CI^{\dagger}_{A_L}} A_L \oplus X_1
    \xrightarrow{\CI_{A_L}}n\cdot1 \oplus L_1
    \xrightarrow{J}n'\cdot1 \oplus y_1
    \xrightarrow{\CI^{\dagger}_{B_L}}n'\cdot B_L \oplus L_2\in\CB_2\, ,
\end{equation}
where $x_1$, $y_1$, $X_1$, $L_1$ and $L_2$ are (possibly non-simple) lines satisfying $\dim \Hom_{\mc B_2}(-,1)=0$ such that $x_1$ may be non-genuine, while $A_L$ and $B_L$ are have the structures of condensable algebras. Moreover, note that $n'\ge n\ge n_{A_L}\ge2$, where $A_L\to n_{A_L}\cdot1 \oplus \cdots$ under condensation. Therefore, there are at least $n_{A_L}\ge2$ junctions arising from $1\in\CB_2$ forming a junction with lines in $B_L$ on $\widehat\CA_S$ (as in Fig.~\ref{fig:A_S action from I}, we are ignoring cases with trivalent junctions involving non-trivial lines living on $\widehat\CA_S$). However, this statement is only consistent with \eqref{ASoutcome} if $B_L\cong1$, and $n_{A_L}=2$. Therefore, we see that $A_L \cong 1 \oplus b$ for some simple boson, $b\in\CB_1$. In particular, we see that when \eqref{ASoutcome} corresponds to a gaugeable symmetry, it implements a 0-form symmetry.\footnote{We will discuss the concrete examples of gauging the symmetry corresponding to $\widehat\CA_S \simeq S_1 \boxplus S_e$, which takes $\TC$ to $\mc D(\mathbb D_6)$, and gauging the symmetry corresponding to $\widehat\CA_S \simeq S_1 \boxplus S_{\psi}$, which takes $\TC$ to $\CZ({\msf{Ising}})$, in Sec.~\ref{ToricCode}.}
We have therefore arrived at the following observation:

\bigskip
\noindent
{\bf Constraint 4.2:} {\it In an Abelian TQFT, a gaugeable symmetry corresponding to an algebra of surfaces  of the form $\widehat\CA_S \simeq S_1\boxplus S_{A'_L}$, is a 0-form symmetry. The dual 1-form symmetry corresponds to a condensable algebra of the form $A_L \cong 1 \oplus b$ with $b$ a simple boson.}

\bigskip
\noindent
In fact, we can immediately generalize this statement as follows. In a non-Abelian TQFT, we also have $n_{S_{A'_L}(1)}^{1}=1$ \cite{Buican:2023bzl}. If we also assume that $n_{S_{A'_L}(1)}^{x}\in\left\{0,1\right\}$ for $x \ncong 1$, then the above constraint also generalizes:

\bigskip
\noindent
    {\bf Constraint 4.3:} {\it In a TQFT, a gaugeable symmetry corresponding to an algebra of surfaces of the form $\widehat\CA_S \simeq S_1\boxplus S_{A'_L}$ is a 0-form symmetry if $n_{S_{A'_L}(1)}^{x}\in\left\{0,1\right\}$ for all genuine lines $x \ncong 1$. The dual 1-form symmetry corresponds to a condensable algebra of the form $A_L \cong 1 \oplus b$ with $b$ a simple boson.}

\bigskip \noindent
This result is a partial converse to one in Sec.~V of \cite{Huston:2022utd}, where the authors showed that (in our language) if $\widehat\CA_S$ is dual to 1-form symmetry gauging, then the trivial surface appears in $\widehat\CA_S$.

Next let us consider if and when a gaugeable symmetry can correspond to
\begin{equation}
\widehat\CA_S \simeq S_1\boxplus S_{\widetilde A_L}\, ,
\end{equation}
where $\widetilde A_L$ is a separable but non-commutative algebra. A priori, such a scenario is compatible with all of our observations above. While we will not answer this question in full generality, we will explain when such surface algebras do and do not exist in certain theories.

To that end, consider an untwisted Abelian Dijkgraaf--Witten theory with input MTC $\mc D(H) \equiv \mc Z(\Vect_H)$, where $H$ is an Abelian gauge group. For simplicity, let us consider the case that $|H|$ is a product of distinct primes.\footnote{In other words, the gauge group is $H = \bigoplus_{i=1}^M \mathbb Z_{p_i}$, where the $p_i$ are distinct primes, and we take the trivial twist, $[0]\in H^3(H,\rU(1))$.
} To simplify matters further, we assume that $\widetilde A_L$ is built from summing over lines in the whole TQFT. In other words, we consider
\begin{equation}\label{DWng}
\widehat\CA_S \simeq S_1\boxplus S_{\mc D(H),\psi}\, ,
\end{equation}
where, to construct $S_{\mc D(H),\psi}$, we have higher-gauged all of the theory. Here, $[\psi]\in H^2(H\oplus H^\vee, U(1))$ is a choice of discrete torsion. Note that we can always choose a gauge in which the $F$-symbols of $\mc D(H)$ are trivial and so we are free to higher-gauge the full theory.\footnote{For example, we can use the gauge in (2.17) of \cite{Lee:2018eqa} and choose generators of the 1-form symmetry corresponding to the electric and magnetic generators of each $\mathbb{Z}_{N_i}\le H$.} In some sense, the $S_{\mc D(H),\psi}$ surfaces are the ``furthest'' from the $S_{A_L}$ surfaces discussed in constraint 4.1, since they are built from lines with a ``maximal'' amount of 1-form anomaly.

We will show the following:
\begin{enumerate}
\item When $S_{\mc D(H),\psi}$ is non-invertible, \eqref{DWng} does not correspond to a gaugeable symmetry.
\item On the other hand, there are examples of invertible $S_{\mc D(G),\psi}$ where \eqref{DWng} corresponds to a gaugeable symmetry.
\end{enumerate}
To understand the second statement, consider the explicit example of untwisted $H=\mathbb{Z}_3$ gauge theory and take
\begin{equation}\label{InvertEx}
\widehat\CA_S \simeq S_1\boxplus S_{\mc D(\mathbb{Z}_3),1}\, ,
\end{equation}
where $S_{\mc D(\mathbb{Z}_3),1}$ is charge conjugation. If we fix the (generalized) discrete torsion and fractionalization such that the $\widehat\CA_S$ fusion 1-category is an appropriate $\mathbb{Z}_3\oplus \mathbb{Z}_3$ Tambara--Yamagami (TY) category,\footnote{Since we higher gauge all of $\CD(\mathbb{Z}_3)$, by the discussion in App.~\ref{FPabelian}, we have a single boundary condition, $\rho$, for $S_{\CD(\mathbb{Z}_3),1}$; this boundary condition corresponds to the non-trivial TY line, and $\CA_S$ has the structure of a $\mathbb{Z}_2$-crossed graded fusion 1-category with $\rho$ being the only simple line in the non-trivial part of the grading.} we can insert a network of $\widehat\CA_S$ and produce $\mc D(\mathbb D_6)$.

To understand the first statement, suppose $S_{\CD(H),\psi}$ is non-invertible. These surfaces are in one-to-one correspondence with \lq\lq sandwiches" in Fig.~\ref{fig:sandwich} with $J$ taken to be the trivial intrinsic surface and $B_L$, $\widetilde B_L$ corresponding to Lagrangian subgroups satisfying 
\begin{equation}\label{AllLag}
B_L=\boxtimes_i\CL_i(x_i)\, , \q \widetilde B_L=\boxtimes_i\CL_i(y_i)\, ,\q x_i\not\cong y_i\, ,\q x_i,y_i\cong e_i\,\, {\rm or}\,\, m_i\, .
\end{equation}
Here $i$ runs over the different prime factors in $H\cong \prod_{i=1}^M\mathbb{Z}_{p_i}$, and $\CL_i(x_i)$ is the Lagrangian subgroup generated by an electric ($e_i$) or magnetic ($m_i$) line in each corresponding $\CD(H_i)$ factor of $\CB_2\simeq\CD(H)$ (note that the theory factorizes as $\CD(H)\simeq\boxtimes_i\CD(H_i)$). Said differently, for each $S_{\CD(H),\psi}$, there is a choice of $x_1,\cdots, x_M$ and $y_1,\cdots, y_M$ such that
\begin{equation}
S_{\CD(H),\psi} \equiv |x_1\rangle \langle y_1 | \otimes | x_2 \rangle \langle y_2 | \otimes \cdots \otimes | x_M \rangle \langle y_M |  \, ,\q x_i\not\cong y_i\, ,
\end{equation}
with $|x_i\rangle$ and $\langle y_i|$ corresponding to gapped boundaries for $\CD(H_i)$. From this construction, it is clear that each non-invertible $S_{\CD(H),\psi}$ maps the trivial line to genuine lines. If we take $S_1$ to be the completely trivial surface (i.e., it does not map untwisted lines to twisted lines), then $\widehat\CA_S$ maps the trivial line to genuine lines.

Now, given the above construction, we will find that there is no consistent fusion 1-category structure we can put on \eqref{DWng}.\footnote{In the language of Sec.~\ref{sec:simple lines after gauging} we will find a contradiction, because we will show that there is no consistent generalized symmetry fractionalization class.} As in the invertible case, we know from the discussion in App.~\ref{FPabelian} that $S_{\mc D(H),\psi}$ has a single bounding line (corresponding to the Dirichlet boundary condition), which we label as $\rho$. Combined with the $\mc D(H)$ lines bounding $S_1$, we expect the structure of a near-group fusion 1-category \cite{siehler2003near}.\footnote{More generally, in any Abelian TQFT, we expect algebra objects of the form $\CA_S \simeq S_1 \boxplus S_{A'}$ for $A'\ne1$ to be given by ``generalized'' near-group fusion 1-categories (these categories have been discussed in \cite{thornton2012generalized}). \label{GenNG}} In fact, this statement must hold up to Morita equivalence, because we have seen in Constraint 4.2 that \eqref{DWng} should correspond to a (non-)invertible 0-form symmetry with dual 1-form symmetry algebra $A_L\in\CB_2$. Then, the fusion 1-category of lines on $\CI$ is just $\CC_{\CI}\simeq(\CB_1)_{A_L}$ (with $\CD(H)\simeq\CB_2\simeq(\CB_1)_{A_L}^{\rm loc}$).

Since our contradiction below will follow from an analysis of quantum dimensions, we loose no generality in assuming the non-trivial fusion rule for $\widehat\CA_S$ is
\begin{equation}
\rho\otimes\rho \cong \bigoplus_{m\in H
,e\in \widehat H}(m,e) \oplus n\cdot\rho \, , \q  n\in \mathbb N\, ,
\end{equation}
where the $(m,e)$ lines are Abelian and realize the $H \oplus \widehat H \cong H \oplus H$ fusion rules of $\mc D(H)$. In our present setting, we can think of $n$ as corresponding to the generalized symmetry fractionalization. Using similar notation to that in \cite{evans2014near}, we refer to a fusion 1-category with these fusion rules and simple objects as $H \oplus H+n$.\footnote{In principle there can be multiple such categories for a given $H$ and $n$ (i.e., we can have multiple choices of generalized discrete torsion). However, this subtlety will not be important for our argument below (i.e., we show that any choice of generalized discrete torsion is inconsistent), and we therefore do not distinguish these possibilities in our notation.} From this discussion, it immediately follows that 
\begin{equation}\label{qdimsRGGn}
d_{\rho}={n+\sqrt{n^2+4|H|^2}\over2}\, ,\q \mc 
D^2_{\CZ(H\oplus  H+n)}=\left(|H|^2+\left({n+\sqrt{n^2+4|H|^2}\over2}\right)^2\right)^2\, ,
\end{equation}
where $\CZ(H\oplus  H+n)$ is the Drinfeld center of $H\oplus H+n$, and $\CD_{\CZ(H\oplus H+n)}:= \sqrt{\sum_{a\in \CZ(H\oplus H +n)} d_a^2}$. These $H\oplus H+n$ categories are highly constrained. In particular, beyond the $n=0$ Tambara--Yamagami case (which we have already associated with charge conjugation for $H=\mathbb{Z}_3$), other such fusion 1-categories exist only if $n=|H|^2-1$ or $n=m|H|^2$, where $m$ is a positive integer \cite{evans2014near,izumi2015cuntz,reutter2023uniqueness}. 

To further constrain our problem, we use associativity of interface fusion as in Sec.~\ref{OPE} to write
\begin{equation}
\begin{split}
    \CI(\CA_S(a))
    &\cong  \CI(S_{A_L}(a)) \cong \CI(\CI^{\dagger}\otimes\CI(a)) \cong (\CI\otimes\CI^{\dagger}\otimes\CI)(a) \cong (\CI\otimes\CI^{\dagger})(\CI(a))
    \\
    &\cong \widehat\CA_S(\CI(a)) \cong (S_1\boxplus S_{\mc D(H),\psi})(\CI(a))\, .
\end{split}
\end{equation}
Setting $a=1$ in the above equation, we obtain
\begin{equation}\label{ATGmap}
    \CI(A_L) \cong (S_1\boxplus S_{\mc D(H),\psi})(1) \cong 1 \oplus S_{\mc D(H),\psi}(1)\, .
\end{equation}
Now, by computing the RHS of the above equation, we can infer the quantum dimension of $A_L$ and therefore constrain \eqref{qdimsRGGn}. The main point is that the number of simple lines appearing in $S_{\mc D(H),\psi}(1)$ must be less than $|H|^2$. Otherwise, the theory is not modular \cite{Buican:2023bzl} and so
\begin{equation}
d_{A_L}<|H|^2+1\, .
\end{equation}
To have a non-invertible $S_{\mc D(H),\psi}$, we need to have $d_{A_L}>2$. Combining this result with \eqref{qdimsRGGn}, we arrive at the relation
\begin{equation}
{\mc D^2_{\CZ(H\oplus H +n)}\over|H|^2\cdot d_{A_L}^2}=|H|^2\, ,
\end{equation}
and solving for $n$ yields
\begin{equation}\label{nrange}
0<n<|H|^2-1\, .
\end{equation}
However, from the results in \cite{evans2014near,izumi2015cuntz,reutter2023uniqueness}, we require that $n=|H|^2-1$ or $n=m|H|^2$ with $m\in\mathbb{N}$. This is a contradiction. In our language, we see there is no consistent choice of generalized symmetry fractionalization. Therefore, we arrive at the following constraint:

\bigskip
\noindent
{\bf Constraint 4.4:} In untwisted Abelian discrete gauge theories with $|H|$ a product of distinct primes, there is no gaugeable symmetry corresponding to a surface algebra of the form $\widehat\CA_S \simeq S_1\boxplus S_{\mc D(H),\psi}$, when $S_{\mc D(H),\psi}$ is non-invertible.\footnote{Recall that our surfaces are defined as in the discussion around \eqref{AllLag}.}

\bigskip
\noindent
Next, we consider additional constraints on gaugeable 0-form symmetries corresponding to algebras of surfaces, $\widehat\CA_S$, whose underlying object is built from higher-gauging condensable 1-form symmetries. More precisely, consider
\begin{equation}
\widehat\CA_S\supset S_{B_L},S_{C_L}\, ,
\end{equation}
where $B_L$ and $C_L$ are separable commutative algebras. We can define these symmetries as in Fig.~\ref{fig:condensation surface A_S from slab theory} and the discussion around \eqref{SlabOut}. Clearly, they map the identity line to genuine lines and so too does $\widehat\CA_S$. First, note that $B_L$ and $C_L$ must braid trivially with themselves
\begin{equation}\label{BCtriv}
\msf S_{B_L,B_L} = {1\over \mc D}d_{B_L}d_{B_L}\, , \q
\msf S_{C_L,C_L}= {1\over \mc D}d_{C_L}d_{C_L}\, .
\end{equation}
Indeed, following similar logic to that around \eqref{ATGmap}
\begin{equation}
\CI(A_L) \cong \widehat\CA_S(1) \cong B_L \oplus C_L \oplus \cdots\, .
\end{equation}
Therefore unless \eqref{BCtriv} holds, $A_L$ braids non-trivially with itself. By the discussion in Sec.~\ref{IntExt}, we know that $\widehat\CA_S(1)$ must contain twisted sector lines, but this is a contradiction. The same argument also shows that we cannot have non-trivial mutual braiding
\begin{equation}
\msf S_{A_L,B_L}\ne {1\over \mc D}d_{A_L}d_{B_L}\, .
\end{equation}
We summarize our discussion as follows:

\bigskip
\noindent
{\bf Constraint 4.5:} Consider an algebra of surfaces, $\widehat\CA_S$, corresponding to a gaugeable 0-form symmetry such that all $S_{B_L}\in\widehat\CA_S$ correspond to condensable algebras and are built as in Fig.~\ref{fig:condensation surface A_S from slab theory}. Then for any $S_{B_L},S_{C_L}\in \widehat\CA_S$,  $B_L$ and $C_L$ must braid trivially with themselves and with each other.

\bigskip \noindent
In Sec.~\ref{ToricCode}, we apply the results of Sec.~\ref{OPE} and Sec.~\ref{GenConstraints} to the gauging of various symmetries of the toric code (or, more precisely, the low energy limit of the toric code) and check that the constraints we have just described are obeyed. However, before getting to these examples, we analyze conditions under which non-invertible gaugings can be broken up into sequential invertible gaugings.

\newsec{Resolving non-invertible gauging in terms of sequences of invertible gaugings}\label{sec:oneshotNshot}

In Sec.~\ref{InvertRev} we reviewed the gauging of invertible 0-form symmetries, and, in Sec.~\ref{OPE}, we discussed the gauging of their non-invertible cousins. However, we have not dwelt much on the relation between these operations. Given that symmetries can be gauged sequentially (for example, we can gauge $\mathbb D_6$ by first gauging $\mathbb{Z}_3$ and then gauging $\mathbb{Z}_2$), a natural question that arises is: can gaugings of non-invertible 0-form symmetries be resolved in terms of sequential gaugings of invertible 0-form symmetries?

As we will see in the explicit example of Sec.~\ref{TCDD8}, the answer is sometimes ``yes''. Although, as we will see in the case of the example of Sec.~\ref{TCDS3}, the answer can also be ``no.'' In this section, we will analyze the landscape of possibilities more generally.\footnote{Although we do not pursue it further in this paper, one motivation for considering this question is to understand if a single non-invertible gauging might make it possible to more simply and efficiently realize non-Abelian topological order for quantum computation (e.g., as in the setups related to those described in \cite{Iqbal:2023wvm,Lyons:2024fsk,Iqbal:2024drh}).}

Let us first explore this question from the dual perspective of gauging 1-form symmetries. Therefore, let us begin by considering an MTC $\mc B$ and a condensable algebra $A_L$ in $\CB$ corresponding to a 1-form symmetry we gauge to produce $\CB_{A_L}^{\rm loc}$. In order for the dual 0-form symmetry, whose gauging implements $\mc B^{\rm loc}_{A_L} \to \mc B$, to be non-invertible, we know from Sec.~\ref{InvertRev} that $A_L$ must contain bosons that condense partially. In that spirit, the $0$-form symmetry gauging, $\mc B^{\rm loc}_{A_L} \to \mc B$, admits a decomposition into a sequential $0$-form gauging (i.e. 0-form gauging in steps as $\mc B^{\rm loc}_{A_L} \to \mathcal{B}_{1} \to \cdots \to \mathcal{B}_{n-1}\to  \mathcal{B}_{n} \equiv \mc B$) if and only if there is a corresponding dual sequence of $1$-form symmetry gaugings. Using the results of Sec.~\ref{InvertRev}, it is clear that each 0-form symmetry we gauge in the sequence will be invertible if and only if the corresponding 1-form symmetry corresponds to an algebra $\widetilde A_i$ in $\CB_{n-i+1}$, such that, under the map $\CB_{n-i+1}\to\CB_{n-i}$, we have $\widetilde A_{L_i}\to d_{\widetilde A_{L_i}}\cdot1$. Summarizing this discussion, we find:

\bigskip
\noindent
{\bf Claim 5.1:} Consider a condensable algebra $A_L$ in $\mc B$, and gauge the corresponding 1-form symmetry to produce $\CB_{A_L}^{\rm loc}$. Suppose that the dual 0-form symmetry in $\CB_{A_L}^{\rm loc}$ is non-invertible and that there exists a sequence of $n\ge2$ 0-form gaugings satisfying
\begin{equation}
\mc B^{\rm loc}_{A_L} \to \mathcal{B}_{1} \to \cdots \to \mathcal{B}_{n-1}\to  \mathcal{B}_{n} \equiv \mc B\, .
\end{equation}
This decomposition involves gauging invertible 0-form symmetries at each step (i.e., we can ``break up'' the non-invertible $\CB_{A_L}^{\rm loc}\to\CB$ gauging into $n$ invertible steps) if and only if the dual 1-form symmetries satisfy
\begin{equation}
\widetilde A_{L_i}\to d_{\widetilde A_{L_i}}\cdot1\, ,
\end{equation}
under the map $\CB_{n-i+1}\to\CB_{n-i}$ induced by gauging $\widetilde A_{L_i}$.

\bigskip \noindent
The above claim involves the data of $n$ distinct theories. However, using the results of \cite{DMNO09}, we can reformulate the above claim in terms of algebras, $A_{L_k}\in\CB$:

\bigskip
\noindent
\textbf{Claim 5.2:} \label{nshotclaim} The 1-form gauging $\mc B \to \mc B^{\rm loc}_{A_L}$ can be formulated as an $n$-step sequential gauging if and only if $A_L\in\CB$ admits an increasing sequence of condensable sub-algebras in $\CB$ with algebra homomorphisms (see Sec. 3.6 of \cite{DMNO09}) $f^{k+1}_{k}:A_{L_k} \to A_{L_{k+1}}$ for $k \in  \left\{0,\cdots,n-1\right\}$, where $A_{L_n} \equiv A_L$, and $A_{L_0} :=1$.

\bigskip \noindent
Let us justify this claim by induction. Let $n=2$.\footnote{For a concrete example illustrating this case, see Sec.~\ref{TCDD8}.} We first prove the ``if'' direction. To that end, consider the condensable algebra $A_{L_2} \equiv A_L\in\CB$ and another condensable algebra, $A_{L_1}$, with homomorphism $f_1^2: A_{L_1} \to A_{L_2}$. Given these algebras, there are two corresponding 1-form symmetry gaugings. They result in $\mc B^{\rm loc}_{A_{L_2}}$ and $\mc B^{\rm loc}_{A_{L_1}}$ respectively.

We wish to know if there is a 1-form symmetry gauging $\mc B^{\rm loc}_{A_{L_1}}\to \mc B^{\rm loc}_{A_{L_2}}$. In fact, by the discussion in Sec. 3.6 of \cite{DMNO09}, we know that $A_{L_2}$ is an algebra over $A_{L_1}$. Therefore, $A_{L_2}$ defines a local right $A_{L_1}$-module, and any local right $A_{L_2}$-module has the structure of a local right $A_{L_1}$-module. As a result, $A_{L_2}$ defines an algebra $\widetilde A_{L_2}$ in $\CB_{A_{L_1}}^{\rm loc}$ and can be employed to perform a two-step sequential gauging so as to produce $\CB_{A_{L_2}}^{\rm loc} \equiv \mc B^{\rm loc}_{A_L}$. In symbols, 
\begin{equation}\label{seqGauge}
    {(\mc B^{\rm loc}_{A_{L_1}})}^{\rm loc}_{\widetilde A_{L_2}} \simeq \mc B^{\rm loc}_{A_{L_2}}\, ,
\end{equation}
which establishes the ``if'' direction of the $n=2$ case. It remains to establish the ``only if'' direction. There must be algebras $A_{L_1}$, $\widetilde A_{L_2}$, and $A_{L_2}$ satisfying \eqref{seqGauge}. By construction, $A_{L_2}$ is an algebra over $A_{L_1}$ and the discussion in \cite{DMNO09} Sec. 3.6 implies that $A_{L_1}$ is a subalgebra of $A_{L_2}$ (more precisely that there exists an algebra homomorphism $f_1^2$ of the type described above). This concludes the study of the $n=2$ case.

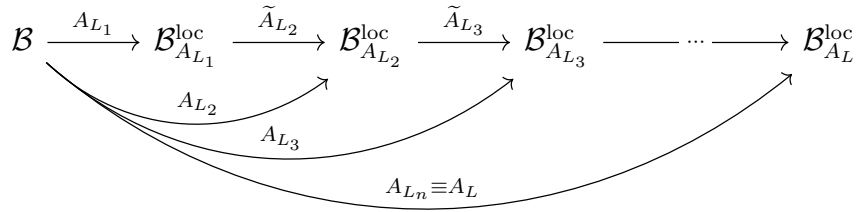
\begin{figure}[h]
\centering
\begin{align*}
\begin{tikzcd}[ampersand replacement=\&, column sep=3em, row sep=2em]
|[alias=A1]| \mc B
\&
|[alias=A2]| \mc B_{A_{L_1}}^\text{loc}
\&
|[alias=A3]| \mc B_{A_{L_2}}^\text{loc}
\&
|[alias=A4]| \mc B_{A_{L_3}}^\text{loc}
\&\&
|[alias=A5]| \mc B_{A_{L}}^\text{loc}
\arrow[from=A1,to=A2,"A_{L_1}"]
\arrow[from=A2,to=A3,"\widetilde A_{L_2}"]
\arrow[from=A3,to=A4,"\widetilde A_{L_3}"]
\arrow[from=A4, to=A5,  "\cdots" description]
\arrow[from=A1,to=A3, pos=.45,"A_{L_2}", bend right=40]
\arrow[from=A1,to=A4, pos=.45,"A_{L_3}", bend right=40]
\arrow[from=A1,to=A5, pos=.45,"A_{L_n} \equiv A_L", bend right=40]    
\end{tikzcd}
\end{align*}

\caption{Decomposing a 1-form symmetry gauging into $n$ steps.}
\label{fig:nshotanyoncondensation}
\end{figure}

Next, let us assume that the claim holds for $n \geq 
2$, and let us show it for $n+1$. As before, we begin by the ``if'' direction. By construction, we have an increasing sequence of condensable algebras $A_{L_1}\to A_{L_2}\to \cdots\to A_{L_{n-1}}\to A_{L_{n}}\to A_{L_{n+1}}$. From the induction step, we know that we can resolve the gauging $\CB\to\CB_{A_{L_n}}^{\rm loc}$ gauging into $n$ steps. From the $n=2$ case, we also know that we can resolve the gauging $\CB\to\CB_{A_L}^{\rm loc}$ into two steps via $\CB\to\CB_{A_{L_n}}^{\rm loc}\to {(\CB_{A_{L_n}}^{\rm loc})}_{\widetilde A_{L_{n+1}}}^{\rm loc}\simeq\CB_{A_L}^{\rm loc}$. Putting these results together, we can resolve the $\CB\to\CB_{A_L}^{\rm loc}$ gauging in $n+1$ steps (see Fig.~\ref{fig:nshotanyoncondensation} for the general case of $n$ steps). The ``only if'' direction proceeds similarly.

\bigskip\noindent
Now, let us consider a simple condition for a non-invertible $\CB_{A_L}^{\rm loc}\to\CB$ 0-form gauging, with $A_L$ a condensable algebra in $\mc B$, to admit a decomposition into $n$ invertible steps. From the discussion above, we know that $A_L$ must admit a sequence of subalgebras $A_{L_0}:=1,A_{L_1},\cdots,A_{L_n}:=A_L$. Given that the 0-form symmetry in question is non-invertible, we know there is  a boson $x \in A_L$ that condenses partially under gauging the symmetry corresponding to $A_L$. Upon condensing $A_L$, we have
\begin{equation}
\label{eq:partially condensing x}
x \to n_{x} \cdot1 \oplus X\, , \q d_{X} = d_{x}-n_{x}>0\, ,
\end{equation}
where $n_x$ is the multiplicity of $x$ in $A_L$.
For the dual gauging process to be sequentially invertible, we must demand that $x$ splits during some step, $\mc B^{\rm loc}_{A_{i-1}}\to \mc B^{\rm loc}_{A_{i}}$, with $x \notin A_{i}$. Otherwise, for some condensable sub-algebra containing $x$, say $A_j$, in the sequence of algebras $A_{L_1}, A_{L_2}, \cdots$, the dual gauging $\CB_{A_{L_{j}}}^{\rm loc}\to\CB_{A_{L_{j-1}}}^{\rm loc}$ would be non-invertible. This is because on condensing $A_j$, $x$ cannot completely condense to identity (otherwise, it will contradict \eqref{eq:partially condensing x}). This logic would then imply, according to the results of Sec.~\ref{InvertRev}, that the gauging $\CB_{A_{L_{j}}}^{\rm loc}\to\CB_{A_{L_{j-1}}}^{\rm loc}$ is non-invertible. Said differently, if $x \in A_L$ is a non-trivial boson that partially condenses, then $x$ must split under the gauging of a sub-algebra $A^{\prime}<A_L$ with $x \notin A^{\prime}$, to exclude non-invertible $0$-form gauging in the dual gauging sequence. 

\bigskip\noindent
\textbf{Claim 5.3:} Consider a condensable algebra $A_L$ in $\CB$ such that the dual 0-form symmetry in $\CB_{A_L}^{\rm loc}$ is non-invertible. The 0-form gauging $\CB_{A_L}^{\rm loc}\to\CB$ can be resolved into a sequence of invertible gaugings only if all bosons $x \in A_L$ that partially condense in $\CB_{A_L}^{\rm loc}$ split under the gauging of a symmetry corresponding to a proper subalgebra $A'<A_L$ with $x\not\in A'$.

\bigskip
\noindent
Given this result, we are motivated to describe a condition for $x$ to split under condensation of a sub-algebra not containing $x$. To that end, recall that for $x$ to split it must be non-simple in $\mc B_{A^{\prime}}^{\rm loc}$. Therefore, we require $\dim\Hom_{\mc B_{A^{\prime}}^{\rm loc}}(x,x)>1$. We have 
\begin{equation}\label{eq:Nshotcondition}
\begin{split}
1 < \dim\Hom_{\mc B_{A^{\prime}}^{\rm loc}}(x,x) &= \dim\Hom_{\mc B}(x,A^{\prime}\otimes x)\q, \q x \notin A^{\prime}\\
&= \sum_{c}~N_{c}~\dim\Hom_\mc B\left(x,c\otimes x\right)\q,\q A^{\prime} = \oplus_{c}N_{c}~c < A\\
& \iff x \in d \otimes x\, .
\end{split}
\end{equation}
for some non-identity line operator $d \in A'$. We have therefore arrived at the following necessary condition for the sequential invertibility of a non-invertible 0-form gauging:\footnote{As we will see in the examples section, this criterion implies that the non-invertible gauging, $\TC\to \mathcal{D}(\mathbb D_6)$, cannot be resolved into invertible steps. This theorem is also consistent with the $\TC\to \mathcal{D}(\mathbb{D}_8)$ non-invertible gauging that we will see can be resolved into two invertible steps.}

\bigskip\noindent
\textbf{Claim 5.4:} Consider a condensable algebra $A_L$ in $\CB$ such that the dual 0-form symmetry in $\CB_{A_L}^{\rm loc}$ is non-invertible. If the 0-form gauging $\CB_{A_L}^{\rm loc}\to\CB$ is sequentially invertible, then for all partially condensing bosons $x \in A_L$, there must be at least one non-trivial boson $y\neq x$ and $y \in A_L$ such that $x \in x\otimes y$ (i.e., $N^{x}_{x\,y} \geq 1$).

\bigskip

Let us revisit this question directly from the perspective of 0-form symmetry. Consider the configuration of Fig.~\ref{fig:nshotanyoncondensation}. For every 1-form symmetry gauging associated with algebras $\widetilde{A}_{L_i}$, with $i \in \{1,\cdots,n\}$ (keeping in mind that $\widetilde{A}_{L_1} \equiv A_{L_1}$ and $A_{L_n}=A_L$) we have a corresponding dual 0-form symmetry gauging associated with an algebra of surface 
\begin{equation}
    \label{SurfaceLineAlgs}
\widetilde\CA_{S_i}\leftrightarrow\widetilde A_{L_i}\, , 
\end{equation}
so that gauging $\widetilde{\mc A}_{S_i}$ maps $\mc B^{\rm loc}_{A_{L_i}}$ to $\mc B^{\rm loc}_{A_{L_{i-1}}}$. 
Let $\widehat \CA_S$ correspond to a non-invertible 0-form symmetry dual to $A_L$ whose gauging we wish to resolve into $n$ steps. For this purpose, we further require, for each 1-form symmetry gauging associated with $A_{L_i}$ in Fig.~\ref{fig:nshotanyoncondensation}, with $i \in \{1,\cdots,n\}$, the corresponding dual 0-form symmetry gauging associated with a surface algebra $\widehat{\mc A}_{S_i}$ in $\Mod(\mc B_{A_L}^{\rm loc})$, so that gauging $\widehat{\mc A}_{S_i}$ maps $\mc B^{\rm loc}_{A_L}$ to $\mc B^{\rm loc}_{A_{L_{n-i}}}$. Note that $\widehat \CA_{S_n} \equiv \widehat \CA_S$.

Recall that as part of the definition of a surface algebra $\widehat{\mc A}_S$, one needs to specify a fusion 1-category also denoted by $\widehat{\mc A}_S$, which encodes the fusion of the lines bounding indecomposable surfaces appearing in $\widehat{\mc A}_S$. In the terminology of Sec.~\ref{sec:simple lines after gauging}, this data specifies the generalized discrete torsion and symmetry fractionalization. In this spirit, we write 
\begin{equation}
\widehat\CA_{S_{k-1}}<\widehat\CA_{S_k}\, ,
\end{equation}
to mean that the fusion 1-category underlying the surface algebra $\widehat\CA_{S_{k-1}}$ is a fusion subcategory of that associated with $\widehat\CA_{S_k}$. We can then think of $\widehat\CA_{S_{k-1}}<\widehat\CA_{S_k}$ as a sub-surface algebra. Given this groundwork, we make the following claim: 

\bigskip
\noindent
\textbf{Claim 5.5}: \label{nshotclaim} Given a condensable algebra, $A_L\in\CB$, the gauging of the dual 0-form symmetry that takes us from $\mc B^{\rm loc}_{A_L}$ to $\CB$ can be formulated as an $n$-step sequential 0-form gauging if and only if the corresponding surface algebra $\widehat\CA_S$ in $ \Mod(\CB_{A_L}^{\rm loc})$ admits an increasing sequence of sub-algebras in $\Mod(\CB_{A_L}^{\rm loc})$,  $\widehat\CA_{S_k} < \widehat\CA_{S_{k+1}}$ for $k \in  \left\{0,\cdots,n-1\right\}$. Here $\widehat\CA_{S_n} \equiv \widehat \CA_S$, and the surface algebra $\widehat\CA_{S_0}$ is trivial and encodes, in particular, the fusion of all the (genuine) lines in $\CB_{A_L}^{\rm loc}$.

\bigskip \noindent
Let us begin by considering the case $n=2$ and suppose that $\CB_{A_L}^{\rm loc}\to\CB$ admits a 2-step 0-form gauging. By construction, $\CB\to\CB_{A_L}^{\rm loc}$ admits a 2-step 1-form gauging, and Claim 5.2 applies. By \cite[Remark 4.12]{DMNO09}, we see that Fig.~\ref{fig:nshotanyoncondensation} and the discussion around \eqref{SurfaceLineAlgs} implies the existence of a surface algebra $\widehat\CA_{S_1}$ such that
\begin{equation}
\label{incAs}
\widehat\CA_{S_0}<\widehat\CA_{S_1}<\widehat\CA_{S_2} \equiv \widehat\CA_{S}\, .
\end{equation}
From these algebras we have the sequential gauging
\begin{equation}
\CB_{A_L}^{\rm loc}
\xrightarrow[]{\;\widetilde{\mc A}_{S_2}\;}
\CB^{\rm loc}_{A_{L_1}}
\xrightarrow[]{\;\widetilde{\mc A}_{S_1}\;}
\CB_{A_{L_0}}^{\rm loc}\simeq\CB\, .
\end{equation}
Conversely, if we have an increasing sequence of surface algebras as in \eqref{incAs}, then Claim 5.2 again applies, and the same conclusion follows. 

Next, let us assume that the claim holds for $n \geq 
2$, and let us show it for $n+1$. By the induction step, we have the result for the sequence of gaugings
\begin{equation}
\CB_{A_L}^{\rm loc}\to\CB^{\rm loc}_{A_{L_{n}}}\to\cdots\to\CB^{\rm loc}_{A_{L_{2}}}\to\CB\, .
\end{equation}
Combining this result with the case $n=2$ finally yields 
\begin{equation}
\CB_{A_L}^{\rm loc}\to\CB^{\rm loc}_{A_{L_{n}}}\to\cdots\to\CB^{\rm loc}_{A_{L_{2}}}\to\CB^{\rm loc}_{A_{L_{1}}}\to\CB\, ,
\end{equation}
as desired. 
\begin{figure}
\begin{align*}
\begin{tikzcd}[ampersand replacement=\&, column sep=3em, row sep=2em]
|[alias=A1]| \mc B
\&
|[alias=A2]| \mc B_{A_{L_1}}^\text{loc}
\&
|[alias=A3]| \mc B_{A_{L_2}}^\text{loc}
\&
|[alias=A4]| \mc B_{A_{L_3}}^\text{loc}
\&\&
|[alias=A5]| \mc B_{A_{L}}^\text{loc}
\arrow[from=A2,to=A1,"\widetilde{\mc A}_{S_1}"']
\arrow[from=A3,to=A2,"\widetilde{\mc A}_{S_2}"']
\arrow[from=A4,to=A3,"\widetilde{\mc A}_{S_3}"']
\arrow[from=A5, to=A4,"\cdots" description]
\arrow[from=A5,to=A1, pos=.55,"\widehat{\mc A}_{S}"', bend left=40]
\arrow[from=A5,to=A2, pos=.55,"\widehat{\mc A}_{S_{n-1}}"', bend left=40]
\arrow[from=A5,to=A3, pos=.55,"\widehat{\mc A}_{S_{n-2}}"', bend left=40]
\arrow[from=A5,to=A4, pos=.55,"\widehat{\mc A}_{S_{n-3}}"', bend left=40] 
\end{tikzcd}
\end{align*}
\caption{Decomposing $0$-form gauging into $n$ steps\label{fig:nshotanyoncondensation0form}}
\end{figure}
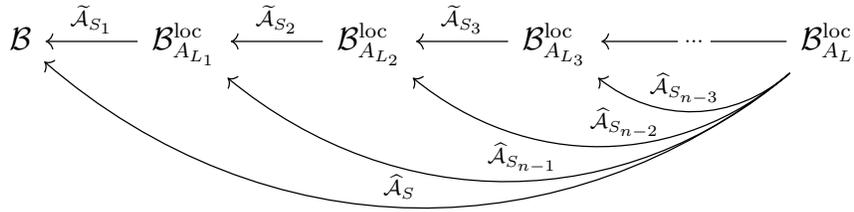
We summarize our above discussion in Fig.~\ref{fig:nshotanyoncondensation0form}.

\bigskip
Now let us focus more specifically on when a non-invertible 0-form gauging gauging can be resolved into invertible gauging in some theory $\CB$. At this point, it is convenient to introduce the notion $\mc B_{/ \widehat{\mc A}_S} := \overline{\mc Z^{\mc B}(\widehat{\mc A}_S)}$ for the result of gauging the algebra of surfaces $\widehat{\mc A}_S$ in $\Mod(\mc B)$. Suppose we have two gaugeable surface algebras $\widehat\CA_{S_1}$ and  $\widehat\CA_{S} \equiv \widehat\CA_{S_2}$ in $\Mod(\CB)$. Let us further suppose that, as fusion 1-categories, we have 
\begin{equation}
\widehat\CA_S\simeq\boxplus_{g\in G}\widehat\CA_{S,g} \, ,
\end{equation}
for some finite group $G$ such that the trivially graded part is $\widehat\CA_{S,1} \equiv \widehat\CA_{S_1}$. 
Note that $\widehat\CA_{S_1}$ is not modular and so $\widehat\CA_S$ is not a $G$-crossed extension of an MTC even though it does have a $G$-grading. Indeed, the $\widehat\CA_{S_1}$ component is the identity component of the grading (it has $g=1$).  Therefore, the gauging that takes us from $\CB\to\CB_{/ \widehat\CA_S}$ is non-invertible. Suppose the gauging operation associated with the separable algebra $\widehat{A}_{S_1}$ is invertible. Then, under the above conditions, we claim $\CB_{/ \widehat\CA_{S_1}}\to\CB_{/ \widehat\CA_{S}}$ exists and is invertible (i.e., we can resolve the original non-invertible gauging in terms of two invertible gaugings). Indeed, by Theorem 4.10 of \cite{DMNO09}, we know that 
\begin{equation}
\CZ(\widehat\CA_S)_{A_{L_1}}\simeq\CZ_{\widehat\CA_{S_1}}(\widehat\CA_S)\, ,
\end{equation}
where $A_{L_1}$ is a condensable algebra in $\CZ(\widehat\CA_S)$ whose 1-form symmetry we gauge to produce the so-called relative center $\CZ_{\widehat\CA_{S_1}}(\widehat\CA_S)$ of $\widehat{\mc A}_S$ with respect to $\widehat{\mc A}_{S_1}$ where $L_1$ and $\widehat A_{S_1}$ are related as in Fig. \ref{fig:nshotanyoncondensation0form}.\footnote{See \cite{DMNO09} for a definition. Roughly speaking, we consider all objects $X$ in $\widehat\CA_S$ and then restrict the half-braiding to those between $X$ and objects $V$ in $\widehat\CA_{S_1}$.} Now, following Sec. 3.1 of \cite{gelaki2009centersgradedfusioncategories}, we know that the $G$-grading of $\widehat\CA_S$ extends to a $G$-crossed braiding on the relative Drinfeld center and that the trivial component of the $G$-grading is $\CZ(\widehat\CA_{S_1})$. In other words, 
\begin{equation}
\CZ(\widehat\CA_S)_{A_{L_1}}\simeq\boxplus_{g\in G}\widehat\CB_g\, , \q \CB_1\simeq\CZ(\widehat\CA_{S_1})\, .
\end{equation}
To see that this result gives us the gauging we want, let us invoke some of the results of Sec.~\ref{sec:Morita2}. In particular, note that
\begin{equation}
\CZ(\widehat\CA_{S_1})\simeq \CB\boxtimes\overline{\CB_{/ \widehat\CA_{S_1}}}\, ,\q \CZ(\widehat\CA_{S})\simeq \CB\boxtimes\overline{\CB_{/ \widehat\CA_{S}}}\, .
\end{equation}
Since $\widehat{\CA}_{S_1}<\widehat{\CA}_S$, we know there is a gauging between the second factors in the above Drinfeld centers (more precisely, between their orientation reversals). The condensable $A_{L_1}$ algebra is then of the form $A_{L_1}\cong 1\boxtimes A_{L_1}'$, where $A_{L_1}'$ is a condensable algebra in $\CB_{/ \widehat\CA_S}$. Therefore, $\boxplus_{g\in G}\widehat\CB_g\simeq\CB\boxtimes\left(\boxplus_{g\in G}\widetilde\CB_{g}\right)$ with $\widetilde\CB_1\simeq\overline\CB_{/ \widehat\CA_{S_1}}$. In particular, we see that there is an invertible gauging by $G$ that takes us from $\CB_{/ \widehat\CA_{S_1}}$ to $\CB_{/ \widehat\CA_{S}}$.

This discussion can be generalized via induction to an $n\ge2$-step process of invertibly gauging a sequence of symmetries that otherwise would be non-invertible in a single gauging. In Sec.~\ref{TCDD8}, we will see an example with $n=2$.

\bigskip
\noindent
{\bf Claim 5.6:} Consider the categorical data in Claim 5.5. If each $\widehat\CA_{S_{k+1}}$ corresponds to an $G_{k+1}$-graded fusion 1-category with trivial component $\widehat\CA_{S_k}$, then a non-invertible gauging can be split into $n$ invertible steps.

\newsec{Gauging Symmetries in $\TC$}\label{ToricCode}
In this section, we illustrate our general discussion above for the particular case of $\TC$ and some of its gaugings. Along the way, we discuss concrete examples of choosing generalized discrete torsion and its imprint on the Dijkgraaf--Witten twists in the gauged theory.

Let us begin by introducing some notation that will be useful in the examples below. First of all, the symmetry structure of the toric code is captured by the fusion 2-category $\Mod(\TC)$, where $\TC := \mc Z(\Vect_{\mathbb Z_2})$. Writing $\mathbb Z_2 \cong \{+1,-1\}$, simple objects in $\Vect_{\mathbb Z_2}$ are the one-dimensional vector spaces $\mathbb C_{+1}$ and $\mathbb C_{-1}$. As is customary, we make use of the shorthand $1 := \mathbb C_{+1}$ and $m := \mathbb C_{-1}$. Similarly, we refer to the two simple objects in $\Rep(\mathbb Z_2) \simeq \Vect_{{\mathbb Z_2}^\vee}$ as $1$ and $e$ respectively. Simple objects in $\TC$ are then denoted by $1 :=(1,1), \ m :=(m,1), \ e :=(1,e), \ f := (m,e) \in \Vect_{\mathbb Z_2 \oplus {\mathbb Z_2}^\vee}$, where we remember that ${\mathbb Z_2}^\vee$ corresponds to the half-braidings (i.e., the electric charges). Finally, we denote $\mathbb Z_2 \oplus 0$, $0 \oplus {\mathbb Z_2}^\vee$, and the diagonal subgroup of $\mathbb Z_2 \oplus {\mathbb Z_2}^\vee$ as $\mathbb Z_2^m$, $\mathbb Z_2^e$, and $\mathbb Z_2^f$ respectively.

Using this notation, the six simple objects in $\Mod(\TC)$ are
\begin{equation}\label{TCsurfaces}
\begin{split}
S_1 &\cong \mc M(0,0) \, ,  \q
S_e \equiv S_{1 \oplus e} \cong \mc M(\mathbb Z_2^e,0) \, , 
\\
S_m &\equiv S_{1 \oplus m} \cong \mc M(\mathbb Z_2^m,0) \, , \q
S_f \equiv S_{1 \oplus f} \cong \mc M(\mathbb Z_2^f,0) \, ,
\\
S_{me}&\equiv S_{\TC} \cong \mc M(\mathbb Z_2^m \oplus \mathbb Z_2^e,0) \, , \q
\tilde S_{me} \equiv S_{\TC,1} \cong \mc M(\mathbb Z_2^m \oplus \mathbb Z_2^e,1) \, ,
\end{split}
\end{equation}
where the second argument of $\CM(G,i)$ corresponds to the discrete torsion (only for $G=\mathbb Z_2 \oplus \mathbb Z_2$ is there such a choice, with $H^2(\mathbb Z_2 \oplus \mathbb Z_2, \rU(1)) \cong \mathbb Z_2$). In writing \eqref{TCsurfaces}, we have used shorthand for the corresponding 1-form symmetry that we higher gauge to build the above surfaces (note that ``$\TC$'' in the above subscripts corresponds to the algebra $A_L=1 \oplus e \oplus m \oplus f$). Next, we give the fusion 2-structure of $\Mod(\TC)$ as determined by the relative Deligne product over $\TC$:
\begin{equation}
\begin{array}{c|cccccc}
\boxtimes_{\TC} & S_1 & S_m & S_e & S_f & S_{me} & \tilde S_{me} \\ \hline
S_1 & S_1 & S_m & S_e & S_f & S_{me} & \tilde S_{me} \\
S_m & S_m & S_m \boxplus S_m & S_{me} & S_{me}& S_{me} \boxplus S_{me} & S_m \\
S_e & S_e & \tilde S_{me} & S_e \boxplus S_e & \tilde S_{me} & S_e & \tilde S_{me} \boxplus \tilde S_{me} \\
S_f & S_f & \tilde S_{me} & S_{me}& S_1 & S_e & S_m \\
S_{me} & S_{me} & S_m& S_{me} \boxplus S_{me} & S_m & S_{me} & S_m \boxplus S_m \\
\tilde S_{me} & \tilde S_{me} & \tilde S_{me} \boxplus \tilde S_{me} & S_e & S_e & S_e \boxplus S_e & \tilde S_{me} 
\end{array} \, .
\end{equation}
In particular, note that the monoidal unit, $S_1$, is equivalent to $\Vect_{\mathbb Z_2 \oplus {\mathbb Z_2}^\vee}$ as a category and that the condensation defect $S_f$ generates a $\mathbb Z_2$ 0-form symmetry (the electric-magnetic duality).

As discussed in Sec.~\ref{sec:Morita2}, fusion 1-categories, $\widehat\CA_S$, together with braided tensor functors $\TC \to \mc Z(\widehat\CA_S)$ specify separable algebras of surfaces in $\Mod(\TC)$. As in the previous sections, we will slightly abuse notation and use $\widehat\CA_S$ to refer both to the fusion 1-category and to the corresponding separable algebra of surfaces in $\Mod(\TC)$. Also, in keeping with our discussion in Sec.~\ref{OPE} and Sec.~\ref{GenConstraints}, we will sometimes refer to $\CA_S$ through the corresponding sum of surfaces.\footnote{As discussed previously, this notation can sometimes be ambiguous, because the same sum of surfaces can potentially have multiple fusion structures imposed on it. We will see examples of this phenomenon in Sec.~\ref{TCDS3alpha} and Sec.~\ref{TCDZ4}.}

\bigskip
\noindent
{\bf Roadmap for the examples:} Our analysis of the examples in Sec.~\ref{TCIsing}-Sec.~\ref{TCDD8} follows much the same pattern, although each example exhibits different phenomena. One unifying theme is that all these examples involve 0-form gauging in $\CB_2\simeq\TC$ and dual 1-form symmetry gauging in a ``larger'' theory, $\CB_1$. Our recipe is as follows: 
\begin{enumerate}
\item We start in the larger theory and identify an algebra, $A_L\in\CB_1$, corresponding to a 1-form symmetry we can gauge in order to produce $\CB_2\simeq\TC$.
\item By the formalism in Sec.~\ref{sec:Morita2} and the discussion in Sec.~\ref{Int1form}, we can identify an algebra of surfaces $\CA_S\simeq S_{A_L}$ in $\Mod(\CB_1)$ corresponding to the 1-form symmetry we gauge.
\item This discussion also allows us to identify the corresponding dual algebra of surfaces $\widehat\CA_S$ in $\Mod(\mc B_2)$.
\item As a straightforward check, we follow Sec.~\ref{OPEdetails} and verify the indecomposable surfaces appearing in the decomposition $\CI^{\dagger}\otimes\CI \simeq \CA_S$ and $\CI\otimes\CI^{\dagger} \simeq \widehat\CA_S$ from the $W$ and $W^{\dagger}$ matrices.
\end{enumerate}
\noindent
The example in Sec.~\ref{TCVec} is also quite similar, but, in this case, $\TC$ plays the role of the larger theory with surface algebra $\CA_S$, and $\Vect$ plays the role of the smaller theory with surface algebra $\widehat\CA_S$. Finally, in Sec.~\ref{TCDZ3} we apply similar logic to a simple example that involves gauging a trivial (split) 2-group.

\subsection{Example: $\mc D(\mathbb Z_2) \to \Vect$}\label{TCVec}
Let us first discuss the simplest gaugings we can perform: $\mathbb{Z}_2^e$ or $\mathbb{Z}_2^m$ 1-form symmetry gauging to produce the trivial theory. As we know from our discussion above, these cases arise when $\mc A_S$ is the fusion 1-category $\Vect_{\mathbb Z_2}$ so that $\CZ(\CA_S)\simeq\TC\boxtimes\overline{\Vect}\simeq\TC$. 

Let us understand this point more explicitly, identify the intrinsic surfaces we can associate with this category, and  determining what happens when we gauge the corresponding symmetry. To that end, note that there are two inequivalent choices of braided tensor functor $\TC \to \TC$: the identity functor or the functor implementing electromagnetic self-duality (i.e., the $e\leftrightarrow m$ exchange fixing $1$ and $f$). Consider the former scenario. Recalling our convention on the naming of objects in $\TC$, the composition $\TC \to \TC \to \Vect_{\mathbb Z_2}$ endows $\Vect_{\mathbb Z_2}$ with a $\TC$-module structure, which we identity as that of $S_e$. Physically, this statement arises from the fact that the two boundary conditions of $S_e$ absorb $1$ and $e$ but are exchanged under fusion with $m$ and $f$. Now, $\mc Z^{\TC}(\Vect_{\mathbb Z_2})$ is clearly equivalent to $\msf{2Vec}$ since only the identity centralizes everything in $\TC$. Putting this logic together, we have the following Morita equivalence of fusion 2-categories
\begin{equation}
{\Mod(\TC)}^*_{\Mod(\CA_S\simeq S_e)} \simeq \TVect \, ,
\end{equation}
which we interpret as the statement that gauging the algebra $\CA_S\in\Mod(\TC)$ generated by the condensation defect $S_e$ amounts to condensing the Lagrangian algebra $1 \oplus e$ in $\TC$ (see also the discussion in the introduction). In particular, one should think of the simple objects in the category $S_e \simeq \Vect_{\mathbb Z_2}$ as the local module, $1 \oplus e$, and the non-local module, $m \oplus f$, over $1 \oplus e \in \TC$.\footnote{Very concretely, this gauging procedure amounts to creating a fine network of so-called Cheshire strings, which are (extrinsic) line defects along which anyonic excitations of type $e$ condense \cite{Kong:2014qka,PhysRevB.96.045136,Kong:2020wmn,Delcamp:2022sjf}.}

Next let us instead choose the braided tensor functor $\TC \to \TC$ that implements electromagnetic duality. Clearly, we now have instead $\CA_S\simeq S_m$ and\footnote{Physically, $1$ and $m$ are now absorbed by the two boundary conditions of $S_m$, and $e$ and $f$ exchange the two.}
\begin{equation}
{\Mod(\TC)}^*_{\Mod(\CA_S\simeq S_m)} \simeq \TVect \, .
\end{equation}
Notice that in these two examples the associator of the fusion 1-category $\Vect_{\mathbb Z_2}$ is chosen to evaluate to the identity. Physically this is the only consistent possibility. Indeed, in the case of $\CA_S\simeq S_m$, with twisted sector lines $c_1,c_2$, we can choose $c_1$ to be the identity of the fusion 1-category. Then $c_1$ clearly has trivial associator. Now, the fusion rule $e \otimes c_1=c_2$, along with the fact that the lines in toric code have trivial associator implies that $c_2$ must have trivial associator. Similar comments apply with $m\leftrightarrow e$ in the previous case. Therefore, in this case both the generalized symmetry fractionalization and the generalized discrete torsion are fixed. Indeed, we could instead consider the generalized discrete torsion $\Vect_{\mathbb Z_2}^\alpha$ whose associator evaluates to a normalized representative of the non-trivial element $[\alpha]$ in $H^3(\mathbb Z_2,\rU(1)) \cong \mathbb Z_2$, but there is no braided tensor functor $\TC \to \mc Z(\Vect_{\mathbb Z_2}^\alpha)$, and thus it cannot be endowed with the structure of a separable algebra in $\Mod(\TC)$.\footnote{In the case of the twisted $\mathbb{Z}_2$ discrete gauge theory, $\mc D^\alpha(\mathbb Z_2)$, this logic is reversed. Then we will have a non-trivial bulk associator and the identification $S_{1 \oplus s\bar s}\simeq\Vect_{\mathbb{Z}^{\alpha}_2}$, where $s$ and $\bar s$ are the semion and anti-semion respectively. Note that $s$ has a non-trivial associator and exchanges the boundary conditions of this surface via fusion. Moreover, the generalized symmetry fractionalization and generalized discrete torsion is again fixed.} This discussion is consistent with our general analysis in Sec.~\ref{Int1form}. Indeed, there we showed that, for twisted sector lines arising from a surface obtained from higher-gauging a commutative algebra, the generalized symmetry fractionalization and the generalized discrete torsion are completely fixed by the commutative algebra.

Next, let us discuss the dual 0-form gauging in the trivial theory to get back to $\TC$ (similar comments apply to gauging the symmetry corresponding to $\CA_S \simeq S_m$ instead). This is the simplest illustration of \eqref{eq:trivialGauging}. Consider the fusion 1-category $\Vect_{\mathbb Z_2}$. As a $\Vect$-module 1-category it decomposes as $S_1 \boxplus S_1$, where $S_1 \simeq \Vect$, so that
\begin{equation}
\Mod(\Vect)^*_{\Mod(S_1\boxplus S_1)}\simeq\Mod(\TC)\, .
\end{equation}
Finally, let us discuss potential indecomposable algebras of surfaces related to the other surfaces in \eqref{TCsurfaces}. As we know from our discussion in Sec.~\ref{sec:Morita2} and Sec.~\ref{GenConstraints}, we should not be able to construct such algebras (in particular, they violate Constraint 4.1 in Sec.~\ref{GenConstraints}).  Indeed, it is clear that neither $S_{me}$ nor $\tilde S_{me}$ can be endowed with the structure of an algebra in $\Mod(\TC)$. What about $S_f$? As a category, $S_f$ is equivalent to $\Vect_{\mathbb Z_2}$, which one can endow with the usual fusion structure. However, there is no braided tensor functor $\TC \to \TC$ that recovers the module structure of $S_f$ when postcomposing it with the forgetful functor. This statement corresponds to the physical fact that one cannot gauge the fermionic $\mathbb{Z}_2$ 1-form symmetry in toric code without turning on a spin structure.

\subsec{Example: $\TC \to \mc Z(\msf{Ising})$}\label{TCIsing}
As we saw in Sec.~\ref{TCVec}, the anyonic excitations $e$ and $m$ play symmetric roles in $\mc D(\mathbb Z_2)$. This phenomenon is the electromagnetic self-duality of the model. Enriching the (2+1)d toric code so as to lift the electromagnetic self-duality to an internal global 0-form $\mathbb Z_2$ symmetry, and subsequently gauging this (invertible) $\mathbb Z_2$ symmetry recovers a string-net model with topological order $\CB_1\simeq\mc Z(\msf{Ising}) \simeq \msf{Ising} \boxtimes \overline{\msf{Ising}}$, where $\msf{Ising}$ is the unitary modular tensor category encoding the Verlinde lines of the Ising conformal field theory.

Let us revisit this gauging procedure within our framework. Denoting the three simple objects of $\msf{Ising}$ by $1$, $\psi$, and $\sigma$, the non-trivial fusion rules read
\begin{equation}
\psi \otimes \psi \cong 1 \, , \q \psi \otimes \sigma \cong \sigma \, , \q \sigma \otimes \sigma \cong 1 \oplus \psi \, .
\end{equation}
Both the associator and braiding are non-trivial, and $\psi$ is a fermion so $R_{\psi,\psi} = -1$. Now, consider the condensable algebra  $A_L \cong 1 \boxtimes \overline 1 \oplus \psi \boxtimes \overline \psi\in\mc Z(\msf{Ising})$ whose multiplication is provided by the monoidal structure. Let us compute $\mc Z(\msf{Ising})_{A_L}$ in close analogy with \cite{PhysRevB.102.045139}. We can think of this fusion 1-category as the domain wall theory, $\CC_{\CI}$, on the interface, $\CI$, separating $\CZ(\msf{Ising})$ from $\CZ(\msf{Ising})_{A_L}^{\rm loc}$.

The simple local modules over $A_L\in\mc Z(\msf{Ising})$ are provided by $1 \to 1 \boxtimes \overline 1 \oplus \psi \boxtimes \overline \psi$, $e \to \sigma \boxtimes \overline \sigma$, $m \to (\sigma \boxtimes \overline \sigma) \otimes (\psi \boxtimes \overline \psi)$, and $f \to \psi \boxtimes \overline 1 \oplus 1 \boxtimes \overline \psi$. Notice that $e$ and $m$ are isomorphic as objects, but their local module structures differ.  On the other hand, the two simple non-local modules are given by $c_1 \to (1 \boxtimes \overline \sigma) \oplus (\psi \boxtimes \overline \sigma)$ and $c_2 \to (\sigma \boxtimes \overline 1) \oplus (\sigma \boxtimes \overline \psi)$. Therefore, $\mc Z(\msf{Ising})_{A_L}$ admits six simple objects, and the action of  the tensor functor $\mc Z(\msf{Ising}) \to \mc Z(\msf{Ising})_{A_L}$ explicitly reads: $1 \boxtimes \overline 1 \to 1$, $1 \boxtimes \overline \psi \to f$, $1 \boxtimes \overline \sigma \to c_1$, $\psi \boxtimes \overline 1 \to f$, $\psi \boxtimes \overline \psi \to 1$, $\psi \boxtimes \overline \sigma \to c_1$, $\sigma \boxtimes \overline 1 \to c_2$, $\sigma \boxtimes \overline \psi \to c_2$, $\sigma \boxtimes \overline \sigma \to e \oplus m$. We can think of this as the map taking bulk $\CZ(\msf{Ising})$ lines to $\CI$. Note that, since $A_L\to2\cdot1$ completely condenses, we know from the discussion around \eqref{invCond2}, that the dual symmetry is invertible.

Let us now explicitly compute the monoidal structure of $\mc Z(\msf{Ising})_{A_L}^\text{loc}$, which is given by the relative Deligne tensor product, $\otimes_{A_L}$, over the algebra $A_L$. From the definition of the relative tensor product and the fusion rules in $\msf{Ising}$, we obtain 
\begin{equation}\label{eeFusion}
e \otimes e \cong (\sigma \boxtimes \overline \sigma) \otimes_{A_L} (\sigma \boxtimes \overline \sigma) \cong 1 \boxtimes \overline 1 \oplus \psi \boxtimes \overline \psi \cong 1 \, .
\end{equation}
Similarly, one finds that $m \otimes m \cong 1\in\mc Z(\msf{Ising})_{A_L}^\text{loc}$. To fix $e\otimes m$, consider the right $A_L$-module $(\sigma \boxtimes \overline \sigma) \otimes_{A_L} (A_L \otimes (\sigma \boxtimes \overline \sigma))$. On the one hand, we have
\begin{equation}
\begin{split}
(\sigma \boxtimes \overline \sigma) \otimes_{A_L} (A_L \otimes (\sigma \boxtimes \overline \sigma))
&\cong (\sigma \boxtimes \overline \sigma) \otimes_{A_L} (\sigma \boxtimes \overline \sigma \oplus (\sigma \boxtimes \overline \sigma) \otimes (\psi \boxtimes \overline \psi))
\\ &\cong 1 \oplus (e \otimes m) \, .
\end{split}
\end{equation}
On the other hand, we have
\begin{equation}
\begin{split}
(\sigma \boxtimes \overline \sigma) \otimes_{A_L} (A_L \otimes (\sigma \boxtimes \overline \sigma))
&\cong (\sigma \boxtimes \overline \sigma) \otimes (\sigma \boxtimes \overline \sigma ) = 1 \boxtimes \overline 1 \oplus 1 \boxtimes \overline \psi \oplus \psi \boxtimes \overline 1 \oplus \psi \boxtimes \overline \psi \\
&\cong 1 \oplus f \, .
\end{split}
\end{equation}
It follows that $e \otimes m \cong f\in\mc Z(\msf{Ising})_{A_L}^\text{loc}$. Putting everything together, one verifies that $\mc Z(\msf{Ising})^\text{loc}_{A_L} \simeq \TC\simeq\CB_2$. Moreover, the following fusion rules
\begin{equation}
\begin{gathered}
c_{1/2} \otimes c_{1/2} \cong 1 \oplus f \, , \q
c_{1/2} \otimes c_{2/1} \cong e \oplus m \, , 
\\
c_{1/2} \otimes e \cong e \otimes c_{1/2} \cong m \otimes c_{1/2} \cong  c_{1/2} \otimes m \cong c_{2/1} \, ,
\end{gathered}
\end{equation}
are easily derived and complete the fusion structure of $\mc Z(\msf{Ising})_{A_L}$. For example, let us work out $c_2 \otimes c_2$, $c_2 \otimes c_1$, and $c_2 \otimes e$. From our definitions, we have 
\begin{equation}
\begin{split}
c_2 \otimes c_2 &\cong ((\sigma \boxtimes \overline 1) \oplus (\sigma \boxtimes \overline \psi)) \otimes _{A_L} ((\sigma \boxtimes \overline 1) \oplus (\sigma \boxtimes \overline \psi)) 
\\ &\cong ((\sigma \boxtimes \overline 1) \otimes A_L) \otimes_{A_L} ((\sigma \boxtimes \overline 1) \otimes A_L) \cong (\sigma \boxtimes \overline 1) \otimes (\sigma \boxtimes \overline 1) \otimes A_L
\\
&\cong (1 \oplus \psi) \boxtimes \overline 1 \otimes A_L \cong 1 \boxtimes \overline 1 \oplus \psi \boxtimes \overline 1 \oplus \psi \boxtimes \overline \psi \oplus 1 \boxtimes \overline \psi \cong 1 \oplus f \, ,
\end{split}
\end{equation}
\begin{equation}
\begin{split}
c_2 \otimes c_1 &\cong ((\sigma \boxtimes \overline 1) \oplus (\sigma \boxtimes \overline \psi)) \otimes _{A_L} ((1 \boxtimes \overline \sigma) \oplus (\psi \boxtimes \overline \sigma)) 
\\ &\cong ((\sigma \boxtimes \overline 1) \otimes A_L) \otimes_{A_L} ((1 \boxtimes \overline \sigma) \otimes A_L) \cong (\sigma \boxtimes \overline 1) \otimes (1 \boxtimes \overline \sigma) \otimes A_L
\\
&\cong (\sigma \boxtimes \overline \sigma) \otimes A_L \cong e \oplus m \, ,
\end{split}
\end{equation}
and 
\begin{equation}
\begin{split}
c_2 \otimes e &\cong ((\sigma \boxtimes \overline 1) \oplus (\sigma \boxtimes \overline \psi)) \otimes_{A_L} ((\sigma \boxtimes \overline \sigma)
\cong ((\sigma \boxtimes \overline 1) \otimes A_L) \otimes_{A_L} (\sigma \boxtimes \overline \sigma)
\\
&\cong (1 \oplus \psi) \boxtimes \overline \sigma \cong c_1 \, ,
\end{split}
\end{equation}
as required. Notice that $\mc Z(\msf{Ising})_{A_L}$ is precisely a $\mathbb Z_2$-graded fusion 1-category, which upon $\mathbb Z_2$-equivariantization, produces $\mc Z(\msf{Ising})$ \cite{Bombin:2010xn,Barkeshli:2014cna}.

Let us now apply the formalism of Sec.~\ref{sec:Morita2}. By construction, we have $\mc Z(\mc Z(\msf{Ising})_{A_L}) \simeq \mc Z(\msf{Ising}) \boxtimes \overline{\TC}$, and thus the fusion 1-category $\mc Z(\msf{Ising})_{A_L}$ has the structure of a separable algebra, $\CA_S$, in the fusion 2-category $\Mod(\mc Z(\msf{Ising}))$. From the discussion in Sec.~\ref{Int1form}, we know that, as a surface $\CA_S \simeq S_{A_L}$. This surface has six boundary conditions corresponding to each of the $A_L$-modules we have computed above, and it appears when we pinch the corresponding $\CI$ as in Fig.~\ref{fig:pinching I}. Moreover, the corresponding fusion 1-category satisfies $\CC_{\CI}\simeq\CZ(\msf{Ising})_{A_L}$.

Note that $\mc Z(\msf{Ising})_{A_L}^\text{op}$ has the structure of a separable algebra $\widehat\CA_S$ in $\Mod(\mc Z(\msf{Ising})_{A_L}^\text{loc}) \simeq \Mod(\TC)$. 
Indeed, recall from Sec.~\ref{sec:Morita2} that the $\TC$-module structure of $\mc Z(\msf{Ising})^\text{op}$ is obtained by postcomposing the braided tensor functor, $\mc Z(\msf{Ising})_{A_L}^\text{loc} \to \mc Z(\mc Z(\msf{Ising})_{A_L}^\text{op})$, provided by the above equivalence with the forgetful functor $\mc Z(\mc Z(\msf{Ising})_{A_L}^\text{op}) \to \mc Z(\msf{Ising})_{A_L}^\text{op}$. Recalling that $S_{f}$ is the indecomposable $\TC$-module 1-category such that $\text{Irr}(S_{f})$ is isomorphic to $(\mathbb Z_2 \oplus \mathbb Z_2) / \mathbb Z_2^{f}$, it readily follows from the identification $\mc Z(\msf{Ising})_{A_L}^\text{loc} \simeq \TC$ and the fusion rules of $\mc Z(\msf{Ising})_{A_L}$ that $\mc Z(\msf{Ising})_{A_L}^\text{op}$ decomposes as $S_1 \boxplus S_f$ as a $\TC$-module 1-category. In particular, this is the surface that appears when we pinch $\CI$ to the right (the opposite direction to the one in Fig.~\ref{fig:pinching I}). When we perform this maneuver, the simple local lines in $\mc Z(\msf{Ising})_{A_L}^\text{op}$ are pulled directly into the $\TC$ bulk while the non-local simple lines have a trailing $S_f$ surface attached when they are pulled into the $\TC$ bulk (e.g., See \cite{Kitaev:2011dxc}).
Our discussion is also clearly consistent with associativity of interface fusion since
\begin{equation}
2\cdot1 \cong \CI(A_L) \cong (S_1\boxplus S_f)(1) \cong 1 \oplus S_f(1)=2\cdot 1\, .
\end{equation}
Putting everything together, one finds that
\begin{equation}
{\Mod(\TC)}^*_{\Mod(S_1 \boxplus S_{f})} \simeq \Mod(\mc Z(\msf{Ising})) \, ,
\end{equation}
which formalizes the notion that gauging the electromagnetic self-duality of $\TC$ yields $\mc Z(\msf{Ising})$. Note that $\widehat{\mc A}_S \simeq S_1 \boxplus S_f$ is a separable algebra in $\Mod(\TC)$ even though $S_f$ on its own is not.\footnote{We can think of this fact as the Poincar\'e dual of the statement that we must sum over all values of the $\mathbb{Z}_2$-valued electromagnetic duality gauge field when we gauge the corresponding symmetry.} This logic is consistent with the more general analysis around \eqref{Apnoncomm} (and also with Constraint 4.2 in Sec.~\ref{GenConstraints}).\footnote{Another example in the same spirit comes from considering the quantum double model with topological order $\mc D(\mathbb Z_3)$. Indeed, gauging the (invertible) charge conjugation symmetry recovers the quantum double model with topological order $\mc D(\mathbb D_6)$. This is the strategy followed by the authors of \cite{Verresen:2021wdv} to prepare a non-Abelian topological order from an Abelian one. Within the framework of Sec.~\ref{sec:Morita2}, we recover this result by considering the algebra $\mathbb C^{\mathbb D_6 / \mathbb Z_3}\in\Rep(\mathbb D_6)$ of functions on $\mathbb D_6 / \mathbb Z_3$ together with pointwise multiplication, which extends to a separable algebra in $\QDDih$. The fusion 1-category of modules over $\mathbb C^{\mathbb D_6 / \mathbb Z_3}$ in $\Rep(\mathbb D_6)$ has the structure of a separable algebra in $\Mod(\TC)$. By postcomposing the braided tensor functor $\TC \to \mc Z(\QDDih_{\mathbb C^{\mathbb D_6 / \mathbb Z_3}})$ with the forgetful functor, one obtains that, as a module 1-category, it decomposes as $S_1 \boxplus S_\text{cc}$, where $S_\text{cc}$ is the invertible condensation defect generating the charge conjugation symmetry. The equivalence
\begin{equation}
{\Mod(\mc D(\mathbb Z_3))}^*_{\Mod(S_1 \boxplus S_\text{cc})} \simeq \Mod(\mc D(\mathbb D_6)) 
\end{equation}
states that gauging the separable algebra in $\Mod(\mc D(\mathbb Z_3))$ with underlying object $S_1 \boxplus S_\text{cc}$ amounts to gauging the charge conjugation symmetry.  As in the case of $\TC\to\CZ(\msf{Ising})$, $S_\text{cc}$ cannot be gauged on its own. Indeed, we have already come across this example \eqref{InvertEx}, where we noted that $S_\text{cc} \equiv S_{\mc D(\mathbb{Z}_3),1}$ is a surface built from an algebra which has a ``maximal'' 1-form symmetry anomaly.}

As a simple check of the above logic, we can determine $\CA_S$ and $\widehat\CA_S$ by specifying the action of $\CI$ on genuine lines (recall the discussion in Fig.~\ref{fig:A_S action from I} and the surrounding text of Sec.~\ref{OPEdetails}). Indeed, using the maps in the paragraph above the one containing \eqref{eeFusion}, we can easily construct the associated $W$ matrix discussed in \eqref{Wdefn}. Then, computing $W^{\dagger}W$ and $WW^{\dagger}$ allows us to confirm that $\CA_S \simeq S_{A_L}$ and $\widehat\CA_S \simeq S_1\boxplus S_f$ (recall that $S_f(e)=m$, $S_f(m)=e$, and all other simple lines are invariant).

\subsection{Example: $\mc D(\mathbb Z_2) \to \mc D(\mathbb D_6)$}\label{TCDS3}
Let us now consider the simplest example of non-invertible 0-form gauging in $\TC$. As in Sec.~\ref{TCIsing}, we start with the larger TQFT. In this case, it is the DW theory with input group the dihedral group $\mathbb D_6$ of order six (we can equivalently write $S_3\cong \mathbb D_6$, where $S_3$ is the group of permutations on three leters) with presentation
\begin{equation}
\mathbb D_6 = \langle r,s \, | \, r^3=s^2=(sr)^2=1 \rangle\, .
\end{equation}
Bulk operators and anyonic excitations of the corresponding quantum double model are both encoded in $\mc D(\mathbb D_6) := \mc Z(\Vect_{\mathbb D_6})$. Its symmetry structure is provided by the fusion 2-category $\Mod(\QDDih)$.

Let us recall the set of lines in this TQFT. To that end, note that, as a set, $\mathbb D_6 = \{1,r,r^2,s,sr,sr^2\}$ is the disjoint union of conjugacy classes $[1]=\{1\}$, $[r]=\{r,r^2\}$, and $[s] =\{s,sr,sr^2\}$ (which describe the trivial permutation, the 3-cycle permutations, and the 2-cycle permutations of $S_3\cong\mathbb D_6$ respectively). The centralizer subgroups of these conjugacy classes are isomorphic to $\mathbb D_6$, $\mathbb Z_3$, and $\mathbb Z_2$ respectively. Irreducible representations of $\mathbb Z_3$ are denoted $\{1,\omega,\bar \omega\}$ such that $\omega \otimes \omega \cong \bar \omega$ and $\omega \otimes \bar \omega \cong 1$, while irreducible representations of $\mathbb D_6$ are denoted $\{1,e,\pi\}$, such that $e \otimes e \cong 1$, $e \otimes \pi \cong \pi$, and $\pi \otimes \pi \cong 1 \oplus e \oplus \pi$.\footnote{Notice that we use the same notation for the sign representations of both $\mathbb Z_2$ and $\mathbb D_6$ since it should always be clear from the context which group we are referring to.} The eight simple objects of $\QDDih$ are then given by the pairs $([1],1), ([1],e), ([1],\pi)$, $([r],1)$, $([r],\omega)$, $([r],\bar \omega)$, $([s],1)$, and $([s],e)$ respectively. The fusion structure of $\QDDih$ is well known \cite{DIJKGRAAF199160,Beigi2011,Cui2015}. For convenience, we reproduce the non-trivial fusion rules below:
\begin{align}
\nn
([1],e) \otimes ([1],e) &\cong ([1],1) \, , \q 
([1],e) \otimes ([1],\pi) = ([1],\pi)
\, , \q
\\ \nn
([1],\pi) \otimes ([1],\pi) &\cong ([1],1) \oplus ([1],e) \oplus ([1],\pi) \, ,
\\ \nn
([r],-) \otimes ([1],e) &\cong ([r],-) \, , \q
([r],-) \otimes ([r],-) = ([1],1) \oplus ([1],e) \oplus ([r],-) \, ,
\\ \nn
([r],1/\omega/\bar \omega) \otimes ([1],\pi) &\cong ([r],\omega/\bar \omega/1) \oplus ([r],\bar \omega/1/\omega) \, ,
\\
([r],1/\omega/\bar \omega) \otimes ([r],\omega/\bar \omega/1) &\cong ([1],\pi ) \oplus ([r],\bar \omega/1/\omega) \, ,
\\ \nn
([1],e) \otimes ([s],1/e) &\cong ([s],e/1) \, ,
\\ \nn
([s],1/e) \otimes ([s],1/e) &\cong ([1],1) \oplus ([1],\pi) \oplus ([r],1) \oplus ([r],\omega) \oplus ([r],\bar \omega) \, ,
\\ \nn
([s],1) \otimes ([s],e) &\cong ([1],e) \oplus ([1],\pi) \oplus ([r],1) \oplus ([r],\omega) \oplus ([r],\bar \omega) \, ,
\\ \nn
([r],-) \otimes ([s],1/e) &\cong ([1],\pi) \otimes ([s],1/e) = ([s],1) \oplus ([s],e) \, ,
\end{align}
while the topological spins are
\begin{equation}
\begin{split}
\theta_{([1],1)} &= \theta_{([1],e)} = \theta_{([1],\pi)} = \theta_{([s],1)} = \theta_{([r],1)} = 1 \, ,
\\
\theta_{([s],e)} &= -1 \, , \q \theta_{([r],\omega )} = e^\frac{2 \pi i}{3} \, , \q \theta_{([r],\bar \omega)} = e^{-\frac{2 \pi i}{3}} \, .
\end{split}
\end{equation}
Now consider the algebra $A_L$ in $\Rep(\mathbb{D}_6)$ isomorphic to the algebra, $\mathbb C^{\mathbb D_6 / \mathbb Z_2}$, of functions on $\mathbb D_6 / \mathbb Z_2$ with pointwise multiplication. It is equipped with a $\mathbb D_6$ action, and we have $\mathbb C^{\mathbb D_6 / \mathbb Z_2} \cong \text{Ind}^{\mathbb D_6}_{\mathbb Z_2}(1) \cong 1 \oplus \pi$. By extension, it  defines a condensable algebra in $\QDDih$ whose underlying object is isomorphic to $A_L \cong ([1],1) \oplus ([1],\pi)$. Since $d_{([1],\pi)}=2$ is larger than $\dim\Hom_{\mc D(\mathbb D_6)}(A_L,([1],\pi))=1$, we know from the discussion in footnote \ref{extremalA} that the dual 0-form symmetry will be non-invertible.

Let us now compute $\QDDih_{A_L}$ explicitly from the above description. Simple modules over the algebra $A_L$ in $\QDDih$ are given by $1 \to ([1],1) \oplus ([1],\pi)$, $e \to ([1],e) \oplus ([1],\pi)$, $m \to ([s],1)$, and $f \to ([s],e)$, which are all local, as well as $c_1 \to ([s],1) \oplus ([s],e)$ and $c_2 \to ([r],1) \oplus ([r],\omega) \oplus ([r], \bar \omega)$, which are non-local. Note that although $c_1$ is isomorphic to $m \oplus f$ as an object, the $A_L$-module structures differ. It follows that $\QDDih_{A_L}$ admits six simple objects. We can think of this fusion 1-category as describing the theory of lines, $\CC_{\CI}$, on the interface, $\CI$, separating $\QDDih$ from the topological phase described by the local modules.

We can describe bringing the bulk $\QDDih$ lines to $\CI$ via the tensor functor $\QDDih \to \QDDih_{A}$ with action: $([1],1) \to 1$, $([1],e) \to e$, $([1],\pi) \to 1 \oplus e$, $([s],1) \to m \oplus c_1$ $([s],e) \to f \oplus c_1$, $([r],1/\omega/\bar \omega) \to c_2$.
The monoidal structure of $\QDDih_{A_L}^\text{loc}$ is given by the relative tensor product, $\otimes_{A_L}$, over the algebra $A_L$. For example, it readily follows from our definitions that 
\begin{equation}
(([1],e) \otimes A_L) \otimes_{A_L} (([1],e) \otimes A_L) \cong ([1],e) \otimes ([1],e) \otimes A_L \cong A_L\, ,
\end{equation}
which yields $e \otimes e \cong 1$ in $\QDDih_{A_L}^\text{loc}$. Using a similar argument, one finds $e \otimes m \cong f$. Furthermore, one can check that $m \otimes m \cong 1$.
Putting everything together, one verifies that $\QDDih_{A_L}^\text{loc} \simeq \TC$, as a fusion 1-category. This is the statement that gauging the (non-invertible) 1-form symmetry $([1],1) \oplus ([1],\pi)$ in $\mathbb D_6$ recovers $\TC$.

The remaining non-trivial fusion rules are
\begin{align}
e \otimes c_1 &\cong c_1
\, , \q
e  \otimes c_2 \cong c_2
\, , \q
m \otimes c_1 \cong c_2 
\, , \q
m \otimes c_2 \cong c_1 \, ,
\\ \nn
c_1  \otimes c_1 &\cong 1 \oplus e \oplus c_2 \, , \q c_1  \otimes c_2 \cong m \oplus f \oplus c_1 \, , \q 
c_2  \otimes c_2 \cong 1 \oplus e \oplus c_2 \, .
\end{align}
For instance, let us work out $m  \otimes c_1 \cong c_2$ by considering the right $A_L$-module $([s],1) \otimes_{A_L}(A_L \otimes ([s],1)) $. On the one hand
\begin{equation}
\begin{split}
([s],1) \otimes_{A_L} (A_L \otimes ([s],1)) &\cong ([s],1) \otimes_{A_L} (([s],1) \oplus ([s],1) \oplus ([s],e))  
\\
&\cong A_L \oplus ([s],1) \otimes_{A_L} (([s],1) \oplus ([s],e)) \, .
\end{split}
\end{equation}
On the other
\begin{equation}
\begin{split}
([s],1) \otimes_{A_L} (A_L \otimes ([s],1)) &\cong ([s],1) \otimes ([s],1) 
\\& \cong  ([1],1) \oplus ([1],\pi) \oplus ([r],1) \oplus ([r],\omega) \oplus ([r],\bar \omega)
\\
&\cong A_L \oplus ([r],1) \oplus ([r],\omega) \oplus ([r],\bar \omega) \, .
\end{split}
\end{equation}
Comparing these two results implies that $m \otimes c_1 \cong c_2$ in $\QDDih_{A_L}^\text{loc}$. Similarly, one can show that $m \otimes c_2 \cong c_1$, and also derive the remaining fusion rules. 

Let us now apply the formalism of Sec.~\ref{sec:Morita2}. By construction, we have $\mc Z(\QDDih_{A_L}) \simeq \QDDih \boxtimes \overline{\TC}$, and thus the fusion 1-category $\QDDih_{A_L}$ has the structure of a separable algebra, $\CA_S$, in the fusion 2-category $\Mod(\QDDih)$. From the discussion in Sec.~\ref{Int1form}, we have $\CA_S \simeq S_{A_L}$. This surface has six boundary conditions corresponding to the $A_L$-modules we have discussed above. We can produce this surface by pinching $\CI$ to the ``left'' as in Fig.~\ref{fig:pinching I}. Analogously to the previous example, $\CC_{\CI}\simeq \CZ(\QDDih_{A_L})$.

Similarly, $\QDDih_{A_L}^\text{op}$ has the structure of a separable algebra $\widehat\CA_S$ in $\Mod(\QDDih_{A_L}^\text{loc})\simeq\Mod(\TC)$. 
Indeed, as we know from Sec.~\ref{sec:Morita2}, the $\TC$-module structure of $\QDDih_{A_L}^\text{op}$ is obtained by postcomposing the braided tensor functor $\QDDih_{A_L}^\text{loc} \to \mc Z(\QDDih_{A_L}^\text{op})$ provided by $\mc Z(\QDDih_{A_L}) \simeq \QDDih \boxtimes \overline{\TC}$ with the forgetful functor $\mc Z(\QDDih_{A_L}^\text{op}) \to \QDDih_{A_L}^\text{op}$. Recalling that $S_e$ is the indecomposable $\TC$-module 1-category such that $\text{Irr}(S_e)$ is isomorphic to $(\mathbb Z_2 \oplus \mathbb Z_2) / \mathbb Z_2^e \cong \mathbb Z_2^m$, it readily follows from the identification $\QDDih_{A_L}^\text{loc} \simeq \TC$ and the fusion rules of $\QDDih_A$ (in particular recalling $m\otimes c_1 \cong c_2$) that $\QDDih_{A_L}^\text{op}$ decomposes as $S_1 \boxplus S_e$ as a $\TC$-module 1-category. Analogously to the discussion in the previous section, this is the surface that appears when we pinch $\CI$ to the right (the opposite direction to the one in Fig.~\ref{fig:pinching I}). Performing this maneuver results in the simple local lines in $\mc Z(\QDDih_{A_L}^\text{op})$ being pulled directly into $\TC$ while the remaining lines bound a trailing $S_e$ surface when they are pulled into the $\TC$ bulk. Our discussion is also clearly consistent with associativity of interface fusion and
\begin{equation}
2\cdot1 \oplus e \cong \CI(A_L) \cong (S_1\boxplus S_e)(1) \cong 2\cdot1 \oplus e\, .
\end{equation}
In particular, we manifestly see that the non-invertibility of $S_e$ is crucial in reproducing the fact that $A_L$ only partially condenses. Putting everything together, one finds that
\begin{equation}
{\Mod(\TC)}^*_{\Mod(S_1 \boxplus S_e)} \simeq \Mod(\QDDih) \, ,
\end{equation}
which formalizes the notion that gauging the non-invertible symmetry encoded in the algebra $S_1 \boxplus S_e$ in the (2+1)d toric code yields the quantum double model with input group $\mathbb D_6$.\footnote{Note that $\QDDih$ admits a braided auto-equivalence that amounts to swapping the roles of the anyonic excitations $([1],\pi)$ and $([r],1)$. It follows that $([1],1) \oplus ([r],1)$ can also be endowed the structure of a condensable algebra in $\QDDih$. The dual of the corresponding condensation process then amounts to gauging the condensation defect $S_1 \boxplus S_m$ in $\TC$.}

As a simple check of the above discussion, we can determine $\CA_S$ and $\widehat\CA_S$ by specifying the action of $\CI$ on genuine lines (recall the discussion in Fig.~\ref{fig:A_S action from I} and the surrounding text of Sec.~\ref{OPEdetails}). Indeed, using the maps in the paragraph above the one containing \eqref{eeFusion}, we can easily construct the associated $W$ matrix discussed in \eqref{Wdefn}. Then, computing $W^{\dagger}W$ and $WW^{\dagger}$ allows us to confirm that $\CA_S \simeq S_{A_L}$ and $\widehat\CA_S \simeq S_1\boxplus S_e$ (recall that $S_e(1) \cong S_e(e) \cong 1 \oplus e$, while $S_e(m)$ and  $S_e(f)$ do not result into any genuine lines).

So far, we showed that gauging $\widehat\CA_S$ can be used to invert the anyon condensation of $([1],1) \oplus ([1],\pi)$. Given the novel nature of this gauging, let us perform it explicitly to check that we indeed get $\CD(\mathbb D_6)$ from $\TC$ upon gauging the symmetry corresponding to $\CA_S$. To that end, let us consider the action of $\CA_S:=S_1 \boxplus S_e$ on the lines. Using the action of $\CI \otimes \CI^{\dagger}$ on the lines, we have

\begin{equation}
\begin{split}
\widehat{\mc A}_S (1) &\cong 2 \cdot 1 \oplus e \, , \q
\widehat{\mc A}_S (e) \cong 1 \oplus 2 \cdot e \, , \q
\widehat{\mc A}_S (m) \cong m \oplus c_1 \, , \q
\widehat{\mc A}_A (f) \cong f \oplus c_1 \, ,
\\
\widehat{\mc A}_S (c_1) &\cong m \oplus f \oplus 2 \cdot c_1 \, , \q 
\widehat{\mc A}_S (c_2) \cong 3 \cdot c_2 \, .
\end{split}
\end{equation}
In Sec.~\ref{sec:simple lines after gauging} we learned that if a simple line operator, $b$, is contained in the action of $\widehat\CA_S$ on $a$, then $a$ and $b$ can form a junction on $\widehat\CA_S$. Summing over $\widehat\CA_S$ involves trivializing this surface. Therefore, after gauging, $a$ and $b$ are not independent lines but instead have a point operator connecting them. Using \eqref{eq:inverse Frobenius reciprocity relation}, we obtain
\be
\Hom_{\TC_{/\widehat\CA_S}}(b,a) \cong \Hom_{\TC}(b,\widehat\CA_S(a))\, .
\ee
In particular, we have 
\be
\Hom_{\TC_{/ \widehat\CA_S}}(1,1) \cong \Hom_{\TC}(1,\widehat\CA_S(1))\cong \mathbb C^2\, .
\ee
Therefore, the trivial line becomes non-simple in the gauged theory and splits into a sum of two simple lines. We write this splitting as
\be
1 \to 1_1 \oplus 1_2\, .
\ee
Similarly, we find
\begin{equation}
    \begin{split}
        e\to e_1 \oplus e_2\, &, \q m\to m\, , \q  \psi\to \psi\, , \\
        c_1\to c_{11} \oplus c_{12}\, &, \q c_2\to c_{21} \oplus c_{22} \oplus c_{23}\, .  
    \end{split}
\end{equation}
However, these lines are not all independent. For example, we have 
\be
\Hom_{\TC_{/\widehat\CA_S}}(1,e) \cong \Hom_{\TC}(1,\widehat\CA_S(e))\cong \mathbb C\, .
\ee
Similarly, $\Hom_{\TC_{/\widehat\CA_S}}(e,1)\cong\mathbb C$. Therefore, one of the simple lines that $1$ splits into should be identified with one of the lines that $e$ splits into. Taking into account all other non-trivial point operators connecting the lines, we get the following equivalence classes of simple lines in $\TC_{/\widehat{\mc A}_S}$:
\be
    \{1_1\}\, , \q \{e_1\}\, , \q \{1_2,e_2\} \, , \q \{m,c_{11}\}\, , \q \{\psi,c_{12}\}\, , \q \{c_{21}\} \, ,\q \{c_{22}\} \, ,\q \{c_{23}\}\, .
\ee
The quantum dimension of these lines are the sum of quantum dimensions of the constituent lines. We obtain $1,1,2,3,3,2,2,2$, respectively. Moreover, the spin of the first 5 lines are $1,1,1,1,-1$, respectively. This agrees with the lines in untwisted Dijkgraaf--Witten theory with gauge group $\mathbb D_6$.

\subsection{Example: $\TC\to \mc D^{\alpha_p}(\mathbb D_6)$ and generalized discrete torsion}\label{TCDS3alpha}
In this section, we continue with our analysis of the non-invertible 0-form gauging that takes us from toric code to the Dijkgraaf--Witten theory with gauge group $\mathbb D_6$, and we explain how to recover the twisted $\mathbb D_6$ theories from the generalized discrete torsion described in Sec.~\ref{sec:simple lines after gauging}.

To that end, consider the twisted Dijkgraaf--Witten theory $\TQDDih$, where the twist $\alpha_p$ is a normalised representative of a class $[\alpha_p] \in H^3(\mathbb{D}_6,\rU(1))\cong\mathbb Z_6$. Let $(a,b):=s^a r^b$, where $s \in \{0,1\}$, and $r \in \{0,1,2\}$ label the group elements. An explicit formula for the 3-cocycle $\alpha_p$ is given by
\be
\alpha_p((a,b),(c,d),(e,f)):=\text{exp}\bigg(-\frac{2\pi i p}{9} \Big [ (-1)^{c+e} b [(-1)^ed +f-((-1)^ed +f) \text{ mod 3}]+\frac{9}{2} ace \Big ]\bigg )\, ,
\ee
where $p \in \{0,\cdots,5\}$ fixes the 3-cocycle (the previous section involved $p=0$). For any $p$, the simple lines are labeled by $([1],1), ([1],e), ([1],\pi)$, $([r],1)$, $([r],\omega)$, $([r],\bar \omega)$, $([s],1)$ and $([s],e)$, as before. However, the twist modifies the topological spins in the following way: 
\be
\label{eq:TQD top spins}
\msf T=\text{diag} \big[1,1,1,e^{\frac{2\pi i}{9}(p-3)},e^{\frac{2\pi i}{9}(p-6)},e^{\frac{2\pi i}{9}(p-9)},e^{\frac{2 \pi i p}{4}},-e^{\frac{2 \pi i p}{4}}\big]\, ,
\ee
and the modular S-matrix is also modified 
\be
\label{eq:TQD S matrix}
\msf S=\frac{1}{6}\begin{pmatrix}
1 & 1 & 2 & 2 & 2 & 2 & 3 & 3\\
1 & 1 & 2 & 2 & 2 & 2 & -3 & -3\\
2 & 2 & 4 & 4 \cos(\frac{2 \pi i}{3}) & 4 \cos(\frac{2 \pi  2}{3}) & 4\cos(\frac{2 \pi  3}{3}) & 0 & 0\\
2 & 2 & 4\cos(\frac{2 \pi i}{3}) & \gamma_{p,2} & \gamma_{p,3} & \gamma_{p,4} & 0 & 0\\
2 & 2 & 4\cos(\frac{2 \pi 2}{3}) & \gamma_{p,3} & \gamma_{p,4} & \gamma_{p,5} & 0 & 0\\
2 & 2 & 4\cos(\frac{2 \pi 3}{3}) & \gamma_{p,4} & \gamma_{p,5} & \gamma_{p,6} & 0 & 0\\
3 & -3 & 0 & 0 & 0 & 0 & 3(-1)^p & 3(-1)^{p+1}\\  
3 & -3 & 0 & 0 & 0 & 0 & 3(-1)^{p+1} & 3(-1)^{p}\\
\end{pmatrix}\, .
\ee
where $\gamma_{p,q}:= 4 \cos(\frac{2\pi i (2p+q)}{9}) $. In spite of the $S$-matrix depending on $p$, one can use the Verlinde formula to verify that the fusion rules of $([1],1), ([1],e), ([1],\pi)$, $([s],1)$ and $([s],e)$ are not modified by a non-trivial $p$. In particular, the Wilson lines $([1],1)$, $([1],e)$ and $([1],\pi)$ still form a $\Rep(\mathbb{D}_6)$ fusion subcategory. As in the previous example, consider gauging the 1-form symmetry associated with the algebra
\be
A_L \cong ([1],1) \oplus ([1],\pi) \, .
\ee
in $\TQDDih$.
From the fact that $\CD_{\TQDDih}^2=36$ and $\dim A_L=3$, we find that the total quantum dimension of the theory, $({\TQDDih})_{A_L}^{\rm loc}$, obtained after gauging $A_L$ is $\CD^2_{{\TQDDih})_{A_L}^{\rm loc}}=4$. 
Moreover, as in the previous section, we find
\be
([1],1)\to 1\, , \q ([1],e) \to  e\, , \q ([1],\pi)\to 1 \oplus e\, .
\ee
for some non-trivial bosonic line operator, $e$, in the gauged theory.
Also, using the fusion rules 
\be
A_L \otimes ([s],1) \cong 2 \cdot ([s],1) \oplus ([s],e)\, , \q A_L \otimes ([s],e) \cong  2([s],e) \oplus  ([s],1)\, ,
\ee
we find that both $([s],1)$ and $(([s],e))$ split into two lines after gauging. Moreover, since these two lines admit a trivalent junction with $A_L$, under gauging, they must map to a common non-genuine line operator. Therefore, we have 
\be
([s],1) \to t \oplus c_1\, , \q  ([s],e) \to u \oplus c_1\, , 
\ee
where $t,u$ are genuine lines and $c_1$ is a non-genuine line in the gauged theory. The topological spins of genuine lines must agree under gauging. As a result
\be
\theta_t=\theta_{([s],1)}=e^{\frac{2 \pi i p}{4}}\, , \q \theta_u=\theta_{([s],e)}=-e^{\frac{2 \pi i p}{4}}\, .
\ee
Using $\CD^2_{{\TQDDih})_{A_L}^{\rm loc}}=4$, we find that the simple lines in the gauged theory are $1$, $e$, $u$ and $t$.
For $p=0,2,4$ these are the simple lines of $\TC$ while for $p=1,3,5$ these are the simple lines of $\mc D^\alpha(\mathbb Z_2)$, the twisted $\mathbb{Z}_2$ gauge theory. From the fusion rules of $A_L$ with the lines of the type $([r],-)$, we find that all three lines of this kind map to the same non-genuine line in the gauged theory, namely
\be
([r],-)\to c_2\, ,
\ee
where $c_2$ is a non-genuine line in $({\TQDDih})^{\rm loc}_{A_L}$.

Let $\widehat\CA_S$ be the dual 0-form symmetry in $({\TQDDih})_{A_L}^{\rm loc}$. From the fact that $\widehat\CA_S \simeq \CI \otimes \CI^{\dagger}$ we can compute $WW^{\dagger}$ and obtain the action of $\widehat\CA_S$ on the lines $1,e,u,t$. We find
\be
\widehat\CA_S (1) \cong 2\cdot 1 \oplus e \, , \q \widehat\CA_S (e) \cong 1 \oplus 2\cdot e \, , \q \widehat\CA_S (u)\cong u\, , \q  \widehat\CA_S (t) \cong t\, .  
\ee
As a result, we see that $\widehat{\mc A}_S \simeq S_1 \boxplus S_e$.
Following the discussion in Sec.~\ref{sec:simple lines after gauging}, we know that in order to gauge the symmetry corresponding to $\widehat\CA_S$, the simple lines $\{1,e,u,t,c_1,c_2\}$ bounding surfaces in $\widehat{\mc A}_S$ (i.e. simple objects in the category $\widehat\CA_S$) must form a fusion 1-category where $1,e,u,t$ are the simple lines of $\TC$ or $\mc D^{\alpha}(\mathbb Z_2)$ depending on the spin of $u$ and $t$. Also, we know from the construction of the surface $S_e$ that the simple twisted sector lines $c_1,c_2$ satisfy 
\be
\label{eq:fusion fixed by module 1-category}
e \otimes c_1 \cong c_1\, , \q e\otimes c_2 \cong c_2\, , \q u\otimes c_1 \cong c_2\, .
\ee
Let us assume that $c_1$ and $c_2$ are self-dual lines. Then, using the fact that they have quantum dimension 2, and demanding consistency with the fusion rules \eqref{eq:fusion fixed by module 1-category} requires
\be\label{XYrules}
c_1\otimes c_1 \cong c_2\otimes c_2 \cong  1 \oplus e \oplus c_1 \, , \q c_1\otimes c_2 \cong u \oplus t \oplus c_1\, .
\ee
Up to permutation of $c_1$ and $c_2$, \eqref{XYrules} are the unique consistent fusion rules. Therefore, following the discussion in Sec.~\ref{sec:simple lines after gauging}, we find that there is a unique choice of ``generalized symmetry fractionalization'' in this case. Since gauging $A_L$ preserves fusion rules, we can use fusion across the interface $\CI$ to check that these are precisely the fusion rules that give rise to $\TQDDih$ upon gauging the 0-form symmetry associated with $\widehat\CA_S$.

In order to gauge the symmetry corresponding to $\widehat\CA_S$, we must ensure that the above fusion rules can be promoted to a fusion 1-category. To that end, let us note that the fusion ring with simple objects $\{1,e,u,t,c_1,c_2\}$ can be written as a product of two fusion rings 
\be
\{1,e,c_1\}\otimes \{1,u\}\, .
\ee
Since $u$ is Abelian and has order two, $\{1,u\}$ must form the fusion 1-category $\Vect_{\DZ_2}^{\widehat\alpha}$ for some $\widehat\alpha\in H^3(\DZ_2,\rU(1))\cong \DZ_2$. It follows from our above discussion that $\widehat\alpha$ is trivial for even $p$ and non-trivial for odd $p$. Besides, $\{1,e,c_2\}$ must obey the fusion rules of $\Rep(\mathbb{D}_6)$. From \cite[Theorem 6.3]{etingof2003classificationfusioncategoriesdimension}, we know that there are three fusion 1-categories with the same fusion rules as $\Rep(\mathbb D_6)$. We will denote these categories as $\CC(\mathbb D_6,\DZ_2,\alpha_p)$, where $\DZ_2<\mathbb D_6$, and $\alpha_p \in H^3(\mathbb D_6,\rU(1))$ with $p$ even. The category $\CC(\mathbb D_6,\DZ_2,\alpha_p)$ is constructed by starting with $\Vect_{\mathbb D_6}^{\alpha_p}$ and gauging a $\DZ_2$ subgroup. Note that since $p$ is even, all order-two subgroups of $\mathbb D_6$ are non-anomalous. When $p$ is odd, $\DZ_2$ subgroups of $\mathbb D_6$ are all anomalous. However, we can gauge the diagonal $\DZ_2$ subgroup of $\Vect_{\DZ_2}^{\widehat\alpha =-1} \boxtimes \Vect_{\mathbb D_6}^{\alpha_p}$. This operation produces three other fusion 1-categories with fusion rules given by those of $\{1,e,u,t,c_1,c_2\}$. Therefore, we have three choices of ``generalized discrete torsion'' in this case.

This argument shows that, starting with $\TC$, we can construct three fusion 1-categories from the simple lines $\{1,e,u,t,c_1,c_2\}$. Upon gauging the symmetry associated with $\widehat\CA_S$, we get $\CD^{\alpha_p}(\mathbb D_6)$ for $p$ even. Similarly, in the $\mc D^{\hat \alpha}(\mathbb Z_2)$ theory, the lines $\{1,e,u,t,c_1,c_2\}$ can be promoted to three distinct fusion 1-categories. Gauging $\widehat\CA_S$ in this case leads to $\CD^{\alpha_p}(\mathbb D_6)$ for $p$ odd. This argument shows that the $\mathbb{D}_6$ Dijkgraaf--Witten theory with the 6 possible twists can all be realized by starting from the untwisted and twisted $\mathbb Z_2$ discrete gauge theories and gauging non-invertible symmetries.\footnote{In the context of the twisted $\mathbb Z_2$ theory, we have $\widehat\CA_S \simeq S_1 \boxplus S_{1+s\bar s}$.}

\subsec{Example: $\TC \to \mc D(\mathbb{Z}_4)$ and $\TC \to \mc D(\mathbb D_4)$}\label{TCDZ4}
In Sec.~\ref{TCDS3alpha}, we explored a set of examples in which the algebras of surfaces $\widehat\CA_S\in\Mod(\TC)$ were identical as objects but differed as fusion 1-categories. This difference corresponded to a choice of generalized discrete torsion in the language of Sec.~\ref{sec:simple lines after gauging}. In this section, we explore a basic pair of examples where the (generalized) symmetry fractionalization differs instead. Moreover, this example involves unfaithful symmetry (as in the examples of Sec.~\ref{ZFib} and Sec.~\ref{TCVec}).

To that end, consider $\CB_1\simeq\CD(\mathbb Z_4)$. Mimicking the notation of $\TC$, simple objects in $\Vect_{\mathbb Z_4}$ are denoted by $1$, $m$, $m^2$ and $m^4$, whereas simple objects in $\Rep(\mathbb Z_4)$ are denoted by $1$, $e$, $e^2$, and $e^3$. Simple objects in $\mc D(\mathbb Z_4)$ are named accordingly. Consider the condensable algebra $A_L = 1 \oplus e^2$. The simple local $A_L$-modules are $A_L$, $m^2 \otimes A_L$, $e \otimes A_L$ and $m^2 \otimes e \otimes A_L$, whereas the simple non-local $A_L$-modules are $m \otimes A_L$, $m^3 \otimes A_L$, $m \otimes e \otimes A_L$ and $m^3 \otimes e \otimes A_L$. It is straightforward to compute fusion rules using the relative product, $\otimes_{A_L}$, and to use preservation of spin to see that $\mc D(\mathbb Z_4)_{A_L}^\text{loc} \simeq \TC$ and  $\mc D(\mathbb Z_4)_{A_L} \simeq \Rep(\mathbb Z_2 \oplus \mathbb Z_4)$. This discussion is consistent with the fact that $\CZ(\CD(\mathbb Z_4)_{A_L})\simeq \CD(\mathbb Z_4)\boxtimes\overline{\TC}$ and the fact that $\mc D(\mathbb Z_4)_{A_L}$ has the structure of a separable algebra $\mc A_S$ in $\Mod(\CD(\mathbb Z_4))$ with $\CA_S \simeq S_{A_L}$ by the discussion in Sec.~\ref{Int1form}. This surface has eight boundary conditions corresponding to each of the $A_L$-modules we have discussed, and it appears when we pinch $\CI$ as in Fig.~\ref{fig:pinching I}. Moreover, the corresponding fusion 1-category satisfies $\CD(\CA_S)\simeq \CC_{\CI} \simeq \CZ(\CD(\mathbb Z_4)_{A_L})$.

By construction, $\mc D(\mathbb Z_4)_{A_L}^\text{op}$ has the structure of a separable algebra $\widehat\CA_S\in\Mod(\mc D(\mathbb Z_4)_{A_L}^\text{loc}) \simeq \Mod(\TC)$. Recall that the $\TC$-module structure of $\mathbb D(\mathbb Z_4)_{A_L}^\text{op}$ is given by $\TC \to \mc Z(\Rep(\mathbb Z_2 \oplus \mathbb Z_4)^\text{op}) \to \Rep(\mathbb Z_2 \oplus \mathbb Z_4)^\text{op}$. It readily follows from the fusion rules in $\mc D(\mathbb Z_4)_{A_L}$
that $\Rep(\mathbb Z_2 \oplus \mathbb Z_4)^\text{op}$ decomposes as $S_1 \boxplus S_1$ as a $\TC$-module 1-category. In particular, this is the surface that appears when we pinch $\CI$ to the right (opposite to the direction in Fig.~\ref{fig:pinching I}). When we perform this pinching, the simple local lines in $\CD(\mathbb{Z}_4)_{A_L}^\text{op}$ are pulled directly into the $\TC$ bulk while the non-local simple lines are ``attached'' to the second $S_1$ surface when they are pulled into the $\TC$ bulk.\footnote{We can think of these latter lines as boundary conditions for the second copy of $S_1$, and we can think of this surface as an extrinsic symmetry. See footnote \ref{MicroExt} for further discussion.} Our constuction is also clearly consistent with associativity of interface fusion and, in particular
\begin{equation}
2\cdot1 \cong \CI(A_L) \cong (S_1\boxplus S_1)(1) \cong 2\cdot1\, .
\end{equation}
Putting everything together, one finds that
\begin{equation}
{\Mod(\TC)}^*_{\Mod(S_1 \boxplus S_1)} \simeq \Mod(\CD(\mathbb Z_4)) \, .
\end{equation}

In the case we have just discussed, we found that $S_1\boxplus S_1\in\Mod(\TC)$ has (generalized) symmetry fractionalization corresponding to $\mathbb Z_2\oplus\mathbb Z_4$ fusion rules (this phenomenon, featuring a non-trivial group extension over the original group, is sometimes referred to in the literature as involving non-trivial symmetry fractionalization). To obtain trivial symmetry fractionalization, consider starting from $\CD(\mathbb D_4)$ (where $\mathbb D_4 \cong \mathbb Z_2 \oplus \mathbb Z_2$ is the dihedral group of order 4). Simple objects in $\Vect_{\mathbb D_4}$ are denoted by $1$, $m_1$, $m_2$, and $m_1m_2$, whereas simple objects in $\Rep(\mathbb D_4)$ are denoted by $1$, $e_1$, $e_2$, and $e_1e_2$. Simple objects in $\mc D(\mathbb D_4)$ are named accordingly. Consider the condensable algebra $A_L := 1 \oplus e_1$. The simple local $A_L$-modules are $A_L$, $m_2 \otimes A_L$, $e_2 \otimes A_L$ and $m_2 \otimes e_2 \otimes A_L$, whereas the simple non-local $A_L$-modules are $m_1 \otimes A_L$, $m_1m_2 \otimes A_L$, $m_1 \otimes e_2 \otimes A_L$, and $m_1m_2 \otimes e_2 \otimes A_L$. We immediately see that $\mc D(\mathbb D_4)_{A_L}^\text{loc} \simeq \mc D(\mathbb Z_2)$ and $\mc D(\mathbb Z_2)_{A_L} \simeq \Rep(\mathbb Z_2 \oplus \mathbb D_4)$. The logic of the non-trivial fractionalization class discussion has obvious analogues in this case. For example,  we find that $\mc D(\mathbb D_4)_{A_L}^\text{op}$ has the structure of a separable algebra in $\Mod(\TC)$ and that it also decomposes as $S_1 \boxplus S_1$ as a $\TC$-module 1-category. Putting everything together, this logic shows that 
\begin{equation}
{\Mod(\TC)}^*_{\Mod(S_1 \boxplus S_1)} \simeq \Mod(\CD(\mathbb D_4)) \, .
\end{equation}
Clearly, the condensation defect $S_1 \boxplus S_1$ can be equipped with two distinct structures of a separable algebra $\widehat\CA_S$ in $\Mod(\TC)$ (i.e., two different symmetry fractionalizations) encoded in the fusion 1-categories $\Rep(\mathbb Z_2 \oplus \mathbb Z_4)$ and $\Rep(\mathbb Z_2 \oplus  \mathbb D_4)$ respectively. 

\subsec{Example: $\TC\to \mc D(\mathbb{D}_8)$}\label{TCDD8}

Like the example in Sec.~\ref{TCDS3}, our example in this section involves non-invertible 0-form symmetry gauging. Unlike this previous example, the algebra of surfaces we consider here also involves an unfaithful sub-symmetry (it is therefore, in some sense, a hybrid of the examples in Sec.~\ref{TCDS3} and Sec.~\ref{TCDZ4}). Moreover, as we will see, the non-invertible gauging we describe here can, alternatively, be understood as a sequential two-step invertible gauging.\footnote{It is therefore an example of the more general discussion in Sec.~\ref{sec:oneshotNshot}, with the first step corresponding to the ``trivial'' fractionalization case discussed in the Sec.~\ref{TCDZ4}.}

As in the previous examples, we start with the larger TQFT which, in this case is $\CD(\DD_8)$. $\mathbb D_{8}$ admits a presentation corresponding to the semi-direct product $\mathbb D_{8} = \zz_{4}\rtimes \zz_{2}$, 
\begin{equation}
\DD_{8} = \langle\,r,s \,|\,r^{4} = s^{2} =1 ,\, srs = r^{-1}\,\rangle\, .
\end{equation}
The 22 anyons of $\mc D(\mathbb D_{8})$ are listed in table~\ref{DD8spins}.
\begin{table}
\begin{center}
\begin{tabular}{|c|c|c|}
\hline
Anyons    & $d$ & $\theta$  \\
\hline
$([1],1) \equiv 1$    & 1 & 1   \\
$([1],e_{1})$    & 1 & 1   \\
$([1],e_{2})$    & 1 & 1   \\
$([1],e_{1}e_{2})$    & 1 & 1   \\
$([1],\pi)$    & 2 & 1   \\
\hline
\hline
$([r^2],1)$    & 1 & 1   \\
$([r^2],e_{1})$    & 1 & 1   \\
$([r^2],e_{2})$    & 1 & 1   \\
$([r^2],e_{1}e_{2})$    & 1 & 1   \\
$([r^2],\pi)$    & 2 & -1   \\
\hline
\end{tabular}
\q 
\begin{tabular}{|c|c|c|}
\hline
Anyons    & $d$ & $\theta$  \\
\hline
$([r],1)$    & 2 & 1   \\
$([r],e)$    & 2 & $i$   \\
$([r],e^{2})$    & 2 & -1   \\
$([r],e^{3})$    & 2 & -$i$   \\
\hline
\hline
$([s],1)$    & 2 & 1   \\
$([s],e_{1})$    & 2 & 1   \\
$([s],e_{2})$    & 2 & -1   \\
$([s],e_{1}e_{2})$    & 2 & -1   \\
\hline
\hline
$([rs],1)$    & 2 & 1   \\
$([rs],e_{1})$    & 2 & 1   \\
$([rs],e_{2})$    & 2 & -1   \\
$([rs],e_{1}e_{2})$    & 2 & -1   \\
\hline
\end{tabular}
\end{center}
\caption{Anyons in $\mc D(\mathbb D_{8})$.}
\label{DD8spins}
\end{table}
There, $\{1,e_{1},e_{2},e_{1}e_{2}\}$ denote irreducible representations of $\zz_{2}\oplus \zz_{2}$ as well as the one-dimensional irreducible representations of $\mathbb D_{8}$, while $\pi$ is the two-dimensional irreducible representation of $\mathbb{D}_{8}$. As before, irreducible representations of $\mathbb Z_4$ (centralizer of $[r]$ in $\mathbb D_{8}$) are denoted by $\{1, e,e^{2},e^{3}\}$.

The details of the condensable algebras in $\CD(\DD_8)$ are worked out in \cite{bhardwaj2024hassediagramsgaplessspt}. Specifically, we study the condensable algebra whose underlying object is $A_L = ([1],1) \oplus ([1],e_{1}) \oplus ([1],\pi)$. As an algebra, this is isomorphic to $\mathbb{C}^{\DD_{8}/\zz_{2}}$, of functions on $\DD_{8}/\zz_{2}$ with pointwise multiplication. We have $\mathbb{C}^{\DD_{8}/\zz_{2}} \cong {\rm Ind}^{\DD_{8}}_{\zz_{2}}(1) \cong 1 \oplus e_{1} \oplus \pi$.\footnote{Notice that $A_L = ([1],1) \oplus ([1],e_{1}) \oplus ([1],\pi)$ has a subalgebra $A_L^{\prime} \cong ([1],1) \oplus ([1],e_{1})$ isomorphic to $\mathbb C^{\zz_{2}\oplus \zz_{2}/\zz_{2}}$. Based our discussion in Sec.~\ref{sec:oneshotNshot}, this implies that the gauging of $A_L$ in $ \CD(\DD_{8})$ can be performed sequentially. Furthermore, since the non-Abelian boson $([1],\pi)$ is fixed by the subalgebra $A_L^{\prime}$, the dual 0-form symmetry gauging satisfies a necessary condition to admit sequential gauging by invertible symmetries. \label{SeqFoot}} Since $d_{([1],\pi)}=2$ is larger than $\dim\Hom_{\CD(\DD_8)}(A_L,([1],\pi))=1$, we know from the discussion in footnote \ref{extremalA} that the dual 0-form symmetry is non-invertible.

The corresponding 1-form symmetry gauging proceeds as follows. First, we write the simple $A_L$-modules in $\CD(\DD_{8})$ 
\begin{equation}
\begin{split}
    1 &\to ([1],1)\oplus ([1],e_{1}) \oplus ([1],\pi)\, , \q m \to ([1],e_{1}e_{2}) \oplus ([1],e_{2})\oplus ([1],\pi)\, ,\\
    e &\to ([s],e_{1}) \oplus ([s],1)\, , \q
    f \to ([s],e_{2})\oplus ([s],e_{1}e_{2})\, ,\\
    d_{1} &\to ([s],1) \oplus ([s],e_{1}e_{2})\, , \q
    d_{2} \to ([r^{2}],e_{1})\oplus ([r^{2}],1) \oplus ([r^2],\pi)\, ,\\
    d_{3} &\to ([r^{2}],e_{2}) \oplus ([r^{2}],e_{1}e_{2}) \oplus ([r^2],\pi)\, , \q 
    d_{4} \to ([s],e_{1}) \oplus ([s],e_{2})\, ,\\
    c_1 &\to ([rs],e_{1})\oplus ([rs],1)\oplus ([rs],e_{1}e_{2}) \oplus([rs],e_{2})\, ,\\
    c_2 &\to ([r],1)\oplus ([r],e^{2}) \oplus ([r],e) \oplus ([r],e^{3})\, .
\end{split}
\end{equation}
The $\{1,e,m,f\}$ simple modules are local, and $\{d_{3},d_{2},d_{4},d_{1},c_1,c_2\}$ are non-local. Analogously to the previous examples, we can think of this fusion 1-category as describing the theory of lines, $\CC_{\CI}$, on the interface $\CI$ separating $\CD(\DD_8)$ from the topological phase described by the local modules.

The monoidal structure of $\mathcal{D}(\DD_{8})_{A_L}$ can be derived by applying the relative product, $\otimes_A$, to the above construction and using the fusion rules in \cite{bhardwaj2024hassediagramsgaplessspt}. One finds
\begin{equation}
\begin{split}
&e\otimes e \cong m\otimes m \cong f\otimes f \cong 1\, ,\\
& e\otimes m \cong f\, ,\\
&d_{3}\otimes d_{3} \cong d_{2}\otimes d_{2} \cong d_{4}\otimes d_{4} \cong d_{1} \otimes d_{1} = 1\, ,\\
&d_{3}\otimes m \cong d_{2} \,, \q d_{3} \otimes e \cong d_{4}\, , \q d_{3} \otimes f \cong d_{1}\, ,\\
&c_{1}\otimes c_{1} \cong c_{2}\otimes c_{2} \cong 1 \oplus m \oplus d_{2} \oplus d_{3}\, ,\\
&m \otimes c_{1} \cong c_{1} \, , \q m \otimes c_{2} \cong c_{2} \, , \q e\otimes c_{1} \cong c_{2}\, .
\end{split}
\end{equation}
From these fusion rules and the topological spins in \eqref{DD8spins}, it is clear that $\CD(\mathbb D_8)_{A_L}^{\rm loc}\cong\TC$. Therefore, gauging $A_L \cong([1],1) \oplus ([1],e_{1}) \oplus ([1],\pi)$ in $\CD(\DD_{8})$ indeed recovers $\TC$.

Let us now once again apply the formalism of Sec.~\ref{sec:Morita2}. To that end, we have $\CZ(\CD(\DD_8)_{A_L})\simeq\CD(\DD_8)\boxtimes\overline{\TC}$, and so the fusion 1-category $\CD(\DD_8)_{A_L}$ has the structure of a separable algebra, $\CA_S$, in the fusion 2-category $\Mod(\CD(\DD_8))$. From the discussion in Sec.~\ref{Int1form}, we know that $\CA_S \simeq S_{A_L}$. This surface has ten boundary conditions corresponding to the $A_L$-modules we have derived above. We can produce this surface by pinching $\CI$ to the ``left” as in Fig.~\ref{fig:pinching I}. Analogously to the previous example, $\CD(\CA_S) \simeq \CC_{\CI} \simeq \CZ(\CD(\DD_8)_{A_L})$.

Similarly, $\CD(\DD_8)_{A_L}^\text{op}$ has the structure of a separable algebra $\widehat\CA_S\in\Mod(\CD(\DD_8)_{A_L}^\text{loc})\simeq\Mod(\TC)$. Indeed, as we know from Sec.~\ref{sec:Morita2}, the $\TC$-module structure of $\CD(\DD_8)_{A_L}^\text{op}$ is obtained by postcomposing the braided tensor functor $\CD(\DD_8)_{A_L}^\text{loc} \to \mc Z(\CD(\DD_8)_{A_L}^\text{op})$ provided by the above equivalence with the forgetful functor $\mc Z(\CD(\DD_8)_{A_L}^\text{op})\to \CD(\DD_8)_{A_L}^\text{op}$. Recalling that $S_m$ is the indecomposable $\TC$-module 1-category such that $\text{Irr}(S_m)$ is isomorphic to $(\mathbb Z_2 \oplus \mathbb Z_2) / \mathbb Z_2^m \cong \mathbb Z_2^e$, it readily follows from the identification $\CD(\DD_8)_{A_L}^\text{loc} \simeq \TC$ and the fusion rules of $\CD(\DD_8)_{A_L}$ (in particular recalling that $e\otimes c_{1} \cong c_{2}$ and that $d_{3}, d_{2}, d_{4},d_{1}$ have the fusion rules of $\TC$) that $\CD(\DD_8)_{A_L}^\text{op}$ decomposes as $S_1 \boxplus S_1\boxplus S_m$ as a $\TC$-module 1-category. Analogously to the discussion in the previous section, this is the surface that appears when we pinch $\CI$ to the right (the opposite direction to the one in Fig.~\ref{fig:pinching I}). Performing this maneuver results in the genuine lines in $\CD(\DD_8)_{A_L}^\text{op}$ being pulled directly into $\TC$ while the Abelian simple lines (i.e., $d_{3}, d_{2}, d_{4}, d_{1}$) bound another copy of $S_1$ (corresponding to an unfaithful symmetry), and the non-Abelian non-local lines (i.e., $c_{1}$ and $c_{2}$) bound a trailing $S_m$ surface when they are pulled into the $\TC$ bulk. Our discussion is also clearly consistent with associativity of interface fusion and, in particular
\begin{equation}
3\cdot1 \oplus m \cong \CI(A_L) \cong (S_1\boxplus S_1\boxplus S_m)(1) \cong 3\cdot 1 \oplus m\, .
\end{equation}
As in the example in Sec.~\ref{TCDS3}, we again see that the non-invertibility of $S_m$ is crucial in reproducing the fact that $A_L$ only partially condenses. Putting everything together, one finds that
\begin{equation}
{\Mod(\TC)}^*_{\Mod(S_1 \boxplus S_1 \boxplus S_m)} \simeq \Mod(\CD(\DD_8)) \, ,
\end{equation}
which formalizes the notion that gauging the non-invertible symmetry encoded in the algebra $S_1 \boxplus S_1\boxplus S_m$ in $\TC$ yields $\CD(\DD_8)$.

As a simple check of the above discussion, we can determine $\CA_S$ and $\widehat\CA_S$ by specifying the action of $\CI$ on genuine lines (recall the discussion in Fig.~\ref{fig:A_S action from I} and the surrounding text of Sec.~\ref{OPEdetails}). Indeed, using the maps in the paragraph above the one containing \eqref{eeFusion}, we can easily construct the associated $W$ matrix discussed in \eqref{Wdefn}. Then, computing $W^{\dagger}W$ and $WW^{\dagger}$ allows us to confirm that $\CA_S \simeq S_{A_L}$ and $\widehat\CA_S \simeq S_1\boxplus S_1\boxplus S_m$, recalling that $S_m(1) \cong S_m(m) \cong 1 \oplus m$ while the action of $S_m$ on $e$ and $f$ does not result in any genuine lines in $\TC$.

From our discussion in Sec.~\ref{sec:oneshotNshot} and footnote \ref{SeqFoot}, we know that the above non-invertible gauging that we have done in ``one shot'' can instead be performed via two invertible shots. Indeed, we notice that $\CD(\DD_{8})_{A_L}$ contains the fusion subcategories $\Vect_{\mathbb Z_2 \oplus \mathbb Z_2}$ and $\Vect_{\mathbb Z_2 \oplus \mathbb Z_2 \oplus \mathbb Z_2}$, with simple objects $\{1, e, m, f\}$ and $\{1,  e,  m,  f,  d_{3},  d_{2},  d_{4},  d_{1}\}$, respectively, such that
\begin{equation}
    \Vect_{\mathbb Z_2 \oplus \mathbb Z_2} < \Vect_{\mathbb Z_2 \oplus \mathbb Z_2 \oplus \mathbb Z_2} < \CD(\DD_{8})_{A_L} \, .   
\end{equation}
Furthermore, from the fusion structure of $\mc D(\mathbb D_8)_{A_L}$, one sees that $\Vect_{\mathbb Z_2 \oplus \mathbb Z_2 \oplus \mathbb Z_2}$ and $\mc D(\mathbb D_8)_{A_L}$ are $\mathbb Z_2$ extensions of $\Vect_{\mathbb Z_2 \oplus \mathbb Z_2}$ and $\Vect_{\mathbb Z_2 \oplus \mathbb Z_2 \oplus \mathbb Z_2}$ respectively.
Applying Claim 5.5 in Sec.~\ref{sec:oneshotNshot}, we know that gauging the symmetry associated with $\CD(\DD_{8})^{\rm op}_{A_L}\simeq S_1\boxplus S_1\boxplus S_m$ admits a decomposition as a 2-step gauging of 0-form symmetries, which consists in first performing the gauging associated with the separable algebra $\Vect_{\mathbb Z_2 \oplus \mathbb Z_2 \oplus \mathbb Z_2}^{\rm op}$ before performing that associated with the separable algebra ${\mc Z_{\Vect_{\mathbb Z_2 \oplus \mathbb Z_2 \oplus \mathbb Z_2}}(\mc D(\mathbb D_8)_{A_L}^{\rm op})}^{\rm op}$, which is given by the relative center of $\mc D(\mathbb D_8)_{A_L}^{\rm op}$ over $\Vect_{\mathbb Z_2 \oplus \mathbb Z_2 \oplus \mathbb Z_2}$. Since $\mc Z^{\mc D(\mathbb Z_2)}(\Vect_{\mathbb Z_2 \oplus \mathbb Z_2 \oplus \mathbb Z_2}^{\rm op}) \simeq \mc D(\mathbb D_4)$, by Claim 5.6 of Sec.~\ref{sec:oneshotNshot}, gauging the symmetry associated with $S_{1}\boxplus S_{1}\boxplus S_{m}$ in $\Mod(\TC)$ admits a decomposition as sequential gauging of invertible $\zz_{2}$-symmetries: 
\begin{equation}
    \TC \longrightarrow \mc D(\mathbb D_4) \longrightarrow \mc D(\mathbb D_8) \, ,
\end{equation}
where the intermediate step is through the second theory appearing in Sec.~\ref{TCDZ4}.

Let us conclude with some comments regarding the fusion structure of $\mc D(\mathbb D_8)_{A_L}$. So far, we have computed its fusion ring, which is equivalent to a tensor product of the fusion rings of $\Rep(\zz_{2})$ and $\msf{TY}(\zz_{2}\oplus \zz_{2},\chi,\epsilon)$, respectively, where $\msf{TY}(\zz_{2}\oplus \zz_{2},\chi,\epsilon)$ is one of the $\mathbb Z_2 \oplus \mathbb Z_2$ Tambara--Yamagami fusion categories\cite{Tambara:1998vmj}. In fact, we have an equivalence of fusion categories $\mc D(\mathbb{D}_{8})_{A_{L}}\simeq \Rep(\zz_{2})\boxtimes \msf{TY}(\zz_{2}\oplus \zz_{2},\chi,\epsilon)$ for
\begin{equation}
\chi(g,h) := g_{1}g_{2}+h_{1}h_{2}\, , \q \epsilon := 1 \, ,
\end{equation}
for every $g,h \in \zz_{2}\oplus \zz_{2}$. In fact, for this choice of $(\chi,\epsilon)$, we have $\msf{TY}(\zz_{2}\oplus \zz_{2},\chi,\epsilon) \simeq \Rep(\DD_{8})$. A similar analysis can be used to yield a theory with symmetry $\Mod(\CD(\mathbb{Q}_{8}))$, where $\mathbb{Q}_{8}$ is the quaternion group of order 8. In this case, we choose the same fusion ring for $\CA_S \simeq S_{1}\boxplus S_{1}\boxplus S_{m}$, but we change the associator via $\chi(g,h) := g_{1}g_{2}+h_{1}h_{2}$, and $\epsilon := -1$.

\subsec{Example: $\TC\to\CD(\mathbb{Z}_3)$}\label{TCDZ3}

In this section, we give a toy example of a gauging that is not pure 0-form or pure 1-form. Unlike the case of pure 1-form symmetry gauging, neither $\CA_S$ nor $\widehat\CA_S$ is indecomposable as an object. Moreover, unlike the case of pure 0-form symmetry gauging, the trivial surface is not part of $\CA_S$ or $\widehat\CA_S$. Instead, we will see that the surface algebra is built from multiple copies of non-trivial surfaces.

Note that there are no 0-form symmetries of $\TC$ that can be gauged to obtain $\CD(\zz_{3})$ or vice-versa. This is because the order of the theories are coprime to each other. However, we can sequentially gauge a 1-form symmetry of $\TC$ so as to obtain the trivial theory before gauging an unfaithful 0-form $\mathbb Z_3$ symmetry. Both operations were derived above and are encoded into the Morita equivalence of fusion 2-categories
\begin{equation}
{\Mod(\TC)}^*_{\Mod(S_e)} \simeq \msf{2Vec} \q \text{and} \q {\msf{2Vec}}^*_{\Mod(S_1 \boxplus S_1 \boxplus S_1)} \simeq \Mod(\mc D(\mathbb Z_3)) \, ,
\end{equation}
where the surfaces $S_e$ and $S_1 \boxplus S_1 \boxplus S_1$ are endowed with the separable algebra structures encoded in the fusion 1-categories $\Vect_{\mathbb Z_2}$ and $\Vect_{\mathbb Z_3}$ respectively. These two operations can be combined in the following way: consider the fusion category $\Vect_{\mathbb Z_2} \boxtimes \Vect_{\mathbb Z_3}^{\rm op}$. It has the structure of a separable algebra $\mc A_S$ in $\Mod(\TC)$ from the fact that $\TC$ is a braided fusion subcategory of $\mc Z(\Vect_{\mathbb Z_2} \boxtimes \Vect_{\mathbb Z_3}^{\rm op})$. Moreover, one can readily check that $\mc A_S$ decomposes as $S_e \boxtimes_{\TC} S_1^{\boxplus 3} \simeq S_e^{\boxplus 3}$ as a $\TC$-module category. Putting everything together, one obtains
\begin{equation}
{\Mod(\TC)}^*_{\Mod(S_e \boxplus S_e \boxplus S_e)} \simeq \Mod(\mc D(\mathbb Z_3)) \, .
\end{equation}
Essentially, this procedure amounts to defining the interface $\mc I$ as in Fig.~\ref{fig:Gensandwich} choosing the Lagrangian algebras $1 \oplus e$ and $1 \oplus \omega \oplus \bar \omega$ in $\TC$ and $\mc D(\mathbb Z_3)$, respectively. The interface $\mc I$ can be thought of as gauging a (trivial) split 2-group symmetry $\mathbb G := (\mathbb Z_3^{[0]},\mathbb Z_2^{[1]},1,1)$ where $\mathbb Z_3^{[0]}$ is an unfaithful 0-form symmetry and $\mathbb Z_2^{[1]}$ is the 1-form symmetry included in $\Mod(\TC)$ associated with $1 \oplus e$. By similar logic, we can think of $\CI^{\dagger}$ as implementing the gauging of a dual split 2-group, $\widetilde{\mathbb G} :=(\zz_{2}^{[0]},\zz^{[1]}_{3},1,1)$ where $\mathbb Z_2^{[0]}$ is an unfaithful 0-form symmetry, and $\mathbb Z_3^{(1)}$ is the 1-form symmetry included in $\Mod(\mc D(\mathbb Z_3))$ associated with $1 \oplus \omega \oplus \bar \omega$. This dual process is associated with a surface algebra $\widehat{\mc A}_S$ that decomposes as $S_\omega \boxplus S_\omega$ as a $\mc D(\mathbb Z_3)$-module category.

\subsec{Constraints on $\TC$ gaugings and target theories}\label{TCconstraints}
In this section, we illustrate some of the implications of the constraints discussed in Sec.~\ref{GenConstraints} on gaugings of toric code. To that end, we first note that Constraint 4.1 immediately rules out gauging symmetries associated with algebras $\mc A_S$ whose underlying surfaces are $S_{me}$, $\tilde S_{me}$, or $S_f$.
Indeed, the corresponding higher-gauged 1-form symmetries are all non-commutative (although we will return to the case of $\CA_S$ with underlying object $S_f$ in the discussion section). These basic facts were discussed from a slightly different perspective at the end of Sec.~\ref{TCVec}.

From Constraint 4.2 (or Constraint 4.3), we know that inserting networks built from $\widehat{A}_S$ with underlying surfaces $S_1 \boxplus S_e$, $S_1 \boxplus S_m$, or $S_1 \boxplus S_f$ should correspond to (non-)invertible 0-form symmetry gauging and that the dual 1-form symmetry should be encoded in an algebra object, $A_L$, built from two simple lines. Indeed, we have explicitly gauged these symmetries in the previous subsections and shown the dual 1-form symmetry has the predicted form. 

Next, from Constraint 4.4 we know that there is no way to gauge a symmetry associated with a surface algebra $\widehat{\mc A}_S$ with underlying surfaces $S_1 \boxplus S_{me}$ or $S_1 \boxplus \tilde S_{me}$,
where these surfaces are of the type discussed around \eqref{AllLag}. We can also apply Constraint 4.5 and conclude that any surface algebra containing both $S_e$ and $S_m$,
where we defined $S_e$ and $S_m$ as in Fig.~\ref{fig:condensation surface A_S from slab theory} cannot correspond to a gaugeable 0-form symmetry. Finally, it is also interesting to note that, as fusion 1-categories, the surface algebras of the form $\widehat \CA_{S} \simeq S_1 \boxplus S_{\widetilde A}$ in the examples of the previous parts of Sec.~\ref{ToricCode} are indeed generalized near group fusion 1-categories (as discussed in  Footnote \ref{GenNG}). For example, in the case of $\TC\to \mc D^{\alpha_p}(\mathbb D_6)$, the surface algebras are different $(\mathbb{Z}_2\oplus \mathbb{Z}_2,\mathbb{Z}_2,0,1)$ categories.

\newsec{Discussion and Conclusion}
In this paper, we have described how to generalize various aspects of invertible 0-form gauging to the world of non-invertible symmetries in (2+1)d. Clearly, many open questions remain, among them:

\begin{itemize}
\item We have discussed the case of gauging non-invertible symmetries in bosonic topological orders in (2+1)d. One natural generalization is to spin TQFTs. For example, in the case of $\TC$, we expect that it is possible to make sense of $\CA_S \simeq S_f$ as fermionic 1-form symmetry gauging in the presence of a spin structure (e.g., we can construct an analog of \eqref{gaugeSeT3} by turning on an appropriate spin structure). It would be interesting to understand generalizations of the formalism in Sec.~\ref{OPE} for this and other spin-structure-dependent theories. 
\item In the same spirit as the previous item, it would be interesting to generalize our results to non-topological QFTs. We expect that much of what we have written in this paper generalizes to this setting, but it would be useful to describe the generalization of Sec.~\ref{OPE} more precisely. 
\item In \cite{Barkeshli:2014cna}, the authors defined and made use of $G$-crossed modularity to derive interesting properties of TQFTs with 0-form symmetry group $G$ on arbitrary 2d spatial slices. It would be worthwhile to understand how to generalize $G$-crossed modularity to the non-invertible case. Already in Sec.~\ref{sec:number of twisted sectors A_S and hat A_S}, we found a non-invertible analog of the fixed point theorem relating $G$-symmetry action on lines with twisted sector states that was proven in \cite{Barkeshli:2014cna} via $G$-crossed modularity. However, our approach did not make direct use of ``generalized'' modularity, and we did not touch upon generalizations of other results in \cite{Barkeshli:2014cna} to the non-invertible setting.
\item We have presented notions of generalized symmetry fractionalization and discrete torsion. It could be useful to develop a corresponding cohomology theory. Perhaps this idea is related to Davydov--Yetter cohomology \cite{Kitaev:2005hzj,etingof2016tensor,cui2019generalized}. In this context it would also be interesting to make more direct contact with the constructions in \cite{Carqueville:2018sld,Mulevicius:2020tgg,Mulevicius:2022gce,Heinrich:2025wkx,Carqueville:2025kqs}.
\item It would be interesting to further constrain the gaugeability of generalized symmetries. In particular, it would be interesting to understand the structure of anomalies in this case and the associated cohomology theory (See \cite{Hsin:2025ria} for some interesting recent work in this direction). It would also be useful to better understand which collections of surfaces correspond to gaugeable 0-form symmetries (e.g., to generalize Constraints 4.2 and 4.3) and which correspond to gaugeable generalized higher structures. In principle, the corresponding $\CA_S$ can be defined by using the discussion in Sec.~\ref{OPEdetails} and the relative Deligne product over more general ``sandwiches'' of TQFTs.
\item In light of recent advances \cite{Verresen:2021wdv,PhysRevX.14.021040,PRXQuantum.4.020339,Ren:2024ayb}, it could be fruitful to revisit the complexity theory for producing non-Abelian topological orders from Abelian ones if one includes the non-invertible 0-form symmetry gaugings we have described.
\item Our discussion in this paper has been mostly in terms of defects. It would be interesting to understand our constructions from the ``Poincar\'e dual'' perspective of generalized gauge fields.
\end{itemize}

\ack{We thank Zhihao Duan, Terry Gannon, Roman Geiko, Milo Moses, Pavel Putrov, Okasha Uddin, Nati Seiberg, Sahand Seifnashri, and Bowen Shi for discussions. M.~B. thanks the Simons Center for Geometry and Physics, the Institute for Advanced Study, and the International Centre for Mathematical Sciences for hospitality during various stages of this work. R.~R. gratefully acknowledges the hospitality of the Simons Center for Geometry and Physics, International Centre for Mathematical Sciences, University of California Los Angeles, Kavli Institute for Theoretical Physics at UC Santa Barbara, and the International Centre for Theoretical Sciences, where part of this work was carried out. M.~B.'s and M.~K.~N.~B.'s work is supported in part by STFC under the grant, \lq\lq Amplitudes, Strings and Duality.” M.~B.'s work was also partly supported by the Royal Society under the grant, “Relations, Transformations, and Emergence in Quantum Field Theory.” M.~K.~N.~B. thanks International Center for Theoretical Sciences and IISER Bhopal for the hospitality, where part of this work was carried out. No new data were generated or analyzed in this study.}

\newpage
\begin{appendices}

\newsec{1-category symmetry, gauging, and Morita duality\label{sec:Morita1}}
We begin by reviewing some of the mathematical formalism behind gauging 0-form symmetries in (1+1)d as well as gauging 1-form symmetries along co-dimension one submanifolds in (2+1)d and in the (2+1)d bulk itself. This appendix also serves to motivate our perspective on 2-category symmetries in the main text. We encourage the reader to consult \cite{etingof2016tensor} for more complete mathematical definitions.

In (1+1)d,  (non-)invertible symmetry should be encoded in a (pivotal unitary) fusion 1-category \cite{Frohlich:2006ch,Bhardwaj:2017xup,Chang:2018iay,Thorngren:2019iar,PhysRevResearch.2.043086}. Given such a category, $\mc C$, we denote representatives of the finitely many isomorphism classes of simple objects in  $\mc C$ by $x_i \in \text{Irr}(\mc C)$. We write their respective quantum dimensions as $d_{x_i}\in \mathbb{R}_+$. The fusion structure of $\mc C$ is described by the triple, $(\otimes, 1_\mc C, \alpha)$, where $\otimes : \mc C \otimes \mc C \to \mc C$ is the tensor product bifunctor, $1_\mc C \in \text{Irr}(\mc C)$ is the unit object, and $\alpha : (- \otimes -) \otimes - \xrightarrow{\sim} - \otimes (- \otimes -)$ is the monoidal associator (recall that $\alpha$ satisfies a ``pentagon axiom'' \cite{etingof2016tensor}).

Twisted local operators charged under the fusion 1-category symmetry, $\mc C$, are encoded in the Drinfeld center, $\mc Z(\mc C)$, of $\mc C$. Recall that an object in $\mc Z(\mc C)$ is a pair $(X,R_{-,X})$ consisting of an object $X\in\mc C$ and a natural isomorphism $R_{-,X} : - \otimes X \xrightarrow{\sim} X \otimes -$, which is required to satisfy a ``hexagon axiom'' involving the monoidal associator, $\alpha$, of $\mc C$. Crucially, for every $\mc C$, its Drinfeld center, $\mc Z(\mc C)$, is a braided fusion 1-category, whose braiding is inherited from the natural isomorphisms $R$ entering the definition of the objects. Furthermore, it is non-degenerate, in the sense that the centralizer $\mc Z^\mc C(\mc C)$ of $\mc Z(\mc C)$ in itself---which consists  of the full subcategory of objects $(X,R_{-,X})$ in $\mc Z(\mc C)$ such that $R_{X_0,X} \circ R_{X,X_0} = \text{id}_{X_0 \otimes X_1}$, for every $X_0 \in \text{Irr}(\mc C)$---is equivalent to the category $\Vect$ of finite-dimensional complex vector spaces.

In a (1+1)d physical theory with a fusion 1-category symmetry $\mc C$, the various ways to gauge $\mc C$ are in one-to-one correspondence with (left) indecomposable finite semisimple module 1-categories over $\mc C$ \cite{Bhardwaj:2017xup}. Given such a $\mc C$-module 1-category, $\mc M$, we denote representatives of the finitely-many isomorphism classes of simple objects in $\mc M$ by $M_a \in \text{Irr}(\mc M)$. The $\mc C$-module structure of $\mc M$ is encapsulated in a pair, $(\act, \alpha^\act)$, where $\act : \mc C \otimes \mc M \to \mc M$ is the action bifunctor, and $\alpha^{\act} : (- \otimes - ) \act - \xrightarrow{\sim} - \act (-  \act -)$ is the module associator (which is required to satisfy a pentagon axiom involving $\alpha$).

The symmetry of the theory resulting from performing the gauging associated with $\mc M$ is encoded in the Morita dual, $\mc C^*_\mc M$, of $\mc C$ with respect to $\mc M$. $\mc C^*_{\CM}$ is defined as the fusion 1-category, $\msf{Fun}_\mc C(\mc M,\mc M)$, of $\mc C$-module endofunctors of $\mc M$. Recall that an object in $\msf{Fun}_\mc C(\mc M,\mc M)$ is a pair $(F,\omega)$ consisting of a functor, $F : \mc M \to \mc M$, and a natural isomorphism, $\omega : F(- \act -) \xrightarrow{\sim} - \act F(-)$ (which is required to satisfy a pentagon axiom involving $\alpha^{\act}$). In this situation, we refer to $\mc C$ and $\mc C^*_\mc M$ as being Morita equivalent. Importantly, the Drinfeld center is an invariant of Morita equivalence in the sense that $\mc Z(\mc C^*_\mc M) \simeq \mc Z(\mc C)$, for every indecomposable finite semisimple $\mc C$-module 1-category $\mc M$. 

The various ways of gauging a fusion 1-category symmetry, $\mc C$, can equivalently be expressed in terms of algebras in $\mc C$. Given such an algebra $A_L$ in $\mc C$, we write the unital associative structure as $(\mu_{A_L},\eta_{A_L})$, where $\mu_{A_L} : A_L \otimes A_L \to A_L$ is the multiplication morphism, and $\eta_{A_L} : 1 \to A_L$ is the unit morphism (these quantities are required to satisfy a ``pentagon axiom'' involving $\alpha$ and ``square axioms''  involving the unit, $1_\mc C$, respectively). Recall that a right module over $A$ consists of an object $M\in\mc C$ and a morphism, $\mu_M : M \otimes A_L \to M$ (which is required to satisfy a ``pentagon axiom'' involving $\alpha$ and the multiplication $\mu_A$).

For every indecomposable finite semisimple (left) $\mc C$-module 1-category $\mc M$, one can find an algebra $A_L$ in $\mc C$ such that the category $\mc C_{A_L}:=\Mod_\mc C(A_L)$ of right $A_L$-modules in $\mc C$ is equivalent to $\mc M$ \cite{Ostrik:2002ohv,ostrik2003module,etingof2016tensor}. Note that Morita equivalent algebras yield equivalent module 1-categories via this construction. Given a $\mc C$-symmetric theory, the actual gauging operation amounts---after endowing $A_L$ with a $\Delta$-separable symmetric Frobenius structure \cite{Fuchs_2002,fuchs2009,kong2019}---to inserting a network of lines labeled by $A_L$ \cite{Bhardwaj:2017xup}.\footnote{In the context of inserting this network along co-dimension one manifolds of a (2+1)d bulk, this procedure is referred to in the high-energy theory literature as ``higher gauging'' \cite{Roumpedakis_2023}.} Alternatively, it can be performed directly in terms of the $\mc C$-module 1-category $\mc M$ as in \cite{Lootens:2021tet,Lootens:2022avn,Lootens:2023wnl,Delcamp:2024cfp}.

Conversely, given a finite semisimple $\mc C$-module 1-category, $\mc M$, a corresponding algebra object $A_L$ in $\mc C$ can be extracted using the internal hom construction. Given $M_1,M_2 \in \text{Irr}(\mc M)$, the internal hom object associated with $M_1$ and $M_2$ is an object, $\underline{\Hom}(M_1,M_2)\in\mc C$, such that $\Hom_\mc C(X \act M_1 , M_2) \cong \Hom_\mc M(X,\underline{\Hom}(M_1,M_2))$ for every $X \in \text{Irr}(\mc C)$. Importantly, for any $M_1,M_2 \in \text{Irr}(\mc M)$, $\underline{\Hom}(M_1,M_1)$ can always be equipped with the structure of an algebra in $\mc C$, and $\underline{\Hom}(M_1,M_2)$ can be equipped with the structure of a right $\underline{\Hom}(M_1,M_1)$-module in $\mc C$ \cite{ostrik2003module,SCHAUMANN2013382}. Furthermore, provided $\mc M$ is indecomposable, the functor $\mc M \ni M_2 \to \underline{\Hom}(M_1,M_2) \in \Mod_\mc C(\underline{\Hom}(M_1,M_1))$ is an equivalence of $\mc C$-module 1-categories. Physically, one can interpret this result as follows: given a finite symmetry $\mc C$, consider gauging the $\Delta$-separable symmetric Frobenius algebra $A_L$ in $\mc C$ within a strip. Denoting by $\mc M$ the corresponding $\mc C$-module 1-category, $\Mod_\mc C(A_L)$, the resulting interfaces can be encoded in the categories $\msf{Fun}_\mc C(\mc C,\mc M) \simeq \mc C$ and $\msf{Fun}_\mc C(\mc M,\mc C) \simeq \mc M^\text{op}$, respectively. Fusing these interfaces result in a line defect in $\mc C$ according to the composition of $\mc C$-module functors $\msf{Fun}_\mc C(\mc C,\mc M) \otimes \msf{Fun}_\mc C(\mc M,\mc C) \to \msf{Fun}_\mc C(\mc C,\mc C)$. In particular, choosing the interfaces associated with the same simple object, $M \in \text{Irr}(\mc M)$, results in a topological line labeled by $\underline{\Hom}(M,M)$, which is Morita equivalent to $A_L$.

Let us consider an explicit example. Let $\Vect_G \simeq \Mod(\mathbb C^G)$ be the fusion 1-category of $G$-graded vector spaces. Given a subgroup, $K \le G$, and a normalised representative, $\phi$, of a cohomology class in $H^2(K,{\rm U}(1))$, the twisted group algebra, $\mathbb C[K]^\phi$, has the structure of an algebra in $\Vect_G$ such that the grading is induced by $K \le G$. We denote by $\mc M(K,\phi)$ the category $\Mod_{\Vect_G}(\mathbb C[K]^\phi)$. It is the indecomposable finite semisimple $\Vect_G$-module 1-category such that $\text{Irr}(\mc M(K,\psi)) \cong G / K$ with module associator defined via $H^2(G,\text{Fun}(G/K,\rU(1))) \cong H^2(K,\rU(1))$ (as follows from Shapiro's lemma). One can check that every indecomposable finite semisimple $\Vect_G$-module 1-category is of the form $\mc M(K,\phi)$ for some subgroup, $K \le G$, and normalised representative, $\phi$, of a cohomology class in $H^2(K,\rU(1))$ \cite{Ostrik:2002ohv,ostrik2003module}. Physically, one interprets the gauging operation of the symmetry $\Vect_G$ associated with $\mc M(K,\phi)$ as the $\phi$-twisted gauging of the sub-symmetry $\Vect_K$ (e.g., See \cite{Brunner:2014lua,Bhardwaj:2017xup,Lootens:2021tet,Lootens:2024gfp,Lu:2024ytl,Lu:2024lzf,Putrov:2024uor}). More generally, given any finite group, $K$, equipped with a group homomorphism, $K \to G$, and normalised representative $\phi$ of a cohomology class in $H^2(K,{\rm U}(1))$, the twisted group algebra $\mathbb C[K]^\phi$ has the structure of a unital associative algebra in $\Vect_G$. In particular, choosing $K = G$ and $\phi=1$, one obtains the $\Vect_G$-module 1-category $\mc M(G,1) \simeq \Vect$. The Morita dual, $(\Vect_G)^*_\Vect$, of $\Vect_G$ with respect to $\Vect$ turns out to be equivalent to the category, $\Rep(G) \simeq \Mod(\mathbb C[G])$, of $G$ representations. It then follows from the Morita equivalence between $\Vect_G$ and $\Rep(G)$ that indecomposable module 1-categories over $\Rep(G)$ are in one-to-one correspondence with those over $\Vect_G$. Concretely, given a pair $(K \leq G ,\phi)$, the corresponding $\Rep(G)$-module 1-category is equivalent to the category $\Rep^\phi(K)$ of $\phi$-projective representations of $K$. 

An interesting aspect of the axiomatisation of a finite semisimple non-invertible symmetry in (1+1)d in terms of a pivotal unitary fusion 1-category $\mc C$ is that---after endowing it with a spherical structure---the same data serves as input to a three-dimensional state-sum topological quantum field theory via the Turaev--Viro--Barrett--Westbury construction \cite{Turaev:1992hq, Barrett:1993ab}. In particular, this correspondence underlies the so-called symmetry topological field theory (SymTFT) proposal \cite{Apruzzi:2021nmk,Freed:2022qnc,Kaidi:2022cpf,Kaidi:2023maf,Delcamp:2024cfp}, and earlier incarnations thereof \cite{Fuchs_2002,KONG201762,PhysRevResearch.2.043086,PhysRevResearch.2.033417,Ji:2021esj,Chatterjee:2022kxb,Huang:2023pyk,Huang:2024ror,PhysRevLett.112.247202,PhysRevLett.121.177203}. In this context, the Drinfeld center, $\mc Z(\mc C)$, which is the quantum invariant the partition function assigns to the circle, encodes the bulk topological defects, or equivalently, the so-called (deconfined) anyonic excitations of the Hamiltonian realization of the theory. 

Given the Hamiltonian realization of a (2+1)d TQFT associated with spherical fusion 1-category $\mc C$, we can sometimes gauge 1-form symmetries associated with these anyons. This gauging procedure is controlled by a choice of condensable algebra in $\mc Z(\mc C)$ \cite{bais2009condensate,Kong:2013aya}. Recall that a condensable algebra in $\mc Z(\mc C)$ is defined as a connected commutative separable algebra $A_L$ in $\mc Z(\mc C)$, where connectedness supposes that $\Hom_{\mc Z(\mc C)}(1_{\mc Z(\mc C)},A) \cong \mathbb C$, commutativity is the statement that $\mu_{A_L} = \mu_{A_L} \circ R_{A_L,A_L}$, while separability requires that the multiplication map $\mu_{A_L}$ splits as a morphism of $(A_L,A_L)$-bimodules \cite{etingof2016tensor}.

Gauging the 1-form symmetry corresponding to $A$ results in a new collection of deconfined anyonic excitations encoded in the MTC, $\mc Z(\mc C)^\text{loc}_{A_L}$, of local $A$-modules in $\mc Z(\mc C)$, where the braiding is inherited from that of $\mc Z(\mc C)$. Recall that a local $A_L$-module is an $A_L$-module $(M,\mu_M)$ such that $\mu_M \circ R_{A_L,M} \circ R_{M,A_L} = \mu_M$.  On the other hand, non-local modules encode confined anyonic excitations. Such a condensation produces a gapped one-codimensional defect separating the topological theories with bulk defects $\mc Z(\mc C)$ and $\mc Z(\mc C)^\text{loc}_{A_L}$, respectively. 
Topological operators along the defect are encoded in the fusion 1-category, $\mc Z(\mc C)_{A_L}$, of (not necessarily local) $A_L$-modules in $\mc Z(\mc C)$. The bulk-boundary correspondence then imposes that $\mc Z(\mc Z(\mc C)_A) \simeq \mc Z(\mc C) \boxtimes \overline{\mc Z(\mc C)_{A_L}^\text{loc}}$, where $\overline{\mc Z(\mc C)_{A_L}^\text{loc}}$ is the braided fusion 1-category with inverse braiding. In particular, this construction realises the Witt equivalence between $\mc Z(\mc C)$ and $\mc Z(\mc C)_{A_L}^\text{loc}$. A special situation occurs whenever the condensable algebra $A_L$ in $\mc Z(\mc C)$ is a  Lagrangian algebra, i.e. $d_{A_L}^2 = \CD^2_{\mc Z(\mc C)}$. In this case, $\mc Z(\mc C)^\text{loc}_{A_L} \simeq \Vect$, and $\mc Z(\mc C)_{A_L}$ is Morita equivalent to $\mc C$. Therefore, it implies that there is a module 1-category, $\mc M$ over $\mc C$, such that $\mc C^*_\mc M \simeq \mc Z(\mc C)_{A_L}$. One can also realize $\mc M$ as the category of modules over an algebra in $\mc C$, in which case the Lagrangian algebra $A_L$ in $\mc Z(\mc C)$ is obtained as its full center \cite{DAVYDOV2010319}. 

Consider for instance the fusion 1-category $\Vect_G$. Simple objects in $\mc Z(\Vect_G)$ are labeled by a choice of conjugacy class and an irreducible representation of the centralizer in $G$ of one of its representatives. We identify several canonical families of condensable algebras \cite{DAVYDOV2017149}. First, let $N \unlhd G$ be a normal subgroup of $G$. Then, $\mathbb C[N]$, with grading  induced by $N \unlhd G$, is a condensable algebra in $\mc Z(\Vect_G)$. As an object in $\mc Z(\Vect_G)$ it is equivalent to $\bigoplus_{[g] \in \text{Cl}(N)}([g],1)$, where $\text{Cl}(N)$ denotes the set of conjugacy classes of $G$ contained in $N$, in such a way that $\mc Z(\Vect_G)^\text{loc}_{\mathbb C[N]} \simeq \mc Z(\Vect_{G/N})$. In particular, choosing $N=G$ yields a Lagrangian algebra, which is that associated with the $\Vect_G$-module 1-category $\Vect$. Second, for any subgroup $K \le G$, $\mathbb C^{G/K}$ defines an algebra in $\Rep(G)$ and, by extension, in $\mc Z(\Vect_G)$ (since $\Rep(G)$ is a fusion subcategory of $\mc Z(\Vect_G)$). As an object in $\mc Z(\Vect_G)$, $\mathbb C^{G/K}$ is equivalent to $([\mathbb 1_G],\text{Ind}_K^G(1))$, where $\text{Ind}_K^G(1)$ designates the induced representation in $G$ of the trivial representation of $K$, in such a way that $\mc Z(\Vect_G)^\text{loc}_{\mathbb C^{G/K}} \simeq \mc Z(\Vect_K)$. In particular, choosing $K=\{\mathbb 1_G\}$ yields another Lagrangian algebra, which is that associated with the module 1-category $\Vect_G$ over itself.

\newsec{Determining the indecomposable surfaces in $\widehat \CA_S$}\label{DetSimple}

In Sec.~\ref{OPE} we determined various properties of the surfaces $\CA_S$ and $\widehat \CA_S$. In particular, we learned from Fig.~\ref{fig: local operators on A_S} that a basis of the vector space of local operators, $V_{\widehat \CA_S}$, on $\widehat \CA_S$ is  in one-to-one correspondence with lines that can end on both $\CI$ and $\CI^{\dagger}$. If the vector space of local operators is $N$-dimensional, then $\widehat\CA_S$ decomposes into $N$ indecomposable surfaces. Let us assume that the interface $\CI$ corresponds to gauging the 1-form symmetry corresponding to the anyon $A_L$.

\begin{figure}[h!]
\centering

\tikzset{every picture/.style={line width=0.75pt}} 

\begin{tikzpicture}[x=0.75pt,y=0.75pt,yscale=-1,xscale=1]

\draw  [color={rgb, 255:red, 0; green, 0; blue, 0 }  ,draw opacity=1 ][fill={rgb, 255:red, 192; green, 188; blue, 188 }  ,fill opacity=1 ] (60.01,146.17) -- (60.03,64.4) -- (136.22,27.05) -- (136.21,108.82) -- cycle ;
\draw  [color={rgb, 255:red, 0; green, 0; blue, 0 }  ,draw opacity=1 ][fill={rgb, 255:red, 192; green, 188; blue, 188 }  ,fill opacity=1 ] (167.17,147) -- (167.18,65.23) -- (243.38,27.89) -- (243.36,109.66) -- cycle ;
\draw  [color={rgb, 255:red, 139; green, 6; blue, 24 }  ,draw opacity=1 ][fill={rgb, 255:red, 139; green, 6; blue, 24 }  ,fill opacity=1 ] (96.93,83.8) .. controls (96.93,83.14) and (97.46,82.61) .. (98.12,82.61) .. controls (98.77,82.61) and (99.3,83.14) .. (99.3,83.8) .. controls (99.3,84.45) and (98.77,84.98) .. (98.12,84.98) .. controls (97.46,84.98) and (96.93,84.45) .. (96.93,83.8) -- cycle ;
\draw  [color={rgb, 255:red, 139; green, 6; blue, 24 }  ,draw opacity=1 ][fill={rgb, 255:red, 139; green, 6; blue, 24 }  ,fill opacity=1 ] (204.27,83.44) .. controls (204.27,82.79) and (204.8,82.26) .. (205.46,82.26) .. controls (206.11,82.26) and (206.64,82.79) .. (206.64,83.44) .. controls (206.64,84.1) and (206.11,84.63) .. (205.46,84.63) .. controls (204.8,84.63) and (204.27,84.1) .. (204.27,83.44) -- cycle ;
\draw [color={rgb, 255:red, 139; green, 6; blue, 24 }  ,draw opacity=1 ]   (98.12,84.15) -- (205.27,83.44) ;
\draw [color={rgb, 255:red, 139; green, 6; blue, 24 }  ,draw opacity=1 ]   (82.63,109.18) -- (189.79,108.48) ;
\draw  [color={rgb, 255:red, 139; green, 6; blue, 24 }  ,draw opacity=1 ][fill={rgb, 255:red, 139; green, 6; blue, 24 }  ,fill opacity=1 ] (188.6,108.83) .. controls (188.6,108.18) and (189.13,107.65) .. (189.79,107.65) .. controls (190.44,107.65) and (190.97,108.18) .. (190.97,108.83) .. controls (190.97,109.49) and (190.44,110.02) .. (189.79,110.02) .. controls (189.13,110.02) and (188.6,109.49) .. (188.6,108.83) -- cycle ;
\draw  [color={rgb, 255:red, 139; green, 6; blue, 24 }  ,draw opacity=1 ][fill={rgb, 255:red, 139; green, 6; blue, 24 }  ,fill opacity=1 ] (81.45,109.18) .. controls (81.45,108.53) and (81.98,108) .. (82.63,108) .. controls (83.29,108) and (83.82,108.53) .. (83.82,109.18) .. controls (83.82,109.84) and (83.29,110.37) .. (82.63,110.37) .. controls (81.98,110.37) and (81.45,109.84) .. (81.45,109.18) -- cycle ;
\draw  [color={rgb, 255:red, 0; green, 0; blue, 0 }  ,draw opacity=1 ][fill={rgb, 255:red, 192; green, 188; blue, 188 }  ,fill opacity=1 ] (399.47,150.17) -- (399.48,68.4) -- (475.68,31.05) -- (475.66,112.82) -- cycle ;
\draw  [color={rgb, 255:red, 0; green, 0; blue, 0 }  ,draw opacity=1 ][fill={rgb, 255:red, 192; green, 188; blue, 188 }  ,fill opacity=1 ] (517.62,146) -- (517.64,64.23) -- (593.83,26.89) -- (593.82,108.66) -- cycle ;
\draw  [color={rgb, 255:red, 139; green, 6; blue, 24 }  ,draw opacity=1 ][fill={rgb, 255:red, 139; green, 6; blue, 24 }  ,fill opacity=1 ] (447.39,82.8) .. controls (447.39,82.14) and (447.92,81.61) .. (448.57,81.61) .. controls (449.23,81.61) and (449.76,82.14) .. (449.76,82.8) .. controls (449.76,83.45) and (449.23,83.98) .. (448.57,83.98) .. controls (447.92,83.98) and (447.39,83.45) .. (447.39,82.8) -- cycle ;
\draw  [color={rgb, 255:red, 139; green, 6; blue, 24 }  ,draw opacity=1 ][fill={rgb, 255:red, 139; green, 6; blue, 24 }  ,fill opacity=1 ] (554.73,82.44) .. controls (554.73,81.79) and (555.26,81.26) .. (555.91,81.26) .. controls (556.57,81.26) and (557.1,81.79) .. (557.1,82.44) .. controls (557.1,83.1) and (556.57,83.63) .. (555.91,83.63) .. controls (555.26,83.63) and (554.73,83.1) .. (554.73,82.44) -- cycle ;
\draw [color={rgb, 255:red, 139; green, 6; blue, 24 }  ,draw opacity=1 ]   (448.57,83.15) -- (462,99) ;
\draw [color={rgb, 255:red, 139; green, 6; blue, 24 }  ,draw opacity=1 ]   (433.09,108.18) -- (462,99) ;
\draw  [color={rgb, 255:red, 139; green, 6; blue, 24 }  ,draw opacity=1 ][fill={rgb, 255:red, 139; green, 6; blue, 24 }  ,fill opacity=1 ] (539.06,107.83) .. controls (539.06,107.18) and (539.59,106.65) .. (540.24,106.65) .. controls (540.9,106.65) and (541.43,107.18) .. (541.43,107.83) .. controls (541.43,108.49) and (540.9,109.02) .. (540.24,109.02) .. controls (539.59,109.02) and (539.06,108.49) .. (539.06,107.83) -- cycle ;
\draw  [color={rgb, 255:red, 139; green, 6; blue, 24 }  ,draw opacity=1 ][fill={rgb, 255:red, 139; green, 6; blue, 24 }  ,fill opacity=1 ] (430.72,108.18) .. controls (430.72,107.53) and (431.25,107) .. (431.9,107) .. controls (432.56,107) and (433.09,107.53) .. (433.09,108.18) .. controls (433.09,108.84) and (432.56,109.37) .. (431.9,109.37) .. controls (431.25,109.37) and (430.72,108.84) .. (430.72,108.18) -- cycle ;
\draw [color={rgb, 255:red, 139; green, 6; blue, 24 }  ,draw opacity=1 ]   (462,99) -- (531,99) ;
\draw [color={rgb, 255:red, 139; green, 6; blue, 24 }  ,draw opacity=1 ]   (531,99) -- (557.1,82.44) ;
\draw [color={rgb, 255:red, 139; green, 6; blue, 24 }  ,draw opacity=1 ]   (531.53,99.09) -- (540.43,107.83) ;

\draw (62.91,62.21) node [anchor=north west][inner sep=0.75pt]    {$\CI^{\dagger }$};
\draw (170.9,63.04) node [anchor=north west][inner sep=0.75pt]    {$\CI$};
\draw (6.54,36.74) node [anchor=north west][inner sep=0.75pt]    {$\Mod( \CB_{2})$};
\draw (246.46,35.95) node [anchor=north west][inner sep=0.75pt]    {$\Mod( \CB_{2})$};
\draw (135.79,35.07) node [anchor=north west][inner sep=0.75pt]    {$\Mod(\CB_{1})$};
\draw (140.97,82.9) node [anchor=north west][inner sep=0.75pt]    {$x _{1}$};
\draw (130.08,109.24) node [anchor=north west][inner sep=0.75pt]    {$x _{2}$};
\draw (413.37,61.21) node [anchor=north west][inner sep=0.75pt]    {$\CI^{\dagger }$};
\draw (521.36,62.04) node [anchor=north west][inner sep=0.75pt]    {$\CI$};
\draw (357,35.74) node [anchor=north west][inner sep=0.75pt]    {$\Mod(\CB_{2})$};
\draw (596.91,34.95) node [anchor=north west][inner sep=0.75pt]    {$\Mod( \CB_{2})$};
\draw (484.25,34.07) node [anchor=north west][inner sep=0.75pt]    {$\Mod( \CB_{1})$};
\draw (486.42,98.9) node [anchor=north west][inner sep=0.75pt]    {$x _{3}$};
\draw (456.97,74.9) node [anchor=north west][inner sep=0.75pt]    {$x _{1}$};
\draw (531.97,69.9) node [anchor=north west][inner sep=0.75pt]    {$x _{1}$};
\draw (439.54,104.99) node [anchor=north west][inner sep=0.75pt]    {$x _{2}$};
\draw (526.54,105.99) node [anchor=north west][inner sep=0.75pt]    {$x _{2}$};
\draw (305,68.4) node [anchor=north west][inner sep=0.75pt]    {$=\sum\limits_{x _{3}}$};

\end{tikzpicture}
\caption{The fusion rules of topological local operators on $\CA_S$ is given by the vertical fusion of lines $x_1,x_2 \in A_L$ that can end on both $\CI$ and $\CI^{\dagger}$. The trivalent junctions are evaluated using the multiplication of the algebra $A_L$.}
\label{fig:OPE of local operators on A_S}
\end{figure}
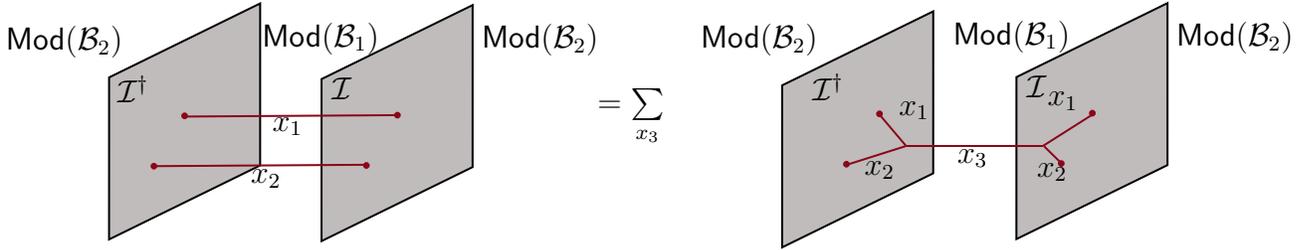

To better understand the above discussion, first note that the fusion of local operators on $\widehat\CA_S$ can be computed using the vertical fusion of lines that can end on both $\CI$ and $\CI^{\dagger}$ (see Fig.~\ref{fig:OPE of local operators on A_S}) \cite{Huston:2022utd}.
These fusion rules defines a simisimple algebra structure on $V_{\widehat\CA_S}$. Let $m$ be the multiplication of this algebra. It is clear that this algebra must be commutative, because topological local operators can be taken around each other in (1+1)d. Let $P_i$, $i=1,\cdots , \text{dim}(V_{A_S})$ be the central idempotents of this algebra. 
These idempotents project $\CA_S$ onto the indecomposable surfaces that form $\CA_S$. In particular, $\CA_S$ with the idempotent $P_i$ on it gives the indecomposable surface $S_i$. The fusion rules of these indecomposable surfaces are given by the product of idempotents
$P_i \otimes P_j$.
Note that the $P_i$ are not projectors with respect to this product. 

Consider the special case when $\CB_2$ describes a TQFT in which all simple lines are invertible. In this case, the action of the indecomposable surfaces in $\widehat\CA_S$ on the lines of $\CB_2$ can be explicitly identified. To that end, consider two lines $a, b\in\CB_2$ forming a junction on $\widehat\CA_S$. Then, the line $(a,\bar b)$ is a boson in the folded theory $\CB_2 \boxtimes \overline{\CB}_2$. Moreover, upon folding, $\CA_S$ becomes a non-simple gapped boundary of $\CB_2 \boxtimes \overline{\CB}_2$, and $(a,\bar b)$ can end on this boundary. The indecomposable surfaces that make up $\CA_S$ can be identified by finding the different maximal subgroups of bosonic lines that can end on this gapped boundary.

In particular, this logic allows us to create non-Abelian anyons in arbitrary Dijkgraaf--Witten theories from Abelian anyons. To that end, consider the category $\Rep(G)$ of Wilson lines corresponding to the gauge group $G$. There is a 1-form symmetry gauging from $\Rep(G)$ to $\Rep(K)$ for any subgroup $K \leq G$ given by the restriction of representations of $G$ to the subgroup $K$. The gauging corresponds to 
\be
A_K \cong \bigoplus_{\pi \in \text{Irr}(\Rep(G))} n_{\pi} \cdot \pi \, ,
\ee
where $n_{\pi}$ satisfies
\be
\pi|_K \cong  n_{\pi} ~ 1_H \oplus \cdots\, ,
\ee
and $1_K$ is the trivial representation of $K$.  Note that when $K$ is a normal subgroup of $G$, then the set of simple lines in $A_K$ are closed under fusion. 

Consider the Dijkgraaf--Witten theory $\mc D^\alpha(G)$ for some 3-cocycle, $\alpha \in H^3(G,{\rm U}(1))$. This theory has a $\Rep(G)$ Lagrangian subcategory of lines. Consider gauging the 1-form symmetry corresponding to $A_K\in\Rep(G)$. Under this operation, we have
\be
\mathcal{I}: \mc D^\alpha(G) \to \CB\, ,
\ee
where $\CB$ is some MTC. In fact, the resulting Dijkgraaf--Witten theory is $\mc D^{\alpha'}(K)$ for some 3-cocycle, $\alpha' \in H^3(K,{\rm U}(1))$. To understand this claim, consider a line $a$ in $\mc \mc D^\alpha(G)$ but not in $\Rep(G)$. If $a$ braids non-trivially with some lines in $A_K$, then it becomes a non-genuine line in $\CB$. If $a$ braids trivially with all lines in $A_K$, then it survives as a line in $\CB$. In fact, if $a$ braids trivially with all lines in $A_K$, it must braid non-trivially with some lines in $\Rep(G)$ that are not in $A_K$. Therefore, $\CI(a)$ braids non-trivially with some line in the subcategory $\Rep(K)$ of $\CB$. As a result, $\Rep(K)$ is a Lagrangian subcategory of $\CB$ implying $\CB$ is equivalent to a Dijkgraaf--Witten theory, $\mc D^{\alpha'}(K)$, for some 3-cocycle $\alpha' \in H^3(K,{\rm U}(1))$. 

The above discussion implies that all Dijkgraaf--Witten theories with non-Abelian gauge groups can be obtained from gauging symmetries of those with Abelian gauge groups. Indeed, consider $\mc \mc D^\alpha(G)$, and let $g\in G$ have order $n$. Then, $g$ generates a cyclic subgroup, $H\cong\DZ_n$, and we have a 1-form symmetry gauging 
\be
\CI: \mc D^\alpha(G) \to \mc D^{\alpha'}(\DZ_n)\, ,
\ee
for some 3-cocycle, $\alpha' \in H^3(\DZ_n,{\rm U}(1))$. Moreover, for any $\alpha'$, the TQFT $\mc D^{\alpha'}(\DZ_n)$ only contains Abelian anyons. Therefore, this condensation shows that all discrete gauge theories with non-Abelian anyons can be related to discrete gauge theories with Abelian anyons under condensation. As a result, there is an inverse gauging of a (generically non-invertible) 0-form symmetry acting on $\mc D^{\alpha'}(\DZ_n)$ that produces $\mc D^\alpha(G)$. Moreover, from $\CI$ and $\CI^{\dagger}$, we can calculate the dual 0-form surface algebra, $\widehat\CA _S$, to be summed over to invert this condensation. 

If we choose $g$ to be the trivial element of the group $G$ in the above discussion, we recover the well-known statement that all discrete gauge theories $\mc D^\alpha(G)$ can be obtained from gauging symmetries of the trivial TQFT. In this case, the 0-form gauging involves gauging an invertible symmetry that acts unfaithfully on the trivial TQFT.

\newsec{Alternate proof of the fixed point theorem for Abelian TQFTs}\label{FPabelian}
In this appendix, we give an alternate proof, for arbitrary Abelian TQFTs, of the more general fixed point theorem proven in Sec.~\ref{sec:number of twisted sectors A_S and hat A_S}. Recall that the general fixed point theorem equates the number of twisted sector lines (defined as the number of boundary conditions of a condensation surface) with the number of lines with self-junctions on the condensation surface (i.e., the number of lines that have a channel for ``passing through'' the surface). Our proof in this appendix gives these quantities in terms of the group-theoretical data that defines the surface in an Abelian TQFT.

In general, the input data for an Abelian TQFT is a pointed MTC of the form ${\Vect}^{\alpha,R}_{H}$, where $H$ is an Abelian finite group, $\alpha\in H^3(H,{\rm U}(1))$ is the associator, and $R$ is the $R$-matrix. We write 
\begin{equation}
    R_{\mathbb C_{h_1},\mathbb C_{h_2}} = R(h_1,h_2) \cdot \text{swap} \, ,
\end{equation}
for every $h_1,h_2 \in H$, and define $B(h_1,h_2) := R(h_1,h_2)R(h_2,h_1)$.
We shall prove the fixed point theorem for all condensation surfaces in Abelian theories.

To that end, we recall that the algebras we can sum over along a co-dimension one surface in spacetime to construct a condensation surface (as in \cite{Roumpedakis_2023}) are in bijection with the pair $(K,\phi)$, where $K \le  H$, and $\phi \in C^{2}(K,{\rm U}(1))$ trivializes the associator, $\alpha|_{K} = d\phi$. The number of simple objects in the corresponding module category is the cardinality of the coset $\left| H/K\right|$ \cite{etingof2009fusioncategorieshomotopytheory}. This is the number of twisted sector lines for the surface $S_{\mathbb C[K]^\phi}$ built from higher-gauging the corresponding algebra $\mathbb C[K]^\phi$ in ${\Vect}^{\alpha,R}_{H}$.

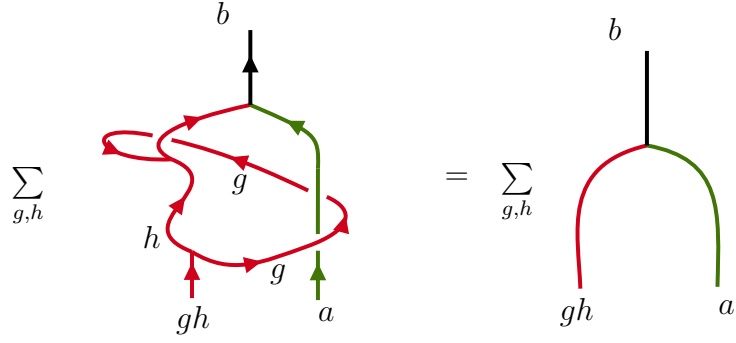
\begin{figure}[h]
\centering

\tikzset{every picture/.style={line width=0.75pt}} 

\begin{tikzpicture}[x=0.75pt,y=0.75pt,yscale=-0.75,xscale=0.75]

\draw [color={rgb, 255:red, 208; green, 2; blue, 27 }  ,draw opacity=1 ][line width=1.5]    (180.13,194.73) .. controls (223.95,212.4) and (145.47,238.07) .. (193.86,257.01) ;
\draw [shift={(189.3,219.81)}, rotate = 125.64] [fill={rgb, 255:red, 208; green, 2; blue, 27 }  ,fill opacity=1 ][line width=0.08]  [draw opacity=0] (11.61,-5.58) -- (0,0) -- (11.61,5.58) -- cycle    ;
\draw [color={rgb, 255:red, 208; green, 2; blue, 27 }  ,draw opacity=1 ][line width=1.5]    (193.86,257.01) -- (193.86,288.15) ;
\draw [shift={(193.86,264.28)}, rotate = 90] [fill={rgb, 255:red, 208; green, 2; blue, 27 }  ,fill opacity=1 ][line width=0.08]  [draw opacity=0] (11.61,-5.58) -- (0,0) -- (11.61,5.58) -- cycle    ;
\draw [color={rgb, 255:red, 208; green, 2; blue, 27 }  ,draw opacity=1 ][line width=1.5]    (284.5,220) .. controls (304.77,228.84) and (301.11,242.7) .. (271.26,253.33) .. controls (241.41,263.97) and (220.44,273.5) .. (193.86,257.01) ;
\draw [shift={(297.19,232.49)}, rotate = 105.37] [fill={rgb, 255:red, 208; green, 2; blue, 27 }  ,fill opacity=1 ][line width=0.08]  [draw opacity=0] (11.61,-5.58) -- (0,0) -- (11.61,5.58) -- cycle    ;
\draw [shift={(240.66,263.52)}, rotate = 167.77] [fill={rgb, 255:red, 208; green, 2; blue, 27 }  ,fill opacity=1 ][line width=0.08]  [draw opacity=0] (11.61,-5.58) -- (0,0) -- (11.61,5.58) -- cycle    ;
\draw [color={rgb, 255:red, 65; green, 117; blue, 5 }  ,draw opacity=1 ][line width=1.5]    (278.55,201.04) -- (278.5,245) ;
\draw [color={rgb, 255:red, 65; green, 117; blue, 5 }  ,draw opacity=1 ][line width=1.5]    (278.55,256.22) -- (278.55,289.47) ;
\draw [shift={(278.55,264.55)}, rotate = 90] [fill={rgb, 255:red, 65; green, 117; blue, 5 }  ,fill opacity=1 ][line width=0.08]  [draw opacity=0] (11.61,-5.58) -- (0,0) -- (11.61,5.58) -- cycle    ;
\draw [color={rgb, 255:red, 208; green, 2; blue, 27 }  ,draw opacity=1 ][line width=1.5]    (180.13,181.68) .. controls (218.71,190.52) and (250.76,205.67) .. (272.01,215.77) ;
\draw [shift={(219.56,193.33)}, rotate = 20.16] [fill={rgb, 255:red, 208; green, 2; blue, 27 }  ,fill opacity=1 ][line width=0.08]  [draw opacity=0] (11.61,-5.58) -- (0,0) -- (11.61,5.58) -- cycle    ;
\draw [color={rgb, 255:red, 208; green, 2; blue, 27 }  ,draw opacity=1 ][line width=1.5]    (233.1,158.12) .. controls (184.05,168.64) and (155.94,182.52) .. (180.13,194.73) ;
\draw [shift={(200.73,166.56)}, rotate = 161.08] [fill={rgb, 255:red, 208; green, 2; blue, 27 }  ,fill opacity=1 ][line width=0.08]  [draw opacity=0] (11.61,-5.58) -- (0,0) -- (11.61,5.58) -- cycle    ;
\draw [color={rgb, 255:red, 65; green, 117; blue, 5 }  ,draw opacity=1 ][line width=1.5]    (233.1,158.12) .. controls (276.91,177.05) and (279.53,177.47) .. (278.55,201.04) ;
\draw [shift={(257.25,168.75)}, rotate = 26.17] [fill={rgb, 255:red, 65; green, 117; blue, 5 }  ,fill opacity=1 ][line width=0.08]  [draw opacity=0] (11.61,-5.58) -- (0,0) -- (11.61,5.58) -- cycle    ;
\draw [color={rgb, 255:red, 208; green, 2; blue, 27 }  ,draw opacity=1 ][line width=1.5]    (180.13,194.73) .. controls (127.16,197.25) and (117.35,168.64) .. (167.38,178.74) ;
\draw [shift={(146.15,191.06)}, rotate = 203.08] [fill={rgb, 255:red, 208; green, 2; blue, 27 }  ,fill opacity=1 ][line width=0.08]  [draw opacity=0] (11.61,-5.58) -- (0,0) -- (11.61,5.58) -- cycle    ;
\draw [line width=1.5]    (233.1,107.76) -- (233.1,158.12) ;
\draw [shift={(233.1,124.64)}, rotate = 90] [fill={rgb, 255:red, 0; green, 0; blue, 0 }  ][line width=0.08]  [draw opacity=0] (11.61,-5.58) -- (0,0) -- (11.61,5.58) -- cycle    ;
\draw [color={rgb, 255:red, 208; green, 2; blue, 27 }  ,draw opacity=1 ][line width=1.5]    (455.02,281.78) .. controls (453.68,243.48) and (441.66,199) .. (499.78,185.41) ;
\draw [color={rgb, 255:red, 65; green, 117; blue, 5 }  ,draw opacity=1 ][line width=1.5]    (547.21,280.54) .. controls (548.55,229.27) and (551.22,194.68) .. (499.78,185.41) ;
\draw [color={rgb, 255:red, 0; green, 0; blue, 0 }  ,draw opacity=1 ][line width=1.5]    (499.78,121.78) -- (499.78,185.41) ;

\draw (181.21,291.39) node [anchor=north west][inner sep=0.75pt]    {$gh$};
\draw (276.6,292.03) node [anchor=north west][inner sep=0.75pt]    {$a$};
\draw (244.35,263.34) node [anchor=north west][inner sep=0.75pt]    {$g$};
\draw (158.9,237.84) node [anchor=north west][inner sep=0.75pt]    {$h$};
\draw (208.13,86.77) node [anchor=north west][inner sep=0.75pt]    {$b$};
\draw (218.56,202.79) node [anchor=north west][inner sep=0.75pt]    {$g$};
\draw (361.79,201.51) node [anchor=north west][inner sep=0.75pt]    {$=$};
\draw (439.18,286.29) node [anchor=north west][inner sep=0.75pt]    {$gh$};
\draw (545.86,287.56) node [anchor=north west][inner sep=0.75pt]    {$a$};
\draw (471.75,97.61) node [anchor=north west][inner sep=0.75pt]    {$b$};
\draw (70.32,197.96) node [anchor=north west][inner sep=0.75pt]    {$\sum\limits_{g,h}$};
\draw (399.23,195.41) node [anchor=north west][inner sep=0.75pt]    {$\sum\limits_{g,h}$};

\end{tikzpicture}
\caption{Condition for the existence of a junction of the line $a$ with a condensation surface.}
\label{Diag1}
\end{figure}

To complete the proof of fixed point theorem, we need to show that the number of lines with self-junction on the surface is also $\left|H/K\right|$. Now, any two lines labelled by $a,\,b \in H$ can form a junction with the condensation surface $S_{\mathbb C[K]^\phi}$ if and if only if the condition given in Fig.~\ref{Diag1} is obeyed (See \cite{Buican:2023bzl,Fuchs_2002}). This condition asks for the point junction between $a$, $b$ and $S_{(K,\phi)}$ to remain invariant under changes of triangulation of the surface (See Figs. 19 and 20 in \cite{Buican:2023bzl}).

We resolve the LHS of Fig.~\ref{Diag1} (see Fig.~\ref{Diag2}) using the $\alpha$, $R$, and $\phi$ data available to us, before invoking the separability of the algebra. Putting everything together, we obtain 
\begin{equation}
\begin{split}
    &\frac{1}{|K|} \bigoplus_{g,h \in K}\frac{\phi(h,g)}{\phi(g,h)}\,R(g,gh)\,B(g,a)\,\alpha(h^{-1},g^{-1},ghg)\,\Hom_{{\Vect}^{\alpha,R}_{H}}(\mathbb C_{gh}\otimes \mathbb C_a, \mathbb C_{b}) 
    \\
    & \q = \bigoplus_{g,h \in K}\Hom_{{\Vect}^{\alpha,R}_{H}}(\mathbb C_{gh}\otimes \mathbb C_a, \mathbb C_{b})\,.
\end{split}
\end{equation}
From \cite{Buican:2023bzl}, we know that in an Abelian theory, the multiplicity of lines acted on by surfaces is $n^{b}_{S_{\mathbb C[K]^\phi}(a)} \in \{0,1\}$. Furthermore, we are only interested in self-junctions so we impose $a=b$, which we restrict the summation to $g = h^{-1}$. These two considerations simplify the equation to
\begin{equation}
\frac{1}{|K|}\,\sum_{k \in K}\left(\frac{\phi(k,k^{-1})}{\phi(k^{-1},k)}\,R(k^{-1},k^{-1}k)\,B(k^{-1},a)\,\alpha(k^{-1},k,k^{-1}kk^{-1}) \right)= 1\, .
\end{equation}
Substituting $R(k^{-1},k^{-1}k) = R(k,1) = 1$, and $\alpha(k^{-1},k,k^{-1}) = d\phi(k^{-1},k,k^{-1}) = \frac{\phi(k^{-1},k)}{\phi(k,k^{-1})}$ yields
\begin{equation}
\frac{1}{|K|}\sum_{k \in K}\,B(k^{-1},a) =1\, ,
\end{equation}
for every $a \in H$.
This argument shows that only lines that braid trivially with the algebra object $\mathbb C[K]^\phi \cong \bigoplus_{k \in K}\, \mathbb C_k$ can have self-junctions. Furthermore, by running this computation in reverse, any line, $x$, that braids trivially with the algebra object is going to have self-junction of dimension $\dim \Hom_{S_{\mathbb C[K]^\phi}}(x,x) = 1$. Therefore, we have shown that the number of lines that have self-junctions is the cardinality of centralizer of the algebra object, which is equal to ${|H|\over |K|}$
\cite{mueger2002structuremodularcategories}. This ratio is equal to the number of representatives in the coset $H/K$, and we have completed the proof.

\begin{figure}[h]
\centering

\tikzset{every picture/.style={line width=0.75pt}} 

\begin{tikzpicture}[x=0.75pt,y=0.75pt,yscale=-0.75,xscale=1]

\draw [color={rgb, 255:red, 208; green, 2; blue, 27 }  ,draw opacity=1 ][line width=1.5]    (128.42,206.55) .. controls (173.53,228.65) and (92.74,260.74) .. (142.56,284.41) ;
\draw [shift={(137.17,239.13)}, rotate = 120.86] [fill={rgb, 255:red, 208; green, 2; blue, 27 }  ,fill opacity=1 ][line width=0.08]  [draw opacity=0] (11.61,-5.58) -- (0,0) -- (11.61,5.58) -- cycle    ;
\draw [color={rgb, 255:red, 208; green, 2; blue, 27 }  ,draw opacity=1 ][line width=1.5]    (142.56,284.41) -- (142.56,323.34) ;
\draw [shift={(142.56,295.57)}, rotate = 90] [fill={rgb, 255:red, 208; green, 2; blue, 27 }  ,fill opacity=1 ][line width=0.08]  [draw opacity=0] (11.61,-5.58) -- (0,0) -- (11.61,5.58) -- cycle    ;
\draw [color={rgb, 255:red, 65; green, 117; blue, 5 }  ,draw opacity=1 ][line width=1.5]    (229.75,214.44) -- (229.75,271.78) ;
\draw [color={rgb, 255:red, 65; green, 117; blue, 5 }  ,draw opacity=1 ][line width=1.5]    (229.75,279.67) -- (229.75,321.23) ;
\draw [shift={(229.75,292.15)}, rotate = 90] [fill={rgb, 255:red, 65; green, 117; blue, 5 }  ,fill opacity=1 ][line width=0.08]  [draw opacity=0] (11.61,-5.58) -- (0,0) -- (11.61,5.58) -- cycle    ;
\draw [color={rgb, 255:red, 208; green, 2; blue, 27 }  ,draw opacity=1 ][line width=1.5]    (128.42,190.24) .. controls (168.14,201.29) and (174.8,249.69) .. (142.56,284.41) ;
\draw [shift={(161.68,224.23)}, rotate = 80.21] [fill={rgb, 255:red, 208; green, 2; blue, 27 }  ,fill opacity=1 ][line width=0.08]  [draw opacity=0] (11.61,-5.58) -- (0,0) -- (11.61,5.58) -- cycle    ;
\draw [color={rgb, 255:red, 208; green, 2; blue, 27 }  ,draw opacity=1 ][line width=1.5]    (182.96,160.78) .. controls (132.46,173.94) and (103.51,191.3) .. (128.42,206.55) ;
\draw [shift={(147.4,172.23)}, rotate = 156.66] [fill={rgb, 255:red, 208; green, 2; blue, 27 }  ,fill opacity=1 ][line width=0.08]  [draw opacity=0] (11.61,-5.58) -- (0,0) -- (11.61,5.58) -- cycle    ;
\draw [color={rgb, 255:red, 65; green, 117; blue, 5 }  ,draw opacity=1 ][line width=1.5]    (182.96,160.78) .. controls (228.06,184.46) and (230.76,184.98) .. (229.75,214.44) ;
\draw [shift={(210.22,175.47)}, rotate = 31.94] [fill={rgb, 255:red, 65; green, 117; blue, 5 }  ,fill opacity=1 ][line width=0.08]  [draw opacity=0] (11.61,-5.58) -- (0,0) -- (11.61,5.58) -- cycle    ;
\draw [color={rgb, 255:red, 208; green, 2; blue, 27 }  ,draw opacity=1 ][line width=1.5]    (128.42,206.55) .. controls (73.89,209.71) and (63.79,173.94) .. (115.29,186.56) ;
\draw [shift={(92.2,201.42)}, rotate = 209.11] [fill={rgb, 255:red, 208; green, 2; blue, 27 }  ,fill opacity=1 ][line width=0.08]  [draw opacity=0] (11.61,-5.58) -- (0,0) -- (11.61,5.58) -- cycle    ;
\draw [line width=1.5]    (182.96,97.84) -- (182.96,160.78) ;
\draw [shift={(182.96,121.01)}, rotate = 90] [fill={rgb, 255:red, 0; green, 0; blue, 0 }  ][line width=0.08]  [draw opacity=0] (11.61,-5.58) -- (0,0) -- (11.61,5.58) -- cycle    ;
\draw [color={rgb, 255:red, 208; green, 2; blue, 27 }  ,draw opacity=1 ][line width=1.5]    (226.72,258.53) .. controls (216.73,258.53) and (218.4,263.75) .. (221.13,270.49) .. controls (223.85,277.23) and (232.39,278.84) .. (237.74,271.28) .. controls (243.09,263.73) and (235.9,258.53) .. (232.71,259.05) ;
\draw [color={rgb, 255:red, 208; green, 2; blue, 27 }  ,draw opacity=1 ][line width=1.5]    (479.71,204.33) .. controls (535.36,226.69) and (435.7,259.16) .. (497.15,283.11) ;
\draw [shift={(490.21,237.72)}, rotate = 126.18] [fill={rgb, 255:red, 208; green, 2; blue, 27 }  ,fill opacity=1 ][line width=0.08]  [draw opacity=0] (11.61,-5.58) -- (0,0) -- (11.61,5.58) -- cycle    ;
\draw [color={rgb, 255:red, 208; green, 2; blue, 27 }  ,draw opacity=1 ][line width=1.5]    (497.15,283.11) -- (497.15,322.5) ;
\draw [shift={(497.15,294.51)}, rotate = 90] [fill={rgb, 255:red, 208; green, 2; blue, 27 }  ,fill opacity=1 ][line width=0.08]  [draw opacity=0] (11.61,-5.58) -- (0,0) -- (11.61,5.58) -- cycle    ;
\draw [color={rgb, 255:red, 65; green, 117; blue, 5 }  ,draw opacity=1 ][line width=1.5]    (619.85,192.01) -- (619.85,237.25) ;
\draw [color={rgb, 255:red, 65; green, 117; blue, 5 }  ,draw opacity=1 ][line width=1.5]    (619.85,237.25) -- (619.85,327.33) ;
\draw [shift={(619.85,273.99)}, rotate = 90] [fill={rgb, 255:red, 65; green, 117; blue, 5 }  ,fill opacity=1 ][line width=0.08]  [draw opacity=0] (11.61,-5.58) -- (0,0) -- (11.61,5.58) -- cycle    ;
\draw [color={rgb, 255:red, 208; green, 2; blue, 27 }  ,draw opacity=1 ][line width=1.5]    (472.89,184.52) .. controls (521.89,195.69) and (536.93,247.98) .. (497.15,283.11) ;
\draw [shift={(516.07,218.19)}, rotate = 70.68] [fill={rgb, 255:red, 208; green, 2; blue, 27 }  ,fill opacity=1 ][line width=0.08]  [draw opacity=0] (11.61,-5.58) -- (0,0) -- (11.61,5.58) -- cycle    ;
\draw [color={rgb, 255:red, 208; green, 2; blue, 27 }  ,draw opacity=1 ][line width=1.5]    (577.67,150.53) .. controls (568.31,152.53) and (528.86,162.12) .. (520.84,164.29) .. controls (512.81,166.46) and (480.19,181.72) .. (472.89,184.52) .. controls (465.59,187.31) and (466.99,197.94) .. (479.71,204.33) ;
\draw [shift={(556.99,155.36)}, rotate = 166.45] [fill={rgb, 255:red, 208; green, 2; blue, 27 }  ,fill opacity=1 ][line width=0.08]  [draw opacity=0] (11.61,-5.58) -- (0,0) -- (11.61,5.58) -- cycle    ;
\draw [shift={(504.07,170.91)}, rotate = 156.48] [fill={rgb, 255:red, 208; green, 2; blue, 27 }  ,fill opacity=1 ][line width=0.08]  [draw opacity=0] (11.61,-5.58) -- (0,0) -- (11.61,5.58) -- cycle    ;
\draw [shift={(469.13,187.8)}, rotate = 91.6] [fill={rgb, 255:red, 208; green, 2; blue, 27 }  ,fill opacity=1 ][line width=0.08]  [draw opacity=0] (11.61,-5.58) -- (0,0) -- (11.61,5.58) -- cycle    ;
\draw [color={rgb, 255:red, 65; green, 117; blue, 5 }  ,draw opacity=1 ][line width=1.5]    (577.67,150.53) .. controls (618.59,169.21) and (620.77,168.76) .. (619.85,192.01) ;
\draw [shift={(600.19,160.9)}, rotate = 26.87] [fill={rgb, 255:red, 65; green, 117; blue, 5 }  ,fill opacity=1 ][line width=0.08]  [draw opacity=0] (11.61,-5.58) -- (0,0) -- (11.61,5.58) -- cycle    ;
\draw [color={rgb, 255:red, 208; green, 2; blue, 27 }  ,draw opacity=1 ][line width=1.5]    (479.71,204.33) .. controls (412.44,207.53) and (457.3,151.52) .. (520.84,164.29) ;
\draw [shift={(453.72,178.22)}, rotate = 322] [fill={rgb, 255:red, 208; green, 2; blue, 27 }  ,fill opacity=1 ][line width=0.08]  [draw opacity=0] (11.61,-5.58) -- (0,0) -- (11.61,5.58) -- cycle    ;
\draw [line width=1.5]    (577.67,100.86) -- (577.67,150.53) ;
\draw [shift={(577.67,117.39)}, rotate = 90] [fill={rgb, 255:red, 0; green, 0; blue, 0 }  ][line width=0.08]  [draw opacity=0] (11.61,-5.58) -- (0,0) -- (11.61,5.58) -- cycle    ;

\draw (129.98,329.42) node [anchor=north west][inner sep=0.75pt]  [font=\small]  {$gh$};
\draw (227.92,330.21) node [anchor=north west][inner sep=0.75pt]  [font=\small]  {$a$};
\draw (211.89,274.08) node [anchor=north west][inner sep=0.75pt]  [font=\small]  {$g$};
\draw (106.74,262.48) node [anchor=north west][inner sep=0.75pt]  [font=\small]  {$h$};
\draw (158.29,73.64) node [anchor=north west][inner sep=0.75pt]  [font=\small]  {$b$};
\draw (168.16,218.66) node [anchor=north west][inner sep=0.75pt]  [font=\small]  {$g$};
\draw (70.23,199.54) node [anchor=north west][inner sep=0.75pt]  [font=\small]  {$g$};
\draw (130.82,141.23) node [anchor=north west][inner sep=0.75pt]  [font=\small]  {$gh$};
\draw (484.3,328.74) node [anchor=north west][inner sep=0.75pt]  [font=\small]  {$gh$};
\draw (604.71,319.28) node [anchor=north west][inner sep=0.75pt]  [font=\small]  {$a$};
\draw (454.37,261.02) node [anchor=north west][inner sep=0.75pt]  [font=\small]  {$h$};
\draw (553.35,78.56) node [anchor=north west][inner sep=0.75pt]  [font=\small]  {$b$};
\draw (550.67,225.86) node [anchor=north west][inner sep=0.75pt]  [font=\small]  {$g$};
\draw (453.43,150.63) node [anchor=north west][inner sep=0.75pt]  [font=\small]  {$g$};
\draw (23.85,220.48) node [anchor=north west][inner sep=0.75pt]    {$\sum\limits_{g,h}$};
\draw (451.11,182.06) node [anchor=north west][inner sep=0.75pt]  [font=\small]  {$gh$};
\draw (501.61,176.34) node [anchor=north west][inner sep=0.75pt]  [font=\small,rotate=-1.92]  {$ghg$};
\draw (519.66,133.57) node [anchor=north west][inner sep=0.75pt]  [font=\small,rotate=-1.92]  {$hg$};
\draw (338.44,220.3) node [anchor=north west][inner sep=0.75pt]  [font=\small]  {$B(g,a)$};
\draw (382.88,220.95) node [anchor=north west][inner sep=0.75pt]  [font=\small]  {$R(g,gh)$};
\draw (265.02,223.81) node [anchor=north west][inner sep=0.75pt]    {$=$};
\draw (301.97,217.54) node [anchor=north west][inner sep=0.75pt]    {$\sum\limits_{g,h}$};

\end{tikzpicture}
\caption{Working out the hom-space condition.}
\label{Diag2}
\end{figure}

\end{appendices}

\newpage
\bibliography{refs_gauging}
\nocite{*}

\end{document}